\documentclass[preprint,notoc]{JHEP3}
\usepackage{slashed}
\usepackage{enumerate}
\usepackage{graphics}

\newcommand{\Ord}{\mathcal{O}}
 
\newcommand{\nn}{\nonumber \\}
\newcommand{\el}{\nonumber \\ &&}

\newcommand{\elale}{\nonumber \\ &=&}

\newcommand{\mintd}[2]{\int {d^{#2} #1 \over (2 \pi)^{#2}}}

\newcommand{\M}{\mathcal{M}}
\newcommand{\T}{\theta}
\newenvironment{changemargin}[2]{%
  \begin{list}{}{%
    \setlength{\topsep}{0pt}%
    \setlength{\leftmargin}{#1}%
    \setlength{\rightmargin}{#2}%
    \setlength{\listparindent}{\parindent}%
    \setlength{\itemindent}{\parindent}%
    \setlength{\parsep}{\parskip}%
  }%
  \item[]}{\end{list}}

\def\eqn#1{eq.~(\ref{#1})}

\newcommand{\be}{\begin{equation}} 
\newcommand{\ee}{\end{equation}} 
\newcommand{\cmb}{\begin{changemargin}}
\newcommand{\cme}{\end{changemargin}}
\newcommand{\bea}{\begin{eqnarray}} 
\newcommand{\eea}{\end{eqnarray}} 
\newcommand{\D}{\delta}
\newcommand{\g}{\gamma}
\newcommand{\A}{\alpha}
\newcommand{\B}{\beta}
\newcommand{\G}{\Gamma}

\newcommand{\mc}{\mathcal}
\newcommand{\mb}{\mathbb}
\def\Tr{\mathop{\rm Tr}\nolimits}

\def\spa#1.#2{\langle#1\,#2\rangle}
\def\spb#1.#2{[#1\,#2]}
\def\sandmm#1.#2.#3{%
\left\langle\smash{#1}{\vphantom1}\right|{#2}%
\left|\smash{#3}{\vphantom1}\right]}
\def\spab#1.#2.#3{\sandmm#1.#2.#3}
\def\spba#1.#2.#3{\sandpp#1.#2.#3}
\def\spaa#1.#2.#3.#4{\sandmp#1.{#2#3}.#4}
\def\spbb#1.#2.#3.#4{\sandpm#1.{#2#3}.#4}
\def\spash#1.#2{\spa{\smash{#1}}.{\smash{#2}}}
\def\spbsh#1.#2{\spb{\smash{#1}}.{\smash{#2}}}

\def\ksl{\not{\hbox{\kern-2.3pt $k$}}}

\def\e{\epsilon}

\def\Ord{{\cal O}}
\def\a{{\cal A}}

\def\Nc{N_c}
\def\Nsym{{\cal N}=4}

\def\li#1{{\mathop{\rm Li}\nolimits}_#1}
\def\Li{\mathop{\rm Li}\nolimits}

\def\pol{\varepsilon}

\preprint{IFT-UAM/CSIC-11-11}
\title{One-Loop $\mathcal{N}=4$ Super Yang-Mills Scattering Amplitudes to All Orders in the Dimensional Regularization Parameter}

\author{Robert M. Schabinger\\
	Instituto de F\'{i}sica Te\'{o}rica UAM/CSIC\\
	Universidad Aut\'{o}noma de Madrid\\
        Cantoblanco, E-28049 Madrid, Espa\~{n}a}
        
\abstract{In this paper we discuss in detail computational methods and new results for one-loop virtual corrections to $\mathcal{N}=4$ super Yang-Mills scattering amplitudes calculated to all orders in $\e$, the dimensional regularization parameter. It is often the case that one-loop gauge theory computations are carried out to $\Ord(\e^0)$, since higher order in $\e$ contributions vanish in the $\e \rightarrow 0$ limit. We will show, however, that the higher order contributions are actually quite useful. In the context of maximally supersymmetric Yang-Mills, we consider two examples in detail to illustrate our point. First we will concentrate on computations with gluonic external states and argue that $\mathcal{N}=4$ supersymmetry implies a simple relation between all-orders-in-$\e$ one-loop $\mathcal{N}=4$ super Yang-Mills amplitudes and the first and second stringy corrections to analogous tree-level superstring amplitudes. For our second example we will derive a new result for the all-orders-in-$\e$ one-loop superamplitude for planar six-particle NMHV scattering, an object which allows one to easily obtain six-point NMHV amplitudes with arbitrary external states. We will then discuss the relevance of this computation to the evaluation of the ratio of the planar two-loop six-point NMHV superamplitude to the planar two-loop six-point MHV superamplitude, a quantity which is expected to have remarkable properties and has been the subject of much recent investigation. To make the presentation as self-contained as possible, we extensively review the prerequisites necessary to understand the main results of this work.}
\begin{document}
\tableofcontents
\section{Overview}
In recent years, tremendous progress has been made towards a more complete understanding of the scattering amplitudes in $\mathcal{N}=4$ super Yang-Mills theory~\cite{origN=4} (hereafter simply $\Nsym$). Lovingly referred to as the ``harmonic oscillator'' of quantum field theory, $\Nsym$ has more symmetry than any other gauge theory, especially in its so-called planar limit~\cite{origtHooft,origMald}. Although the theory's S-matrix has been under investigation for nearly 30 years~\cite{origGreenSchwBrink}, the last five have been particularly exciting. Numerous ground-breaking discoveries have been made (like the application of the AdS/CFT correspondence~\cite{origMald} to gluon scattering at strong coupling~\cite{origAldayMald}, a hidden dual superconformal symmetry of the planar theory~\cite{DHKSdualconf}, and a dual description of the leading singularities of the S-matrix as integrals over periods in a Grassmann manifold~\cite{dualSmat} to name just a few) and there is no reason to believe that we have learned everything $\Nsym$ has to teach us. 

One of our main goals in this work is to further develop existing tools for the calculation of one-loop $\Nsym$ amplitudes to all orders in the dimensional regularization~\cite{origtHooftVelt} parameter. This parameter, $\e$, is introduced to cut off the IR divergences that appear in massless gauge theory calculations (we encourage readers less familiar with the structure of IR divergences in gauge theory to peruse Appendix \ref{dimregs}). We will illustrate our methods by considering examples where our results find useful application. At times we will develop aspects of $\Nsym$ S-matrix theory that appear to be of purely academic interest, but, in fact, a significant part of the computational machinery discussed in this paper can be applied to calculations in any quantum field theory. When techniques are applicable only in $\Nsym$ we will try to emphasize this. Before delving into the details of the problems we want to solve, a few words of historical introduction are in order.

$\mathcal{N}=4$ SYM is a very special four dimensional quantum field theory and its S-matrix has a number of unusual properties, many of which were unknown until very recently. We begin by reviewing some of its better-known features. The field content of the model consists of a gauge field $A_{\mu}$, four Majorana fermions $\psi_i$, three real scalars $X_p$, and three real pseudo-scalars $Y_q$. All fields are in the adjoint representation of a compact gauge group, $G$. The Lagrange  density of $\mathcal{N}=4$ is given by \cite{myfirst}\footnote{Here we use the conventions of Peskin and Schroeder~\cite{PeskinSchroeder} for the Lagrange density, which differ somewhat from the conventions of~\cite{myfirst}. Throughout this paper, when not explicitly defined, the reader may assume that our conventions for perturbation theory coincide with those of Peskin and Schroeder.} 
\bea
\label{LagDens}
\mathcal{L} &&= \textrm{tr} \bigg\{ -\frac{1}{2} F_{\mu \nu} F^{\mu \nu} + \bar{\psi_i} \slashed{D}  \psi_i + D^{\mu} X_p D_{\mu} X_p + D^{\mu} Y_q D_{\mu} Y_q \\ \nonumber
&& - i g \bar{\psi_i} \alpha^p_{i j}[X_p,\psi_j]+g \bar{\psi}_i \gamma_5 \beta^q_{i j} [Y_q,\psi_j] \\ \nonumber
&& +\frac{g^2}{2}\bigg([X_l,X_k][X_l,X_k]+[Y_l,Y_k][Y_l,Y_k]+2[X_l,Y_k][X_l,Y_k]\bigg )\bigg\},
\eea
where the $4 \times 4$ matrices $\alpha^p$ and $\beta^q$ are given by  \footnote{$\sigma_0$ is the $2 \times 2$ identity matrix.}
\bea
&& \alpha^1 = \left(\begin{array}{cc} i \sigma_2 & 0 \\ 0 & i \sigma_2 \end{array}\right),~~\alpha^2 = \left(\begin{array}{cc} 0 & - \sigma_1 \\ \sigma_1 & 0 \end{array}\right),~~\alpha^3 = \left(\begin{array}{cc} 0 & \sigma_3 \\ -\sigma_3 & 0 \end{array}\right),  \\ \nonumber
&& \beta^1 = \left(\begin{array}{cc} -i \sigma_2 & 0 \\ 0 & i \sigma_2 \end{array}\right),~~\beta^2 = \left(\begin{array}{cc} 0 & -i \sigma_2 \\ -i \sigma_2 & 0 \end{array}\right),~~\beta^3 = \left(\begin{array}{cc} 0 & \sigma_0 \\ -\sigma_0 & 0 \end{array}\right).
\eea
Once the gauge group and coupling constant $g$ are fixed, the theory is uniquely specified. It turns out that in scattering amplitude calculations it is somewhat more typical to pair up the scalars and pseudoscalars and work with three complex scalar fields. The presence of four supercharges means that there is an $SU(4)$ R-symmetry acting on the fields.  This symmetry acts on the state space as well and dictates selection rules for $\Nsym$ scattering amplitudes.

One of the first remarkable discoveries made about the $\Nsym$ model is that the
scale invariance of the classical Lagrange density (\ref{LagDens}) remains a symmetry at
the quantum level~\cite{SohniusWest}, implying that the $\beta$ function vanishes to
all orders
in perturbation theory. It follows~\cite{HoweStelleWest,HoweST,BrinkLN1,BrinkLN2,Lemes} that the
theory is UV finite in
perturbation theory (it turns out that the $\beta$ function remains
zero non-perturbatively
as well, but this is trickier to prove~\cite{Seiberg}). The classical superconformal
invariance
of the classical Lagrange density (\ref{LagDens}) (see Appendix B for a brief
discussion of
the $\Nsym$ superconformal group) continues to be a quantum mechanical
symmetry of
all correlation functions of gauge-invariant operators.

Most of the work on $\Nsym$ scattering amplitudes focuses on the massless, superconformal $\Nsym$ model described above but we note in passing that it is also possible to construct an $\Nsym$ model with both massive and massless fields~\cite{Fayet}. One can give some of the scalar fields in (\ref{LagDens}) vacuum expectation values  (VEVs) at the cost of superconformal invariance and some of the generators of $G$. Formally, the fact that the six scalar fields can acquire VEVs without breaking supersymmetry implies that the theory has a six-dimensional moduli space of vacua. The $\Nsym$ model where some, but not all, of the gauge group generators are broken by scalar VEVs is called the Coulomb phase of the theory. While most of the literature has focused on the massless, conformal phase of the theory, the Coulomb phase is also quite interesting and is starting to attract the attention it deserves~\cite{myfirst,Higgsreg1,Higgsreg2,Higgsreg3,Higgsreg4}. Unfortunately, a proper discussion of the Coulomb phase is beyond the scope of this paper and we focus our attention exclusively on the S-matrix in the conformal phase of the theory, using dimensional regularization to regulate the IR divergences.

To better understand what makes $\Nsym$ so much simpler than garden-variety quantum field theories, it is instructive to compare  the form of the one-loop virtual corrections in $\Nsym$ to, say, those in ordinary Yang-Mills theory. To be concrete, consider the four-gluon scattering amplitude in both models. Na\"{i}vely, one might think that the final results in the $\Nsym$ model are naturally expressed in terms of the Feynman integral basis for pure Yang-Mills, modulo UV divergent contributions. In fact, the basis of Feynman integrals that one needs for one-loop $\Nsym$ calculations form an even smaller subset. To understand this point, we must take a closer look at the integral basis for four-point scattering in pure Yang-Mills theory~\cite{PassarinoVeltman}, pictured in Figure \ref{4pttops}.

\FIGURE{
\resizebox{0.95\textwidth}{!}{\includegraphics{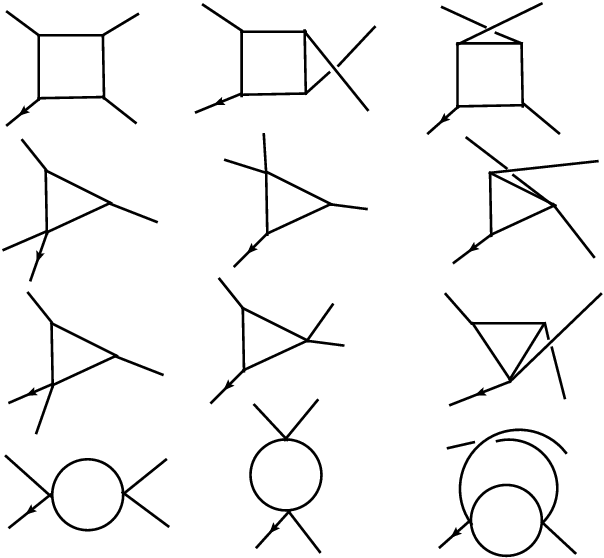}}
\caption{The possible integral topologies that could enter into the calculation of the one-loop four-gluon amplitude in pure Yang-Mills theory. The external particles are all taken to be outgoing and they are understood to be labeled clockwise beginning with the decorated leg.}
\label{4pttops}}

The integrals in each line of Figure \ref{4pttops} have a different topology. From top to bottom, we have scalar box integrals, triangle integrals, and bubble integrals. For instance, the 1st integral in the top row is defined as
\bea
-i(4 \pi)^{2-\e}\mintd{p}{4-2\e} {1\over p^2 (p-k_1)^2 (p-k_1-k_2)^2(p+k_4)^2}
\eea
and the third integral in the bottom row is defined as
\bea
-i(4 \pi)^{2-\e}\mintd{p}{4-2\e} {1\over p^2 (p-k_1-k_3)^2}
\eea
After a moment's thought it is clear that the $D = 4 - 2\e$ bubble integrals are UV divergent and the $D = 4 - 2\e$ triangle and box integrals are UV finite (let us stress that they are {\bf NOT} IR finite (see {\it e.g.} Appendix \ref{dimregs})).\footnote{It's worth pointing out that, in general, the basis integrals in spacetime dimensions other than four can appear as well. As explained in the \ref{GUD} the UV/IR behavior of such integrals is rather different.}

If we follow the above line of reasoning, we might guess that the one-loop four-gluon amplitude in $\Nsym$ is built out $D = 4 - 2\e$ triangles and boxes, but not $D = 4 - 2\e$ bubbles. Remarkably, this turns out not to be the case; the one-loop four-gluon $\Nsym$ amplitude is built out of box integrals only. What is even more remarkable is that, with the caveat that we drop contributions $\Ord(\e)$ and higher, this conclusion holds~\cite{1001lessons} for $n$-gluon scattering amplitudes\footnote{As we will be clear later, it is now known that this conclusion holds for one-loop $\Nsym$ scattering amplitudes with arbitrary external states.}.

For our purposes, the result in the above paragraph will not suffice; we are interested in studying $\Nsym$ amplitudes to all orders in $\e$ and we therefore need to modify the integral basis. Actually, this is not too hard. It has been clear at least since the work of~\cite{oneloopdimreg} that all one has to do is add scalar pentagon integrals to the basis. Then one can express any one-loop $\Nsym$ scattering amplitude in terms of pentagons and boxes to all orders in $\e$. These ideas will be explained in much more detail in Section \ref{revcomp} after the necessary framework has been developed. 

Another main theme of this paper is a novel relation between one-loop scattering amplitudes in $\Nsym$ gauge theory and tree-level scattering amplitudes in open superstring theory\footnote{Tree-level amplitudes of massless particles in open superstring constructions compactified to four dimensions have a universal form~\cite{ST1}.}. With a bit of inspiration, the relationships to be discussed can be derived from the existing string theory literature. To the best of our knowledge, however, they are unknown at the time of this writing. We shall test our relationships explicitly in the simplest non-trivial case to establish confidence that they are correct. 

What do we mean by ``the simplest non-trivial case?'' It turns out that there is a natural organizing principle for the S-matrix of $\Nsym$. If we label all external momenta as outgoing, as is conventional, then $\Nsym$ amplitudes can be organized according to a natural isomorphism between their little-group transformation properties and their $SU(4)_R$ transformation properties. In particular, we can assign a set of $SU(4)_R$ indices to each external state according to whether it is a positive helicity gluon, negative helicity fermion etc. Then there is a natural map~\cite{DHKSdualconf,SUSYBCFW,origElvang} from each external state of the theory to a subset\footnote{The map given in (\ref{operatormap}) is obviously not unique. Any consistent permutation of the flavor and $SU(4)_R$ labels for a given type of field would define an equally valid map.} of $\{1,2,3,4\}$. In what follows $g^\pm(p_i)$ is a positive or negative helicity gluon of momentum $p_i$, $\phi_a^{\pm}(p_i)$ is a positive or negative helicity fermion of flavor $a$ and momentum $p_i$, and $S^\pm_a(p_i)$ is a complex scalar of flavor $a$ and momentum $p_i$. A scalar has no helicity so the assignment of ``$+$'' and ``$-$'' is arbitrary (but useful) for scalar particles.
\bea
&&g^+(p_i)\leftrightarrow p_i 
\nn
\phi_1^+(p_i) \leftrightarrow p_i^1 ~~~~~~~~~ \phi_2^+(p_i) &&\leftrightarrow p_i^2 ~~~~~~~~~ \phi_3^+(p_i) \leftrightarrow p_i^3 ~~~~~~~~~ \phi_4^+(p_i) \leftrightarrow p_i^4
\nn
S_1^+(p_i) \leftrightarrow p_i^{12}  ~~~~~~~~~~~~~~ &&S_2^+(p_i) \leftrightarrow p_i^{23} ~~~~~~~~~~~~~ S_3^+(p_i) \leftrightarrow p_i^{13}
\nn
S_1^-(p_i) \leftrightarrow p_i^{34} ~~~~~~~~~~~~~~ &&S_2^-(p_i) \leftrightarrow p_i^{14} ~~~~~~~~~~~~~ S_3^-(p_i) \leftrightarrow  p_i^{24}
\nn
\phi_1^-(p_i) \leftrightarrow p_i^{234} ~~~~~~~~~ \phi_2^-(p_i) &&\leftrightarrow p_i^{134} ~~~~~~~~~\phi_3^-(p_i) \leftrightarrow p_i^{124} ~~~~~~~~~ \phi_4^-(p_i) \leftrightarrow p_i^{123} 
\nn
&& g^-(p_i) \leftrightarrow p_i^{1234}
\label{operatormap}
\eea
The only {\it a priori} non-zero scattering amplitudes are those that respect the R-symmetry; it must be possible to collect $k$ complete copies of $\{1,2,3,4\}$, where $k$ is a non-negative integer.

 To make this rather abstract discussion more concrete, we consider examples for $k = 0,~1,~{\rm and}~2$. For $k = 0$ we have, for example, the all-positive helicity amplitude $\mathcal{A}\left(p_1,p_2,p_3,p_4\right)$. For $k = 1$ a good example is the four-positive helicity fermion amplitude $\mathcal{A}\left(p_1^1,p_2^2,p_3^3,p_4^4\right)$. Finally, an example for $k = 2$ is the four-point amplitude with a positive-negative helicity gluon pair and a positive-negative helicity fermion pair $\mathcal{A}\left(p_1,p_2^{1234},p_3^{1},p_4^{234}\right)$. In fact, it turns out that supersymmetry forces all scattering amplitudes with $k = 0~ {\rm or}~1$ to be equal to zero (see Appendix \ref{SWI} for a discussion of the supersymmetric Ward identities responsible for this). This implies that the first non-zero amplitudes have $k = 2$. Such amplitudes are called MHV amplitudes for historical reasons\footnote{MHV stands for maximally helicity violating. The $n$-point MHV amplitude describes, for example, a scattering experiment where two negative helicity gluons go in and $n-4$ positive helicity gluons and 2 negative helicity gluons come out. Such an outcome violates helicity as much as is possible at tree-level in QCD.}. 
 
At the outset of the author's investigations, Stieberger and Taylor had recently discovered~\cite{ST1} a relation between the one-loop gluon $\Nsym$ MHV amplitudes and the tree-level gluon open superstring MHV amplitudes for which they had no explanation. Our work demystifies the relation they found and generalizes it as much as possible. Since Stieberger and Taylor showed explicitly that all MHV amplitudes satisfy the relation, it is of some interest to look at the simplest uncalculated example as an explicit test of our proposed generalization of the simpler Stieberger-Taylor relation. In other words, we ought to calculate the all-orders-in-$\e$ one-loop six-gluon\footnote{It is straightforward to check that at least six external particles need to participate in order to get an NMHV amplitude.} next-to-MHV (NMHV) amplitudes in $\Nsym$. Fortunately, Stieberger and Taylor have already tabulated all independent six-gluon NMHV amplitudes in open superstring theory~\cite{ST4} compactified to four dimensions. The existence of these results will make it significantly easier to check our proposed relations. Furthermore, our relations shed some light in a non-obvious way on an old result in pure Yang-Mills. In a nutshell, we are able to explain why $A^{1-{\rm loop}}_{1;\,\mathcal{N}=0}(k_1,k_2,\cdots,k_n)$ vanishes when $n > 4$ and three of the gluons are replaced by photons. 

The precise statement of our relations between gauge and string theory is somewhat technical and we postpone further discussion of it to Section \ref{gsrel}. Suffice it to say that the gauge theory side of our relation requires one-loop $\Nsym$ amplitudes calculated to all orders in $\e$. At the outset of our investigations, it was not clear precisely what computational strategy was most appropriate. Therefore the entirety of Section \ref{gluoncomp} and part of Section \ref{supercomp} will be devoted to the explicit calculation of all-orders-in-$\e$ one-loop amplitudes in $\Nsym$.

Before we can discuss what is perhaps our nicest result for all-order one-loop $\Nsym$ amplitudes, we have to review several exciting recent developments in the theory of $\Nsym$ scattering amplitudes. For what follows the planar limit of the $\Nsym$ theory will be indispensable. The planar limit will be defined more carefully in Section \ref{revcomp}, but for now we discuss the simple example of four-gluon scattering to get across the main idea and to illustrate why taking this limit simplifies the S-matrix. We remind the reader that the building blocks for one-loop four-point scattering amplitudes in $\Nsym$ are the three box integrals in the top row of Figure \ref{4pttops}. Only the first of the three can be drawn in a plane without self-intersections. Operationally, the planar limit is reached if one computes the complete $\Nsym$ amplitude and then throws away all basis integrals that cannot be drawn in a plane without self-intersections. The reduction in the number of basis integrals (three to one) is modest at the one-loop four-point level. However, as you add more loops and legs, working in the planar limit dramatically reduces the complexity of the final results.  

We now specialize to the planar limit and discuss some of the remarkable features of the $\Nsym$ S-matrix in this limit. Particularly exciting is the fact that, in the planar limit, it is possible to completely solve the perturbative S-matrix (up to momentum independent pieces) for the scattering of either four gluons or five gluons (and, by $\Nsym$ supersymmetry, all four- and five-point amplitudes). Starting from the work of~\cite{ABDK}, Bern, Dixon, and Smirnov (BDS) made an all-loop, all-multiplicity proposal for the finite part of the MHV amplitudes in $\Nsym$. In this paper~\cite{BDS}, BDS explicitly demonstrated that their ansatz was valid for the four-point amplitude through three loops. Subsequent work demonstrated that the BDS ansatz holds for the five-point amplitude through two-loops~\cite{TwoLoopFive} and that the strong coupling form of the four-point amplitude calculated via the AdS/CFT correspondence (the unfamiliar reader should consult Appendix \ref{ADS/CFT} for a brief description of this important result) has precisely the form predicted by BDS~\cite{origAldayMald}.  

In fact,~\cite{origAldayMald} sparked a significant parallel development. Motivated by the fact that the strong coupling calculation proceeded by relating the four-point gluon amplitude to a particular four-sided light-like Wilson loop, the authors of~\cite{DKS4pt} were able to show that the finite part of the four-point light-like Wilson loop at one-loop matches the finite part of the planar one-loop four-gluon scattering amplitude. The focus of~\cite{DKS4pt} was on the planar four-gluon MHV amplitude, but it was shown in~\cite{BHTWLn} that this MHV amplitude/light-like Wilson loop correspondence holds for all one-loop MHV amplitudes  in $\Nsym$. As will be made clear in Section \ref{WL/MHV}, an arbitrary $n$-gluon light-like Wilson loop should be conformally invariant in position space.\footnote{Strictly speaking, the conformal symmetry is anomalous due to the presence of divergences at the cusps in the Wilson loop. If one regulates these divergences and subtracts the conformal anomaly, then the finite part of what remains will be conformally invariant.} What was not at all obvious before the discovery of the amplitude/Wilson loop correspondence is that $\Nsym$ scattering amplitudes must be (dual) conformally invariant in {\it momentum} space. 

It turns out that this hidden symmetry (referred to hereafter as dual conformal invariance) has non-trivial consequences for the $\Nsym$ S-matrix. Assuming that the MHV amplitude/light-like Wilson loop correspondence holds to all loop orders, the authors of~\cite{DHKSward} were able to prove that dual conformal invariance  fixes the (non-perturbative) form of all the four- and five- point gluon helicity amplitudes (recall that non-MHV amplitudes first enter at the six-point level) in $\Nsym$. Up to trivial factors, they showed that the functional form of the (dual) conformal anomaly coincides with that of the BDS ansatz. Subsequently, work was done at strong coupling~\cite{BM,BRTW} that provides evidence for the assumption made in~\cite{DHKSward} that the MHV amplitude/light-like Wilson loop correspondence holds to all orders in perturbation theory. Quite recently, the symmetry responsible for the correspondence was understood from a perspective that bears on the results seen at weak coupling as well~\cite{DrummondFerro}.

The idea is that, due to the fact that non-trivial conformal cross-ratios can first be formed at the six-point level, one would na\"{i}vely expect the four- and five-point amplitudes to be momentum-independent constants to all orders in perturbation theory. It is well-known, however, that gluon loop amplitudes have IR divergences. These IR divergences explicitly break the dual conformal symmetry and it is precisely this breaking which allows four- and five- gluon loop amplitudes to have non-trivial momentum dependence. In fact, the arguments of~\cite{DHKSward} allowed the authors to predict the precise form that the answer should take and they found (up to trivial constants) complete agreement with the BDS ansatz to all orders in perturbation theory.

At this stage, it was unclear whether the appropriate hexagon Wilson loop would still be dual to the six-point MHV amplitude at the two-loop level. This question was decisively settled in the affirmative by the work of~\cite{BDKRSVV} on the scattering amplitude side and~\cite{DHKS2nd2L6pt} on the Wilson loop side. Another issue settled by the authors of~\cite{BDKRSVV} and~\cite{DHKS2nd2L6pt} was the question of whether the BDS ansatz fails at two loops and six points. It had already been pointed out by Alday and Maldacena in~\cite{AMlargen} that the BDS ansatz must fail to describe the analytic form of the $L$-loop $n$-gluon MHV amplitude for sufficiently large $L$ and $n$, but it had not yet been conclusively proven until the appearance of~\cite{BDKRSVV} and~\cite{DHKS2nd2L6pt} that $L = 2$ and $n = 6$ was the simplest possible example of BDS ansatz violation. The difference between the full answer and the BDS ansatz is called the remainder function and it is invariant under the dual conformal symmetry. 

Since full two-loop six-point calculations are extremely arduous, one might hope that there is a smoother route to proving the above fact. In fact, Bartels, Lipatov, and Sabio Vera~\cite{BLSV1,BLSV2} derived an approximate formula for the imaginary part of the two-loop remainder function in a particular region of phase-space and multi-Regge kinematics. For some time, this formula was the subject of controversy, due to subtleties associated with analytical continuation of two-loop amplitudes. In~\cite{mysecond} the present author confirmed the controversial result\footnote{To appreciate the subtlety here the reader may wish to read the discussion at the end of~\cite{mysecond,BNST} as it relates to the erratum at the end of~\cite{DDG}.} of BLSV for the imaginary part of the remainder by explicitly continuing the full results of~\cite{DHKS2nd2L6pt} into the Minkowski region of phase-space in question.

The authors of~\cite{DHKS2nd2L6pt}, Drummond, Henn, Korchemsky, and Sokatchev (DHKS), recently discovered an even larger symmetry of the planar S-matrix. In~\cite{DHKSdualconf} DHKS found that there is actually a full dual $\Nsym$ superconformal symmetry acting in momentum space, which they appropriately christened dual superconformal symmetry. One of the main ideas utilized in ~\cite{DHKSdualconf} is that all of the scattering amplitudes with the same value of $n$ are unified into a bigger object called an on-shell $\Nsym$ superamplitude. This superamplitude can be further expanded into $k$-charge sectors and we will often refer to the $k$-charge sectors of a given superamplitude as superamplitudes as well. For example, the $n = 6$, $k = 2$ superamplitude would contain component amplitudes like $\mathcal{A}\left(p_1^{1234},p_2^{1234},p_3,p_4,p_5,p_6\right)$ and $\mathcal{A}\left(p_1^{1},p_2^{1234},p_3^{234},p_4,p_5,p_6\right)$ among others. 

In~\cite{DHKSdualconf} DHKS made an intriguing conjecture for the ratio of the $k = 3$ and $k = 2$ six-point superamplitudes. They argued that the $k = 3$ superamplitude is naturally written as the $k = 2$ superamplitude times a function invariant under the dual superconformal symmetry. DHKS explicitly demonstrated that their proposal holds in the one-loop approximation. This ratio function has recently been the subject of intensive investigation and there are strong arguments in favor of it~\cite{BHMPYangian2}. Nevertheless, it would be nice to see explicitly that the two-loop ratio function is invariant under dual superconformal symmetry and this was done quite recently by Kosower, Roiban, and Vergu for an appropriately defined even part of the ratio function~\cite{KRV}. It turns out that the all-orders-in-$\e$ one-loop six-point $\Nsym$ NMHV superamplitude is necessary to explicitly test the dual superconformal invariance of the ratio function at two loops in dimensional regularization. In Section \ref{WL/MHV} we review the dual superconformal symmetry and explain how the all-orders-in-$\e$ one-loop formula we present in Section \ref{supercomp} can be rewritten to manifest this hidden symmetry as much as possible.

To summarize, the structure of this article is as follows. In Section \ref{revcomp} we review the modern computational techniques prerequisite to the topics discussed later in the paper. In Section \ref{gluoncomp} we discuss a new, efficient approach to the calculation of all-orders-in-$\e$ one-loop $\Nsym$ amplitudes, with the one-loop six-point gluon NMHV amplitude as our main non-trivial example. In Section \ref{gsrel} we discuss a novel relation between one-loop $\Nsym$ gauge theory and tree-level open superstring theory and illustrate its usefulness by solving an old puzzle in pure Yang-Mills. In Section \ref{supercomp} we discuss $\Nsym$ on-shell supersymmetry and extend our results for all-orders-in-$\e$ one-loop six-gluon NMHV amplitudes to the full one-loop $\Nsym$ NMHV superamplitude. In Section \ref{WL/MHV} we elaborate on the light-like Wilson loop/MHV amplitude correspondence, on dual superconformal invariance and on the relevance of the results of Section \ref{supercomp} to testing the dual superconformal invariance of the two-loop NMHV ratio function. Finally, in Section \ref{sum}, we summarize the main results of our work. In addition, we provide several appendices where we discuss important topics that deserve some attention but would be awkward to include in the main text. In Appendix \ref{dimregs} we discuss dimensional regularization, its usefulness in the regularization of IR divergences, and the structure of these divergences in planar $\Nsym$ gauge theory at the one-loop level. In Appendix \ref{sconf} we give a brief introduction to the $\Nsym$ superconformal group and present in considerable detail the $\Nsym$ superconformal and dual superconformal algebras. Appendix \ref{sconf} contains the complete conformal and dual superconformal algebras and corrects various misprints existing in the literature. In Appendix \ref{SWI} we give a brief introduction to the supersymmetric Ward identities, consequences of supersymmetry for the S-matrix that result in linear relations between many of the components of $\Nsym$ superamplitudes. Finally, in Appendix \ref{ADS/CFT}, we explain the AdS/CFT correspondence in general terms and then apply it to the calculation of the strong coupling form of the four-gluon amplitude in $\Nsym$.
\section{Review of Computational Technology}
\label{revcomp}
In this section we review some of the tools that make state-of-the-art gauge theory computations possible. In \ref{color} we discuss color decompositions, useful procedures that allow one to isolate the independent color structures that appear in the final results at the very beginning of a calculation. In \ref{planar} we define the planar limit of Yang-Mills theory. In \ref{SH} we introduce the spinor helicity formalism, a very convenient way of dealing with the external wave-functions of fermions and gauge bosons. In \ref{BCFW} we introduce the BCFW recursion relation and discuss its main applications. In \ref{GU4} we introduce the four dimensional generalized unitarity method at the one-loop level in the context of $\Nsym$ and discuss the integral basis, valid through $\Ord(\e^0)$, needed to use it. Finally, in \ref{GUD} we generalize the results of \ref{GU4} to $4 - 2\e$ dimensional spacetime ($D$ dimensional generalized unitarity). 
\subsection{Color Decompositions}
\label{color}
In non-Abelian gauge theories one has to deal with the color, helicity, and kinematic degrees of freedom separately. Otherwise the resulting expressions for loop-level virtual corrections in scattering processes become much too complicated. In this work, we will only need to deal with color adjoints (all fields in $\Nsym$ live in the adjoint representation of the gauge group, which we take to be $SU(\Nc)$) and therefore restrict ourselves to discussing methods applicable to the situation where all of the external particles are in the adjoint representation. This should be contrasted to the situation in QCD. There one needs to deal with external particles that are color fundamentals as well (the fermions). Decoupling the color degrees of freedom in QCD is possible as well (see {\it e.g.}~\cite{BDKW} for an example at the one-loop level). The resulting decomposition, however, is much less simple. 

To begin, we illustrate the concept of color decomposition by analyzing a contribution to four-gluon scattering at tree-level. In what follows, we deviate from the standard normalization (described in textbooks like~\cite{PeskinSchroeder}) and replace 
$${\rm Tr}\{T^a T^b\} = {\D^{a b} \over 2}$$
with  
$${\rm Tr}\{T^a T^b\} = \D^{a b} ~{\rm .}$$
If this alternative normalization convention is not adopted the color decomposition results in objects that have an annoying $2^n$ out front for an $n$-point scattering process (see {\it e.g.}~\cite{Yasui} for some sample calculations with the standard conventions). We will see that the usual Feynman rules can be split up into simpler rules that only contribute to specific color structures. As a first example, we consider the $s$-channel Feynman diagram drawn in Figure \ref{4gluecolor}. For this diagram we need only the gluon propagator and the three-gluon vertex and we work in 't Hooft-Feynman gauge using the conventions of~\cite{PeskinSchroeder}. 

\FIGURE{
\resizebox{0.75\textwidth}{!}{\includegraphics{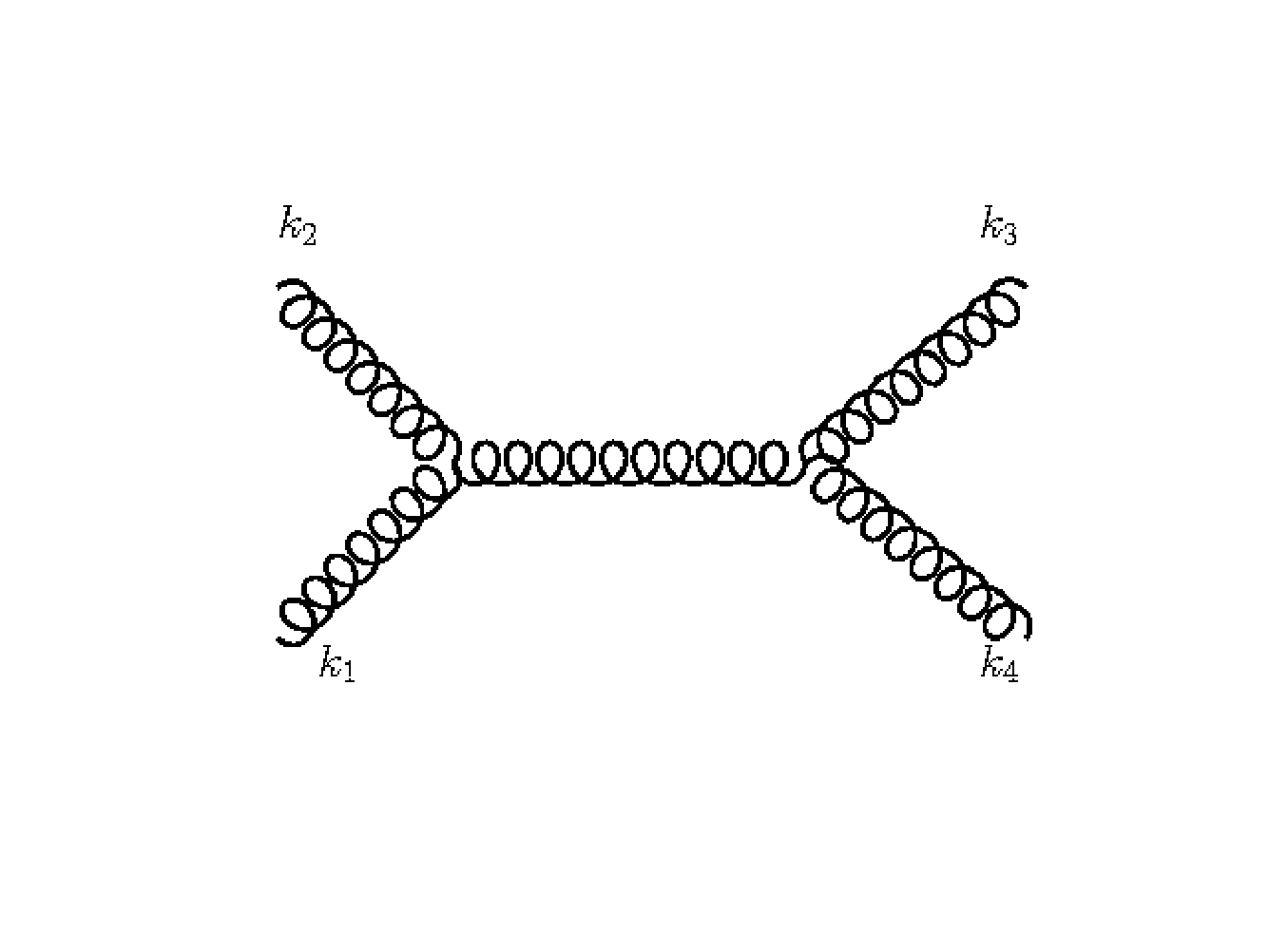}}
\caption{The $s$-channel diagram contributing to the tree-level four-gluon amplitude.}
\label{4gluecolor}}
\vspace{.75 in}
Denote the graph of Figure \ref{4gluecolor} by $\mathcal{A}_S$. We then have
\bea
\a_S &=& \left(g f^{a_1 a_2 b}\big[\pol^{h_1}(k_1)\cdot \pol^{h_2}(k_2)(k_1-k_2)_\mu+\pol_\mu^{h_2}(k_2)\pol^{h_1}(k_1)\cdot (2 k_2+k_1)+\pol_\mu^{h_1}(k_1)\pol^{h_2}(k_2)\cdot(-2 k_1-k_2)\big]\right)\times
\el
\times \left({-i g^{\mu \nu} \delta^{b c} \over (k_1 + k_2)^2}\right) \times
\el
\times \left(g f^{a_3 a_4 c}\big[\pol^{h_3}(k_3)\cdot \pol^{h_4}(k_4)(k_3-k_4)_\nu+\pol_\nu^{h_4}(k_4)\pol^{h_3}(k_3)\cdot (2 k_4+k_3)+\pol_\nu^{h_3}(k_3)\pol^{h_4}(k_4)\cdot(-2 k_3-k_4)\big]\right) \,{\rm ,}\nn
\label{raw4glue}
\eea
where we have kept the polarizations of the gluons arbitrary. For now we ignore the dependence of $\a_S$ on everything except color. This leaves us with 
\be
f^{a_1 a_2 b}f^{a_3 a_4 b}
\label{initcolorcont}
\ee
To push further, we exploit a few simple facts about the Lie algebra of $SU(\Nc)$. The structure constants of the group can be written as
\be
f^{a b c} = -{i \over \sqrt{2}} {\rm Tr}\{T^{a}T^{b}T^{c}-T^{a}T^{c}T^{b}\}
\ee
and the structure constants with one index contracted up with a fundamental representation generator matrix can be rewritten by using the algebra itself:
\be
- {i \over \sqrt{2}}[T^a,~T^b] = f^{a b c}T^c \,{\rm .}
\ee
These relations allow us to write (\ref{initcolorcont}) completely in terms of the $T^a$. After massaging the color factors into the desired form
\bea
f^{a_1 a_2 b}f^{a_3 a_4 b} &=& -{i \over \sqrt{2}} {\rm Tr}\{T^{a_1}T^{a_2}T^{b}-T^{a_1}T^{b}T^{a_2}\} f^{a_3 a_4 b} 
\el
= -{1\over 2}{\rm Tr}\{T^{a_1}T^{a_2}[T^{a_3},T^{a_4}]-T^{a_1}[T^{a_3},T^{a_4}]T^{a_2}\}
\elale
-{1\over 2}{\rm Tr}\{T^{a_1}T^{a_2}T^{a_3}T^{a_4}-T^{a_1}T^{a_2}T^{a_4}T^{a_3}-T^{a_1}T^{a_3}T^{a_4}T^{a_2}+T^{a_1}T^{a_4}T^{a_3}T^{a_2}\} \nn
\eea
we can now go back to eq. (\ref{raw4glue})
\bea
\a_S &=& i g^2 {\rm Tr}\{T^{a_1}T^{a_2}T^{a_3}T^{a_4}-T^{a_1}T^{a_2}T^{a_4}T^{a_3}-T^{a_1}T^{a_3}T^{a_4}T^{a_2}+T^{a_1}T^{a_4}T^{a_3}T^{a_2}\} \times
\el
\times {1\over 2}\big[\pol^{h_1}(k_1)\cdot \pol^{h_2}(k_2)(k_1-k_2)_\mu+\pol_\mu^{h_2}(k_2)\pol^{h_1}(k_1)\cdot (2 k_2+k_1)+\pol_\mu^{h_1}(k_1)\pol^{h_2}(k_2)\cdot(-2 k_1-k_2)\big]
 \times
\el
\times {g^{\mu \nu} \over (k_1 + k_2)^2} \big[\pol^{h_3}(k_3)\cdot \pol^{h_4}(k_4)(k_3-k_4)_\nu+\pol_\nu^{h_4}(k_4)\pol^{h_3}(k_3)\cdot (2 k_4+k_3)+\pol_\nu^{h_3}(k_3)\pol^{h_4}(k_4)\cdot(-2 k_3-k_4)\big]\nn
\label{fincoldemo}
\eea
and we see explicitly that we've achieved our goal. This amplitude contributes to four different color structures and four different {\it color-ordered partial amplitudes}. A color-ordered partial amplitude is defined as the collection of all terms from all the different diagrams that have the same color structure. The reason that this decomposition is so useful in practice is that the color structures are independent and therefore, by construction, each color-ordered partial amplitude must be gauge invariant. Since a color structure is only defined up to cyclic permutations (because the trace is cyclicly symmetric), we choose representatives for them with the convention that $T^{a_1}$ is always the first generator matrix to appear in any trace structure. To make sure that the notation is clear, we remind the reader that $a_1$ denotes the color label, valued in $\{1,\cdots,\Nc\}$, of the first gluon. However, due to cyclic symmetry, what we decide to call the first gluon is completely arbitrary. Finally, we note that this technique can easily be used to expand the color factors of the four-gluon vertex as well.

Although the $n$-point generalization of the above tree-level color decomposition is typically explained in a field theory context, for us it is useful to follow the route that was taken historically. In~\cite{Manganocolor,BerendsGiele88} it was pointed out that an elegant way to derive color decomposition formulae in field theory is to first derive the formulae in an open superstring theory and then take a particular limit (the infinite string tension limit) in which the unwanted modes in the superstring theory decouple and $\Nsym$ gauge theory falls out. This approach benefits us because we will need the tree-level color decomposition for open superstring theory in Section \ref{gsrel} when we discuss our non-trivial relations between one-loop $\Nsym$ amplitudes and tree-level open superstring amplitudes.

It has been known since the early days of superstring theory, that one can write a open superstring theory amplitude as
\bea
&& \a^{tree}_{str}\left(k_1^{h_1},~k_2^{h_2},~\cdots,~k_n^{h_n}\right) = 
\el
g^{n-2} \sum_{\sigma \in S_n/\mathcal{Z}_n} {\rm Tr}\{T^{a_{\sigma(1)}}T^{a_{\sigma(2)}}~\cdots T^{a_{\sigma(n)}} \} A^{tree}_{str}\left(k_{\sigma(1)}^{h_{\sigma(1)}},~k_{\sigma(2)}^{h_{\sigma(2)}},~\cdots,~k_{\sigma(n)}^{h_{\sigma(n)}}\right)
\elale
g^{n-2} \sum_{\sigma \in S_n/\mathcal{Z}_n} {\rm Tr}\{T^{a_{\sigma(1)}}T^{a_{\sigma(2)}}~\cdots T^{a_{\sigma(n)}} \} A^{tree}\left(k_{\sigma(1)}^{h_{\sigma(1)}},~k_{\sigma(2)}^{h_{\sigma(2)}},~\cdots,~k_{\sigma(n)}^{h_{\sigma(n)}}\right) + \Ord(\alpha'^2)\,{\rm ,}\nn
\label{fintreecols}
\eea
where partial amplitudes are associated to appropriate Chan-Paton factors and one sums over all distinguishable permutations\footnote{String amplitudes can be expressed as a power series in the inverse string tension, $\alpha'$, and (\ref{fintreecols}) is the zeroth order term. More discussion of the perturbation theory of the massless modes in open superstring is given in \ref{gsrel}.}. In the infinite string tension limit, the formula above reduces to an analogous one for $U(\Nc)$ $\Nsym$ gauge theory:
\bea
&& \a^{tree}\left(k_1^{h_1},~k_2^{h_2},~\cdots,~k_n^{h_n}\right) = 
\el
g^{n-2} \sum_{\sigma \in S_n/\mathcal{Z}_n} {\rm Tr}\{T^{a_{\sigma(1)}}T^{a_{\sigma(2)}}~\cdots T^{a_{\sigma(n)}} \} A^{tree}\left(k_{\sigma(1)}^{h_{\sigma(1)}},~k_{\sigma(2)}^{h_{\sigma(2)}},~\cdots,~k_{\sigma(n)}^{h_{\sigma(n)}}\right) \,{\rm .}\nn
\label{fintreecol}
\eea
At this stage, the alert reader may be wondering whether the $U(\Nc)$ written above should really be $SU(\Nc)$. Actually, this is not the case. Locally, one can always write $U(\Nc) \simeq SU(\Nc)\times U(1)$ and, since $U(1)$ is Abelian, we are effectively in $SU(\Nc)$ because any partial amplitude containing the $U(1)$ particle (photon) must vanish. This cancellation is simply due to the fact that such a particle has no way to couple to states that live in the adjoint of $SU(\Nc)$. In fact, this vanishing yields the following identity
\be
0 = A^{tree}\left(k_1^{h_1},~k_2^{h_2},~\cdots,~k_n^{h_n}\right)+ A^{tree}\left(k_2^{h_2},~k_1^{h_1},~\cdots,~k_n^{h_n}\right)+\cdots+ A^{tree}\left(k_2^{h_2},~k_3^{h_3},~\cdots,~k_1^{h_1}\right)
\label{photdecoup}
\ee
for the partial amplitudes, commonly referred to as the photon decoupling identity. Eq. (\ref{photdecoup}) is nothing but the expression for an (vanishing) amplitude with a photon with momentum $k_1$ and helicity $h_1$ and $n-1$ $SU(\Nc)$ adjoint particles. The reason that there is a sum with $k_1^{h_1}$ inserted in all possible positions is that the photon is not ordered with respect to the $n-1$ adjoint particles and therefore we have to symmetrize with respect to the insertion of the photon. Finally, we remark that there is another feature of (\ref{fintreecol}) that should bother the alert reader: it looks like there are $(n-1)!$ independent partial amplitudes that need to be computed to determine the full amplitude. In practice, the situation is much better; recently, the authors of~\cite{BCJ} showed that, in fact, there are really only $(n-3)!$ independent partial amplitudes.

To summarize, the claim is that, at tree-level, $\Nsym$ amplitudes\footnote{All that we have really assumed is that our gauge group is $SU(\Nc)$ and that all of the fields transform in the adjoint representation.} organized in terms of their color structures look identical to open superstring amplitudes organized in terms of their Chan-Paton~\cite{ChanPaton} factors. It is also very important to note that all we had to assume about the superstring construction to get this to work at tree-level is that it approaches $\Nsym$ field theory in a particular limit. In a nutshell, this is why the tree-level S-matrix of massless modes in superstring theory is model independent. This model independence disappears at loop level and one has to work much harder. Nevertheless an analogous program can be carried out at one-loop~\cite{BKcolor} and it has been shown that one can use string constructions to derive useful one-loop color decomposition formulae in field theory. It is to this topic that we now turn.

Unfortunately, there are many more color structures at one-loop than at tree-level and this complicates things. Since a description of the necessary heterotic string construction would take us much too far afield, we refer the interested reader to~\cite{BKcolor} and simply state the main result of their work. In $\Nsym$, one-loop scattering amplitudes can be decomposed into color-ordered partial amplitudes using the formula
\begin{changemargin}{-.4 in}{0 in}
\bea
&& \a^{1-{\rm loop}}_{\Nsym}\left(k_1^{h_1},~\cdots,~k_n^{h_n}\right) = 
g^{n-2}{g^2 \Nc \mu^{2\e} e^{-\gamma_E \e}\over (4 \pi)^{(2-\e)}} \sum_{\sigma \in S_n/\mathcal{Z}_n} {\rm Tr}\{T^{a_{\sigma(1)}}~\cdots T^{a_{\sigma(n)}} \} \times
\el
 \times  A^{1-{\rm loop}}_{1;\,\Nsym}\left(k_{\sigma(1)}^{h_{\sigma(1)}},~\cdots,~k_{\sigma(n)}^{h_{\sigma(n)}}\right)+ g^{n-2}{g^2 \mu^{2\e} e^{-\gamma_E \e} \over (4 \pi)^{2-\e}} \sum_{m = 2}^{[{n \over 2}]+1}\Bigg( \sum_{\sigma \in~ S_n/(\mathcal{Z}_{m-1}\times\mathcal{Z}_{n-m+1})} {\rm Tr}\{T^{a_{\sigma(1)}}~\cdots T^{a_{\sigma(m-1)}}\}\times
\el \times{\rm Tr}\{T^{a_{\sigma(m)}}~\cdots T^{a_{\sigma(n)}} \} A^{1-{\rm loop}}_{2;\,\Nsym}\left(k_{\sigma(1)}^{h_{\sigma(1)}},~\cdots,~k_{\sigma(n)}^{h_{\sigma(n)}}\right)\Bigg)\,{\rm ,}\nn
\label{finloopcol}
\eea
\end{changemargin}
where 
\be
\Big[{n \over 2}\Big] \equiv {\rm Floor}\left({n \over 2}\right)\,{\rm .}
\ee
In this work we will very often use the notation $A^{1-{\rm loop}}_{1}$ as a somewhat abbreviated version of $A^{1-{\rm loop}}_{1;\,\Nsym}$. The appearance of $e^{-\gamma_E \e}/(4 \pi)^{2-\e}$ in (\ref{finloopcol}) is necessary for technical reasons and will be explained in Subsection \ref{GU4}. The factor $\mu^{2\e}$, the unit of mass, is explained in Appendix \ref{dimregs} and is important in theories where the interesting observables are infrared finite. Also, in contrast to what happened at tree-level, we have both single-trace color structures and double-trace color structures. Actually, as we will see in the next Subsection, the double-trace structures will be irrelevant for us because they are sub-leading in the number of colors, $\Nc$. In any case, at one-loop, there are analogs of the photon decoupling relations at tree-level and these allow one to express the coefficients of the double-trace color structures in terms of the coefficients of the single-trace color structures~\cite{BKcolor}. The coefficients of the single-trace structures are commonly referred to as the leading color-ordered partial amplitudes, again because of their dominance at large $\Nc$. 

\subsection{Planar Limit}
\label{planar}
Long ago, 't Hooft observed that non-Abelian gauge theories simplify dramatically~\cite{origtHooft} in a particular limit, in which one fixes the combination $\lambda = 2 \Nc ~g^2$, eliminates $g$ in favor of $\Nc$ and $\lambda$, and then takes $\Nc$ to infinity ($\lambda$ is referred to as the 't Hooft coupling in his honor). One thing 't Hooft conjectured was that large $\Nc$ gauge theory ought to have a stringy description. This idea was given new life by Maldacena in his ground-breaking work~\cite{origMald} on the near-horizon geometry of $AdS_5 \times S_5$ (see Appendix \ref{ADS/CFT}). In brief, Maldacena showed that type IIB superstring theory in an $AdS_5 \times S_5$ background is dual to a $\Nsym$ SYM gauge theory. Maldacena's duality was incredibly novel because it related planar $\Nsym$ at strong coupling to classical type IIB superstring theory at weak coupling. In this paper, we will see that unexpected simplicity also emerges in the planar limit of {\it weakly} coupled $\Nsym$.  

For our purposes, the advantage of working in the planar limit is that the simple tree-level color decomposition formula of eq. (\ref{fintreecol}) actually generalizes to multi-loop amplitudes. This is not hard to guess at the one-loop level from the decomposition formula (\ref{finloopcol}). In this formula, the single-trace color structures have an explicit factor of $\Nc$ out front that the double-trace structures do not. It follows that the single-trace structures dominate in the large $\Nc$ limit. To be explicit, the planar $L$-loop color decomposition formula is 
\bea
&&{\cal A}_1^{L-{\rm loop}}\left(k_1^{h_{\sigma(1)}}, k_2^{h_{\sigma(2)}}, ~\ldots, k_n^{h_{\sigma(n)}}\right) = 
\el
 g^{n-2}\left({g^2 \Nc \mu^{2\e} e^{-\gamma_E \e}\over (4 \pi)^{2-\e}}\right)^L
 \sum_{\sigma \in~ S_n/Z_n}
 \, \Tr\{T^{a_{\sigma(1)}} T^{a_{\sigma(2)}}
   \ldots T^{a_{\sigma(n)}}\}
      A_{1}^{L-{\rm loop}}\left(k_1^{h_{\sigma(1)}}, k_2^{h_{\sigma(2)}}, ~\ldots, k_n^{h_{\sigma(n)}}\right)\,{\rm .} \nn
\label{LLoopDecomposition}
\eea
Clearly, this is going to be much easier to work with than a full $L$-loop color decomposition.

Although $\Nsym$ supersymmetry by itself is very powerful and puts highly non-trivial constraints on the perturbative S-matrix, $\Nsym$ supersymmetry together with the planar limit is even more powerful. In section \ref{WL/MHV} we will discuss a so-called hidden symmetry of the $\Nsym$ S-matrix that emerges in the large $\lambda$ limit. This symmetry, called dual superconformal invariance is like a copy of the ordinary superconformal invariance of the $\Nsym$ that acts in {\it momentum} space\footnote{The Lagrange density of $\Nsym$ is manifestly superconformally invariant in position space. We encourage the reader unfamiliar with superconformal symmetry to peruse Appendix \ref{sconf}.}. 
  
\subsection{Spinor Helicity Formalism}
\label{SH}
In this Subsection we review the spinor helicity formalism, a computational framework suitable for research-level helicity amplitude computations in non-Abelian gauge theories. Although the formalism has been fully developed for more than twenty years~\cite{Calkul1,Calkul2,Calkul3,Calkul4,Calkul5,Calkul6}, it has only recently begun to replace the traditional techniques in mainstream textbooks (see {\it e.g.}~\cite{Srednicki}). The underlying assumption of the method is that the correct way to deal with the S-matrix is to compute helicity amplitudes. This is a useful approach in practice because many helicity amplitudes are protected by supersymmetry or related by discrete symmetries (entering either from parity invariance or the color decomposition). 

The spinor helicity formalism is designed to streamline the computation of helicity amplitudes and to allow one to express the results obtained in as simple a form as possible. In the spinor helicity formalism, {\it everything} that enters into the computation of a helicity amplitude is expressed in terms of the same set of building blocks. For example, all dot products involving the polarization vectors of the gluons (or photons) in the scattering process under consideration are expressed in terms of spinor products. This is also true for invariants built out of scalar products of external four-momenta. Another nice feature of the method is directly related to the way in which external polarization vectors are dealt with in the spinor helicity framework. In a moment, we will see that on-shell gauge invariance is built into the spinor helicity polarization vectors and this can be exploited to cancel numerous terms at the beginning of the calculation. In this subsection we closely follow the conventions and discussion of~\cite{Dixon96rev}, though we will ultimately rewrite everything in more modern notation.

Consider free particle solutions to the massless Dirac equation. The positive and negative energy solutions, $u^\pm(p)$ and $v^\mp(p)$ are equivalent up to phase conventions~\cite{Srednicki}. In particular, it is possible to make a choice where $v^\mp(p) = u^\pm(p)$ and we can eliminate $v^\mp(p)$ in favor of $u^\pm(p)$. In this language, a momentum invariant, $s_{ij} \equiv (p_i+p_j)^2$, is easily expressed in terms of spinor products as
\be
s_{i j} = \bar{u}^-(p_i)u^+(p_j)\bar{u}^+(p_j)u^-(p_i) \,{\rm .}
\label{sijsh}
\ee
The above identity can be verified by turning the right-hand side of (\ref{sijsh})into a trace over gamma matrices and projection operators (see {\it e.g.}~\cite{PeskinSchroeder} if unfamiliar). The polarization vectors $\e^+(p_i)$ and $\e^-(p_i)$ are expressed as
\bea
&&\pol^+_\mu(p_i) = {\bar{u}^-(q_i)\gamma_\mu u^-(p_i)\over \sqrt{2}~\bar{u}^-(q_i)u^+(p_i)}
\label{polvecs+} \\
&&
\pol^-_\mu(p_i) = {\bar{u}^-(p_i)\gamma_\mu u^-(q_i)\over \sqrt{2}~\bar{u}^+(p_i)u^-(q_i)}\,{\rm ,}
\label{polvecs-}
\eea 
where the four-momentum $q_i$ is called the reference momentum associated to $p_i$. Though it is probably not clear how we arrived at eqs. (\ref{polvecs+}) and (\ref{polvecs-}) (see~\cite{TUPhoton,XZC} and, more recently,~\cite{BFLWard} for details), it should be plausible that one can build a $\bf (1/2, 1/2)$ wavefunction out of a $\bf (1/2, 0)$ wavefunction and a $\bf (0, 1/2)$ wavefunction\footnote{Here we are using the representation theory of $SU(2) \times SU(2)$ to label the states of definite helicity. This is possible because the complexified proper orthochronous Lorentz group and $SU(2) \times SU(2)$ have isomorphic Lie algebras. The complexification is necessary to obtain the desired isomorphism between the Lie algebras but doesn't affect the labeling of representations. See {\it e.g.}~\cite{WiedemannKirstenMuller} for more details.}. In fact, using the massless Dirac equation, it is straightforward to show that eqs. (\ref{polvecs+}) and (\ref{polvecs-}) are solutions of the Fourier transformed Yang-Mills field equations in the appropriate limit. Furthermore, they are normalized in the standard way
\be
\pol^{+}(p_i)\cdot\left(\pol^+(p_i)\right)^* = \pol^{+}(p_i)\cdot \pol^-(p_i) = \left(\pol^{-}(p_i)\right)^*\cdot\pol^-(p_i) = -1
\ee
and satisfy the completeness relation appropriate for a light-like axial gauge~\cite{Dixon96rev}:
\be
\sum_{\lambda = \pm} \pol^{\lambda}_\mu(p_i, q_i) \left(\pol^{\lambda}_\nu(p_i, q_i)\right)^* = -g_{\mu \nu} + {p_{i\,\mu} q_{i\,\nu}+p_{i\,\nu} q_{i\,\mu}\over p \cdot q}\,.
\ee
The reference momentum associated to $p_i$ is present because the polarization vector of a gluon state is only defined up to a term proportional to $p_i$; it is always permissible to add a term $\alpha \,p_\mu$ to $\pol^\pm_\mu(p)$ since the Ward identities of gauge theory guarantee that any such term will drop out of gauge invariant quantities.

The arbitrariness introduced into the definitions of the polarization vectors ($\pol^\pm(p_i) \equiv \pol^\pm(p_i, q_i)$) by the $q_i$ can be effectively exploited because of the following identities:
\bea
&&\bar{u}^+(p_j) \slashed{\pol}^-(p_i, p_j) = \bar{u}^-(p_j)\slashed{\pol}^+(p_i, p_j) = 0
\el
\slashed{\pol}^+(p_i, p_j) u^+(p_j) = \slashed{\pol}^-(p_i, p_j) u^-(p_j) = 0
\el
\pol^+(p_i, p_j)\cdot\pol^-(p_j, q) = \pol^+(p_i, q)\cdot \pol^-(p_j, p_i) = 0
\el
\pol^+(p_i, q) \cdot \pol^+(p_j, q) = \pol^-(p_i, q)\cdot \pol^-(p_j, q) = 0
\el
\pol^+(p_i, q) \cdot q = \pol^-(p_i, q) \cdot q = 0\,{\rm .}
\label{polidents}
\eea

Thankfully, the traditional, clunky notation for spinors is no longer used. It makes much more sense to define
\be
u^+(p_i) \equiv  |i~\rangle ~~~~~~ u^-(p_i) \equiv |i~] ~~~~~~ \bar{u}^+(p_i) \equiv [i~| ~~~~~~ \bar{u}^-(p_i) \equiv \langle i~| \,{\rm .}
\ee
Then the eqs. (\ref{sijsh}), (\ref{polvecs+}), and (\ref{polvecs-}) discussed above can be rewritten as
\bea
&&s_{i j} = \spa{i}.j\spb{j}.i \label{modsij}\\
&&\pol^+_\mu(p_i) = {\spab{q_i}.\gamma_\mu.i \over \sqrt{2}\, \spa{q_i}.i} \label{+p}\\
&&\pol^-_\mu(p_i) = {\langle i |\, \gamma_\mu \,| q_i ] \over \sqrt{2}\, \spb{i}.{q_i}}
\label{-p}
\eea
It is also conventional to suppress explicit slashes: $p^\mu_j \spab1.\gamma_\mu.2$ is written as $\spab1.j.2$. Finally, for reference, we translate the Fierz identity, the Schouten identity, and the identity $\sum_{i = 1}^n k_i = 0$ (in an $n$ particle scattering process) into modern spinor language:
\bea
&&\spab{i}.\gamma^\mu.j \spab{k}.\gamma_\mu.\ell = 2\,\spb{i}.k \spa{\ell}.j
\el
\spa{i}.j \spa{k}.\ell + \spa{i}.k \spa{\ell}.j + \spa{i}.\ell \spa{j}.k = 0
\el
\sum_{i = 1}^n \spb{j}.i \spa{i}.k = 0
\label{shidents}
\eea

To gain some familiarity with the ideas of this Subsection, we calculate the tree-level four-gluon amplitude, $A^{tree}\left(k_1^{1234},k_2^{1234},k_3,k_4\right)$. We started evaluating the $s$-channel diagram in (\ref{fincoldemo}). Actually, this is the {\it only} diagram we have to evaluate since the $t$-channel and contact diagrams are identically zero if we choose $q_1 = q_2 = k_4$ and $q_3 = q_4 = k_1$, as can easily be verified using eqs. (\ref{polidents}). This is a remarkable feature of the spinor helicity formalism. Many terms that would be present in a traditional calculation vanish if one makes a judicious choice for the reference momenta. In fact, it is now known (and will be made clear in the next subsection) that one {\it never} needs to deal with four-gluon vertices in the evaluation of gluonic tree amplitudes. Continuing with our computation, we evaluate the $s$-channel diagram using eqs. (\ref{modsij}), (\ref{+p}), and (\ref{-p}) and rewrite all invariants in spinor language
\begin{changemargin}{-.5 in}{0 in}
\bea
&&A^{tree}\left(k_1^{1234},k_2^{1234},k_3,k_4\right) = {i\over 2}\big[\pol^{-}(k_1)\cdot \pol^{-}(k_2)(k_1-k_2)_\mu+\pol_\mu^{-}(k_2)\pol^{-}(k_1)\cdot (2 k_2+k_1)
\el
+\pol_\mu^{-}(k_1)\pol^{-}(k_2)\cdot(-2 k_1-k_2)\big] {g^{\mu \nu} \over (k_1 + k_2)^2} \big[\pol^{+}(k_3)\cdot \pol^{+}(k_4)(k_3-k_4)_\nu+\pol_\nu^{+}(k_4)\pol^{+}(k_3)\cdot (2 k_4+k_3)
\el+\pol_\nu^{+}(k_3)\pol^{+}(k_4)\cdot(-2 k_3-k_4)\big]
\elale 
{i\over 2 \spa1.2\spb2.1} \big[ \pol^-(k_1)\cdot\pol^-(k_2)\pol^+(k_3)\cdot\pol^+(k_4)(k_1-k_2)\cdot(k_3-k_4)
\el+\pol^-(k_1)\cdot\pol^-(k_2)(k_1-k_2)\cdot\pol^+(k_4)(k_3 + 2 k_4)\cdot \pol^+(k_3)
+\pol^-(k_1)\cdot\pol^-(k_2)(k_1-k_2)\cdot\pol^+(k_3)(-2 k_3-k_4)\cdot\pol^+(k_4)
\el+\pol^+(k_3)\cdot\pol^+(k_4)(2 k_2+k_1)\cdot\pol^-(k_1)(k_3-k_4)\cdot\pol^-(k_2)
+\pol^-(k_2)\cdot\pol^+(k_4)(2 k_2+k_1)\cdot \pol^-(k_1)(2 k_4+k_3)\cdot\pol^+(k_3)
\el+\pol^-(k_2)\cdot\pol^+(k_3)(2 k_2+k_1)\cdot \pol^-(k_1)(-2 k_3-k_4)\cdot\pol^+(k_4)
\el
+\pol^+(k_3)\cdot\pol^+(k_4)(-2 k_1-k_2)\cdot\pol^-(k_2)(k_3-k_4)\cdot\pol^-(k_1)
\el
+\pol^-(k_1)\cdot\pol^+(k_4)(-2 k_1-k_2)\cdot\pol^-(k_2)(2 k_4+k_3)\cdot\pol^+(k_3)
\el
+\pol^-(k_1)\cdot\pol^+(k_3)(-2 k_1-k_2)\cdot\pol^-(k_2)(-2 k_3-k_4)\cdot\pol^+(k_4)\big]
\elale {i\over 2 \spa1.2\spb2.1} \big[-4\pol^-(k_2)\cdot\pol^+(k_3)k_2\cdot\pol^-(k_1)k_3\cdot\pol^+(k_4)\big]
\elale {i\over 2 \spa1.2\spb2.1} \left[-4 {\spab2.{\gamma_\mu}.4\over \sqrt{2}\,\spb2.4}{\spab1.{\gamma^\mu}.3\over \sqrt{2}\,\spa1.3}{\spab1.2.4\over\sqrt{2}\,\spb1.4}{\spab1.3.4\over\sqrt{2}\,\spa1.4} \right] \,{\rm .}
\label{4ptsimp}
\eea
\end{changemargin}
In the above, all but one (~$\pol^-(k_2) \cdot \pol^+(k_3)$~) of the dot products of polarization vectors vanished as a consequence of our choice of reference momenta. Using eqs. (\ref{shidents}), we arrive at the final result
\be
A^{tree}\left(1^{1234},2^{1234},3,4\right) = {i \spa1.2^4\over \spa1.2 \spa2.3 \spa3.4 \spa4.1}{\rm .}
\ee
In fact, this result generalizes to arbitrarily many gluons~\cite{ParkeTaylor} and we take this opportunity to define the $n$-gluon tree-level MHV amplitude (Parke-Taylor formula):
\be
A_{n;\,\spa{i}.{j}}^{\textrm{\scriptsize{MHV}}}\equiv A^{tree}\left(1,...,i^{1234},...,j^{1234},...,n\right)=i { \langle i j \rangle^4 \over \langle 1 2 \rangle \langle 2 3 \rangle...\langle n 1 \rangle} \,{\rm .}
\ee 
The identities of eqs. (\ref{shidents}) are useful in simple situations (like the calculation above) but, in practice, it usually makes more sense to simplify spinor strings using complex deformations of spinor variables and the analyticity properties of scattering amplitudes. This will be discussed more in the next subsection. 
\subsection{BCFW Recursion}
\label{BCFW}
In this Subsection we review the recursion relation of Britto, Cachazo, Feng, and Witten (BCFW recursion), a powerful tool used primarily for the calculation of tree-level helicity amplitudes in massless gauge theories. Shortly after BCFW recursion was developed in~\cite{origBCFW} it was realized, probably by the authors of~\cite{allNMHV} and perhaps also by the authors of the original paper, that BCFW recursion is also useful when confronted with the problem of simplifying messy linear combinations of rational functions of spinor products. This is because, in many situations of practical interest, physical linear combinations of spinor products are tightly constrained by their singularity structure. 

Although, it is possible~\cite{SUSYBCFW,DHsuperBCFW} to write down a manifestly $\Nsym$ supersymmetric version of BCFW recursion, we present the recursion relation in its original incarnation. Suppose that all tree-level $(n-1)$ and lower-point gluon helicity amplitudes are known. Britto, Cachazo, Feng, and Witten showed that one can write any $n$-gluon tree amplitude in terms of particular deformations of the known lower-point gluon amplitudes. The algorithm will be easy to understand once we explain the concept.

The main idea is that scattering amplitudes are analytic functions of all their inputs. To exploit this analyticity it makes sense to complexify all four-momenta in the problem before going any further. Now, imagine factorizing the amplitude we wish to calculate on a collinear or multi-particle pole\footnote{If these notions are not familiar, see~\cite{Dixon96rev} for an elementary discussion.}, say the one associated to the invariant $K^2 \equiv (k_i +\cdots+k_{i+j})^2$:
\bea
&&A^{tree}\left(k_1^{h_1},\cdots,k_i^{h_i},\cdots,k_{i+j}^{h_{i+j}},\cdots,k_n^{h_n}\right) \stackrel{K^2 \rightarrow~ 0}{\longrightarrow} \nn
&& \sum_h A^{tree}\left(k_i^{h_i},\cdots,k_{i+j}^{h_{i+j}},K^h\right){-i\over K^2}~A^{tree}\left(-K^{-h},k_{i+j}^{h_{i+j}},\cdots,k_n^{h_n},k_1^{h_1},\cdots,k_{i-1}^{h_{i-1}}\right) \,{\rm .}\nn
\label{factor}
\eea
This is the picture that one should have in mind. Intuitively, BCFW recursion is based on the observation that the set of all such limits of a particular $n$-point gluon tree amplitude actually carry all the information necessary to reconstruct the complete tree amplitude. In general, one should only expect this approach to work for tree amplitudes; we shall see later that factorization properties are typically not sufficient to fix the analytical structure of amplitudes at the one-loop level and higher.

To try and realize this intuitive picture of reversing collinear and multi-particle factorization limits more concretely, BCFW found it convenient to consider a particular analytical continuation applicable to general gluon tree amplitudes (under appropriate assumptions). Consider the following deformation of the holomorphic spinor associated to $k_\ell$ and the anti-holomorphic spinor associated to $k_m$:
\bea
\lambda_\ell &&\rightarrow \lambda_\ell(z) = \lambda_\ell + z \lambda_m\nn
\tilde{\lambda}_m &&\rightarrow \tilde{\lambda}_m(z) = \tilde{\lambda}_m - z \tilde{\lambda}_\ell \,{\rm ,}
\label{BCFWs}
\eea
where $z$ is a complex parameter. At the level of spinors it is not even obvious that this complex deformation is well-defined. The corresponding relations for $k_\ell$ and $k_m$
\bea
k_\ell^\mu &&\rightarrow k_\ell^\mu(z) = k_\ell^\mu + {z \over 2}\spab{m}.\gamma^\mu.\ell \nn
k_m^\mu &&\rightarrow k_m^\mu(z) = k_m^\mu - {z \over 2}\spab{m}.\gamma^\mu.\ell
\eea
make it clear that the BCFW shift (eq. (\ref{BCFWs})) preserves overall momentum conservation
\be
\sum_{i=1}^n k_i = 0
\ee
and, furthermore, a small calculation using (\ref{shidents}) makes it clear that
\be
k_\ell(z)^2 = k_m(z)^2 = 0 \,{\rm .}
\ee
So the BCFW deformation is well-defined after all.

We now evaluate the integral
\be
{1\over 2 \pi i} \oint_C dz {A^{tree}\left(k_1^{h_1},\cdots,k_m^{h_m}(z),\cdots,k_\ell^{h_\ell}(z),\cdots,k_n^{h_n}\right)\over z}
\ee
in two different ways, assuming that $C$ is a very large circle in the complex $z$-plane that encloses all poles of the integrand. Of course, we know the answer must be zero
\be
{1\over 2 \pi i} \oint_C dz {A^{tree}\left(k_1^{h_1},\cdots,k_m^{h_m}(z),\cdots,k_\ell^{h_\ell}(z),\cdots,k_n^{h_n}\right)\over z} = 0
\label{BCFW0}
\ee
by virtue of the choice of contour and Cauchy's theorem. On the other hand\footnote{Here we can proceed only under the assumption that the integrand goes to zero fast enough that $C$ can be safely taken to infinity. This assumption is justified for a large class of shifts~\cite{origBCFW}.}, we can write
\bea
&&{1\over 2 \pi i} \oint_C dz {A^{tree}\left(k_1^{h_1},\cdots,k_m^{h_m}(z),\cdots,k_\ell^{h_\ell}(z),\cdots,k_n^{h_n}\right)\over z} = A^{tree}\left(k_1^{h_1},\cdots,k_m^{h_m},\cdots,k_\ell^{h_\ell},\cdots,k_n^{h_n}\right) \el
+ \sum_\alpha {\rm Res}_{z = z_\alpha}\Bigg\{ {A^{tree}\left(k_1^{h_1},\cdots,k_m^{h_m}(z),\cdots,k_\ell^{h_\ell}(z),\cdots,k_n^{h_n}\right)\over z}\Bigg\}
\label{BCFWrel}
\eea
where $\alpha$ is indexing the poles of the amplitude in $z$ induced in particular factorization channels by the BCFW shift. Though it is not at all obvious, it can be shown~\cite{origBCFW} that it is always possible to find {\it some} shift (and associated pair $(k_m^{h_m},k_\ell^{h_\ell})$) for which (\ref{BCFWrel}) is valid (we focus on pure glue for now). Combining eqs. (\ref{BCFW0}) and (\ref{BCFWrel}), we see that the amplitude at the origin of $z$-space (which is what we want) is given by a sum of residues of the shifted amplitude divided by $z$:
\bea
&&A^{tree}\left(k_1^{h_1},\cdots,k_m^{h_m},\cdots,k_\ell^{h_\ell},\cdots,k_n^{h_n}\right) = - \sum_\alpha {\rm Res}_{z = z_\alpha}\Bigg\{ {A^{tree}\left(k_1^{h_1},\cdots,k_m^{h_m}(z),\cdots,k_\ell^{h_\ell}(z),\cdots,k_n^{h_n}\right)\over z}\Bigg\} \,{\rm .}\nn
\label{BCFWfin}
\eea
Since the physical poles that amplitudes can develop must all be of the form
\be
{1\over (k_i +\cdots+k_{i+j})^2}
\ee
for various $i$ and $j$, it is possible to develop a recursive algorithm based on eqs. (\ref{factor}) and (\ref{BCFWfin}).

Specifically, an $n$-point gluon amplitude can be expressed as a sum over factorization channels such that $k_m$ and $k_\ell$ are not both on the same side of the intermediate state\footnote{This is because no residue in $z$ can arise unless there is non-trivial $z$ dependence in $(k_i +\cdots+k_{i+j})^2$. The fact that BCFW shifts respect momentum conservation makes such non-trivial $z$ dependence impossible if both shifted particles are on the same side of the intermediate state.} in (\ref{factor}). Each factorization channel should be evaluated after the chosen BCFW shift has been made, with the value of $z$ fixed by solving the equation $(k_i +\cdots+k_{i+j})^2(z) = 0$. The technique is best illustrated through a simple example. 

As such, we derive the six-gluon tree amplitude $A^{tree}\left(k_1^{1234},k_2^{1234},k_3^{1234},k_4,k_5,k_6\right)$ using the deformation
\bea
&&\lambda_4(z) = \lambda_4 + z \lambda_3\nn
&&\tilde{\lambda}_3(z) = \tilde{\lambda}_3 - z \tilde{\lambda}_4 \,{\rm .}
\eea
Possible contributions for this choice of shift are shown pictorially in Figure \ref{BCFWfig}.

\FIGURE{
\resizebox{.75\textwidth}{!}{\includegraphics{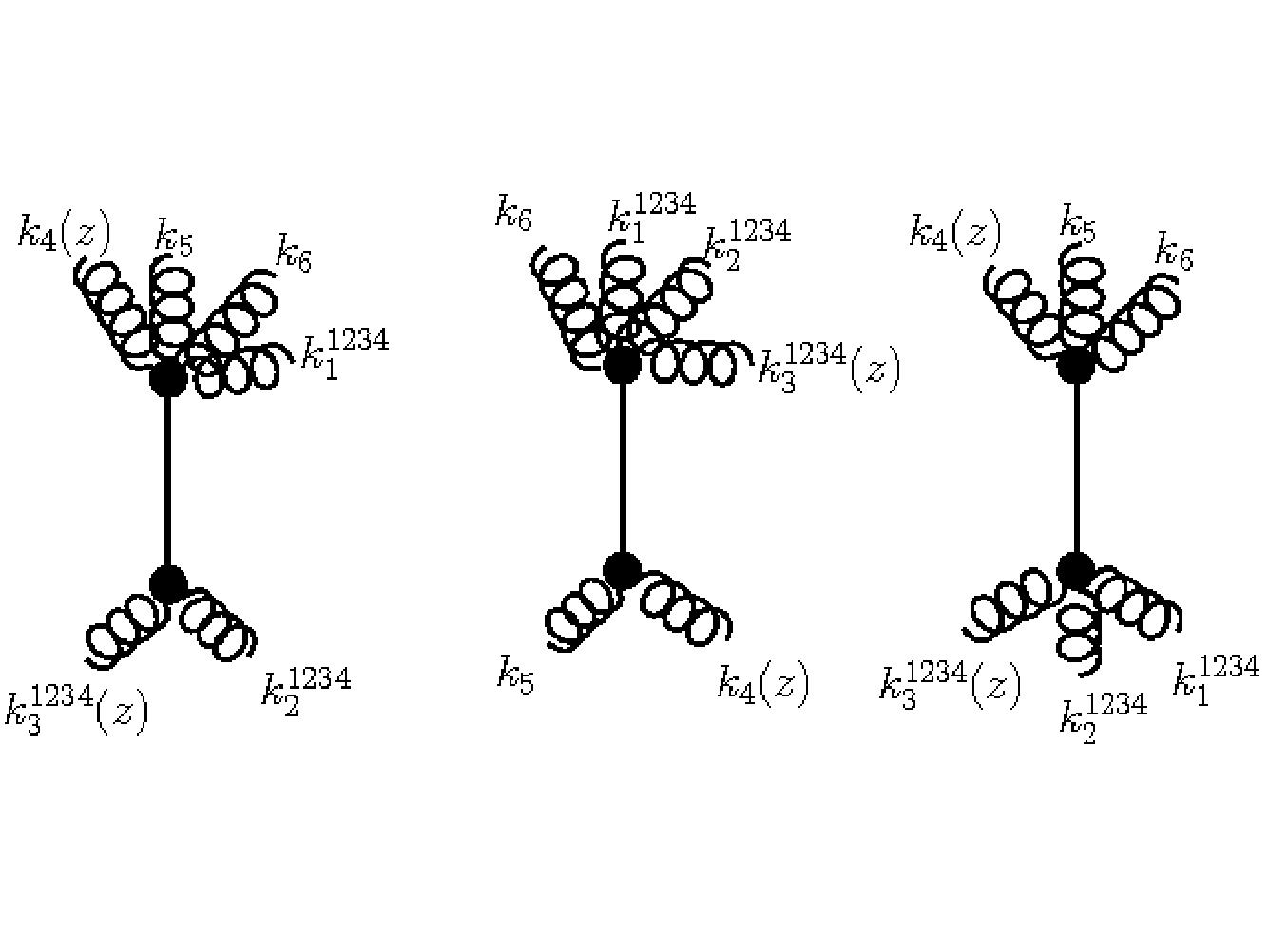}}
\caption{The three BCFW diagrams that could potentially contribute to $A^{tree}\left(k_1^{1234},k_2^{1234},k_3^{1234},k_4,k_5,k_6\right)$.}
\label{BCFWfig}}

Clearly the multi-particle channel gives zero contribution, due to the fact that gluon amplitudes with zero or one negative helicity (and all the rest positive) are protected by supersymmetry (see Appendix \ref{SWI} for a discussion). The first and second graphs in Figure \ref{BCFWfig} are non-zero and, if we label them $\a_1$ and $\a_2$, we have
\bea
\a_1 &=& \sum_h A^{tree}\left(k_2^{1234},k_3^{1234}(z),-(k_2+k_3)^{-h}(z)\right){-i\over (k_2+k_3)^2}A^{tree}\left((k_2+k_3)^h(z),k_4(z),k_5,k_6,k_1^{1234}\right)
\elale A^{tree}\left(k_2^{1234},k_3^{1234}(z),-K_{23}(z)\right){-i\over s_{23}}A^{tree}\left(K_{23}^{1234}(z),k_4(z),k_5,k_6,k_1^{1234}\right)
\elale {i \spa2.{\hat{3}}^4\over \spa2.{\hat{3}} \spa{\hat{3}}.{-\hat{K}_{23}} \spa{-\hat{K}_{23}}.2} {-i\over s_{23}} {i \spa{\hat{K}_{23}}.1^4\over \spa{\hat{K}_{23}}.{\hat{4}}\spa{\hat{4}}.5\spa5.6\spa6.1\spa1.{\hat{K}_{23}}}
\eea
and
\bea
\a_2 &=& \sum_h A^{tree}\left(k_4(z),k_5,-(k_4+k_5)^{-h}(z)\right){-i\over (k_4+k_5)^2}A^{tree}\left((k_4+k_5)^h(z),k_6,k_1^{1234},k_2^{1234},k_3^{1234}(z)\right)
\elale A^{tree}\left(k_4(z),k_5,-K_{45}^{1234}(z)\right){-i\over s_{45}}A^{tree}\left(K_{45}(z),k_6,k_1^{1234},k_2^{1234},k_3^{1234}(z)\right)
\elale {i \spb{\hat{4}}.5^4\over \spb{\hat{4}}.5 \spb5.{-\hat{K}_{45}} \spb{-\hat{K}_{45}}.{\hat{4}}}{-i\over s_{45}} {i \spb{\hat{K}_{45}}.6^4\over \spb{\hat{K}_{45}}.6 \spb6.1 \spb1.2 \spb2.{\hat{3}} \spb{\hat{3}}.{\hat{K}_{45}}} \,{\rm .}
\eea
In the above, a hatted spinor variable reminds us which spinors have been shifted. At this stage, we see why complexified momenta are necessary; if we worked with real momenta the three point amplitudes $A^{tree}\left(k_2^{1234},k_3^{1234}(z),-K_{23}(z)\right)$ and $A^{tree}\left(k_4(z),k_5,-K_{45}^{1234}(z)\right)$ in the above would vanish identically. In order to further simplify the above equations we need
\bea
\spa{a}.{\hat{K}_{23}} &=& {\spab{a}.{K_{23}}.4 \over \spb{\hat{K}_{23}}.4} ~~~~{\rm and} \nn
\spb{\hat{K}_{45}}.{b} &=& {\spab{3}.{K_{45}}.b \over \spa{3}.{\hat{K}_{23}}}
\label{BCFWid}
\eea
together with the solutions of $(k_2+k_3)^2(z)=0$ and $(k_4+k_5)^2(z)=0$:
\bea
z_{23} &=& {s_{23}\over \spab3.2.4} ~~~~{\rm and} \nn
z_{45} &=& -{s_{45}\over \spab3.5.4} \,{\rm .}
\eea
It may concern the reader that we are not able to express all quantities appearing in $\a_1$ and $\a_2$ explicitly in terms of unshifted spinors. Actually, this will turn out not to be a problem; all factors of $\spb{\hat{K}_{23}}.4$ and $\spa{3}.{\hat{K}_{23}}$  cancel out of the final result as can easily be verified by counting how many times they will appear in the numerator and in the denominator of each expression. Applying identities (\ref{BCFWid}) to $\a_1$ and $\a_2$, we finally obtain
\bea
\a_1 &=& {i \spa2.3^4 \spab1.{K_{23}}.4^4 \over \spa2.3 \spab3.{K_{23}}.4 \spab2.{K_{23}}.4 s_{23} (\spab4.{K_{23}}.4+z_{23}\spab3.{K_{23}}.4)(\spa4.5+z_{23}\spa3.5) \spa5.6 \spa6.1 \spab1.{K_{23}}.4}\nn
\a_2 &=& {i \spb4.5^4 \spab3.{K_{45}}.6^4 \over \spb4.5 \spab3.{K_{45}}.5 \spab3.{K_{45}}.4 s_{45} \spab3.{K_{45}}.6  \spb6.1 \spb1.2 (\spb2.3-z_{45}\spb2.4)(\spab3.{K_{45}}.3-z_{45}\spab3.{K_{45}}.4)} \nn
\eea
and
\bea
A^{tree}(k_1^{1234},k_2^{1234},k_3^{1234},k_4,k_5,k_6) = \a_1+\a_2
\eea
which can be confirmed (numerically using {\it e.g.} S@M~\cite{S@M}) by comparing to the result given in~\cite{ManganoParke}.

Before moving on to loop-level calculations, we need to say a few words about the application of BCFW recursion to the simplification of messy rational linear combinations of spinor products. This works well when there is reason to believe that an expression for which you have an ugly formula naturally collapses down to a single term. BCFW then allows you to systematically guess the form of the allegedly simple common denominator. To better understand this we consider the following thought experiment. Suppose that instead of choosing $q_1 = q_2 = k_4$ and $q_3 = q_4 = k_1$ to evaluate $A^{tree}\left(k_1^{1234},k_2^{1234},k_3,k_4\right)$ in eq. (\ref{4ptsimp}) we instead made an unintelligent choice of reference momenta that resulted in more than one Feynman diagram making a non-zero contribution to the amplitude. Now imagine making a table of all possible BCFW shifts, numerically evaluated at a randomly chosen non-degenerate phase-space point. Initially, we would find that all shifts produce several poles in $z$. We could deduce that everything should be put over the common denominator $\spa1.2\spa2.3\spa3.4\spa4.1$ by systematically multiplying the unsimplified expression for $A^{tree}\left(k_1^{1234},k_2^{1234},k_3,k_4\right)$ by each invariant in the problem and then checking to see if our table of all BCFW shifts has a simpler $z$-pole structure or not. In short order, we would be able to deduce that
\be
A^{tree}\left(k_1^{1234},k_2^{1234},k_3,k_4\right) = {C\over \spa1.2 \spa2.3\spa3.4\spa4.1} \,{\rm .}
\ee
This is powerful because evaluating BCFW shifts numerically is a lot less labor intensive than attempting an analytic simplification. Once the denominator is determined, it is then a simple matter to use dimensional analysis and the known little group rescaling properties\footnote{The little group of the Lorentz group in four dimensions for a massless external state is $SO(2)$. Keeping in mind $k_i = \lambda_i \tilde{\lambda}_i$, we expect scattering amplitudes to transform covariantly under the rescaling $\lambda_i \rightarrow t_i \lambda,\,\tilde{\lambda}_i \rightarrow t_i^{-1} \tilde{\lambda}_i$. The precise transformation law of course depends on the helicities of the massless external particles in the scattering process: we have $A(t_i \lambda_i,\,t_i^{-1} \tilde{\lambda}_i; h_i) = \left(\prod_{i=1}^n t_i^{-2 h_i}\right) A(\lambda_i,\,\tilde{\lambda}; h_i)$.} of the amplitude to fix $C = \spa1.2^4$. This thought experiment might seem somewhat contrived, but, at least in $\Nsym$, many loop-level calculations result in objects that are naturally put over a single denominator. While more non-trivial amplitudes have constituents like $\spab2.{5+3}.6$ and $\langle6|4+5|2+3|1\rangle$, it is straightforward to try a large number of guesses in a fraction of a second using a computer. Once the denominator is determined, it is typically possible to apply the dual constraints of little group covariance and correct dimensionality to great effect. All new results in this work were simplified using some variant of this technique.

Finally, we should emphasize that arbitrary tree-level scattering processes in $\Nsym$ can be generated using BCFW recursion~\cite{SUSYBCFW,DHsuperBCFW} provided that the above discussion is supersymmetrized and described in the language of the $\Nsym$ on-shell superspace introduced in Section \ref{supercomp}. 
\subsection{Generalized Unitarity in Four Dimensions}
\label{GU4}
We now turn to loop-level calculations. Most of the calculations in this paper are at the one-loop level, but the ideas reviewed in this subsection and the next, with appropriate modifications, have been applied to multi-loop calculations as well. The program of {\it generalized unitarity} pioneered by Bern, Dixon, Dunbar, and Kosower~\cite{BDDKMHV,BDDKNMHV} was developed to replace the traditional Feynman diagram based approach to loop-level calculations. For most loop-level applications it is much more efficient to use generalized unitarity diagrams because they are built out of on-shell tree amplitudes and the number of contributions scales roughly like number of topologies times the number of particle species allowed to run in the loop. This is a already a big improvement over the usual Feynman diagram expansion. As will be made clear, once you take into account the fact that each diagram is also easier to compute, the generalized unitarity approach is even more attractive.

As was made clear in the introduction, any one-loop planar $\Nsym$ amplitude can be written as
\be
A_1^{1-{\rm loop}}(k_1^{h_1},\cdots,k_n^{h_n}) = \sum_\alpha C_\alpha I^{(\alpha)}_4 + \Ord(\e)\,{\rm ,}
\label{gen1loop}
\ee
where $\alpha$ labels the specific kinematic structure of the box integral (more on our labeling scheme below) and each box integral is evaluated through $\Ord(\e^0)$.  Much of the power of the generalized unitarity technique comes from (\ref{gen1loop}), so it is worth spending some time trying to understand it. It turns out that (\ref{gen1loop}) is very special to $\Nsym$. An equation similar to (\ref{gen1loop}) would hold for generic $\mathcal{N}=2$ and $\mathcal{N}=1$ gauge theories, except that triangle and bubble integrals (discussed briefly in the introduction) would have to be added to the box integrals on the right-hand side~\cite{1001lessons, BDDKMHV}. Less supersymmetry ({\it i.e.} $\mathcal{N}=1$ super Yang-Mills) makes such a relation less powerful and a little more difficult to work  with (see {\it e.g.}~\cite{Forde}). For $\mathcal{N}=0$ it becomes harder still and we really need ideas from the next subsection to make an analogous construction. 

Before we start, we need a convenient way to enumerate the box topologies for a planar $n$-particle scattering process. Consider, as usual, a regular $n$-gon with one external line attached at each vertex. In an approach based on Feynman diagrams this would be the highest-point Feynman integral that could possibly appear prior to integral reduction. There are 
\be
\left(\begin{array}{c} n \\ n-4\end{array}\right) = {n! \over (n-4)! 4!}
\label{noboxes}
\ee
ways to collapse this $n$-gon down to a box. Consequently, it is natural to label each box in the integral basis by an $n-4$-tuple of integers corresponding to the internal lines that need to be erased to produce the box in question\footnote{Our convention will be to start counting with the propagator connecting the 1st and $n$th vertices.}. In this work we will mostly be interested in $n = 6$ for which (\ref{noboxes}) gives 15 boxes. 

This formula gives the largest number of boxes that could possibly appear. Depending on the helicity configuration, certain classes of boxes may make no contribution to the sum in (\ref{gen1loop}). To be less cryptic, we write (\ref{gen1loop}) out explicitly for $A_1^{1-{\rm loop}}\left(k_1^{1234},k_2^{1234},k_3,k_4,k_5,k_6\right)$ and $A_1^{1-{\rm loop}}\left(k_1^{1234},k_2^{1234},k_3^{1234},k_4,k_5,k_6\right)$:\footnote{In eqs. (\ref{6ptgMHV}) and (\ref{6ptgNMHV}) $s_i \equiv s_{i\,i+1}$ and $t_i \equiv s_{i\,i+1\,i+2}$, where indices are mod 6. We will frequently use this notation in our discussions of six-point scattering. The notation can, of course, be generalized to describe a basis of kinematic invariants for arbitrary $n$. For instance, at the eight-point level, invariants like $w_{1} \equiv (k_1+k_2+k_3+k_4)^2$ will enter.}
\begin{changemargin}{-.6 in}{0 in}
\bea
&&A_1^{1-{\rm loop}}\left(k_1^{1234},k_2^{1234},k_3,k_4,k_5,k_6\right) = {A_{n;\,\spa{1}.{2}}^{\textrm{\scriptsize{MHV}}} \over 2}\Big(- s_3 s_4 I_{4}^{(1,2)}- {s_4 s_5} I_{4}^{(2,3)} - {s_5 s_6} I_{4}^{(3,4)} -{s_1 s_6 }I_{4}^{(4,5)} - {s_1 s_2 }I_{4}^{(5,6)}
\el - {s_2 s_3 } I_{4}^{(1,6)}+ (s_3 s_6 - t_2 t_3) I_{4}^{(1,4)} + (s_1 s_4 - t_1 t_3) I_{4}^{(2,5)} + (s_2 s_5 - t_1 t_2) I_{4}^{(3,6)} +  \Ord(\e)\Big) 
\label{6ptgMHV}
\eea
\cme
\begin{changemargin}{-.6 in}{0 in}
\bea
&&A_1^{1-{\rm loop}}\left(k_1^{1234},k_2^{1234},k_3^{1234},k_4,k_5,k_6\right) = -{i\over2}  {\spa1.2 \spa2.3 \spb4.5 \spb5.6 \spab3.{(1+2)}.6 \spab1.{(2+3)}.4 t_1^3 \over s_1 s_2 s_4 s_5 (t_1 t_2-s_2 s_5)(t_1 t_3-s_1 s_4)}\Big( s_4 s_5 I_4^{(2,3)}+s_1 s_2 I_4^{(5,6)} 
\el +s_6 t_1 I_4^{(3,5)} + s_3 t_1 I_4^{(2,6)}\Big)-{i\over2}\Bigg(\bigg({\spab1.{(2+3)}.4\over t_2}\bigg)^4 {\spa2.3 \spa3.4 \spb5.6 \spb6.1 \spab4.{(2+3)}.1 \spab2.{(3+4)}.5 t_2^3 \over s_2 s_3 s_5 s_6 (t_2 t_3-s_3 s_6)(t_2 t_1-s_2 s_5)} 
\el+\bigg({\spa2.3 \spb5.6 \over t_2}\bigg)^4 {\spb2.3 \spb3.4 \spa5.6 \spa6.1 \spab1.{(2+3)}.4 \spab5.{(3+4)}.2 t_2^3 \over s_2 s_3 s_5 s_6 (t_2 t_3-s_3 s_6)(t_2 t_1-s_2 s_5)}\Bigg) \Big( s_5 s_6 I_4^{(3,4)}+s_2 s_3 I_4^{(6,1)}+s_1 t_2 I_4^{(4,6)}+s_4 t_2 I_4^{(1,3)}\Big)
\el-{i\over2}\Bigg(\bigg({\spab3.{(1+2)}.6\over t_3}\bigg)^4 {\spa6.1 \spa1.2 \spb3.4 \spb4.5 \spab2.{(6+1)}.5 \spab6.{(1+2)}.3 t_3^3 \over s_6 s_1 s_3 s_4 (t_3 t_1-s_1 s_4)(t_3 t_2-s_6 s_3)}
\el+\bigg({\spa1.2 \spb4.5 \over t_3}\bigg)^4 {\spb6.1 \spb1.2 \spa3.4 \spa4.5 \spab5.{(6+1)}.2 \spab3.{(1+2)}.6 t_3^3 \over s_6 s_1 s_3 s_4 (t_3 t_1-s_1 s_4)(t_3 t_2-s_6 s_3)}\Bigg) \Big( s_6 s_1 I_4^{(4,5)}+s_3 s_4 I_4^{(1,2)}+s_2 t_3 I_4^{(1,5)}
\el+s_5 t_3 I_4^{(2,4)}\Big) +  \Ord(\e) \,{\rm .}
\label{6ptgNMHV}
\eea
\end{changemargin}

In the six-point MHV amplitude, all of the boxes with two adjacent external masses enter with zero coefficient and in the six-point NMHV amplitude all of the boxes with two diametrically opposed external masses enter with zero coefficient. Boxes with two external masses are special in that they have different analytic structures depending on whether the two external masses are adjacent or diametrically opposed. For historical reasons, two mass box integrals with adjacent external masses are called two mass hard and two mass box integrals with diametrically opposed external masses are called two mass easy (see Figure \ref{boxints}). The six different types of box integrals that appear in planar one-loop $\Nsym$ calculations are summarized in Figure \ref{boxints}.

\FIGURE{
\resizebox{0.65\textwidth}{!}{\includegraphics{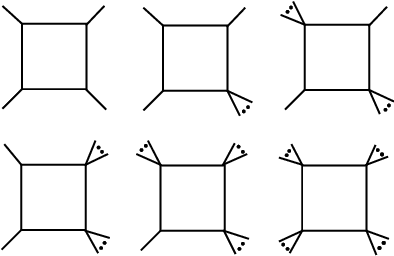}}
\vbox{\vskip 1 in}
\caption{The one-loop basis of scalar integrals in $\Nsym$ through $\Ord(\e^0)$ consists of the: zero mass box integral, one mass box integral, two mass easy box integral, two mass hard box integral, three mass box integral, and four mass box integral.}
\label{boxints}}

Explicit formulae for these integral functions will be provided shortly, but first let us say a few more words about (\ref{gen1loop}). Clearly, the zero mass box will only appear for the special case of four particle scattering. For general $n$, planar $\Nsym$ MHV amplitudes are built out of one mass and two mass easy boxes~\cite{BDDKMHV} and planar $\Nsym$ NMHV amplitudes are built out of one mass, two mass easy, two mass hard, and three mass boxes~\cite{allNMHV}; four mass boxes don't appear until the eight-point ${\rm N}^2$MHV level. In particular, the absence of two mass easy basis integrals in $A_1^{1-{\rm loop}}\left(k_1^{1234},k_2^{1234},k_3^{1234},k_4,k_5,k_6\right)$ does not generalize to higher $n$ NMHV amplitudes. 

Before going further, we need to carefully define the one-loop Feynman integrals which enter into our perturbative calculations. In general, the Feynman integrals that enter into calculations in massless gauge theories have severe IR divergences that need to be regulated. In dimensional regularization~\cite{origtHooftVelt} one regulates the IR divergences by analytically continuing the scattering amplitude under consideration from $D = 4$ to $D = 4 - 2 \e$ and then computing its Laurent expansion about $\e = 0$ (see Appendix \ref{dimregs} for background). We make the definition
\bea
I_n^{D=4-2 \e}
&\equiv& i (-1)^{n+1}(4\pi)^{2-\e} \int 
    {d^{4-2\e} p \over (2\pi)^{4-2\e} }
  { 1 \over p^2 \ldots
    (p-\sum_{i=1}^{n-1} K_i )^2 }
 \nonumber \\   &=& i (-1)^{n+1} (4\pi)^{2-\e} \int 
    {d^{4} p \over (2\pi)^{4} }
    {d^{-2\e} \mu \over (2\pi)^{-2\e} }
  { 1 \over (p^2 - \mu^2) \ldots (
    (p-\sum_{i=1}^{n-1} K_i )^2 - \mu^2) } \,{\rm .}\nn
\label{IntDef}
\eea
The prefactor $i (-1)^{n+1} (4\pi)^{2-\e}$ cancels a factor of $i (-1)^{n} (4\pi)^{\e-2}$ that always arises in the calculation of one-loop Feynman integrals and on the second line we have explicitly separated out the integrations into four dimensional and $-2\e$ dimensional pieces. This second form will be useful in the next subsection and later on. In what follows, we give explicit results for a representative sample of basis integrals that enter into planar one-loop $\Nsym$ scattering amplitudes. For our representative of each integral species (Figure \ref{boxints}), we shall use the kinematics of the point at which the basis integral first enters into the sum in eq. (\ref{gen1loop}). For example, we use five-point kinematics for the one mass box because this integral first appears in planar one-loop five-point MHV amplitudes. A technical point to be aware of is that, in expanding an $L$-loop Feynman integral in $\e$, a factor of $e^{\gamma_E \e\,L}$ is expanded with the Feynman integral in order to prevent a proliferation of factors of $\gamma_E$ that would otherwise occur. This remark and definition (\ref{IntDef}) explain why the prefactors in eq. (\ref{finloopcol}) have the form that they do. Following~\cite{BDDKNMHV} (and checking against the more recent article~\cite{allNMHV}), the epsilon expansions through $\Ord(\e^0)$ of the basis integrals are presented below in the same order that they appeared in Figure \ref{boxints}:
\be
I_4 = {\G(1+\e)\G^2(1-\e)\over s t \,\G(1-2\e)}
\bigg\{
{2 \over \e^2} \Big[ ( -s)^{-\e}+ (-t)^{-\e} \Big]
- \ln^2\left( {-s \over - t} \right) - \pi^2 \bigg\} \,{\rm ,}
\label{box0}
\ee
\bea
 I_{4}^{(5)} &=& { -2 \G(1+\e)\G^2(1-\e) \over s_1 s_2 \,\G(1-2\e)}
 \bigg\{
 -{1\over\e^2} \Big[ (-s_1)^{-\e} +
(-s_2 )^{-\e} - (-s_4)^{-\e} \Big] 
\el + \Li_2\left(1-{s_4 \over s_1}\right)
   + \ \Li_2\left(1-{s_4 \over s_2}\right)
   +{1\over 2} \ln^2\left({-s_1 \over -s_2}\right)\
+ {\pi^2\over6} \bigg\}\,{\rm ,}
\label{box1}
\eea
\bea
 I_{4}^{(2,5)} &=& { -2 \G(1+\e)\G^2(1-\e) \over (t_1 t_3 -s_1 s_4) \,\G(1-2\e)}
 \bigg\{
  - {1\over\e^2} \Big[ (-t_1)^{-\e} + (-t_3)^{-\e}
              - (-s_1)^{-\e} - (-s_4)^{-\e} \Big] 
\el + \Li_2\left(1-{s_1 \over t_1}\right)
  + \Li_2\left(1-{s_1 \over t_3}\right)
  + \Li_2\left(1-{s_4 \over t_1}\right)
\el + \Li_2\left(1-{s_4 \over t_3}\right)
- \Li_2\left(1-{s_1 s_4
\over t_1 t_3}\right)
    + {1\over 2} \ln^2\left({-t_1 \over -t_3}\right) \bigg\}\,{\rm ,}
    \label{box2e}
\eea
\bea
 I_{4}^{(2,4)} &=& { -2 \G(1+\e)\G^2(1-\e) \over s_5 t_3 \,\G(1-2\e)}
 \bigg\{
  - {1\over 2\e^2} \Big[ (-s_5)^{-\e} + 2(-t_3)^{-\e}
              - (-s_1)^{-\e} - (-s_3)^{-\e} \Big] 
\el -{1\over2}\ln\left({-s_1 \over -s_5}\right)\ln\left({-s_3 \over -s_5}\right)+ {1\over 2} \ln^2\left({-s_5 \over -t_3 }\right) + \Li_2\left(1-{ s_1 \over t_3}\right)
   + \Li_2\left(1-{s_3 \over t_3 }\right) \bigg\}\,{\rm ,}\nn
   \label{box2h}
\eea
\bea
 I_{4}^{(3,5,7)} &=& { -2 \G(1+\e)\G^2(1-\e) \over (t_1 t_6-s_2 s_6) \,\G(1-2\e)}
 \bigg\{
 -{1\over 2\e^2} \Big[ (-t_1 )^{-\e} + (-t_6)^{-\e}
     - (-s_2)^{-\e}
     - (-s_6)^{-\e} \Big] 
\el -{1\over2}\ln\left({-s_2 \over -t_6}\right)\ln\left({-s_4 \over -t_6}\right)-{1\over2}\ln\left({-s_4 \over -t_1}\right)\ln\left({-s_6 \over -t_1}\right)+\ {1\over2}\ln^2\left({-t_1 \over -t_6}\right)
\el + \Li_2\left(1-{s_2\over t_1 }\right)
   + \Li_2\left(1-{s_6 \over t_6}\right)
  -  \Li_2
\left(1-{s_2 s_6\over t_1 t_6 }\right) \bigg\}\,{\rm ,}
\label{box3}
\eea
\bea
 I_{4}^{(2,4,6,8)} &=& {-\G(1+\e)\G^2(1-\e)  \over  w_1 w_3 \rho\,\G(1-2\e)}
\bigg\{ - \Li_2\left({1\over2}(1-\lambda_1+\lambda_2+\rho)\right)
  + \Li_2\left({1\over2}(1-\lambda_1+\lambda_2-\rho)\right)
\el - \Li_2\left(
   -{1\over2\lambda_1}(1-\lambda_1-\lambda_2-\rho)\right)
  + \Li_2\left(-{1\over2\lambda_1}(1-\lambda_1-
    \lambda_2+\rho)\right) 
\el - {1\over2}\ln\left({\lambda_1\over\lambda_2^2}\right)
   \ln\left({ 1+\lambda_1-\lambda_2+\rho \over 1+\lambda_1
        -\lambda_2-\rho }\right) \bigg\} \,{\rm ,}
        \label{box4}
\eea
where
\be
 \rho \equiv \sqrt{1 - 2\lambda_1 - 2\lambda_2
+ \lambda_1^2 - 2\lambda_1\lambda_2 + \lambda_2^2}\, {\rm ,}
\ee
and
\bea
\lambda_1 &=& {s_1 s_3\over w_1 w_3} \, {\rm ,}
\\
\lambda_2 &=& {s_5 s_7 \over w_1 w_3} \,{\rm .}
\eea

We will never need to use these results directly to compute our amplitudes. By combining a power variant of the optical theorem for Feynman diagrams (if unfamiliar see {\it e.g.}~\cite{PeskinSchroeder}), the fact that the above integrals form a complete basis for one-loop planar $\Nsym$ scattering amplitudes through $\Ord(\e^0)$, and the fact that the above integrals are uniquely determined by how they develop residues when viewed as contour integrals in $\mathbb{C}^4$ (there is a canonical choice for the contours which shall be discussed shortly), we can deduce all of the coefficients in the sum of eq. (\ref{gen1loop}) for any given amplitude without explicitly evaluating a single Feynman integral~\cite{BDDKMHV,SharpLS}. This is the power of the generalized unitarity technique in the context of planar $\Nsym$.\footnote{As mentioned before, similar arguments can be used to greatly simplify calculations in theories with less supersymmetry. However, such calculations are usually harder because the ansatz for the amplitude (right-hand side of (\ref{gen1loop})) will be less tightly constrained and may contain many more terms.} This is a remarkable claim, so we examine it in detail. First let us be clear about one subtle point, in the above formulae all of the basis integrals were evaluated in the Euclidean region after Wick rotation because this is technically easier; in the Euclidean region the Laurent expansions of Feynman loop integrals have real coefficients. In what follows we deal with Feynman integrals before Wick rotation and therefore it makes sense to ignore the prefactor of $i (-1)^{n+1} (4\pi)^{2-\e}$ until we actually have our final answer and are ready to Wick rotate from Minkowski space to Euclidean space. At that point it can be trivially restored. In general, it is useful to think about loop calculations in two phases, the first being the determination of the coefficients in eq. (\ref{gen1loop}) and the second being the actual analytical evaluation of the basis integrals. 

To better understand how generalized unitarity is superior to Feynman diagrams, we compare the two approaches for the simple example of the five-point amplitude \\$A^{1-{\rm loop}}_1(k_1^{1234},k_2^{1234},k_3,k_4,k_5)$. This exercise is perfect for us because the actual answer is well-known and it fits on a page even if written out in gory detail: 
\begin{changemargin}{-.6 in}{0 in}
\bea
&&A_{1}^{1-{\rm loop}}(k_1^{1234},k_2^{1234},k_3,k_4,k_5)  =  i {A_{5;\langle 12 \rangle}^{\textrm{\scriptsize{MHV}}} \over 2} \left( s_{2}s_{3} I_4^{(1)}+s_{3}s_{4} I_4^{(2)}+ s_{4}s_{5} I_4^{(3)}+s_{5}s_{1} I_4^{(4)}+s_{1}s_{2} I_4^{(5)} + \Ord(\e)\right)  \,{\rm ,}
\nn
&&A_{1}^{1-{\rm loop}}(k_1^{1234},k_2^{1234},k_3,k_4,k_5)  = i {A_{5;\langle 12 \rangle}^{\textrm{\scriptsize{MHV}}} \over 2} \bigg( s_{2}s_{3} \int 
    {d^{4-2\e} p \over (2\pi)^{4-2\e} }
  { 1 \over (p-k_1)^2 (p-k_1-k_2)^2 (p+k_4+k_5)^2
    (p+k_5)^2 }
\el+s_{3}s_{4} \int 
    {d^{4-2\e} p \over (2\pi)^{4-2\e} }
  { 1 \over p^2 (p-k_1-k_2)^2 (p+k_4+k_5)^2
    (p+k_5)^2 }+ s_{4}s_{5} \int 
    {d^{4-2\e} p \over (2\pi)^{4-2\e} }
  { 1 \over p^2 (p-k_1)^2(p+k_4+k_5)^2
    (p+k_5)^2 }
\el+s_{5}s_{1} \int 
    {d^{4-2\e} p \over (2\pi)^{4-2\e} }
  { 1 \over p^2 (p-k_1)^2 (p-k_1-k_2)^2
    (p+k_5)^2 }+s_{1}s_{2} \int 
    {d^{4-2\e} p \over (2\pi)^{4-2\e} }
  { 1 \over p^2 (p-k_1)^2 (p-k_1-k_2)^2 (p+k_4+k_5)^2
     } 
     \el+ \Ord(\e)\bigg) \,{\rm .}
     \label{5ptapprox}
\eea
\end{changemargin}
We now argue that, by complexifying the loop momentum, $p$, and changing the contour from the usual one over all $\mathbb{R}^{1,3}$ to a particular 4-torus $T^4 \cong S^1\times S^1 \times S^1 \times S^1$ embedded in $\mathbb{C}^4$, we can isolate a single coefficient, say $i s_1 s_2 A_{5;\langle 12 \rangle}^{\textrm{\scriptsize{MHV}}}/2$, on the right-hand side of (\ref{5ptapprox}).

To obtain a meaningful relation, we will of course be interested in somehow performing the same sequence of operations on the left-hand side of (\ref{5ptapprox}). This is less obvious and is where the optical theorem for Feynman diagrams comes into play. Ultimately, the left-hand side will be evaluated using a variant of the optical theorem, generalized unitarity, a name coined by in Eden, Landshoff, Olive, and Polkinghorne in their classic text~\cite{AnalyticSmatrix}.\footnote{We should point out that most of the formalism reviewed in this subsection was developed in~\cite{quadcuts}.} For now, one should simply remember that we must at some point return to the question of how to make sense of the ``raw'' expression for the amplitude (that delivered directly from Feynman diagrams) with respect to whatever contours of integration we introduce on the right-hand side of (\ref{5ptapprox}).

The idea proposed above is well-motivated. One of the main reasons loop-level computations (even in UV finite theories like $\Nsym$) are hard is that one has to worry about regulating divergences in the momentum integrals over all $p$-space. It would be nice if there was some meaningful IR-finite data that one could extract by considering the amplitude on contours in $\mathbb{C}^4$ other than $\mathbb{R}^{1,3}$. We now show how this is realized in the present example. We consider eq. (\ref{5ptapprox}) on a contour $\G_p$ defined by
\be
\G_p = \{p \in\mathbb{C}^4: ~~|p^2|<\D,\,|(p-k_1)^2|<\D,\,|(p-k_1-k_2)^2|<\D,\,|(p+k_4+k_5)^2|<\D \}
\ee
for sufficiently small $\D$. On this contour, we can evaluate the right-hand side of eq. (\ref{5ptapprox}),
\begin{changemargin}{-.6 in}{0 in}
\bea
&& i {A_{5;\langle 12 \rangle}^{\textrm{\scriptsize{MHV}}} \over 2} \bigg( s_{2}s_{3} \int_{\G_p} 
    {d^{4} p \over (2\pi i)^{4} }
  { 1 \over (p-k_1)^2 (p-k_1-k_2)^2 (p+k_4+k_5)^2
    (p+k_5)^2 }
\el+s_{3}s_{4} \int_{\G_p} 
    {d^{4} p \over (2\pi i)^{4} }
  { 1 \over p^2 (p-k_1-k_2)^2 (p+k_4+k_5)^2
    (p+k_5)^2 }+ s_{4}s_{5} \int_{\G_p} 
    {d^{4} p \over (2\pi i)^{4} }
  { 1 \over p^2 (p-k_1)^2(p+k_4+k_5)^2
    (p+k_5)^2 }
\el+s_{5}s_{1} \int_{\G_p} 
    {d^{4} p \over (2\pi i)^{4} }
  { 1 \over p^2 (p-k_1)^2 (p-k_1-k_2)^2
    (p+k_5)^2 }+s_{1}s_{2} \int_{\G_p} 
    {d^{4} p \over (2\pi i)^{4} }
  { 1 \over p^2 (p-k_1)^2 (p-k_1-k_2)^2 (p+k_4+k_5)^2
     } \bigg) \,{\rm ,}\nn
     \label{5ptaprxrhs}
\eea
\end{changemargin}
using a multidimensional generalization of Cauchy's residue theorem. Note that we have set $\e$ to zero. Very soon we will see that, on $\G_p$, eq. (\ref{5ptapprox}) is perfectly well-defined in $D = 4$. $\G_p$ is a product of four tiny circles that wrap all of the singularities of the integrand
$${ 1 \over p^2 (p-k_1)^2 (p-k_1-k_2)^2 (p+k_4+k_5)^2
     }$$
in $\mathbb{C}^4$ and, furthermore, fail to wrap all four of the singularities of any of the other four integrands. The so-called global residue theorem~\cite{GriffithsHarris} allows us to evaluate (\ref{5ptaprxrhs}) in an incredibly straightforward fashion. The first four terms in (\ref{5ptaprxrhs}) give zero contribution because a non-zero contribution can only arise if all singularities of the integrand are wrapped by the contour. The final term,
$$i{A_{5;\langle 12 \rangle}^{\textrm{\scriptsize{MHV}}} \over 2} s_{1}s_{2} \int_{\G_p} 
    {d^{4} p \over (2\pi i)^{4} }
  { 1 \over p^2 (p-k_1)^2 (p-k_1-k_2)^2 (p+k_4+k_5)^2
     }\,{\rm ,}$$
evaluates to
\be
i{A_{5;\langle 12 \rangle}^{\textrm{\scriptsize{MHV}}} \over 2} s_{1}s_{2}\, {\rm det}^{-1}\left({\partial\over \partial p_\mu} ~(p-K_i)^2\right)\Bigg|_{p_\mu =\, p_\mu^*}\,{\rm ,}
\ee
where the new factor is just the Jacobian that results if one changes the integration variables from the $p_\mu$ to the propagator denominators, $(p-K_i)^2$, themselves. This Jacobian is evaluated on the solution of the four equations $(p^*-K_i)^2 = 0$. Thus, it appears that our strategy to isolate the integral coefficient $i s_1 s_2 A_{5;\langle 12 \rangle}^{\textrm{\scriptsize{MHV}}}/2$ was almost successful. The only problem is that the definition of $\G_p$ does not specify a unique contour of integration; the system $(p^*-K_i)^2 = 0$ has two solutions, $p^{*\,(1)}$ and $p^{*\,(2)}$. The question of which contour is the ``right'' one to use is an important technical one that we return to later; it will be much easier to address this point once we've explained how to interpret the left-hand side of eq. (\ref{5ptapprox}) evaluated on $\G_p$.

The optical theorem for Feynman diagrams is usually presented in text books (see {\it e.g.}~\cite{PeskinSchroeder}) for individual Feynman diagrams. In its simplest incarnation, it relates the product of two tree-level diagrams integrated over an appropriate Lorentz invariant phase-space (of the external lines of the two tree amplitudes that depend on $p$) to the imaginary part of the one-loop Feynman diagram built by gluing together the two tree diagrams in the kinematic channel under consideration (called the channel ``being cut''). See Figure \ref{ordunitarity} for a cartoon depicting an $s$-channel cut of a diagram that would enter into the calculation of the one-loop virtual corrections to Bhabha scattering.\footnote{Technically speaking, a cut can be implemented by replacing of a set of propagators by delta functions that force the momenta carried by the replaced propagators on-shell.} 

\FIGURE{
\resizebox{0.9\textwidth}{!}{\includegraphics{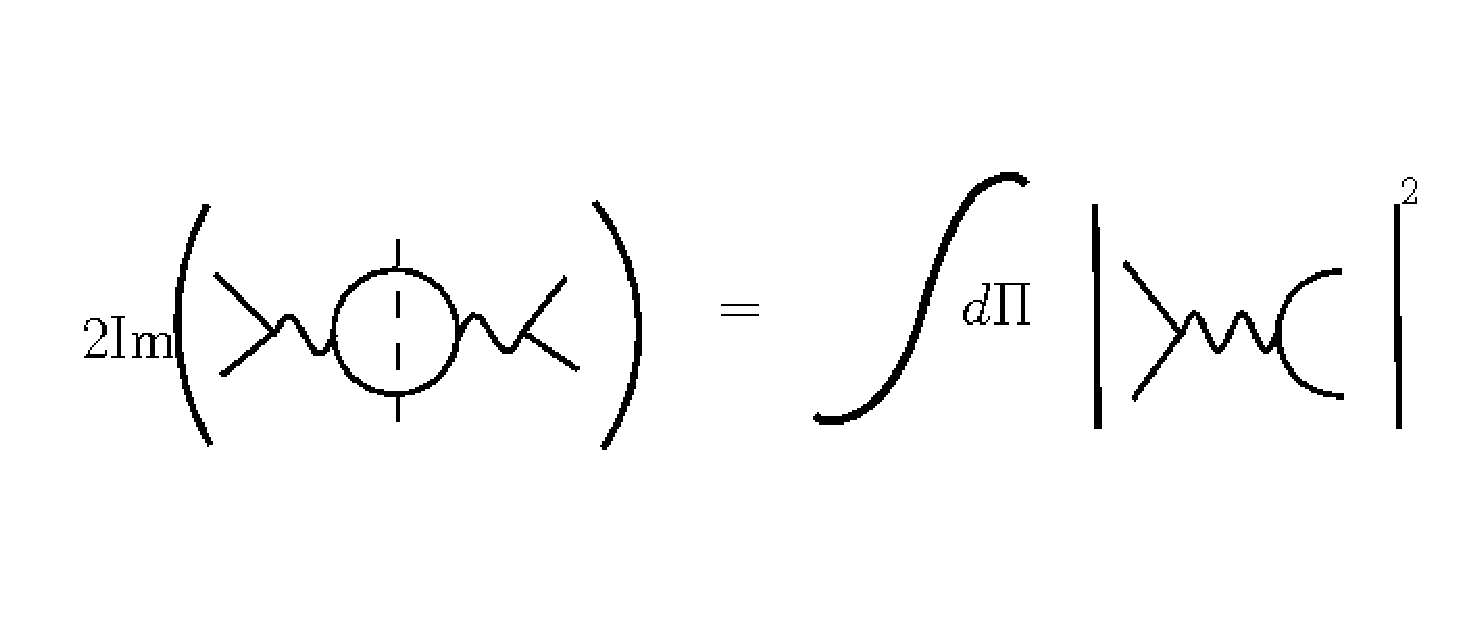}}
\vbox{\vskip 1 in}
\caption{An example of the optical theorem for Feynman diagrams for an $s$-channel vacuum polarization graph that contributes to the one-loop amplitude for Bhabha scattering. On the left-hand side we have twice the imaginary part of a one-loop Feynman diagram and on the right-hand side we have a product of two tree-level Feynman diagrams integrated over the Lorentz invariant phase-space of the cut propagators.}
\label{ordunitarity}}

From this point of view, the optical theorem is only useful as a cross-check on individual Feynman diagrams, not as a calculational tool. However, the generalized version of the optical theorem is much more powerful because it relates {\it on-shell} tree amplitudes integrated over an appropriate Lorentz invariant phase-space to the imaginary parts of pieces of complete one-loop amplitudes. Furthermore, one might hope that $\Nsym$ is a theory where the integral basis is such that the analytic structure of each basis element can be deduced from its imaginary part without any ambiguities. Indeed, as was shown in~\cite{BDDKMHV}, $\Nsym$ is in the class of so-called cut-constructible theories. One might guess that this is the case by looking at the explicit formulae of eqs. (\ref{box0})-(\ref{box4}). 

To better understand this discussion, we introduce our variant of the optical theorem and apply it to the $I_4^{(5)}$ topology of $A_{1}^{1-{\rm loop}}(k_1^{1234},k_2^{1234},k_3,k_4,k_5)$. Consider the complete set of Feynman diagrams that have gluons running in the loop\footnote{In the particular case of $A_{1}^{1-{\rm loop}}(k_1^{1234},k_2^{1234},k_3,k_4,k_5)$, it turns out that, once internal lines are put on-shell, the $SU(4)_R$ symmetry does not allow non-zero contributions from fermions or scalars. This is a general phenomenon that will occur in pure-glue one-loop $\Nsym$ amplitudes whenever multiple external lines are attached to a corner of a generalized unitarity diagram and all of these lines are positive (or negative) helicity gluons.} and the topology of $I_4^{(5)}$. These diagrams are drawn in Figure \ref{BFGdiags}. It is worth pointing out that the set of Feynman diagrams in Figure \ref{BFGdiags} are only valid if we work in background field gauge. It was first understood in~\cite{BDDKMHV} that there are enormous practical advantages to working in background field gauge when one is faced with the task of computing a one-loop $\Nsym$ scattering amplitude in which all the external states are gluons. The reason for this is that the use of background field gauge reduces the degree of the loop momentum polynomial in the numerator of each Feynman integrand. In the calculation of an $n$-point scattering amplitude, one expects a loop momentum polynomial of degree $n-4$ in the numerator of an $n$-point contribution to the amplitude. For us, this means that the use of background field gauge will allow us to express $A_{1}^{1-{\rm loop}}(k_1^{1234},k_2^{1234},k_3,k_4,k_5)$ in such a way that there will be at most one power of the loop momentum in any given term in the numerator of the Feynman integral with five propagator denominators (pentagon topology) and no powers of the loop momentum in the numerators of the five daughter integrals (the notion of daughter integral is explained in \ref{muint}) with four propagator denominators (box topology). 

\FIGURE{
\resizebox{0.75\textwidth}{!}{\includegraphics{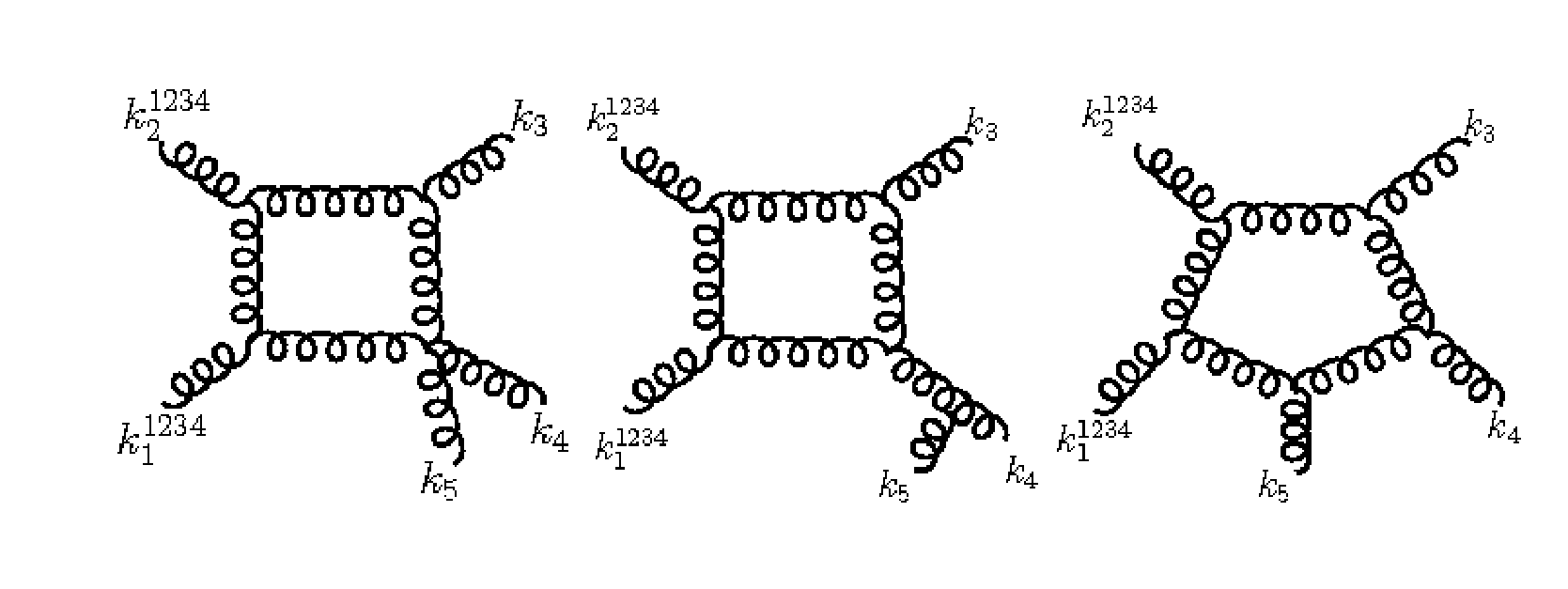}}
\vbox{\vskip 1 in}
\caption{Background field gauge Feynman diagrams that contribute to the $I_4^{(5)}$ topology of $A_{1}^{1-{\rm loop}}(k_1^{1234},k_2^{1234},k_3,k_4,k_5)$.}
\label{BFGdiags}}

First let us think about the pentagon diagram. It is possible to exploit that fact that there is at most one power of the loop momentum in the numerator of the pentagon diagram to reduce the pentagon diagram to a sum over boxes. We can see this directly from results derived in~\cite{oneloopdimreg} truncated to $\Ord(\e^0)$. If we Feynman parametrize the loop momentum in the pentagon diagram using Feynman parameters $x_i$, the following two formulas can be used to write the pentagon diagram as a sum of scalar box integrals:
\bea
I_5 &=& {1\over 2} \sum_{j=1}^5 C_j I_{4}^{(j)} + \Ord(\e) \\
I_5[x_i] &=& {1\over 2}\sum_{j=1}^5 S_{i j}^{-1} I_{4}^{(j)} + \Ord(\e)
\eea
In \ref{muint} we derive these formulae and carefully define the functions $C_j$ and $S_{i j}$. The important point is that there is a piece of the pentagon diagram that has the same topology as the other two diagrams drawn in Figure \ref{BFGdiags} and it is this piece that should be grouped together with those diagrams.

At last, we are set up to explain the principle of generalized unitarity. Suppose we cut through, say, propagators one and three in the two true box diagrams and the relevant piece of the reduced pentagon. The claim is that, if we add up all three pieces, the fact that they share the same topology (and we have added up all possible contributions) guarantees that the sum will be a product of two on-shell tree amplitudes, one with external momenta $-p$, $k_1$, $k_2$, and $p-k_1-k_2$ and one with external momenta $-p+k_1+k_2$, $k_3$, $k_4$, $k_5$ and $p$ , integrated over the appropriate Lorentz invariant phase-space. Since $\Nsym$ is cut-constructible, we can easily invert this process and deduce complete one-loop integrands of a particular topology by calculating two appropriate on-shell tree amplitudes and tacking on a couple of missing propagator denominators. In fact, this discussion motivates replacing Feynman diagrams completely in favor of what we'll call generalized unitarity diagrams. The generalized unitarity diagram (singular) associated to the discussion of this paragraph is drawn in Figure \ref{I45GUdiag}.

\FIGURE{
\resizebox{0.65\textwidth}{!}{\includegraphics{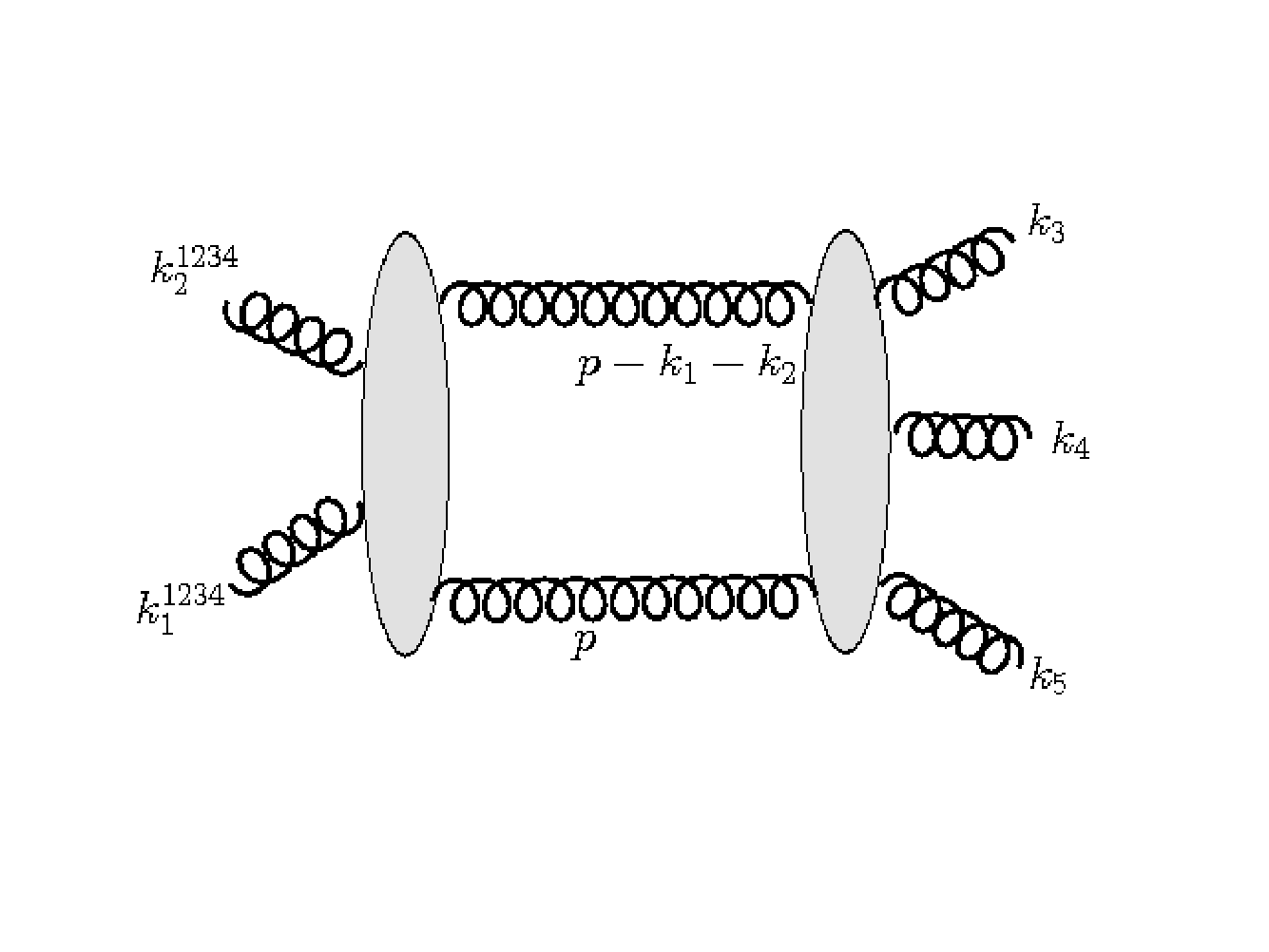}}
\vbox{\vskip 1 in}
\caption{The $s_1$-channel generalized unitarity diagram can be used to reconstruct the coefficient of the $I_4^{(5)}$ topology of $A_{1}^{1-{\rm loop}}(k_1^{1234},k_2^{1234},k_3,k_4,k_5)$ but the information about $I_4^{(5)}$ will have to be disentangled from that pertaining to all the other boxes detected by the cut (all except $I_1^{(1)}$ and $I_1^{(3)}$). The blobs on either side of the lines carrying momenta $p$ and $p - k_1 - k_2$ (the propagators to be cut in evaluating the diagram) represent on-shell tree amplitudes.}
\label{I45GUdiag}}

Using generalized unitarity with double cuts is already very powerful, but we want to do even better. We would like to use the principle of generalized unitarity in a way that meshes well with our earlier discussion of multidimensional contour integrals and the right-hand side of eq. (\ref{5ptapprox}). What this amounts to is cutting all the propagators in all contributions with a particular box topology. For the five-point example of this subsection, we would end up with the diagram of Figure \ref{I45maxGU}.
\FIGURE{
\resizebox{0.75\textwidth}{!}{\includegraphics{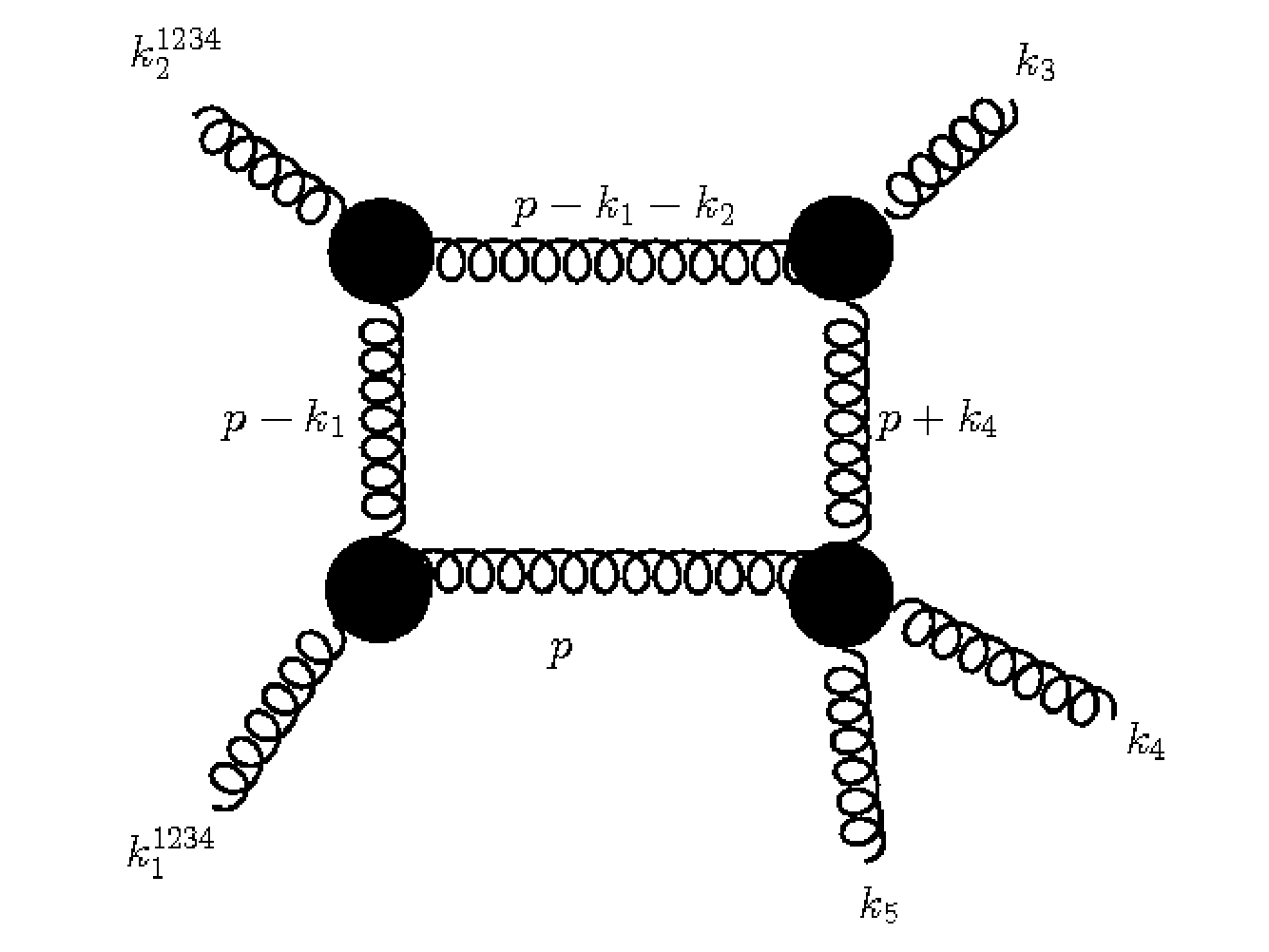}}
\vbox{\vskip 1 in}
\caption{The maximal generalized unitarity diagram that contributes to the $I_4^{(5)}$ topology of $A_{1}^{1-{\rm loop}}(k_1^{1234},k_2^{1234},k_3,k_4,k_5)$. As before, each blob represents an on-shell tree amplitude and each line in the diagram is to be cut in the evaluation of the diagram. This time, however, the cuts {\it only} detect $I_4^{(5)}$.}
\label{I45maxGU}}
In other words, we can reconstruct the integrand of the piece of the one-loop  amplitude $A_{1}^{1-{\rm loop}}(k_1^{1234},k_2^{1234},k_3,k_4,k_5)$ (with gluons running in the loop) and topology $I_4^{(5)}$ by first multiplying four appropriate tree amplitudes together (three three-point amplitudes and one four-point amplitude) and then tacking on the missing propagator denominators:
\begin{changemargin}{-.6 in}{0 in}
\bea
&&A_{1}^{1-{\rm loop}}\left(k_1^{1234},k_2^{1234},k_3,k_4,k_5\right)\Big|_{I_4^{(5)}} = \int {d^{4} p \over (2\pi i)^{4}} A^{tree}\left(-p, k_1^{1234}, (p-k_1)^{1234} \right)
\el A^{tree}\left((-p+k_1), k_2^{1234}, p-k_1-k_2\right) A^{tree}\left((-p+k_1+k_2)^{1234}, k_3, p+k_4+k_5\right) \times
\el \times A^{tree}\left((-p-k_4-k_5)^{1234}, k_4, k_5, p^{1234}\right){1\over p^2 (p-k_1)^2 (p-k_1-k_2)^2 (p+k_4+k_5)^2}+
\el
\el
\el \int {d^{4} p \over (2\pi i)^{4}} A^{tree}\left(-p, k_1^{1234}, p-k_1 \right)A^{tree}\left((-p+k_1)^{1234}, k_2^{1234}, p-k_1-k_2\right)
\el A^{tree}\left((-p+k_1+k_2)^{1234}, k_3, p+k_4+k_5\right) A^{tree}\left((-p-k_4-k_5)^{1234}, k_4, k_5, p^{1234}\right)\times
\el\times{1\over p^2 (p-k_1)^2 (p-k_1-k_2)^2 (p+k_4+k_5)^2} \nn
\label{5pt45piece}
\eea
\end{changemargin}
In this example, there are just two consistent assignments of the internal $SU(4)_R$ indices that would not obviously give a vanishing result by virtue of the SUSY Ward identities $\a(k_1,k_2,\cdots,k_n) = 0$ and $\a(k_1^{1234},k_2,\cdots,k_n) = 0$ or their parity conjugates\footnote{Parity acts on scattering amplitudes as complex conjugation (interchange of angle and square brackets in spinor products).}. In general, one should sum over all the internal configurations allowed by the $SU(4)_R$ symmetry.\footnote{In component language, what we mean is that, usually, the four Majorana fermions running in the loop and the three complex scalars running in the loop would give non-vanishing contributions.} There is, however, a more subtle constraint which forces the first term in eq. (\ref{5pt45piece}) to zero. It turns out that when there are configurations with two on-shell three-point amplitudes next to each other, the adjacent three-point amplitudes must have different $SU(4)_R$ index structures or they vanish~\cite{BDKrev}. Therefore, the product 
\bea
&&A^{tree}\left(-p, k_1^{1234}, (p-k_1)^{1234} \right)A^{tree}\left((-p+k_1), k_2^{1234}, p-k_1-k_2\right) 
\el A^{tree}\left((-p+k_1+k_2)^{1234}, k_3, p+k_4+k_5\right) A^{tree}\left((-p-k_4-k_5)^{1234}, k_4, k_5, p^{1234}\right) \nonumber
\eea
 must vanish because $A^{tree}\left((-p+k_1), k_2^{1234}, p-k_1-k_2\right)$ and $A^{tree}\left((-p+k_1+k_2)^{1234}, k_3, p+k_4+k_5\right)$ have the same $SU(4)_R$ index structure.

All the hard work is now done and we can evaluate the reconstructed one-loop integrand of eq. (\ref{5pt45piece}) as a contour integral (with $\G_p$ as the contour) in exactly the same way that we evaluated the right-hand side of eq. (\ref{5ptaprxrhs}). We find that eq. (\ref{5pt45piece}) gives
\bea
&&A^{tree}\left(-p, k_1^{1234}, p-k_1 \right)A^{tree}\left((-p+k_1)^{1234}, k_2^{1234}, p-k_1-k_2\right)\times
\el \times A^{tree}\left((-p+k_1+k_2)^{1234}, k_3, p+k_4+k_5\right) A^{tree}\left((-p-k_4-k_5)^{1234}, k_4, k_5, p^{1234}\right)\times
\el\times {\rm det}^{-1}\left({\partial\over \partial p_\mu} ~(p-K_i)^2\right)\Bigg|_{p_\mu = \,p_\mu^*} \,{\rm ,}
\label{lhsLS}
\eea
where $p^{*\,(i)}$ represents either of the two solutions of $(p^*-K_i)^2 = 0$. Finally, we can put together the left- and right-hand sides of (\ref{5ptapprox}) evaluated on $\G_p$:
\bea
&&A^{tree}\left(-p, k_1^{1234}, p-k_1 \right)A^{tree}\left((-p+k_1)^{1234}, k_2^{1234}, p-k_1-k_2\right)\times
\el  \times A^{tree}\left((-p+k_1+k_2)^{1234}, k_3, p+k_4+k_5\right) A^{tree}\left((-p-k_4-k_5)^{1234}, k_4, k_5, p^{1234}\right)\Bigg|_{p_\mu = p_\mu^*} 
\elale i {A_{5;\langle 12 \rangle}^{\textrm{\scriptsize{MHV}}} \over 2} s_{1}s_{2}\,{\rm ,}
\label{LSsketch}
\eea
where we have cancelled a Jacobian factor from both sides. The only loose end to tie up is what to do about the fact that there are actual two distinct contours specified by $\G_p$. There is no natural reason to chose one solution of $(p^*-K_i)^2 = 0$ over the other and, to make matters worse, it appears that the equations defined by (\ref{LSsketch}) are not consistent. On one solution to $(p^*-K_i)^2 = 0$, $p^{*\,(1)} = k_1^\mu + 1/2\,\spa2.3  \spab1.{\gamma^\mu}.2/\spa1.3 $, the product of trees on the left-hand side of (\ref{LSsketch}) is $i s_1 s_2 A_{5;\langle 12 \rangle}^{\textrm{\scriptsize{MHV}}}$ and on the other solution, $p^{*\,(2)} = k_1^\mu + 1/2\,\spb2.3  \spab2.{\gamma^\mu}.1/\spb1.3$, the product of trees is 0. To summarize, our two equations read
\bea
0 &=& i {s_1 s_2\over 2} A_{5;\langle 12 \rangle}^{\textrm{\scriptsize{MHV}}}
\label{LS51}\\
i s_1 s_2 A_{5;\langle 12 \rangle}^{\textrm{\scriptsize{MHV}}} &=& i {s_1 s_2\over 2} A_{5;\langle 12 \rangle}^{\textrm{\scriptsize{MHV}}}
\label{LS52}
\eea
The reason for the lack of consistency is that we truncated our basis at $\Ord(\e^0)$. As we will see in the next section, adding a massless pentagon integral to the basis of Feynman integrals fixes this apparent inconsistency in a very nice way. For now, we appeal to symmetry to fix our problem~\cite{SharpLS}. Our two solutions, $p^{*\,(1)}$ and $p^{*\,(2)}$, are related by parity symmetry (complex conjugation of the spinor products). We want our final answer to be parity invariant and one way to ensure this is to simply add eqs. (\ref{LS51}) and (\ref{LS52}). If we do this we find that the left-hand side is equal to the right-hand side, which implies that this procedure works, at least for $A_{1}^{1-{\rm loop}}(k_1^{1234},k_2^{1234},k_3,k_4,k_5)$. It turns out that this ad hoc prescription will work for general one-loop amplitudes if one doesn't care about the higher order in $\e$ pieces of the amplitude. For general multi-loop amplitudes, however, such sloppy analysis is simply not sufficient. We will also need something better for our all-orders-in-$\e$ computations at the one-loop level. It is to this that we turn in the next subsection. 
\subsection{Generalized Unitarity in $D$ Dimensions}
\label{GUD}
The great thing about generalized unitarity is that it works in very general situations (the unitarity of the S-matrix is a consequence of probability conservation in quantum mechanics). In particular, generalized unitarity is compatible with dimensional regularization because dimensional regularization preserves unitarity. As was shown shortly after the seminal papers on the generalized unitarity technique were published, there is no inherent restriction to $\Ord(\e^0)$; if desired, one can compute amplitudes to all orders in $\e$ by working a little harder.~\cite{BernMorgan} We explain how this works in the context of the example of the last subsection and, in the process, explain how one needs to modify the basis of planar one-loop box integrals used so far if one is interested in computing planar $\Nsym$ amplitudes to all orders in the dimensional regularization parameter.

At the five-point level, there is an obvious candidate integral that one could try adding to the basis of scalar boxes: the massless pentagon. Let us try to prove that there is a non-vanishing pentagon contribution to $A^{1-{\rm loop}}_1(k_1^{1234},k_2^{1234},k_3,k_4,k_5)$ that we ignored in the last subsection. Our experience in \ref{GU4} has taught us that it is a bad idea to try and think of the pentagon integrals in the problem as reductive box contributions from the start. Rather, we should make the most general ansatz of scalar basis integrals that makes sense and let the amplitude decide how it wants to be written. To this end, we perform the same generalized unitarity analysis on the left-hand side of (\ref{5ptapprox}) that we did in the last subsection but this time make an ansatz $\sum_i A_i \,I_4^{(i)}+B\,I_5$ for the right-hand side. As before, we can get a non-zero contribution from $A_4 I_4^{(5)}$ but now we will also have a non-zero contribution from $B\,I_5$:
\bea
&&B \int_{\G_p} {d^4p\over (2 \pi i)^4} {1\over p^2 (p - k_1)^2 (p - k_1 - k_2)^2 (p + k_4 + k_5)^2 (p + k_5)^2 } 
\elale B\,{1\over (p + k_5)^2} {\rm det}^{-1}\left({\partial\over \partial p_\mu} ~(p-K_i)^2\right)\Bigg|_{p_\mu = \,p_\mu^*} \,{\rm .}
\eea
This makes all the difference. Instead of eqs. (\ref{LS51}) and (\ref{LS52}) we now have the system 
\bea
0 &=& A_4 + B\,{1\over (p^{*\,(1)} + k_5)^2}
\label{LS5D1}\\
i s_1 s_2 A_{5;\langle 12 \rangle}^{\textrm{\scriptsize{MHV}}} &=& A_4 + B\,{1\over (p^{*\,(2)} + k_5)^2}
\label{LS5D2}
\eea
which is consistent and solvable. We will follow~\cite{SharpLS} and refer to this technique as the leading singularity method (the leading singularities being the left-hand sides of the above equations). We find
\begin{equation}
A_4 = -i A_{5;\langle 12 \rangle}^{\textrm{\scriptsize{MHV}}}
\frac{s_{1}s_{2}\tilde\beta_5}{\beta_5-\tilde\beta_5}, \qquad B =
i A_{5;\langle 12 \rangle}^{\textrm{\scriptsize{MHV}}}\frac{s_{5}s_{1}s_{2}}{\beta_5-\tilde\beta_5} \,{\rm ,}
\label{LS5ptfin}
\end{equation}
where
\begin{equation}
\beta_{5} = \left(1+ \frac{\langle 2 3\rangle [2 5] }{\langle
1 3\rangle [1 5]} \right)^{-1}, \qquad \tilde\beta_{5} =
\left(1+ \frac{\langle 2 5\rangle [2 3] }{\langle 1 5\rangle
[1 3]} \right)^{-1}\,{\rm .}
\end{equation}
The formula for $A_4$ bears no resemblance to $i s_1 s_2 A_{5;\langle 12 \rangle}^{\textrm{\scriptsize{MHV}}}/2$ and, indeed, one can check numerically (using {\it e.g.} S@M~\cite{S@M}) that they are not equal. At first sight, it appears that the leading singularity method fails to reproduce the known result. In actuality there is no contradiction because, secretly, the result obtained in the previous subsection was expressed in terms of a different basis with $4 - 2\e$ dimensional boxes and $6 - 2\e$ dimensional pentagon integrals as opposed to the $4-2\e$ dimensional box and pentagon integrals we used above. The connection between these two bases is the reduction formula derived in \ref{muint} using traditional techniques:\footnote{Note that eq. (\ref{PtoDCB}) employs the original conventions of definition (\ref{IntDef}).} 
\be
I_5^{D=4-2\e} = {1\over 2}\bigg[\sum_{j=1}^5 C_j I_{4}^{(j),\,D=4-2\e} + 2\e C_0 I_5^{D=6-2 \e} \bigg]\,{\rm .}
\label{PtoDCB}
\ee
The answer looks much more compact when expressed in this basis (employed in the last subsection); the higher order in $\e$ terms are more cleanly separated from those that are present through $\Ord(\e^0)$. This is related to the fact that there will always be an explicit $\e$ out front of $I_5^{D=6-2 \e}$ and $I_5^{D=6-2 \e}$ is both UV and IR finite~\cite{oneloopdimreg}. In this paper we will refer to the basis with all elements expanded about $D = 4$ as the geometric basis and the basis with $I_5^{D=6-2 \e}$ pentagons as the dual conformal basis (this notation will be motivated in Section \ref{WL/MHV}). Simplifying (\ref{LS5ptfin}) after projecting the geometric basis onto the dual conformal basis using (\ref{PtoDCB}), we find that 
\be
\tilde{A}_4 = -{s_1 s_2\over 2} A_{5;\langle 12 \rangle}^{\textrm{\scriptsize{MHV}}}
\label{A45ptDCB}
\ee
as before and
\be
\tilde{B} = \e ~\pol(1,2,3,4) A_{5;\langle 12 \rangle}^{\textrm{\scriptsize{MHV}}}\,{\rm ,}
\label{B5ptDCB}
\ee
where we have made the useful definition
\be
 \pol(i,j,m,n) \equiv 4 i \pol_{\mu \nu \rho \sigma} k_i^\mu k_j^\nu k_m^\rho k_n^\sigma = \spb{i}.j \spa{j}.m \spb{m}.n \spa{n}.i - \spa{i}.j \spb{j}.m \spa{m}.n \spb{n}.i \, {\rm .}
 \label{epstensor}
\ee
In above eq. (\ref{A45ptDCB}) is the coefficient of $I_4^{(5),\,D = 4-2\e}$ and eq. (\ref{B5ptDCB}) is the coefficient of $I_5^{D=6-2\e}$ in the conventions of definition (\ref{IntDef}). 

What we have learned is that, in the case of $A^{1-{\rm loop}}_1(k_1^{1234},k_2^{1234},k_3,k_4,k_5)$, we are able to learn all about the higher-order in $\e$ pieces of the amplitude without ever leaving four dimensions. It would be great if the leading singularity method gave us all of the pentagon coefficients for arbitrary $n$ but, unfortunately, life is not so simple. In fact, as we shall see in the next subsection, one needs to develop more machinery to calculate the pentagon integrals already at the six-point level. In a nutshell, what we need to do is to further develop the $D$ dimensional unitarity technique of Bern and Morgan~\cite{BernMorgan} to reconstruct one-loop integrands in $\Nsym$ without dropping any higher order in $\e$ pieces.

We now review the Bern-Morgan approach to $D$-dimensional integrand reconstruction to prepare the reader for the next section where we discuss simple extensions of their results. As an illustration, we  consider the amplitude $A^{1-{\rm loop}}_{1;\,\mathcal{N}=0}(k_1,k_2,k_3,k_4)$ in pure Yang-Mills theory. Following~\cite{BernMorgan}, we remind the reader of the second form for $I_4^{D = 4 - 2\e}$ where we split up the integral over the loop momentum into four dimensional and $-2\e$ dimensional pieces:
\be
I_4^{D=4-2 \e} = -i (4\pi)^{2-\e} \int 
    {d^{4} p \over (2\pi)^{4} }
    {d^{-2\e} \mu \over (2\pi)^{-2\e} }
  { 1 \over (p^2 - \mu^2) ((p-k_1)^2 - \mu^2) ((p-k_1-k_2)^2 - \mu^2)(
    (p+k_4)^2 - \mu^2) } \,{\rm .}\nn
\ee
If we consider an $s$-channel cut of the above zero mass box integral, we find the on-shell conditions
\be
p^2 = \mu^2 \qquad (p-k_1-k_2)^2 = \mu^2\,{\rm .}
\ee
It follows that, to reconstruct the complete one-loop integrand in $D$ dimensions using the principle of generalized unitarity, one should simply imagine that the lines of the tree amplitudes on either side of the unitarity cut(s) (external lines of the trees that have $p$-dependent momenta) have a mass $\mu$. Actually, the procedure of gluing trees together to form loops is a little more complicated in our approach because we do not have in hand an analog of the spinor helicity framework in $-2\e$ dimensions\footnote{It is possible that, with a bit of inspiration, we might be able to profitably make use of some combination of the formalisms worked out in~\cite{Dennen,CheungC,BCDHI}.}. Consequently, the whole process is more closely related to traditional perturbation theory. In particular, summing over internal degrees of freedom inside the loop being reconstructed is much more labor intensive than it is in four dimensions. One trick to try and avoid tedious algebra, which works better in some situations than in others, is to perform a supersymmetric decomposition of the amplitude. For example, if we rewrite a loop of gluons in the following way:
$$A_g = (A_g + 4 A_f + 3 A_s) - 4 (A_f + A_s) + A_s$$
We see that the contribution from a loop of gluons ({\it i.e.} pure Yang-Mills theory) can be derived by summing the answer in $\Nsym$ and the contribution from a loop of complex scalars and then subtracting off the contribution from four $\mathcal{N}=1$ chiral multiplets. For the present application this works beautifully because the first two terms on the right-hand side of the above equation are protected by supersymmetry and vanish (see Appendix \ref{SWI}). It follows that
\be
A^{1-{\rm loop}}_{1;\,\mathcal{N}=0}(k_1,k_2,k_3,k_4) = A^{1-{\rm loop}}_{1;\,{\rm scalar}}(k_1,k_2,k_3,k_4)
\label{all+susydecomp}
\ee
and, in this particular case, we can avoid some numerator algebra by calculating $A^{1-{\rm loop}}_{1;\,{\rm scalar}}(k_1,k_2,k_3,k_4)$ instead of $A^{1-{\rm loop}}_{1;\,\mathcal{N}=0}(k_1,k_2,k_3,k_4)$. 

Generalized unitarity applied to $A^{1-{\rm loop}}_{1;\,{\rm scalar}}(k_1,k_2,k_3,k_4)$ gives\footnote{In what follows, we will very often be interested in amplitudes where some of the external states have definite helicity and some should be thought of as having any of the possible physical polarizations. We label external states with indeterminate polarization as $(q)_{x}$ where $q$ is the momentum carried by the external particle and $x$ denotes the particle type ($s$ or $\bar{s}$ for scalar states, $f$ or $\bar{f}$ for fermion states, and $g$ for gluon states).}
\bea
&&{1\over (4 \pi)^{2-\e}} A^{1-{\rm loop}}_{1;\,{\rm scalar}}(k_1,k_2,k_3,k_4) = \int 
    {d^{4} p \over (2\pi)^{4} }
    {d^{-2\e} \mu \over (2\pi)^{-2\e} }
  \bigg({ i \over p^2 - \mu^2} A^{tree}_{\mu^2}\left((-p)_{s},k_1,k_2,(p-k_1-k_2)_{\bar{s}}\right)
\el{i\over(p-k_1-k_2)^2 - \mu^2} A^{tree}_{\mu^2}\left((-p+k_1+k_2)_s,k_3,k_4,p_{\bar{s}}\right)
  \el+ { i \over p^2 - \mu^2} A^{tree}_{\mu^2}\left((-p)_{\bar{s}},k_1,k_2,(p-k_1-k_2)_s\right) {i\over(p-k_1-k_2)^2 - \mu^2} A^{tree}_{\mu^2}\left((-p+k_1+k_2)_{\bar{s}},k_3,k_4,p_s\right)\bigg) \,{\rm .}\nn
\eea
The massive scalar amplitudes $A^{tree}_{\mu^2}\left((-p)_s,k_1,k_2,(p-k_1-k_2)_{\bar{s}}\right)$ and $A^{tree}_{\mu^2}\left((-p)_{\bar{s}},k_1,k_2,(p-k_1-k_2)_s\right)$ are equal, as are $A^{tree}_{\mu^2}\left((-p+k_1+k_2)_s,k_3,k_4,p_{\bar{s}}\right)$ and $A^{tree}_{\mu^2}\left((-p+k_1+k_2)_{\bar{s}},k_3,k_4,p_s\right)$. Using
\bea
A^{tree}_{\mu^2}\left((-p)_s,k_1,k_2,(p-k_1-k_2)_{\bar{s}}\right) &=& {i \mu^2 \spb1.2 \over \spa1.2 ((p-k_1)^2-\mu^2)}~~~~{\rm and} \\
A^{tree}_{\mu^2}\left((-p+k_1+k_2)_s,k_3,k_4,p_{\bar{s}}\right) &=& {i \mu^2 \spb3.4 \over \spa3.4 ((p+k_4)^2-\mu^2)}\,{\rm ,}
\eea
which can be derived from Feynman diagrams, we find 
\bea
&&{1\over (4 \pi)^{2-\e}} A^{1-{\rm loop}}_{1;\,{\rm scalar}}(k_1,k_2,k_3,k_4) = {1\over (4 \pi)^{2-\e}} {2 \spb1.2 \spb3.4\over \spa1.2 \spa3.4}\times
\el \times \int 
    {d^{4} p \over (2\pi)^{4} }
    {d^{-2\e} \mu \over (2\pi)^{-2\e} }
  { \mu^4 \over (p^2 - \mu^2)((p-k_1)^2-\mu^2)((p-k_1-k_2)^2 - \mu^2)((p+k_4)^2-\mu^2)} \nn
&&= {1\over (4 \pi)^{2-\e}} {2 i \spb1.2 \spb3.4\over \spa1.2 \spa3.4} I_4^{D = 4 - 2\e}[\mu^4] \\
A^{1-{\rm loop}}_{1;\,\mathcal{N}=0}(k_1,k_2,k_3,k_4) &&= A^{1-{\rm loop}}_{1;\,{\rm scalar}}(k_1,k_2,k_3,k_4) = {2 i \spb1.2 \spb3.4\over \spa1.2 \spa3.4} I_4^{D = 4 - 2\e}[\mu^4]\,{\rm .}
\eea
A basis integral with some power of $\mu^2$ inserted in the numerator is usually referred to as a $\mu$-integral and such terms will play a central role in this work. It is often convenient to rewrite $\mu$-integrals in terms of dimensionally shifted integrals. This is easily accomplished by manipulating the $-2\e$ dimensional part of the integration measure in eq. (\ref{IntDef}). Written out, the $-2\e$ dimensional integral is 
\be
\int {d^{-2\e} \mu \over (2\pi)^{-2\e} } f(\mu^2) = \int{d \Omega_{-2\e}\over (2\pi)^{-2\e}}\int_0^\infty d\mu \mu^{-2\e-1}f(\mu^2) = {1\over 2} \int {d \Omega_{-2\e}\over (2\pi)^{-2\e}}\int_0^\infty d\mu^2 (\mu^2)^{-\e-1}f(\mu^2)\, {\rm ,}
\ee
where, as usual,
\be
\int {d \Omega_{-2\e}} = {2 \pi^{-\e} \over \G(-\e)}\,{\rm .}
\ee
Now, if we replace $f(\mu^2)$ with $\mu^{2 r}$ we can absorb the extract factors of $\mu^2$ into the integration measure:
\begin{changemargin}{-.6 in}{0 in}
\be
\int {d^{-2\e} \mu \over (2\pi)^{-2\e} } \mu^{2 r}f(\mu^2) = {(2\pi)^{2 r}\int d\Omega_{-2\e}\over \int d\Omega_{2 r -2\e}} \int {d^{2 r-2\e}\mu \over (2 \pi)^{2 r-2 \e}} f(\mu^2) = -\e(1-\e)(2-\e)\cdots(r-1-\e)(4\pi)^r\int {d^{2 r-2\e}\mu \over (2 \pi)^{2 r-2 \e}} f(\mu^2)\,{\rm .}
\ee
\end{changemargin}
If $r$ is a natural number, this analysis leads to
\be
I_n^{D = 4 - 2\e}[\mu^{2 r}] = -\e(1-\e)(2-\e)\cdots(r-1-\e)I_n^{D = 2 r + 4 - 2\e}
\label{DSmu}
\ee
relating $\mu$-integrals and dimensionally-shifted integrals. Now, a very interesting phenomenon can occur, which we illustrate by applying eq. (\ref{DSmu}) to our result for $A^{1-{\rm loop}}_{1;\,\mathcal{N}=0}(k_1,k_2,k_3,k_4)$. We first rewrite the answer
\be
A^{1-{\rm loop}}_{1;\,\mathcal{N}=0}(k_1,k_2,k_3,k_4) = {2 i \spb1.2 \spb3.4\over \spa1.2 \spa3.4} I_4^{D = 4 - 2\e}[\mu^4] = -{2 \e (1-\e)i \spb1.2 \spb3.4\over \spa1.2 \spa3.4} I_4^{D = 8 - 2\e}
\ee
and then Feynman parametrize it:
\be
A^{1-{\rm loop}}_{1;\,\mathcal{N}=0}(k_1,k_2,k_3,k_4) = -{2 \e (1-\e)i \spb1.2 \spb3.4\over \spa1.2 \spa3.4} \G(\e)\int_0^1 dx\int_0^{1-x}dy\int_0^{1-x-y}dz {1\over \mathcal{D}(x,y,z)^\e}\,{\rm .}
\ee
Remarkably, the $\e$ expansion of the above starts at $\Ord(\e^0)$. Explicitly, we find 
\bea
A^{1-{\rm loop}}_{1;\,\mathcal{N}=0}(k_1,k_2,k_3,k_4) &=& -{2 i \spb1.2 \spb3.4\over \spa1.2 \spa3.4} \int_0^1 dx\int_0^{1-x}dy\int_0^{1-x-y}dz + \Ord(\e)
\elale -{i \spb1.2 \spb3.4\over 3 \spa1.2 \spa3.4}+ \Ord(\e)\,{\rm .}
\eea
At first sight, this result might seem rather puzzling since, without the $\mu^4$ in the numerator, the integral $I_4^{D = 4 -2 \e}$ is UV finite and IR divergent. What has happened is that, in shifting to $D = 8 - 2\e$, we have induced a UV divergence (the integral now has the same number of powers of the loop momenta in the measure of integration as it has in the denominator) and the IR divergences effectively got regulated by the $\mu^2$ factors in the propagator denominators. The explicit $\e$ in the numerator coming from eq. (\ref{DSmu}) is canceling the induced UV pole, which is why the $\e$ expansion of $A^{1-{\rm loop}}_{1;\,\mathcal{N}=0}(k_1,k_2,k_3,k_4)$ starts at $\Ord(\e^0)$.

Although we have been focusing on scalars running in the loop we could equally well have performed the above analysis for a loop of fermions with one obvious additional complication: the need to sum over internal spin states in a Lorentz covariant way. Typical tree amplitudes with a pair of massive fermions will be built out of a string beginning with $\bar{u}^{\pm}(p)$ and ending with $u^\pm(p)$. In order to fuse together two such tree amplitudes across a unitarity cut, we simply use the spin sum identity 
\be
\sum_{s} u^s(p)\bar{u}^s(p) = \slashed{p} + \mu
\label{fermspsum}
\ee
heavily used in traditional perturbation theory~\cite{PeskinSchroeder}. In \ref{effgcomp}, we treat a gluon running in the loop as well. Due to the fact there is no straightforward massive counterpart (with two spin states) to the massless gluon, treating an internal gluon line requires a little more thinking. In the last few years, $D$-dimensional unitarity has been systematized by several different groups~\cite{Rocket1,Rocket2,BadgerGUD,Ossola}.

Also, we wish to remark that there is no reason for us to restrict ourselves to double cuts; as we shall see in the next section, we can profit enormously by using quintuple cuts in $D$ dimensions to determine individual pentagon coefficients one at a time. The idea is conceptually similar to what we did with quadruple cuts and box coefficients in \ref{GU4}, though it is a bit more complicated. It turns out that the leading singularity method supplemented by $D$ dimensional quintuple cuts allows one to efficiently calculate all-orders-in-$\e$ one-loop $\Nsym$ amplitudes.
\section{Efficient Computation and New Results For One-Loop $\Nsym$ Gluon Amplitudes Calculated To All Orders in $\e$}
\label{gluoncomp}
\subsection{Efficient Computation Via $D$ Dimensional Generalized Unitarity}
\label{effgcomp}
In order to harness the power of $D$-dimensional unitarity for the application at hand, we have to extend the results of Bern and Morgan to treat cut internal gluon lines. To be clear, many other authors have thought about extending the Bern-Morgan approach to integrand reconstruction (see {\it e.g.}~\cite{Rocket1,Rocket2,BadgerGUD,Ossola}). All of them either focus on a getting numerical results or isolate terms that would be missed by four dimensional generalized unitarity. There are obviously many applications where it makes sense to follow one of these strategies. In this paper, however, we have a different goal. We further develop the Bern-Morgan approach and show that it is a very efficient way to analytically reconstruct general one-loop integrands. In fact, we expect that our approach will mesh well with the spinor integration reduction technique of~\cite{ABFKM,BFY}, which is applicable to general field theory amplitudes at one-loop. Although these references analyzed a variety of processes, they started with integrands obtained by other means in all cases except that of a complex scalar running in the loop. A general strategy for the analytical reconstruction of one-loop integrands in $D$ dimensions was not discussed. In what follows we fill in this gap.

Now, to illustrate our approach to $D$-dimensional unitarity, we offer an alternative derivation of the massless pentagon coefficient of eq. (\ref{B5ptDCB}) associated to \\$A^{1-{\rm loop}}_1(k_1^{1234},k_2^{1234},k_3,k_4,k_5)$. All we really need to do right now is extend Bern-Morgan to the case of purely gluonic external states with a massless vector running in the loop. Later on we will also treat the case where some of the external gluons are replaced by fermions. It seems likely that so far most researchers have found it expedient to side-step the question of how to properly treat a gluon running in the loop by exploiting supersymmetry decompositions as was done in Subsection \ref{GUD}. We argue that it is no more difficult to calculate directly. 

We warm up by repeating the analysis of the last subsection, but for the pentagon coefficient of $A^{1-{\rm loop}}_1(k_1^{1234},k_2^{1234},k_3,k_4,k_5)$ using quintuple cuts. Using the massive scalar three-point vertices~\cite{BGKSmassive},
\bea
A^{tree}_{\mu^2}\left((-p)_s,k_1,(p-k_1)_{\bar{s}}\right) &&= -i \sqrt{2} p \cdot \pol^+(k_1)~~~~{\rm and}
\\ A^{tree}_{\mu^2}\left((-p)_s,k_1^{1234},(p-k_1)_{\bar{s}}\right) &&= -i \sqrt{2} p \cdot \pol^-(k_1){\rm ,}
\eea
and quintuple $D$ dimensional generalized unitarity cuts we can deduce the pentagon integral coefficient for the scalar loop contribution to the five-point MHV amplitude. In the above, the polarization vectors can be evaluated using eqs. (\ref{polvecs+}) and (\ref{polvecs-}) because we are implicitly using the four dimensional helicity scheme (see \ref{4DHS}) where the external polarization vectors are kept in four dimensions. The result of this calculation is
\begin{changemargin}{-.4 in}{0 in}
\bea
&&A_{1;\,{\rm scalar}}^{1-{\rm loop}}\left(k_1^{1234},k_2^{1234},k_3,k_4,k_5\right)\Big|_{I_5} = A^{tree}_{\mu^2}\left((-p_*)_s,k_1^{1234},(p_*-k_1)_{\bar{s}}\right)  \times
\el \times A^{tree}_{\mu^2}\left((-p_*+k_1)_s,k_2^{1234},(p_*-k_1-k_2)_{\bar{s}}\right) A^{tree}_{\mu^2}\left((-p_*+k_1+k_2)_s,k_3,(p_*+k_4+k_5)_{\bar{s}}\right)\times
\el \times A^{tree}_{\mu^2}\left((-p_*-k_4-k_5)_s,k_4,(p_*+k_5)_{\bar{s}}\right)A^{tree}_{\mu^2}\left((-p_*-k_5)_s,k_5,(p_*)_{\bar{s}}\right) \,{\rm ,}\nn
\label{s5ptpent}
\eea
\end{changemargin}
where $p_*^\nu$ solves the on-shell conditions:
\cmb{-.6 in}{0 in}
\bea
p_*^2 - \mu^2 = 0 \qquad (p_*-k_1)^2 - \mu^2 &=& 0 \qquad (p_*-k_1-k_2)^2 - \mu^2 = 0 \nonumber \\
(p_*+k_4+k_5)^2 - \mu^2 = 0 & \qquad & (p_*+k_5)^2 - \mu^2 = 0 \,{\rm .}
\eea
\cme
It turns out that, in this case, the solution is unique and is given by~\cite{BFY} expanding the four dimensional, massive loop momentum with respect to a basis $K_1$, $K_2$, $K_3$, and $K_4$ of four-vectors:
\be
p^\nu = L_1 K_1^\nu + L_2 K_2^\nu + L_3 K_3^\nu + L_4 K_4^\nu
\label{5ptos1}
\ee
and then solving a system of linear equations for the $L_i$ coefficients. It makes sense to choose the $K$'s to be the four-vectors in the problem; in the present example we set
\be
K_1 = k_1 + k_2 \qquad K_2 = k_1 \qquad K_3 = -k_4-k_5 \qquad K_4 = -k_5{\rm .}
\ee
Explicitly, we have
\be
\left(\begin{array}{c}L_1\\L_2\\L_3\\L_4\end{array}\right) = {1\over2}\left(\begin{array}{cccc}K_1^2 & K_1\cdot K_2 & K_1\cdot K_3 & K_1 \cdot K_4 \\K_2\cdot K_1 & K_2^2 & K_2\cdot K_3 & K_2 \cdot K_4\\K_3\cdot K_1 & K_3\cdot K_2 & K_3^2 & K_3 \cdot K_4\\K_4\cdot K_1 & K_4\cdot K_2 & K_4\cdot K_3 & K_4^2\end{array}\right)^{-1} \left(\begin{array}{c}K_1^2\\K_2^2\\K_3^2\\K_4^2\end{array}\right)\,{\rm .}
\label{5ptos2}
\ee

Now that we are warmed up, we are ready to try the quintuple cut of the fermion loop contribution. The only reason that the fermion loop contribution is more complicated is that we have to sum over internal fermion spin states using eq. (\ref{fermspsum}); the net result of the sum over internal states for the scalar loop contribution is just an overall factor of two. Although Bern and Morgan did not literally give their fermions a mass $\mu$, our procedure is easily deduced from the discussion in their paper~\cite{BernMorgan}.

To reconstruct the one-loop integrand, we need tree amplitudes with two massive fermions and a gluon:
\bea
A^{tree}_{\mu^2}\left(p_{\bar{f}},k_1,(-p-k_1)_f\right) &=& -{i \over \sqrt{2}}\bar{u}(p)\slashed{\pol}^+(k_1) u(p+k_1)\\
A^{tree}_{\mu^2}\left(p_{\bar{f}},k_1^{1234},(-p-k_1)_f\right) &=&  -{i \over \sqrt{2}}\bar{u}(p)\slashed{\pol}^-(k_1) u(p+k_1)
\eea
where we don't worry about specifying the spins of the fermions because we will ultimately sum over them using (\ref{fermspsum}). For the quintuple cut of the fermion loop we find  
\begin{changemargin}{-.6 in}{0 in}
\bea
&&A_{1;\,{\rm fermion}}^{1-{\rm loop}}\left(k_1^{1234},k_2^{1234},k_3,k_4,k_5\right)\Big|_{I_5} = - \left(-{i\over \sqrt{2}}\right)^5  \bar{u}(p_*)\slashed{\pol}^+(k_5) u(p_*+k_5) \bar{u}(p_*+k_5)\slashed{\pol}^+(k_4) u(p_*+k_4+k_5)
\el \bar{u}(p_*+k_4+k_5)\slashed{\pol}^+(k_3) u(p_*-k_1-k_2) \bar{u}(p_*-k_1-k_2)\slashed{\pol}^-(k_2) u(p_*-k_1)\bar{u}(p_*-k_1)\slashed{\pol}^-(k_1) u(p_*) 
\elale - \left({i\over \sqrt{2}}\right)^5 {\Tr}\Big[\slashed{\pol}^+(k_5)(\slashed{p}_*+\slashed{k}_5 + \mu)\slashed{\pol}^+(k_4) (\slashed{p}_*+\slashed{k}_4+\slashed{k}_5 + \mu)
\el \slashed{\pol}^+(k_3) (\slashed{p}_*-\slashed{k}_1-\slashed{k}_2 + \mu)\slashed{\pol}^-(k_2) (\slashed{p}_*-\slashed{k}_1 + \mu)\slashed{\pol}^-(k_1) (\slashed{p}_* + \mu) \Big] \,{\rm .}\nn
\label{f5ptpent}
\eea
\end{changemargin}

In this context, the extra overall minus sign is a result~\cite{BernMorgan} of using three-point amplitudes with spinor strings of the form $\bar{u}(p)\slashed{\pol}^+(k_1) u(p+k_1)$, when really they should have spinor strings of the form $\bar{u}(p)\slashed{\pol}^+(k_1) u(-p-k_1)$. Now that we understand how to deal with a loop of fermions, it is natural to ask what the analogous prescription is for a loop of gluons. Clearly, to start we need to write down three-point gluon amplitudes
\bea
A^{tree}_{\mu^2}\left(-p_{g},k_1,(p-k_1)_g\right) &=&  i \sqrt{2}\left(\pol^+(k_1)\cdot p~ g_{\rho \sigma}+k_{1\,\rho}~\pol^+_\sigma(k_1)-k_{1\,\sigma}~ \pol^+_\rho(k_1)\right)\pol^{*\,\rho}(p)\pol^\sigma(p-k_1)\nn
\\
A^{tree}_{\mu^2}\left(-p_{g},k_1^{1234},(p-k_1)_g\right) &=&   i \sqrt{2}\left(\pol^-(k_1)\cdot p~ g_{\rho \sigma}+k_{1\,\rho}~\pol^-_\sigma(k_1)-k_{1\,\sigma}~ \pol^-_\rho(k_1)\right)\pol^{*\,\rho}(p)\pol^\sigma(p-k_1)\nn
\eea
without committing to a specific choice of polarization for the gluons with $p$-dependent external momenta. These degrees of freedom will eventually be summed over. Actually, the correct summation procedure is fairly obvious~\cite{PeskinSchroeder}. We can use the na\"{i}ve replacement 
\be
\sum_\lambda \pol^\lambda_\rho(k_1)\pol^{*\,\lambda}_\sigma(k_1) \rightarrow -g_{\rho \sigma} 
\ee
valid in Abelian gauge theory, provided that we correct for the fact that we are overcounting states by including the quintuple cut of a ghost loop. This is simple since the contribution from a ghost loop is nothing but the contribution from a complex scalar loop with an extra overall minus sign coming the fact that the ghost field obeys Fermi-Dirac statistics:
\be
A_{1;\,{\rm ghost}}^{1-{\rm loop}}\left(k_1^{1234},k_2^{1234},k_3,k_4,k_5\right)\Big|_{I_5} = -A_{1;\,{\rm scalar}}^{1-{\rm loop}}\left(k_1^{1234},k_2^{1234},k_3,k_4,k_5\right)\Big|_{I_5}\,{\rm .}
\ee
Returning to the quintuple cut of the gluon loop, we have 
\begin{changemargin}{-.6 in}{0 in}
\bea
&&A_{1;\,{\rm gluon}}^{1-{\rm loop}}\left(k_1^{1234},k_2^{1234},k_3,k_4,k_5\right)\Big|_{I_5} = \left(i \sqrt{2}\right)^5  \pol^{*\,\rho_1}(p_*) \Big(\pol^-(k_1)\cdot p_*~ g_{\rho_1 \sigma_1}+k_{1\,\rho_1}~\pol^-_{\sigma_1}(k_1)
\el-k_{1\,\sigma_1}~ \pol^-_{\rho_1}(k_1)\Big)\pol^{\sigma_1}(p_*-k_1) \pol^{*\,\rho_2}(p_*-k_1) \Big(\pol^-(k_2)\cdot (p_*-k_1)~ g_{\rho_2 \sigma_2}+k_{2\,\rho_2}~\pol^-_{\sigma_2}(k_2)
\el-k_{2\,\sigma_2}~ \pol^-_{\rho_2}(k_2)\Big)\pol^{\sigma_2}(p_*-k_1-k_2)  
\pol^{*\,\rho_3}(p_*-k_1-k_2) \Big(\pol^+(k_3)\cdot (p_*-k_1-k_2)~ g_{\rho_3 \sigma_3}+k_{3\,\rho_3}~\pol^+_{\sigma_3}(k_3)
\el-k_{3\,\sigma_3}~ \pol^+_{\rho_3}(k_3)\Big)\pol^{\sigma_3}(p_*+k_4+k_5)
\pol^{*\,\rho_4}(p_*+k_4+k_5) \Big(\pol^+(k_4)\cdot (p_*+k_4+k_5)~ g_{\rho_4 \sigma_4}+k_{4\,\rho_4}~\pol^+_{\sigma_4}(k_4)
\el-k_{4\,\sigma_4}~ \pol^+_{\rho_4}(k_4)\Big)\pol^{\sigma_4}(p_*+k_5)\pol^{*\,\rho_5}(p_*+k_5) \left(\pol^+(k_5)\cdot (p_*+k_5)~ g_{\rho_5 \sigma_5}+k_{5\,\rho_5}~\pol^+_{\sigma_5}(k_5)-k_{5\,\sigma_5}~ \pol^+_{\rho_5}(k_5)\right)\pol^{\sigma_5}(p_*)
\elale \left(i \sqrt{2}\right)^5   \left(\pol^-(k_1)\cdot p_*~ g_{\rho_1 \sigma_1}+k_{1\,\rho_1}~\pol^-_{\sigma_1}(k_1)-k_{1\,\sigma_1}~ \pol^-_{\rho_1}(k_1)\right)\left(-g^{\sigma_1 \rho_2}\right) 
\el\left(\pol^-(k_2)\cdot (p_*-k_1)~ g_{\rho_2 \sigma_2}+k_{2\,\rho_2}~\pol^-_{\sigma_2}(k_2)-k_{2\,\sigma_2}~ \pol^-_{\rho_2}(k_2)\right)\left(-g^{\sigma_2 \rho_3}\right) 
\el\left(\pol^+(k_3)\cdot (p_*-k_1-k_2)~ g_{\rho_3 \sigma_3}+k_{3\,\rho_3}~\pol^+_{\sigma_3}(k_3)-k_{3\,\sigma_3}~ \pol^+_{\rho_3}(k_3)\right)\left(-g^{\sigma_3 \rho_4}\right)
\el\left(\pol^+(k_4)\cdot (p_*+k_4+k_5)~ g_{\rho_4 \sigma_4}+k_{4\,\rho_4}~\pol^+_{\sigma_4}(k_4)-k_{4\,\sigma_4}~ \pol^+_{\rho_4}(k_4)\right)\left(-g^{\sigma_4 \rho_5}\right)
\el\left(\pol^+(k_5)\cdot (p_*+k_5)~ g_{\rho_5 \sigma_5}+k_{5\,\rho_5}~\pol^+_{\sigma_5}(k_5)-k_{5\,\sigma_5}~ \pol^+_{\rho_5}(k_5)\right)\left(-g^{\sigma_5 \rho_1}\right)\,{\rm .}
\label{g5ptpent}
\eea
\end{changemargin}

Finally, we combine together all of the above results with the appropriate multiplicities:
\cmb{-1.0 in}{0 in}
\bea
A_{1}^{1-{\rm loop}}\left(k_1^{1234},k_2^{1234},k_3,k_4,k_5\right)\Big|_{I_5} &=& 3 A_{1;\,{\rm scalar}}^{1-{\rm loop}}\left(k_1^{1234},k_2^{1234},k_3,k_4,k_5\right)\Big|_{I_5}+4 A_{1;\,{\rm fermion}}^{1-{\rm loop}}\left(k_1^{1234},k_2^{1234},k_3,k_4,k_5\right)\Big|_{I_5} 
\el
+\left(A_{1;\,{\rm gluon}}^{1-{\rm loop}}\left(k_1^{1234},k_2^{1234},k_3,k_4,k_5\right)\Big|_{I_5}-A_{1;\,{\rm scalar}}^{1-{\rm loop}}\left(k_1^{1234},k_2^{1234},k_3,k_4,k_5\right)\Big|_{I_5}\right) \,{\rm ,}\nn
\label{5ptpent}
\eea
\cme
where we have dealt with the ghost loop contribution as discussed above. One can straightforwardly check (numerically using {\it e.g.} S@M~\cite{S@M}) that, after projecting (\ref{5ptpent}) onto the dual conformal basis using eq. (\ref{PtoDCB}), the result agrees with that obtained earlier using the leading singularity method (eq. (\ref{B5ptDCB})). Evaluating the numerator algebra becomes slightly more involved for quintuple cuts of one-loop six-gluon amplitudes, but we will still be able to use the above procedure to great effect. 

We are finally in a position to outline the strategy that we will use to solve, say, $A_{1}^{1-{\rm loop}}(k_1^{1234},k_2^{1234},k_3,k_4,k_5,k_6)$ to all orders in $\e$.  This amplitude works well as an example because its full analytical form is known~\cite{oneloopselfdual}:
\cmb{-.6 in}{0 in}
\bea
&&A_{1}^{1-{\rm loop}}(k_1^{1234},k_2^{1234},k_3,k_4,k_5,k_6) ={A_{6;~\langle 12 \rangle}^{\textrm{\scriptsize{MHV}}} \over 2} \bigg(- s_3 s_4 I_{4}^{(1,2),\,D=4-2\e}- {s_4 s_5} I_{4}^{(2,3),\,D=4-2\e} 
\el- {s_5 s_6} I_{4}^{(3,4),\,D=4-2\e} -{s_1 s_6 }I_{4}^{(4,5),\,D=4-2\e}  - {s_1 s_2 }I_{4}^{(5,6),\,D=4-2\e}- {s_2 s_3 } I_{4}^{(1,6),\,D=4-2\e}+ (s_3 s_6 - t_2 t_3) I_{4}^{(1,4),\,D=4-2\e} 
\el+ (s_1 s_4 - t_1 t_3) I_{4}^{(2,5),\,D=4-2\e} + (s_2 s_5 - t_1 t_2) I_{4}^{(3,6),\,D=4-2\e}
+ \e \sum_{i=1}^6 \pol(i+1, i+2, i+3, i+4) I_5^{(i),\,D=6-2\e}
\el+\e~ \textrm{tr}[\slashed{k_1}\slashed{k_2}\slashed{k_3}\slashed{k_4}\slashed{k_5}\slashed{k_6}] I_6^{D=6-2\e}\bigg) \,{\rm .}
\label{6ptMHV}
\eea
\cme
We will, of course, mostly be interested in evaluating all\footnote{In the next subsection we will go through the exercise of determining how many independent NMHV gluon  amplitudes there are (ignoring $\Nsym$ supersymmetry for now).} six-gluon NMHV amplitudes, but the strategy utilized for $A_{1}^{1-{\rm loop}}(k_1^{1234},k_2^{1234},k_3,k_4,k_5,k_6)$ carries over in a completely straightforward fashion to the other six-gluon amplitudes.

The general idea is that, while the leading singularity method does not fix everything to all orders in $\e$ starting at six points, the method is very powerful and {\it does} fix everything up to terms with trivial soft and collinear limits. To illustrate this point let us discuss to what extent the universal factorization properties of $A_{1}^{1-{\rm loop}}(k_1^{1234},k_2^{1234},k_3,k_4,k_5,k_6)$ under soft and collinear limits determine the analytic form of the amplitude, given that we already know $A_{1}^{1-{\rm loop}}(k_1^{1234},k_2^{1234},k_3,k_4,k_5)$ to all orders in $\e$. It turns out that there is only one function in $A_{1}^{1-{\rm loop}}(k_1^{1234},k_2^{1234},k_3,k_4,k_5,k_6)$ that is not constrained in this approach: One can check that 
\be
\textrm{tr}[\slashed{k_1}\slashed{k_2}\slashed{k_3}\slashed{k_4}\slashed{k_5}\slashed{k_6}]
\ee
has no soft or collinear limits in any channel.\footnote{A soft or collinear limit for planar amplitudes is particularly simple because one only has to consider nearest-neighbor pairs of momenta. If unfamiliar, see~\cite{Dixon96rev} for an elementary discussion of planar soft and collinear limits.} Therefore any attempt to deduce the form of the one-loop six-gluon MHV amplitude from that of the one-loop five-gluon amplitude by demanding consistency of the soft and collinear limits will miss terms like that above. 

This ambiguity is reflected in the solution of the leading singularity equations for $A_{1}^{1-{\rm loop}}(k_1^{1234},k_2^{1234},k_3,k_4,k_5,k_6)$. Solving the system of $15\times 2 = 30$ equations in $15 + 6 = 21$ unknowns\footnote{Here we would like to remind the reader that, if desired, the one-loop scalar hexagon integral may be expressed as a linear combination of six one-loop scalar pentagon integrals (see eq. (\ref{hexred})).} determines 20 of the unknown integral coefficients in terms of one of the pentagon coefficients, say that associated to $I_5^{(1),\,4-2\e}$. The point is that if we can evaluate one pentagon coefficient using $D$-dimensional unitarity, then the leading singularity equations, which require only four dimensional inputs, give us everything else. This is a much better strategy than trying to evaluate the quintuple cut of each pentagon independently because it allows one to solve for all the pentagon coefficients with a minimum of effort beyond that required to determine the coefficients of the boxes. 

Before going any further, we should clarify a potentially confusing point about the solution to the leading singularity equations for $A_{1}^{1-{\rm loop}}(k_1^{1234},k_2^{1234},k_3,k_4,k_5,k_6)$. Suppose we let $B_1$ be the coefficient of $I_5^{(1),\,4-2\e}$. Then a generic box coefficient, say that of $I_{4}^{(1,6),\,D=4-2\e}$, will have the form $\alpha_{16}+\beta_{16} B_1$. It may seem strange that the box coefficient associated to $I_{4}^{(1,6),\,D=4-2\e}$ depends on the pentagon coefficient $B_1$. This apparent paradox is resolved by projecting the geometric basis onto the dual conformal basis: the pentagons $I_{5}^{(1),\,D=4-2\e}$ and $I_{5}^{(6),\,D=4-2\e}$ each contribute to the coefficient of $I_{4}^{(1,6),\,D=4-2\e}$ in the dual conformal basis after the formula (\ref{PtoDCB}) is applied to them. Remarkably, these extra contributions conspire to cancel all of the $B_1$ dependence that was present in the coefficient of $I_{4}^{(1,6),\,D=4-2\e}$, considered as an element of the geometric basis.

In solving the leading singularity equations, we were free to choose any pentagon coefficient we wanted as the parameter undetermined by the system. The reason that we chose the coefficient of $I_5^{(1),\,4-2\e}$ is that it is particularly simple to determine this integral coefficient using quintuple cuts. This follows from the fact that $A^{tree}_{\mu^2}\left((p-k_1)_s,k_1^{1234},k_6,(-p-k_6)_{\bar{s}}\right)$, $A^{tree}_{\mu^2}\left((p-k_1)_{\bar{f}},k_1^{1234},k_6,(-p-k_6)_f\right)$, and $A^{tree}_{\mu^2}\left((p-k_1)_{g},k_1^{1234},k_6,(-p-k_6)_g\right)$ can each be represented by a single Feynman diagram:
\cmb{-.6 in}{0 in}
\bea
A^{tree}_{\mu^2}\left((p-k_1)_s,k_1^{1234},k_6,(-p-k_6)_{\bar{s}}\right) &=& -{i \spab1.{p}.6^2 \over s_6 \spab1.p.1}
\label{mu4tree1}\\ A^{tree}_{\mu^2}\left((p-k_1)_{\bar{f}},k_1^{1234},k_6,(-p-k_6)_f\right) &=& {i (p+k_6)\cdot \pol^+(k_6)\over \spab1.p.1}\bar{u}(p+k_6)\slashed{\pol}^-(k_1)u(p-k_1)
\label{mu4tree2}\\ A^{tree}_{\mu^2}\left((p-k_1)_{g},k_1^{1234},k_6,(-p-k_6)_g\right) &=& -{2 i\pol^\rho(p-k_1) \pol^{*\,\sigma}(p+k_6) \over \spab1.p.1}\Big(\pol^-(k_1)\cdot p ~\pol^+(k_6)\cdot p ~g_{\rho \sigma} 
\nonumber\\&&+ \pol^+(k_6)\cdot p ~k_{1\,\sigma} \pol^-_\rho(k_1)-\pol^+(k_6)\cdot p ~k_{1\,\rho}~ \pol^-_\sigma(k_1) + \pol^-(k_1)\cdot p ~k_{6\,\sigma}~ \pol^+_{\rho}(k_6) 
\nonumber\\&&- \pol^-(k_1)\cdot p ~k_{6\,\rho}~ \pol^+_{\sigma}(k_6) - k_1\cdot k_6 \pol^-_\rho(k_1)\pol^+_\sigma(k_6)\Big)
\label{mu4tree3}\eea
\cme
Using the same logic that was employed for the five-point pentagon coefficient calculated above and the results of eqs. (\ref{mu4tree1})-(\ref{mu4tree3}), it is straightforward to compute $B_1$. One subtlety is that the line $p^2-\mu^2$ is left uncut in the evaluation of this integral coefficient. As a result, the expression for $\mu^2$ is not given simply by $p_*^2$, the way that it was in the five-point example worked out in detail above. Instead, one has the relation $p_*^2-2 p_*\cdot k_1 = \mu^2$. Also, to use the framework of eqs. (\ref{5ptos1}) and (\ref{5ptos2}), we have to make the following adjustments: (\ref{5ptos1}) becomes
\be
p^\nu = L_1 K_1^\nu + L_2 K_2^\nu + L_3 K_3^\nu + L_4 K_4^\nu+k_1
\ee
and the $K_i$ four-vectors all need to be shifted by $-k_1$ ({\it i.e.} instead of $K_1 = k_1 + k_2 + k_3$, we have $K_1 = k_2 + k_3$). 

Before presenting the results of our all-orders-in-$\e$ six-gluon NMHV calculations, we make some remarks about how we expect the strategy outlined for six-gluon MHV amplitudes to generalize to higher-multiplicity amplitudes. First of all, we conjecture that the number of unconstrained by the one-loop soft/collinear bootstrap is controlled by kinematics as opposed to dynamics ({\it i.e.} independent of the $k$ in $\rm{N}^k$MHV). We interpret the fact that this is true at the six-point level (in the sense that the leading singularities for both MHV and NMHV amplitudes fix everything up to a single pentagon coefficient) as evidence for this proposal. If this conjecture is indeed correct, it follows that the number of terms in arbitrary $n$-point amplitudes left unconstrained by the soft/collinear bootstrap is equal to the number of unconstrained terms in the $n$-gluon MHV amplitudes. The number of such terms is 6 in the seven-gluon and 21 in the eight-gluon MHV amplitudes and we expect the answer to be the same for their non-MHV counterparts.

Thus, we conclude this subsection by conjecturing that the number of terms left unconstrained by the soft/collinear bootstrap at the $n$-point level (pentagon coefficients undetermined by the leading singularity method) is
\be
\left(\begin{array}{c}n - 1 \\ 5\end{array}\right) = {(n-5)(n-4)(n-3)(n-2)(n-1)\over 120} \,{\rm ,}
\ee
equal to the number of pentagons at the $(n-1)$-point level. Loosely speaking, we can think of this result as the statement that, at the $n$-point level an independent object analogous to $\textrm{tr}[\slashed{k_1}\slashed{k_2}\slashed{k_3}\slashed{k_4}\slashed{k_5}\slashed{k_6}]$ can be constructed for each pentagon integral at the $(n-1)$-point level without spoiling any of the soft/collinear constraints relating $n$-point one-loop planar amplitudes to $(n-1)$-point one-loop planar amplitudes.
\subsection{The All-Orders in $\e$ Planar One-Loop $\Nsym$ NMHV Six-Gluon Amplitudes}
\label{gresults}
In this subsection, we give our formulae for the one-loop planar six-gluon NMHV pentagon coefficients in $\Nsym$ and discuss the structural similarities between our results and certain two-loop planar six-gluon integral coefficients entering into the NMHV amplitudes calculated in~\cite{KRV}. Our first task, of course, is to understand how many independent NMHV gluon amplitudes there are (delaying a discussion of the constraints coming from $\Nsym$ supersymmetry until Subsection \ref{gendisonshell}). Na\"{i}vely, there are a large number of possibilities:
\begin{tabular}{p{0.47\linewidth} p{0.4\linewidth}}
   \begin{enumerate}
\item $A_{1}^{1-{\rm loop}}\left(k_1^{1234},k_2^{1234},k_3^{1234},k_4,k_5,k_6\right)$
\item $A_{1}^{1-{\rm loop}}\left(k_1^{1234},k_2^{1234},k_3,k_4^{1234},k_5,k_6\right)$
\item $A_{1}^{1-{\rm loop}}\left(k_1^{1234},k_2^{1234},k_3,k_4,k_5^{1234},k_6\right)$
\item $A_{1}^{1-{\rm loop}}\left(k_1^{1234},k_2^{1234},k_3,k_4,k_5,k_6^{1234}\right)$
\item $A_{1}^{1-{\rm loop}}\left(k_1^{1234},k_2,k_3^{1234},k_4^{1234},k_5,k_6\right)$
\item $A_{1}^{1-{\rm loop}}\left(k_1^{1234},k_2,k_3^{1234},k_4,k_5^{1234},k_6\right)$
\item $A_{1}^{1-{\rm loop}}\left(k_1^{1234},k_2,k_3^{1234},k_4,k_5,k_6^{1234}\right)$
\item $A_{1}^{1-{\rm loop}}\left(k_1^{1234},k_2,k_3,k_4^{1234},k_5^{1234},k_6\right)$
\item $A_{1}^{1-{\rm loop}}\left(k_1^{1234},k_2,k_3,k_4^{1234},k_5,k_6^{1234}\right)$
\item $A_{1}^{1-{\rm loop}}\left(k_1^{1234},k_2,k_3,k_4,k_5^{1234},k_6^{1234}\right)$
   \end{enumerate} &
   \begin{enumerate}
   \setcounter{enumi}{10}
      \item $A_{1}^{1-{\rm loop}}\left(k_1,k_2^{1234},k_3^{1234},k_4^{1234},k_5,k_6\right)$
\item $A_{1}^{1-{\rm loop}}\left(k_1,k_2^{1234},k_3^{1234},k_4,k_5^{1234},k_6\right)$
\item $A_{1}^{1-{\rm loop}}\left(k_1,k_2^{1234},k_3^{1234},k_4,k_5,k_6^{1234}\right)$
\item $A_{1}^{1-{\rm loop}}\left(k_1,k_2^{1234},k_3,k_4^{1234},k_5^{1234},k_6\right)$
\item $A_{1}^{1-{\rm loop}}\left(k_1,k_2^{1234},k_3,k_4^{1234},k_5,k_6^{1234}\right)$
\item $A_{1}^{1-{\rm loop}}\left(k_1,k_2^{1234},k_3,k_4,k_5^{1234},k_6^{1234}\right)$
\item $A_{1}^{1-{\rm loop}}\left(k_1,k_2,k_3^{1234},k_4^{1234},k_5^{1234},k_6\right)$
\item $A_{1}^{1-{\rm loop}}\left(k_1,k_2,k_3^{1234},k_4^{1234},k_5,k_6^{1234}\right)$
\item $A_{1}^{1-{\rm loop}}\left(k_1,k_2,k_3^{1234},k_4,k_5^{1234},k_6^{1234}\right)$
\item $A_{1}^{1-{\rm loop}}\left(k_1,k_2,k_3,k_4^{1234},k_5^{1234},k_6^{1234}\right)$
   \end{enumerate}\\
\end{tabular}

many of which are obviously related by parity\footnote{Recall that CP is a good symmetry of perturbative scattering amplitudes even in pure $\mathcal{N} = 0$ Yang-Mills.} or cyclic symmetry\footnote{Recall from \ref{planar} that, for example, the amplitudes $A_{1}^{1-{\rm loop}}(k_1^{1234},k_2^{1234},k_3^{1234},k_4,k_5,k_6)$ and $A_{1}^{1-{\rm loop}}(k_2^{1234},k_3^{1234},k_4,k_5,k_6,k_1^{1234})$ are equal by virtue of the color structure in the planar limit.}. In particular, we will now show that all of the above can be related to $A_{1}^{1-{\rm loop}}\left(k_1^{1234},k_2^{1234},k_3^{1234},k_4,k_5,k_6\right)$, $A_{1}^{1-{\rm loop}}\left(k_1^{1234},k_2^{1234},k_3,k_4^{1234},k_5,k_6\right)$, and $A_{1}^{1-{\rm loop}}\left(k_1^{1234},k_2,k_3^{1234},k_4,k_5^{1234},k_6\right)$ (amplitudes 1., 2., and 6. in the above). In fact, we will see in Section \ref{supercomp} that it is possible to derive a beautiful all-orders-in-$\e$ $\Nsym$ supersymmetrization of the six-point NMHV amplitudes using only these three all-orders-in-$\e$ component amplitudes as inputs. To start, we see immediately that amplitudes 11. - 20. are related to 1. - 10. by parity which acts on the amplitudes by reversing the helicities of all gluons. Next, we see that 4. and 10. are related to 1. through a series of cyclic shifts followed by a relabeling of the momenta of the external gluons. Similarly, amplitudes 7. and 8. are related to 2. through cyclic shifts followed by a relabeling. Finally, 3., 5., and 9. are related to 2. through cyclic shifts, a relabeling, {\it and} a parity transformation. Amplitude 6. can't be related to 1. or 2. through some combination of parity and cyclicity, so we need to include it in our basis as well.

Now that we understand why it makes sense to focus on $A_{1}^{1-{\rm loop}}\left(k_1^{1234},k_2^{1234},k_3^{1234},k_4,k_5,k_6\right)$, $A_{1}^{1-{\rm loop}}\left(k_1^{1234},k_2^{1234},k_3,k_4^{1234},k_5,k_6\right)$, and $A_{1}^{1-{\rm loop}}\left(k_1^{1234},k_2,k_3^{1234},k_4,k_5^{1234},k_6\right)$, we present our results for these amplitudes. To begin, let us recall the results of the calculations performed in~\cite{BDDKNMHV}. The authors of that work determined the box coefficients for all NMHV gluon amplitudes in the dual conformal basis. The $6 - 2\e$ dimensional pentagon coefficients, however, were undetermined. It was found that
\bea
\nonumber A_{1}^{1-{\rm loop}}\left(k_1^{1234},k_2^{1234},k_3^{1234},k_4,k_5,k_6\right)&&=-{1\over2}B_1 \Bigg( s_4 s_5 I_4^{(2,3)}+s_1 s_2 I_4^{(5,6)}+s_6 t_1 I_4^{(3,5)}+s_3 t_1 I_4^{(2,6)}\Bigg) \\ \nonumber
&&-{1\over2}B_2 \Bigg( s_5 s_6 I_4^{(3,4)}+s_2 s_3 I_4^{(6,1)}+s_1 t_2 I_4^{(4,6)}+s_4 t_2 I_4^{(1,3)}\Bigg) \\ \nonumber 
&&-{1\over2}B_3 \Bigg( s_6 s_1 I_4^{(4,5)}+s_3 s_4 I_4^{(1,2)}+s_2 t_3 I_4^{(1,5)}+s_5 t_3 I_4^{(2,4)}\Bigg) \\ \nonumber 
&& +K_1 \e I_5^{(1),6-2 \e}+K_2 \e I_5^{(2),6-2 \e}+K_3 \e I_5^{(3),6-2 \e}
\\ 
&& +K_4 \e I_5^{(4),6-2 \e}+K_5 \e I_5^{(5),6-2 \e}+K_6 \e I_5^{(6),6-2 \e} \, ,
\eea
\bea
\nonumber A_{1}^{1-{\rm loop}}\left(k_1^{1234},k_2,k_3^{1234},k_4,k_5^{1234},k_6\right)&&=-{1\over2}G_1 \Bigg( s_4 s_5 I_4^{(2,3)}+s_1 s_2 I_4^{(5,6)}+s_6 t_1 I_4^{(3,5)}+s_3 t_1 I_4^{(2,6)}\Bigg) \\ \nonumber
&&-{1\over2}G_2 \Bigg( s_5 s_6 I_4^{(3,4)}+s_2 s_3 I_4^{(6,1)}+s_1 t_2 I_4^{(4,6)}+s_4 t_2 I_4^{(1,3)}\Bigg) \\ \nonumber 
&&-{1\over2}G_3 \Bigg( s_6 s_1 I_4^{(4,5)}+s_3 s_4 I_4^{(1,2)}+s_2 t_3 I_4^{(1,5)}+s_5 t_3 I_4^{(2,4)}\Bigg) \\ \nonumber 
&& +F_1 \e I_5^{(1),6-2 \e}+F_2 \e I_5^{(2),6-2 \e}+F_3 \e I_5^{(3),6-2 \e}
\\ 
&& +F_4 \e I_5^{(4),6-2 \e}+F_5 \e I_5^{(5),6-2 \e}+F_6 \e I_5^{(6),6-2 \e} \, , 
\eea
and
\bea
\nonumber A_{1}^{1-{\rm loop}}\left(k_1^{1234},k_2^{1234},k_3,k_4^{1234},k_5,k_6\right)&&=-{1\over2}D_1 \Bigg( s_4 s_5 I_4^{(2,3)}+s_1 s_2 I_4^{(5,6)}+s_6 t_1 I_4^{(3,5)}+s_3 t_1 I_4^{(2,6)}\Bigg) \\ \nonumber
&&-{1\over2}D_2 \Bigg( s_5 s_6 I_4^{(3,4)}+s_2 s_3 I_4^{(6,1)}+s_1 t_2 I_4^{(4,6)}+s_4 t_2 I_4^{(1,3)}\Bigg) \\ \nonumber 
&&-{1\over2}D_3 \Bigg( s_6 s_1 I_4^{(4,5)}+s_3 s_4 I_4^{(1,2)}+s_2 t_3 I_4^{(1,5)}+s_5 t_3 I_4^{(2,4)}\Bigg) \\ \nonumber 
&& +H_1 \e I_5^{(1),6-2 \e}+H_2 \e I_5^{(2),6-2 \e}+H_3 \e I_5^{(3),6-2 \e}
\\ 
&& +H_4 \e I_5^{(4),6-2 \e}+H_5 \e I_5^{(5),6-2 \e}+H_6 \e I_5^{(6),6-2 \e} \,.
\eea
All of the spin factors which entered into the box coefficients ($B_i$, $G_i$, and $D_i$) were determined. They are given by
\bea
B_1 &=& B_0 \label{B1}\\
B_2 &=& \bigg({\spab1.{2+3}.4\over t_2}\bigg)^4 B_0\Big|_{j\rightarrow j+1}+\bigg({\spa2.3 \spb5.6 \over t_2}\bigg)^4 B_0^{\langle~\rangle \leftrightarrow [~]}\Big|_{j\rightarrow j+1} \\
B_3 &=& \bigg({\spab3.{1+2}.6\over t_3}\bigg)^4 B_0\Big|_{j\rightarrow j-1}+\bigg({\spa1.2 \spb4.5 \over t_3}\bigg)^4 B_0^{\langle~\rangle \leftrightarrow [~]}\Big|_{j\rightarrow j-1} \, ,
\eea
\bea
G_1 &=& \bigg({\spab5.{4+6}.2\over t_1}\bigg)^4 B_0+\bigg({\spa1.3 \spb4.6 \over t_1}\bigg)^4 B_0^{\langle~\rangle \leftrightarrow [~]} \\
G_2 &=& \bigg({\spab3.{2+4}.6\over t_2}\bigg)^4 B_0^{\langle~\rangle \leftrightarrow [~]}\Big|_{j\rightarrow j+1}+\bigg({\spa5.1 \spb2.4 \over t_2}\bigg)^4 B_0\Big|_{j\rightarrow j+1} \\
G_3 &=& \bigg({\spab1.{2+6}.4\over t_3}\bigg)^4 B_0^{\langle~\rangle \leftrightarrow [~]}\Big|_{j\rightarrow j-1}+\bigg({\spa3.5 \spb6.2 \over t_3}\bigg)^4 B_0\Big|_{j\rightarrow j-1} \, ,
\eea
and
\bea
D_1 &=& \bigg({\spab4.{1+2}.3\over t_1}\bigg)^4 B_0+\bigg({\spa1.2 \spb5.6 \over t_1}\bigg)^4 B_0^{\langle~\rangle \leftrightarrow [~]} \\
D_2 &=& \bigg({\spab1.{2+4}.3\over t_2}\bigg)^4 B_0\Big|_{j\rightarrow j+1}+\bigg({\spa2.4 \spb5.6 \over t_2}\bigg)^4 B_0^{\langle~\rangle \leftrightarrow [~]}\Big|_{j\rightarrow j+1} \\
D_3 &=& \bigg({\spab4.{1+2}.6\over t_3}\bigg)^4 B_0\Big|_{j\rightarrow j-1}+\bigg({\spa1.2 \spb3.5 \over t_3}\bigg)^4 B_0^{\langle~\rangle \leftrightarrow [~]}\Big|_{j\rightarrow j-1} \, ,
\eea
where
\be
B_0 = i{\spa1.2 \spa2.3 \spb4.5 \spb5.6 \spab3.{1+2}.6 \spab1.{2+3}.4 t_1^3 \over s_1 s_2 s_4 s_5 (t_1 t_2-s_2 s_5)(t_1 t_3-s_1 s_4)} \, . \label{B0}
\ee
Using the strategy outlined it \ref{effgcomp}, we reproduce the above and, furthermore, find explicit expressions for the $K_i$, $G_i$, and $H_i$.

Although the raw answers obtained using the method described in the last subsection are already compact enough to fit on a single page, it is clearly desirable to find more compact formulae. In their work on the two-loop planar NMHV gluon amplitudes~\cite{KRV}, Kosower, Roiban, and Vergu derived explicit expressions for all possible $\mu$-integral hexabox coefficients (see Figure \ref{hexabox} for an illustration). Motivated by issues of IR consistency that we will elaborate on in Section \ref{WL/MHV}, we evaluated the answers they obtained numerically and were able to find a straightforward mapping between their results and ours. To explain this relationship, it is useful to consider a concrete example.

We consider the coefficient of the $s_1$-channel hexabox integral (see Figure \ref{hexabox}) that appears in the amplitude $A_{1}^{2-{\rm loop}}\left(k_1^{1234},k_2^{1234},k_3^{1234},k_4,k_5,k_6\right)$ calculated to  all orders in $\e$. 
\FIGURE{
\resizebox{0.6\textwidth}{!}{\includegraphics{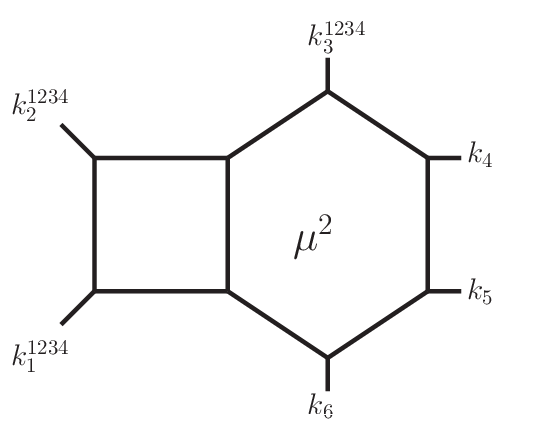}}
\vbox{\vskip 1 in}
\caption{The $s_1$-channel hexabox $\mu$-integral topology of $A_{1}^{2-{\rm loop}}\left(k_1^{1234},k_2^{1234},k_3^{1234},k_4,k_5,k_6\right)$. The factor of $\mu^2$ should be thought of as belonging to the hexagon subdiagram.}
\label{hexabox}}
It turns out that this $\mu$-integral hexabox coefficient and $K_2$ (coefficient of the $s_1$-channel $6 - 2\e$ dimensional pentagon coefficient that appears in $A_{1}^{1-{\rm loop}}\left(k_1^{1234},k_2^{1234},k_3^{1234},k_4,k_5,k_6\right)$ are simply related:
\be
K_2 = {C_2 \over 2 s_1} \mathcal{K}_2 \, {\rm ,}
\label{hexpentrel}
\ee
where we have given the hexabox coefficient the convenient label $\mathcal{K}_2$ and $C_2$ is one of the variables that we used to define the reduction of the one-loop scalar hexagon integral to a sum of six scalar pentagons in \ref{muint}:
\be
I_6 = {1\over 2}\sum_{i = 1}^6 C_i I_5^{(i)}\,{\rm .}
\label{hexred}
\ee
This relation makes a certain amount of sense if we think about collapsing the box in the $\mu$-integral hexabox in Figure \ref{hexabox} to a point. This turns the hexabox $\mu$-integral into a pentagon $\mu$-integral. Evidently, $s_1$ appears because we are working with the $s_1$ channel hexabox and, perhaps, $C_2$ appears because we are relating an object with six external legs to one with five. In any case, the above relation will allow us to exploit extremely simple results found for the NMHV hexabox coefficients~\cite{KRV} to write beautiful formulas for the $K_i$, $F_i$, and $H_i$. We find
\bea
K_1 &=& {i \over 2} C_1 {\Big(\spab2.{1+6}.5 \spab1.{2+3}.4 \spab3.{1+2}.6 +\spab5.{1+6}.2 \spa1.2 \spa2.3 \spb4.5 \spb5.6 \Big)^2\over s_6 s_3 \spab2.{6+1}.5 \spab5.{6+1}.2} \nonumber \\ \\
K_2 &=& {i \over 2} C_2 {\spab3.{1+2}.6^2 \spa1.2^2 \spb4.5^2 t_1^2 \over s_1 s_4 \spab3.{1+2}.6 \spab6.{1+2}.3} \\
K_3 &=& {i \over 2} C_3 {\spab1.{2+3}.4^2 \spa2.3^2 \spb5.6^2 t_1^2 \over s_2 s_5 \spab1.{2+3}.4 \spab4.{2+3}.1} \\
K_4 &=& {i \over 2} C_4 {\Big(\spab2.{1+6}.5 \spab1.{2+3}.4 \spab3.{1+2}.6 +\spab5.{1+6}.2 \spa1.2 \spa2.3 \spb4.5 \spb5.6 \Big)^2\over s_6 s_3 \spab2.{6+1}.5 \spab5.{6+1}.2} \nonumber \\ \\
K_5 &=& {i \over 2} C_5 {\spab3.{1+2}.6^2 \spa1.2^2 \spb4.5^2 t_1^2 \over s_1 s_4 \spab3.{1+2}.6 \spab6.{1+2}.3} \\
K_6 &=& {i \over 2} C_6 {\spab1.{2+3}.4^2 \spa2.3^2 \spb5.6^2 t_1^2 \over s_2 s_5 \spab1.{2+3}.4 \spab4.{2+3}.1} \, ,
\eea
\bea
F_1 &=& {i \over 2} C_1 {\Big(\spab5.{6+1}.2 \spab3.{2+4}.6 \spab1.{3+5}.4 +\spab2.{6+1}.5 \spb6.2 \spb2.4 \spa1.5 \spa3.5 \Big)^2\over s_6 s_3 \spab5.{6+1}.2 \spab2.{6+1}.5} \nonumber \\ \\
F_2 &=& {i \over 2} C_2 {\Big(\spab3.{1+2}.6 \spab1.{3+5}.4 \spab5.{4+6}.2 +\spab6.{1+2}.3 \spa1.3 \spa3.5 \spb2.6 \spb4.6 \Big)^2\over s_1 s_4 \spab3.{1+2}.6 \spab6.{1+2}.3} \nonumber \\ \\
F_3 &=& {i \over 2} C_3 {\Big(\spab1.{2+3}.4 \spab5.{4+6}.2 \spab3.{5+1}.6 +\spab4.{2+3}.1 \spb2.4 \spb4.6 \spa3.1 \spa5.1 \Big)^2\over s_2 s_5 \spab1.{2+3}.4 \spab4.{2+3}.1} \nonumber \\ \\
F_4 &=& {i \over 2} C_4 {\Big(\spab5.{6+1}.2 \spab3.{2+4}.5 \spab1.{3+5}.4 +\spab2.{6+1}.5 \spb2.6 \spb4.2 \spa5.1 \spa5.3 \Big)^2\over s_6 s_3 \spab5.{6+1}.2 \spab2.{6+1}.5} \nonumber \\ \\
F_5 &=& {i \over 2} C_5 {\Big(\spab3.{1+2}.6 \spab1.{3+5}.4 \spab5.{4+6}.2 +\spab6.{1+2}.3 \spa1.3 \spa3.5 \spb2.6 \spb4.6 \Big)^2\over s_1 s_4 \spab3.{1+2}.6 \spab6.{1+2}.3} \nonumber \\ \\
F_6 &=& {i \over 2} C_6 {\Big(\spab1.{2+3}.4 \spab5.{4+6}.2 \spab3.{5+1}.6 +\spab4.{2+3}.1 \spb2.4 \spb4.6 \spa3.1 \spa5.1 \Big)^2\over s_2 s_5 \spab1.{2+3}.4 \spab4.{2+3}.1} \, ,\nonumber \\ 
\eea
and
\bea
H_1 &=& {i \over 2} C_1 {\Big(\spab2.{6+1}.5 \spab1.{2+4}.3 \spab4.{1+2}.6 +\spab5.{6+1}.2 \spa1.2 \spa2.4 \spb3.5 \spb5.6 \Big)^2\over s_6 s_3 \spab2.{6+1}.5 \spab5.{6+1}.2} \nonumber \\ \\
H_2 &=& {i \over 2} C_2 {\spa1.2^2 \Big(\spab3.{1+2}.6 \spab4.{1+2}.3 \spb5.3 +\spab6.{1+2}.3 \spab4.{1+2}.6 \spb5.6\Big)^2\over s_1 s_4 \spab3.{1+2}.6 \spab6.{1+2}.3} \nonumber \\ \\
H_3 &=& {i \over 2} C_3 {\spb5.6^2 \Big(\spab1.{2+3}.4 \spab4.{1+2}.3 \spa2.4 +\spab4.{2+3}.1 \spab1.{2+4}.3 \spa2.1\Big)^2\over s_2 s_5 \spab4.{2+3}.1 \spab1.{2+3}.4} \nonumber \\ \\
H_4 &=& {i \over 2} C_4 {\Big(\spab2.{6+1}.5 \spab1.{2+4}.3 \spab4.{1+2}.6 +\spab5.{6+1}.2 \spa1.2 \spa2.4 \spb3.5 \spb5.6 \Big)^2\over s_6 s_3 \spab2.{6+1}.5 \spab5.{6+1}.2} \nonumber \\ \\
H_5 &=& {i \over 2} C_5 {\spa1.2^2 \Big(\spab3.{(1+2)}.6 \spab4.{1+2}.3 \spb5.3 +\spab6.{1+2}.3 \spab4.{1+2}.6 \spb5.6\Big)^2\over s_1 s_4 \spab3.{1+2}.6 \spab6.{1+2}.3} \nonumber \\ \\
H_6 &=& {i \over 2} C_6 {\spb5.6^2 \Big(\spab1.{2+3}.4 \spab4.{1+2}.3 \spa2.4 +\spab4.{2+3}.1 \spab1.{2+4}.3 \spa2.1\Big)^2\over s_2 s_5 \spab4.{2+3}.1 \spab1.{2+3}.4} \, {\rm .}\nonumber \\
\eea
We also checked these results against a Feynman diagram calculation.

These results have a couple of striking features of which we have only a partial understanding. The numerators of all the spin factors (divided by the appropriate $C_i$) are perfect squares. Furthermore, the pentagon coefficients possess a certain $i \rightarrow i+3$ symmetry:
\be
{K_1 \over C_1} = {K_4 \over C_4} \qquad
{K_2 \over C_2} = {K_5 \over C_5} \qquad
{K_3 \over C_3} = {K_6 \over C_6} \, {\rm .}
\label{mystrel}
\ee
with analogous formulas for the $F_i$ and $H_i$. As we will see in Section \ref{WL/MHV}, this $i \rightarrow i+3$ symmetry is related to the action of parity when the amplitude is written in a way that makes a hidden symmetry\footnote{This hidden symmetry is called dual superconformal invariance and some background and motivation for it is provided in the second part of Appendix \ref{ADS/CFT}.} of the planar S-matrix as manifest as possible. In the next section we explore an interesting connection between all-orders-in-$\e$ one-loop $\Nsym$ amplitudes and the first two non-trivial orders in the $\alpha'$ expansion of tree-level superstring amplitudes. The explicit one-loop results presented so far in this paper will provide us with useful explicit cross-checks on the relations we propose.
\section{New Relations Between One-Loop Amplitudes in $\Nsym$ Gauge Theory and Tree-Level Amplitudes in Open Superstring Theory}
\label{gsrel}
Before reviewing the scattering of massless modes in open superstring theory, we motivate what follows. Stieberger and Taylor~\cite{ST1} calculated the lowest-order, $\Ord(\alpha'^2)$, stringy corrections to $\Nsym$ tree-level gluon MHV amplitudes.\footnote{As mentioned in the introduction, tree-level amplitudes of massless particles in open superstring constructions compactified to four dimensions have a universal form~\cite{ST1}.} They found that their result\footnote{In Stieberger and Taylor's notation, say at the six-point level, $[[1]]_1 = s_1$, $\{\,[[1]]_1\,\} = s_1 + s_2 + s_3 + s_4 + s_5 + s_6$, $[[1]]_2 = t_1$, and $\{\,[[1]]_2\,\} = t_1 + t_2 + t_3$.},
\bea
A^{tree}_{str}\left(k_1^{1234},k_2^{1234},k_3,\cdots,k_n\right)\Bigl|_{\Ord(\alpha'^2)} &=& -{\pi^2\over 12} A_{n;\langle 12 \rangle}^{\textrm{\scriptsize{MHV}}}\Bigg(
 \sum_{k=1}^{\left[\frac{n}{2}-1\right]}\{\,[[1]]_k [[2]]_k\,\}-
\sum_{k=3}^{\left[\frac{n}{2}-1\right]}\{\,[[1]]_k [[2]]_{k-2}\,\}\nonumber\\
&& +~C^{(n)}+\sum_{j<k<\ell<m<n} \pol(j,k,\ell,m)\Bigg)\, ,\\
C^{(n)} &=& \Bigg\{ \begin{array}{ll}
        -\{\,[[1]]_{\frac{n}{2}-2}  [[{n \over 2}+1]]_{\frac{n}{2}-2}\}  
& \mbox{$n>4$, even,}\nonumber \\
         -\big\{\,[[1]]_{\frac{n-5}{2}} [[\frac{n+1}{2}]]_{\frac{n-3}{2}}\big\} & \mbox{$n>5$, odd},\end{array}
         \label{stringMHVsimp}
\eea
was precisely equal to $-6 \zeta(2)$ times the analogous one-loop $\Nsym$ amplitude with a factor of $\mu^4$ inserted into the numerator of each basis integral, $A_{1}^{1-{\rm loop}}\left(k_1^{1234},k_2^{1234},k_3,\cdots,k_n\right)[\mu^4]\Bigl|_{\e\to0}$. This non-obvious connection was actually made by showing that both 
$${A^{tree}_{str}\left(k_1^{1234},k_2^{1234},k_3,\cdots,k_n\right)\Bigl|_{\Ord(\alpha'^2)}\over \spa1.2^4} ~~~~{\rm and}~~~~ {A_{1}^{1-{\rm loop}}\left(k_1^{1234},k_2^{1234},k_3,\cdots,k_n\right)[\mu^4]\Bigl|_{\e\to0} \over \spa1.2^4}$$
 are, apart from trivial constants, equal to the all-plus one-loop amplitude in pure Yang-Mills theory~\cite{oneloopselfdual}, $A_{1;\,\mathcal{N}=0}^{1-{\rm loop}}\left(k_1,k_2,\cdots,k_n\right)$. The only reason an equivalence between 
 $${A_{1}^{1-{\rm loop}}\left(k_1^{1234},k_2^{1234},k_3,\cdots,k_n\right)[\mu^4]\Bigl|_{\e\to0}\over\spa1.2^4} ~~~~{\rm and}~~~~ {A_{1;\,\mathcal{N}=0}^{1-{\rm loop}}\left(k_1,k_2,k_3,\cdots,k_n\right)}$$ is possible is that both have the same manifest invariance under cyclic shifts $i \rightarrow i+1$. It is hard to imagine that additional relationships between $\mathcal{N}=0$ and $\Nsym$ amplitudes could exist because, in general, there is no reason to expect $\mathcal{N}=0$ and $\Nsym$ amplitudes to have similar symmetry properties (for more general amplitudes there is no trick analogous to dividing the one-loop MHV amplitude by $\spa1.2^4$). Indeed, it is incredibly likely that this relation between pure Yang-Mills and $\Nsym$ is purely accidental. However, additional relations between superstring tree amplitudes and $\Nsym$ one-loop amplitudes are a more realistic possibility. It is this possibility that we discuss in this section. The new results presented are based on unpublished work done in collaboration with Lance J. Dixon~\cite{myfourth}.
\subsection{Organization of the Tree-Level Open Superstring S-matrix}
\label{BornInfeld}
For the simple case of a $U(1)$ gauge group, it has been known since the work of Fradkin and Tseytlin~\cite{FT} that the effective action governing the low-energy dynamics of open superstrings ending on a single Dirchlet 3-brane (though the connection between gauge symmetry and D-branes remained hidden until the work of Dai, Leigh, and Polchinski in~\cite{DLP} and Leigh in~\cite{L}) is nothing but a  supersymmetrization of the Born-Infeld action. This action, expressed in terms of the Maxwell field strength tensor,
\bea
\mathcal{L}_{BI} &=& {1 \over (2 \pi g \alpha')^2} \left(1 - \sqrt{{\rm Det}\big(~g_{\mu \nu} + (2 \pi g \alpha') F_{\mu \nu} ~\big)} \right) \nonumber \\
&=& {1 \over (2 \pi g \alpha')^2} \left(1 - \sqrt{1 + {(2 \pi g \alpha')^2\over2}F_{\mu \nu}F_{\mu \nu} - {(2 \pi g \alpha')^4\over16}~ (F_{\mu \nu} \tilde{F}_{\mu \nu})^2} \right) \nonumber \\
&=& {1 \over (2 \pi g \alpha')^2} \left(1 - \sqrt{1 + {(2 \pi g \alpha')^2\over2}F_{\mu \nu}F_{\mu \nu} - {(2 \pi g \alpha')^4\over4}~ \Big(F_{\mu \nu}F_{\nu \rho}F_{\rho \sigma}F_{\sigma \mu} - {1 \over 2}~\big(F_{\mu \nu}F_{\nu \mu}\big)^2\Big)} \right) \nonumber \\
&=& -{1 \over 4} F_{\mu \nu} F_{\mu \nu} + 3~ \zeta(2) g^2 \alpha'^2 \left(F_{\mu \nu}F_{\nu \rho}F_{\rho \sigma}F_{\sigma \mu} - {1 \over 4}~\big(F_{\mu \nu}F_{\nu \mu}\big)^2\right) + \Ord\left(g^4 \alpha'^4 F^6\right)
\label{BI}
\eea
was proposed in~\cite{BI} as an alternative description of electrodynamics. In the context of string scattering, the constant $\alpha'$ is identified with the string tension. A natural generalization to the case of a $U(\Nc)$ gauge group is realized~\cite{T97} when the open superstrings under consideration end on a stack of $\Nc$ coincident $D_3$-branes. This situation is unfortunately much more complicated to describe with an effective action and there is no known analog of \eqn{BI}; the non-Abelian Born-Infeld action, as it is commonly called, is only known up to fourth order~\cite{Wulff,KKNSWrev} in $\alpha'$ (reference~\cite{KKNSWrev} is a review with many more references to the original literature). For us, only the first two non-trivial orders in this expansion play an important role. Due to the fact that there is no $\Nsym$ supersymmetrizable operator of mass dimension six\footnote{In this section we use lower-case $d$ for operator dimensions and upper-case $D$ for spacetime dimensions.} ($d = 6$) that one can write down in terms of non-Abelian field strengths and covariant derivatives, the first two non-trivial orders in the $\alpha'$ expansion are actually $\Ord(\alpha'^2)$ and $\Ord(\alpha'^3)$. In our conventions, the non-Abelian Born-Infeld action is given by
\bea
\mathcal{L}_{NABI} &=& -{1\over4} \Tr \Big\{ F_{\mu \nu} F_{\mu \nu} \Big\} 
\nonumber \\ && + \zeta(2) g^2 \alpha'^2 \Tr \Big\{ {1\over2}F_{\mu \nu} F_{\nu \rho}
                    F_{\rho \sigma} F_{\sigma \mu}
                +   F_{\mu \nu} F_{\nu \rho} 
                    F_{\sigma \mu} F_{\rho \sigma}  
        - {1\over8} F_{\mu \nu} F_{\rho \sigma}
                    F_{\nu \mu} F_{\sigma \rho}
   - {1\over4} F_{\mu \nu} F_{\nu \mu}
                    F_{\rho \sigma} F_{\sigma \rho } \Big\} 
 \nonumber\\ &&  - 8 ~\zeta(3) \alpha'^3 \Tr \Big\{{i g^3\over \sqrt{2}} \Big( F_{\mu \nu} F_{\nu \rho}
                    F_{\rho \sigma} F_{\tau \mu} F_{\sigma \tau} + F_{\mu \nu} F_{\sigma \tau}             F_{\nu \rho} F_{\tau \mu} F_{\rho \sigma} - {1\over2} F_{\mu \nu} F_{\nu \rho}  F_{\sigma \tau}     F_{\rho \mu} F_{\tau \sigma}\Big)
\nonumber \\ && + g^2 \Big( {1\over2} (D_{\mu}F_{\nu \rho}) (D_{\mu} F_{\rho \sigma}) F_{\tau \nu} F_{\sigma \tau} 
 + {1\over2}(D_{\mu} F_{\nu \rho})  F_{\tau \nu} (D_{\mu} F_{\rho \sigma})  F_{\sigma \tau} - F_{\mu \nu} (D_{\mu} F_{\rho \sigma}) (D_{\tau} F_{\nu \rho})  F_{\sigma \tau}
 \nonumber\\ && - {1\over8} (D_{\mu} F_{\nu \rho}) F_{\sigma \tau} (D_{\mu} F_{\rho \nu}) F_{\tau \sigma} + (D_{\tau} F_{\mu \nu}) F_{\rho \sigma} (D_{\mu} F_{\nu \rho})  F_{\sigma \tau} \Big)\Big\} + \Ord\left(\alpha'^4 \Tr F^6\right)\,.
\label{nonABI}
\eea
The form reproduced above is very nearly that given in~\cite{BMM}, but their conventions are slightly different\footnote{It also appears that their overall normalization differs from ours by a factor of two.}. We use the conventions of~\cite{Dixon96rev}, in which
\be
F_{\mu \nu} = \partial_\mu A_\nu - \partial_\nu A_\mu - {i g \over \sqrt{2}} [A_\mu, A_\nu]
\ee
and
\be
D_\mu \Phi = \partial_\mu \Phi - {i g \over \sqrt{2}} [A_\mu, \Phi] \, .
\ee
Results essentially identical to those above appeared in~\cite{KS} (see also the later work of~\cite{Drummond03}) and the derivative terms at $\Ord(\alpha'^3)$ were obtained earlier in~\cite{Bilal}. We have normalized our $\Ord(\alpha'^2)$ and $\Ord(\alpha'^3)$ effective Lagrangians so that they reproduce the appropriate terms in the $\alpha'$ expansion of the string scattering results given in~\cite{ST2}, where a representative leading four-point color-ordered partial amplitude is
\bea
A^{tree}_{str}(k_1^{1234},k_2^{1234},k_3,k_4) &=& A_{4;\langle 12 \rangle}^{\textrm{\scriptsize{MHV}}} {\Gamma(1+\alpha' s)\Gamma(1+\alpha' t)\over \Gamma(1 + \alpha'(s+t))} \nonumber \\
&=& A_{4;\langle 12 \rangle}^{\textrm{\scriptsize{MHV}}} \bigg( 1 - \zeta(2) s t ~\alpha'^2 + \zeta(3) s t (s+t) ~\alpha'^3 
\nonumber \\ && - {\zeta(4) \over 4} s t (4 s^2+s t+4 t^2)~\alpha'^4 +\Ord(\alpha'^5)\bigg) \, .
\label{ST4pt}
\eea
\subsection{New Relations}
\label{results}
We now return to the observed correspondence between the results of~\cite{oneloopselfdual} and~\cite{ST1} discussed briefly at the beginning of this section. By comparing the two references it is easy to see that
\be
A^{tree}_{str}\left(k_1^{h_1},\cdots,k_n^{h_n}\right)\Bigl|_{\Ord(\alpha'^2)} = -6 \zeta(2) A_{1}^{1-{\rm loop}}\left(k_1^{h_1},\cdots,k_n^{h_n}\right)[\mu^4]\Bigl|_{\e\to0} \, {\rm ,}
\label{oldconj}
\ee
where the gluon helicity configuration is MHV and should of course match on both sides of eq. (\ref{oldconj}). 

Since our notation may not be completely obvious, we consider an illustrative example. Specifically, we check that eq. \ref{oldconj} holds for the five-gluon MHV amplitude $A_{1}^{1-{\rm loop}}\left(k_1^{1234},k_2^{1234},k_3,k_4,k_5\right)$. In terms of unevaluated scalar Feynman integrals~\cite{oneloopselfdual}, we have
\bea
 A_{1}^{1-{\rm loop}}\left(k_1^{1234},k_2^{1234},k_3,k_4,k_5\right)  &=&  {- A_{5;\langle 12 \rangle}^{\textrm{\scriptsize{MHV}}} \over 2} \Bigg( s_{2}s_{3} I_4^{(1),~D=4-2\e}+s_{3}s_{4} I_4^{(2),~D=4-2\e}+ s_{4}s_{5} I_4^{(3),~D=4-2\e}\nonumber
\\ &&  +s_{5}s_{1} I_4^{(4),~D=4-2\e}+s_{1}s_{2} I_4^{(5),~D=4-2\e}-2 \e ~\pol(k_1,k_2,k_3,k_4) I_5^{D=6-2\e}\Bigg) \, {\rm .}\nn
\eea
Applying the dimension shift operation of~\cite{oneloopselfdual} to the amplitude sends $\e \to \e-2$ and $I_n^D \to I_n^D [\mu^4]$:
\bea
 A_{1}^{1-{\rm loop}}\left(k_1^{1234},k_2^{1234},k_3,k_4,k_5\right)[\mu^4] &&=  {- A_{5;\langle 12 \rangle}^{\textrm{\scriptsize{MHV}}} \over 2 } \Bigg( s_{2}s_{3} I_4^{(1),~D=4-2\e}[\mu^4] + s_3 s_{4} I_4^{(2),~D=4-2\e}[\mu^4]
\nonumber \\ &&   +  s_{4}s_{5} I_4^{(3),~D=4-2\e}[\mu^4]+s_{5}s_{1} I_4^{(4),~D=4-2\e}[\mu^4] + s_{1}s_{2} I_4^{(5),~D=4-2\e}[\mu^4]
\nonumber \\ && -2 (\e-2) ~\pol(k_1,k_2,k_3,k_4) I_5^{D=6-2\e}[\mu^4] \Bigg) \, .
\eea
Applying eq. (\ref{DSmu}) gives
\bea
 A_{1}^{1-{\rm loop}}\left(k_1^{1234},k_2^{1234},k_3,k_4,k_5\right)[\mu^4] && =  { A_{5;\langle 12 \rangle}^{\textrm{\scriptsize{MHV}}} \e (1-\e)\over 2} \Bigg( s_{2}s_{3} I_4^{(1),~D=8-2\e} + 
 s_{3}s_{4} I_4^{(2),~D=8-2\e} +
 \nonumber \\ &&  s_{4}s_{5} I_4^{(3),~D=8-2\e}+s_{5}s_{1} I_4^{(4),~D=8-2\e} + 
 s_{1}s_{2} I_4^{(5),~D=8-2\e}
 \nonumber \\ &&  -2 (\e-2) ~\pol(k_1,k_2,k_3,k_4) I_5^{D=10-2\e} \Bigg) \, .
\eea
Finally, we take the limit as $\e\to0$. As explained in Subsection \ref{GUD}, the terms which survive are those proportional to the ultraviolet singularities of the dimensionally-shifted basis integrals.
\bea
 A_{1}^{1-{\rm loop}}\left(k_1^{1234},k_2^{1234},k_3,k_4,k_5\right)[\mu^4]\Bigl|_{\e\to0} &&=  { A_{5;\langle 12 \rangle}^{\textrm{\scriptsize{MHV}}} \over 2} \Bigg( s_{2}s_{3} \Big(~{1\over6}~\Big) + 
 s_{3}s_{4} \Big(~{1\over6}~\Big)  +  s_{4}s_{5} \Big(~{1\over6}~\Big)
 \nonumber \\ &&  +s_{5}s_{1} \Big(~{1\over6}~\Big)  + 
 s_{1}s_{2} \Big(~{1\over6}~\Big)  + 4 ~\pol(1,2,3,4) \Big(~{1\over 24}~\Big)  \Bigg)
\nonumber \\ && =  { A_{5;\langle 12 \rangle}^{\textrm{\scriptsize{MHV}}} \over 12} 
  \Big( \big\{ s_{2}s_{3} \big\}  +  \pol(k_1,k_2,k_3,k_4) \Big) \, ,
\eea
where, following \cite{ST1}, $\big\{ s_{2}s_{3} \big\}$ represents the sum of $s_{2}s_{3}$ and its four cyclic permutations.
Finally, plugging this expression into \eqn{oldconj} gives the following prediction for \\ $A^{tree}_{str}\left(k_1^{1234},k_2^{1234},k_3,k_4,k_5\right)\Bigl|_{\Ord(\alpha'^2)}$:
\be
A^{tree}_{str}\left(k_1^{1234},k_2^{1234},k_3,k_4,k_5\right)\Bigl|_{\Ord(\alpha'^2)}=  -\zeta(2) {A_{5;\langle 12 \rangle}^{\textrm{\scriptsize{MHV}}} \over 2 } 
  \Big( \big\{ s_{2}s_{3} \big\}  +  \pol(k_1,k_2,k_3,k_4) \Big) \, .
\ee
By comparing to the all-$n$ result for the $\Ord(\alpha'^2)$ stringy corrections given at the beginning of this section, it is clear that the prediction of the conjecture for the $\Ord(\alpha'^2)$ piece of $A^{tree}_{str}\left(k_1^{1234},k_2^{1234},k_3,k_4,k_5\right)$ is correct. It is obvious from the above analysis that we would have been unsuccessful had we performed the dimension shift operation on the expression usually associated with the five-gluon one-loop MHV amplitude,
\bea
  {- A_{5;\langle 12 \rangle}^{\textrm{\scriptsize{MHV}}} \over 2} \Bigg( s_{2}s_{3} I_4^{(1),~D=4-2\e}+s_{3}s_{4} I_4^{(2),~D=4-2\e}
  + s_{4}s_{5} I_4^{(3),~D=4-2\e} + \\ +s_{5}s_{1} I_4^{(4),~D=4-2\e}+s_{1}s_{2} I_4^{(5),~D=4-2\e}\Bigg) \, , \nonumber
\label{ordepsans}
\eea
illustrating that eq. (\ref{oldconj}) is only applicable if one works to all orders in $\e$ on the field theory side. We wish to stress that, although we find the language of~\cite{oneloopselfdual} convenient, we could have used the coefficients of the UV poles of $\Nsym$ one-loop MHV amplitudes considered in $D = 8 - 2 \e$ to define the right-hand side of \eqn{oldconj} and nothing would have changed, apart from maybe an unimportant overall minus sign.

Now, suppose we want to generalize the Stieberger-Taylor relation. One obvious question is whether we can relax their requirement that the helicity configuration on both sides of (\ref{oldconj}) be MHV. Indeed, we will see that the relation actually holds for general helicity configurations. Fortunately, Stieberger and Taylor calculated all six-point NMHV open superstring amplitudes in~\cite{ST4} (unfortunately not in a form as elegant as eq. (\ref{stringMHVsimp})). As a first check, we verified that $A_{1}^{1-{\rm loop}}\left(k_1^{1234},k_2^{1234},k_3^{1234},k_4,k_5,k_6\right)$, $A_{1}^{1-{\rm loop}}\left(k_1^{1234},k_2^{1234},k_3,k_4^{1234},k_5,k_6\right)$, and $A_{1}^{1-{\rm loop}}\left(k_1^{1234},k_2,k_3^{1234},k_4,k_5^{1234},k_6\right)$ all satisfy (\ref{oldconj}). There exists, in fact, a more general way to argue that relation (\ref{oldconj}) should be helicity-blind. Furthermore, it is possible to show that one can use all-orders-in-$\e$ $\Nsym$ Yang-Mills amplitudes to derive the $\Ord(\alpha'^3)$ stringy corrections as well. It was pointed out in~\cite{DunbarTurner} that the $\Nsym$ theory considered in $D = 8 - 2\e$ has UV divergences and that the requirements that the counterterm Lagrangian respect $\Nsym$ supersymmetry and have $d = 8$ uniquely fix it to be the $\Nsym$ supersymmetrization of ${\rm Tr}\{F^4\}$ (2nd line of eq. (\ref{nonABI})), the {\it same} operator that appears at $\Ord(\alpha'^2)$ in the non-Abelian Born-Infeld action of~\cite{T97}. Now it is clear why we found that, up to a trivial constant, one-loop $\Nsym$ gluon amplitudes dimensionally shifted to $D = 8 - 2\e$ are equal to the $\Ord(\alpha'^2)$ stringy corrections to $\Nsym$ gluon tree amplitudes: The underlying effective Lagrangians are completely determined by dimensional analysis and $\Nsym$ supersymmetry. In other words, there is only one $\Nsym$ supersymmetrizable $d = 8$ operator built out of field-strength tensors and their covariant derivatives. 

This is not, however, the end of the story. That the non-Abelian Born-Infeld action is fixed to order $\Ord(\alpha'^2)$ by $\Nsym$ supersymmetry is perhaps more widely appreciated than the fact that it is fixed to order $\Ord(\alpha'^3)$ by $\Nsym$ supersymmetry. It is highly non-trivial to prove the above claim (see~\cite{KS,Collinucci}), but it is true; there is a unique $\Nsym$ supersymmetrizable linear combination of the available $d = 10$ operators (schematically, there are only two such operators, $D^2 F^4$ and $F^5$) built out of field strength tensors and their covariant derivatives. On the field theory side, Dunbar and Turner showed that $D = 10 - 2\e$ counterterm Lagrangians are built out of (an appropriate $\Nsym$ supersymmetrization of) the $d = 10$ operators $F^5$ and $D^2 F^4$. The results of~\cite{KS,Collinucci} clearly imply that this $\Nsym$ supersymmetric linear combination, being unique, coincides with the $\Ord(\alpha'^3)$ terms in the non-Abelian Born-Infeld action (eq. (\ref{nonABI})). As an additional check, we evaluated $A_{1}^{1-{\rm loop}}\left(k_1^{1234},k_2^{1234},k_3,k_4,k_5,k_6\right)[\mu^6]\Big|_{\e \to 0}$ and $A_{1}^{1-{\rm loop}}\left(k_1^{1234},k_2^{1234},k_3,k_4,k_5,k_6\right)[\mu^6]\Big|_{\e \to 0}$ and observed that, up to an overall factor of $60 \zeta(3)$, the results obtained precisely matched the appropriate stringy corrections (eq. (\ref{stringMHVsimp})). These observations indicate that an analogous relationship, 
\be
A^{tree}_{str}\left(k_1^{h_1},\cdots,k_n^{h_n}\right)\Bigl|_{\Ord(\alpha'^3)} = 60 \zeta(3) A_{1}^{1-{\rm loop}}\left(k_1^{h_1},\cdots,k_n^{h_n}\right)[\mu^6]\Bigl|_{\e\to0}
\label{newconj}
\ee
holds in this case (again for arbitrary helicity configurations).

To summarize, we have seen that quite a bit of non-trivial information about the low-energy dynamics of open superstrings is encoded in all-orders-in-$\e$ one-loop $\Nsym$ amplitudes. At this point, one might hope that the trend continues and the stringy corrections are all somehow encoded in the $\Nsym$ theory considered in some higher dimensional spacetime. Unfortunately, there is no analog of \eqn{oldconj} and \eqn{newconj} at $\Ord(\alpha'^4)$. It is not hard to see this explicitly at the level of four-point amplitudes. 

Based on our experience so far, one might expect the four-point MHV amplitude dimensionally shifted to $D=12-2\e$ to match the $\Ord(\alpha'^4)$ stringy correction given in eq. (\ref{ST4pt}) up to a multiplicative constant. However, a short calculation shows that
\be
A_{1}^{1-{\rm loop}}(k_1^{1234},k_2^{1234},k_3,k_4)[\mu^8]\Big|_{\e \to 0} = {s t (2 s^2+s t+2 t^2)\over 840} A_{4;\langle 12 \rangle}^{\textrm{\scriptsize{MHV}}}
\ee
which does {\it not} have the same $s$ and $t$ dependence as the $\Ord(\alpha'^4)$ stringy correction,
\be
A^{tree}_{str}(k_1^{1234},k_2^{1234},k_3,k_4)\Big|_{\Ord(\alpha'^4)} = - {\zeta(4) \over 4} s t (4 s^2+s t+4 t^2) A_{4;\langle 12 \rangle}^{\textrm{\scriptsize{MHV}}} \, .
\ee

Although it was originally hoped that $\Nsym$ supersymmetry would constrain the non-Abelian Born-Infeld action to all orders in $\alpha'$, it is now clear that this already fails to work at $\Ord(\alpha'^4)$~\cite{Collinucci}. Since it is illuminating, we repeat the argument of~\cite{Collinucci}. One can easily see that there must be more than one independent $\Nsym$ superinvariant at $\Ord(\alpha'^4)$ by comparing the $\Ord(\alpha'^4)$ terms in the Abelian Born-Infeld action to the $\Ord(\alpha'^4)$ terms in the non-Abelian Born-Infeld action responsible for the $\Ord(\alpha'^4)$ piece of the four-point tree open superstring amplitude. It is clear from eq. (\ref{BI}) that the Abelian Born-Infeld action doesn't contain any derivative terms. On the other hand, operators of the form $(D F)^4$ are the only dimension ten operators which can enter into and produce the observed $\Ord(\alpha'^4)$ four-point tree-level superstring amplitude~\cite{Bilal},
\bea
A^{tree}_{str}(k_1^{1234},k_2^{1234},k_3,k_4)\Bigl|_{\Ord(\alpha'^4)} = - {\zeta(4) \over 4} s t (4 s^2+s t+4 t^2) A_{4;\langle 12 \rangle}^{\textrm{\scriptsize{MHV}}} \, .
\label{restateordalp4}
\eea
Since the $\Ord(\alpha'^4)$ terms in the Abelian Born-Infeld action form an $\Nsym$ superinvariant by themselves (since they are present in the Abelian case where no derivative terms are allowed), the linear combination of operators of the form $(D F)^4$ responsible for the above result must be part of an distinct $\Nsym$ superinvariant.

Before leaving this section, we make one further remark about our results at $\Ord(\alpha'^2)$ that is relevant to $n$-gluon scattering. One might expect that the stringy corrections at this order in $\alpha'$ would obey a photon-decoupling relation exactly like the one in pure Yang-Mills at tree level, where replacing a single gluon by a photon produces a vanishing result. This turned out to be too simplistic. The ${\rm Tr}\{F^4\}$ operator that governs the dynamics at this order in $\alpha'$ can, in fact, couple one or even two photons to gluons. However, once you have replaced at least three external photons, the matrix elements do vanish, so long as at least one of the gluons touching the insertion of ${\rm Tr}\{F^4\}$ is off-shell. Figure \ref{ThreePhotonFigure} illustrates how this works. 

For example, replacing, for sake of argument, gluons $k_2^{1234}$, $k_3$, and $k_4$ by photons results in the identity
\bea
&& 0 = A_{str}^{tree}(k_1^{1234},k_2^{1234},k_3,k_4,k_5)\Big|_{\Ord(\alpha'^2)}+A_{str}^{tree}(k_1^{1234},k_2^{1234},k_4,k_3,k_5)\Big|_{\Ord(\alpha'^2)}
\nonumber \\ && + A_{str}^{tree}(k_1^{1234},k_3,k_4,k_2^{1234},k_5)\Big|_{\Ord(\alpha'^2)}+A_{str}^{tree}(k_1^{1234},k_3,k_2^{1234},k_4,k_5)\Big|_{\Ord(\alpha'^2)}
\nonumber \\ && + A_{str}^{tree}(k_1^{1234},k_4,k_3,k_2^{1234},k_5)\Big|_{\Ord(\alpha'^2)}+A_{str}^{tree}(k_1^{1234},k_4,k_2^{1234},k_3,k_5)\Big|_{\Ord(\alpha'^2)}
\nonumber \\ && + A_{str}^{tree}(k_1^{1234},k_2^{1234},k_3,k_5,k_4)\Big|_{\Ord(\alpha'^2)}+A_{str}^{tree}(k_1^{1234},k_2^{1234},k_4,k_5,k_3)\Big|_{\Ord(\alpha'^2)}
\nonumber \\ && + A_{str}^{tree}(k_1^{1234},k_3,k_4,k_5,k_2^{1234})\Big|_{\Ord(\alpha'^2)}+A_{str}^{tree}(k_1^{1234},k_3,k_2^{1234},k_5,k_4)\Big|_{\Ord(\alpha'^2)}
\nonumber \\ && + A_{str}^{tree}(k_1^{1234},k_4,k_3,k_5,k_2^{1234})\Big|_{\Ord(\alpha'^2)}+A_{str}^{tree}(k_1^{1234},k_4,k_2^{1234},k_5,k_3)\Big|_{\Ord(\alpha'^2)}
\nonumber \\ && + A_{str}^{tree}(k_1^{1234},k_2^{1234},k_5,k_3,k_4)\Big|_{\Ord(\alpha'^2)}+A_{str}^{tree}(k_1^{1234},k_2^{1234},k_5,k_4,k_3)\Big|_{\Ord(\alpha'^2)}
\nonumber \\ && + A_{str}^{tree}(k_1^{1234},k_3,k_5,k_4,k_2^{1234})\Big|_{\Ord(\alpha'^2)}+A_{str}^{tree}(k_1^{1234},k_3,k_5,k_2^{1234},k_4)\Big|_{\Ord(\alpha'^2)}
\nonumber \\ && + A_{str}^{tree}(k_1^{1234},k_4,k_5,k_3,k_2^{1234})\Big|_{\Ord(\alpha'^2)}+A_{str}^{tree}(k_1^{1234},k_4,k_5,k_2^{1234},k_3)\Big|_{\Ord(\alpha'^2)}
\nonumber \\ && + A_{str}^{tree}(k_1^{1234},k_5,k_2^{1234},k_3,k_4)\Big|_{\Ord(\alpha'^2)}+A_{str}^{tree}(k_1^{1234},k_5,k_2^{1234},k_4,k_3)\Big|_{\Ord(\alpha'^2)}
\nonumber \\ && + A_{str}^{tree}(k_1^{1234},k_5,k_3,k_4,k_2^{1234})\Big|_{\Ord(\alpha'^2)}+A_{str}^{tree}(k_1^{1234},k_5,k_3,k_2^{1234},k_4)\Big|_{\Ord(\alpha'^2)}
\nonumber \\ && + A_{str}^{tree}(k_1^{1234},k_5,k_4,k_3,k_2^{1234})\Big|_{\Ord(\alpha'^2)}+A_{str}^{tree}(k_1^{1234},k_5,k_4,k_2^{1234},k_3)\Big|_{\Ord(\alpha'^2)}\, {\rm .}
\eea

An immediate out-growth of our three-photon decoupling relation for ${\rm Tr}\{F^4\}$ matrix elements is a plausible explanation of the observation~\cite{allplus} that, for the all-plus helicity configuration at one loop in pure Yang-Mills, replacing three gluons by photons always gives zero for the five- and higher-point amplitudes. Stieberger and Taylor showed that MHV ${\rm Tr}\{F^4\}$ matrix elements are closely related to the all-plus one-loop pure Yang-Mills amplitudes and, therefore, it is reasonable to expect the photon-decoupling identity discussed above for ${\rm Tr}\{F^4\}$ matrix elements to carry over to the all-plus one-loop pure Yang-Mills amplitudes as well.
\FIGURE{
\resizebox{0.70\textwidth}{!}{\includegraphics{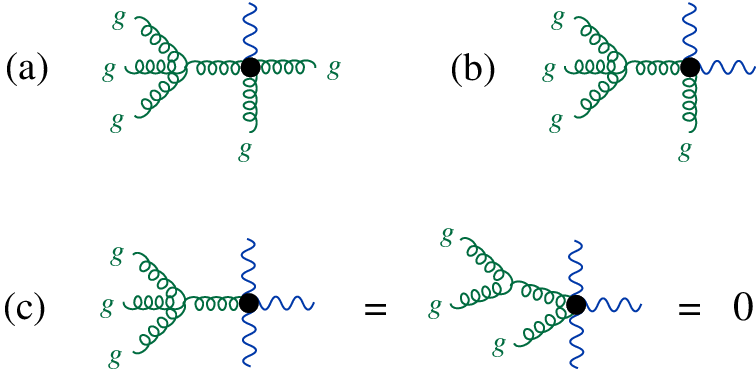}}
\vbox{\vskip 1 in}
\caption{Matrix elements of ${\rm Tr}\{F^4\}$ for a number of gluons and 
(a) a single photon, or (b) two photons, can be non-vanishing, as explained
  above.  On the other hand, matrix elements with (c) three or more
  photons have to vanish (except for the case $n=4$ with four photons).}
\label{ThreePhotonFigure}}
\section{On-Shell Superspace and All-Orders One-Loop $\Nsym$ Superamplitudes}
\label{supercomp}
So far in this paper, $\Nsym$ supersymmetry has played a somewhat peripheral role in that all results have been presented in component form and we have rather arbitrarily focused on the computation of $n$-gluon scattering amplitudes. In this section we discuss a powerful formalism that unifies all amplitudes with a given $k$-charge\footnote{Recall that, so far, we have defined the $k$-charge of an amplitude operationally as how many complete copies of the set $\{1,2,3,4\}$ appear in the helicity labels of the amplitude's external lines ({\it e.g.} $\a(k_1,k_2^{123},k_3^4,k_4,k_5^{12},k_6^{34})$ has $k$-charge two).}. We will see that this is naturally accomplished by introducing auxiliary Grassmann variables. Essentially, the point of introducing these variables is that it facilitates a better understanding of the symmetries of the S-matrix and allows one to make more of them manifest. The goal is to replace the scattering amplitudes studied so far (amplitudes with a particular $SU(4)_R$ index structure) with $\Nsym$ {\it superamplitudes}, $SU(4)_R$ invariant objects that contain as components all $\Nsym$ amplitudes of a given multiplicity, $n$. Writing results in terms of superamplitudes will also allow us to give a less heuristic definition of $k$-charge. Indeed, this section should make it much more clear why we use $SU(4)_R$ indices to label component $\Nsym$ scattering amplitudes. 
\subsection{General Discussion of $\Nsym$ On-Shell Superspace}
\label{gendisonshell}
The usual construction of the massless $\Nsym$ supermultiplet begins with the anticommutator of the supercharges 
\bea
\{Q^a_{~\alpha}, \bar{Q}_{b~\dot{\alpha}}\} &=& \D_b^{\,~a}~ P_{\alpha \dot{\alpha}}\elale
 \D_b^{\,~a}~ \lambda_{\alpha} \tilde{\lambda}_{\dot{\alpha}}\,,
\label{fundsusyalg}
\eea
in which one chooses $p^\mu = ({\bf p},0,0,{\bf p})$ to define a preferred reference frame. It then follows that some of the supercharges anticommute with themselves and everything else in this frame and others act as creation ($\bar{Q}_{b~\dot{1}}$) or annihilation ($Q^a_{~1}$) operators on the space of states. This approach is useful for some purposes ({\it e.g.} determining the particle content of the massless supermultiplet) but to describe scattering it is better to try and build a formalism where the supercharges act as creation and annihilation operators on the space of states in a manifestly Lorentz covariant way.\footnote{This alternative approach is not new~\cite{Nair,Sokatchev}, but its power was not properly appreciated until very recently~\cite{DHKSdualconf,EFK,SUSYBCFW}.} Our goal is readily accomplished if we introduce a set of four Grassmann variables, $\{\eta_\ell^1,\eta_\ell^2,\eta_\ell^3,\eta_\ell^4\}$, for each external four-vector, $p_\ell$, in the problem. Then one can easily see that (suppressing the $\ell$ label for now)
\be
Q^a_{~\alpha}= \lambda_{\alpha} \eta^a \qquad {\rm and} \qquad \bar{Q}_{b\,\dot{\alpha}} = \tilde{\lambda}_{\dot{\alpha}} {\partial \over \partial \eta^b}
\label{susycharges}
\ee
together satisfy (\ref{fundsusyalg}). Furthermore, the introduction of the $\eta^a$ allows one to build a super wavefunction (Grassmann coherent state) for each external line
\begin{eqnarray}
  \Phi(p,\eta) &=& G^{+}(p) + \G_a(p) \eta^a  + \frac{1}{2!2!}\e_{a b c d}S^{a c}(p)\eta^b \eta^d 
  + \frac{1}{3!} \e_{a b c d}\bar{\G}^{a}(p)\eta^b\eta^c\eta^d  \el
    + \frac{1}{4!}\e_{a b c d} G^{-}(p) \eta^a\eta^b\eta^c \eta^d\,.
  \label{supwvfun}
\end{eqnarray}
which makes it possible to consider $\Nsym$ scattering with half of the supersymmetries (the $Q^a_{~\alpha}$ supercharges which we have chosen to implement multiplicatively) made manifest. To convince the reader that (\ref{susycharges}) and (\ref{supwvfun}) make sense, we must construct the covariant analogs of $Q^i_{~1}$ and $\bar{Q}_{j~\dot{1}}$ in the traditional, non-covariant approach ({\it i.e.} we need to identify the relevant creation operators). In fact, given that $Q^a_{~\alpha} \lambda^\alpha = 0$ and $\bar{Q}_{b~\dot{\alpha}}\tilde{\lambda}^{\dot{\alpha}} = 0$ (the supercharges only have components parallel to $\lambda^\alpha$ and $\tilde{\lambda}^{\dot{\alpha}}$), we can read off the analogs of $Q^i_{~1}$ and $\bar{Q}_{j~\dot{1}}$: the annihilation and creation operators are simply the components of the supercharges along the directions of the spinors, $\hat{a}^c \equiv Q^c = \eta^c$ and $\hat{a}_d^{\dagger} \equiv \bar{Q}_{d} = \partial/\partial \eta^d$, and they satisfy the algebra
\be
\{Q^c, \bar{Q}_{d}\} = \D^{~\,c}_{d} \,{\rm .}
\ee
Now that we know what the creation operators are we can act on the super wavefunction (\ref{supwvfun}) in various combinations. All that we have to do to show that (\ref{supwvfun}) is complete and correct is find some combination of creation operators ($\eta$ derivatives) that isolate each term in the super wavefunction. Following~\cite{DHKSdualconf,origElvang}, we have 
\bea
&& \Phi(p,\eta)\Big|_{\eta^n = 0} = G^{+}(p) \qquad \bar{Q}_{a} \Phi(p,\eta)\Big|_{\eta^{n} = 0} = \G_{a}(p) \qquad  {1\over 2!}\bar{Q}_{a} \bar{Q}_{b} \e^{a b c d} \Phi(p,\eta)\Big|_{\eta^n = 0} = S^{c d}(p)
  \el {1\over 3!} \bar{Q}_{a} \bar{Q}_{b} \bar{Q}_{c} \e^{a b c d} \Phi(p,\eta)\Big|_{\eta^n = 0} = \bar{\G}^{d}(p) \qquad 
    {1\over 4!} \bar{Q}_{a} \bar{Q}_{b} \bar{Q}_{c} \bar{Q}_{d} \e^{a b c d} \Phi(p,\eta)\Big|_{\eta^n = 0} = G^{-}(p) \,{\rm .}\nn
\eea

Evidently, our on-shell superspace construction is well-defined and it therefore makes sense to speak about $\Nsym$ on-shell superamplitudes, $\a(\Phi_1,\cdots, \Phi_n)$, that take into account all elements of the planar\footnote{Clearly, at the moment, this is a choice we are making since supersymmetry commutes with color.} S-matrix with $n$ external states simultaneously. The $n$-point superamplitude is naturally expanded into $k$-charge sectors as\footnote{As discussed in Appendix \ref{SWI}, the $k = 0$ and $k = 1$ sectors (and by parity the $k = n - 1$ and $k = n$ sectors) are identically zero for non-degenerate kinematical configurations.}
\begin{equation}
    \a(\Phi_1,\cdots, \Phi_n) = \a_{n;2} + \a_{n;3} + \cdots + \a_{n;\, n-2} \,{\rm .}
    \label{kexpans}
\end{equation}
So far, we have defined $k$-charge at the level of component amplitudes. For example, $A(p_1^{1234},p_2^{1234},p_3,p_4)$ and $A(p_1^{1234},p_2^{123},p_3^{4},p_4)$ both have $k$-charge two because one needs two copies of $\{1,2,3,4\}$ to label their external states. At the level of the superamplitude, the $k$-charge of a given term on the right-hand side of (\ref{kexpans}) is determined by the number of Grassmann parameters that appear in it divided by four\footnote{The $SU(4)_R$ rotates the Grassmann parameters into each other and the superamplitude must be a singlet under R-symmetry transformations. This is impossible unless, for a given term, each $SU(4)$ index, $\{1,2,3,4\}$, appears the same number of times.}. We will refer to $\a_{n;k}$ (a $k$-charge sector of the superamplitude) as a superamplitude since there is usually no possibility of confusion. 

We now turn to the MHV tree-level superamplitude, $\a_{n;2}^{tree}$, which has the simplest superspace structure and can be completely determined by matching onto the Parke-Taylor formula (or any other component amplitude for that matter). Clearly, the simplest possible superspace structure is given by the eight-fold Grassmann delta function itself and corresponds to the first term on the right-hand side of (\ref{kexpansD}),
\be
\a_{n;2}^{tree} = {1\over 16}\prod_{a = 1}^4 \sum_{i,j = 1}^n \spa{i}.j \eta^a_i \eta^a_j~~ \hat{\a}_{n;2}^{tree} \,{\rm ,}
\label{undetMHVsup}
\ee
where we have used the explicit formula for the Grassmann delta function $\D^{(8)}\left(Q^{a\,\A}\right)$ derived in Appendix \ref{SWI}. Suppose we are interested in computing $A^{tree}\left(p_1^{1234},p_2^{1234},p_3,\cdots,p_n\right)$ using $\a_{n;2}^{tree}$. To compute this amplitude one expands (\ref{undetMHVsup}) and extracts the coefficient of $\eta_1^1\eta_1^2\eta_1^3\eta_1^4\eta_2^1\eta_2^2\eta_2^3\eta_2^4$. We will denote this combination as $\eta_1^{1234} \eta_2^{1234}$. The result of this calculation is the numerator of the familiar  Parke-Taylor amplitude times $\hat{\a}_{n;2}^{tree}$:
\be
A^{tree}\left(p_1^{1234},p_2^{1234},p_3,\cdots,p_n\right) = \spa1.2^4~~ \hat{\a}_{n;2}^{tree} \,{\rm .}
\ee
It follows that
\be
\hat{\a}_{n;2}^{tree} = {i \over \spa1.2 \spa2.3 \cdots \spa{n}.1}
\ee
and
\be
\a_{n;2}^{tree} = i{{1\over 16} \prod_{a = 1}^4 \sum_{i,j = 1}^n \spa{i}.j \eta^a_i \eta^a_j \over \spa1.2 \spa2.3 \cdots \spa{n}.1} \,{\rm .}
\label{MHVsup}
\ee
We have successfully given a unified description of all MHV tree amplitudes in $\Nsym$. In fact, for appropriate supersymmetry-preserving variants of dimensional regularization\footnote{We refer the interested reader to Subappendix \ref{4DHS}, where we describe the four dimensional helicity scheme, the particular variant used in most multi-loop studies of $\Nsym$ scattering amplitudes.}, it turns out that the superspace structure in the MHV case is independent of the loop expansion~\cite{oneloopselfdual,dualSmat} and we can write
\be
\a_{n;2} = i{{1\over 16} \prod_{a = 1}^4 \sum_{i,j = 1}^n \spa{i}.j \eta^a_i \eta^a_j \over \spa1.2 \spa2.3 \cdots \spa{n}.1}\Bigg(1+\left({g^2 \Nc \mu^{2\e} e^{-\gamma_E \e} \over (4 \pi)^{2-\e}}\right) M_{1-{\rm loop}}+ \left({g^2 \Nc \mu^{2\e} e^{-\gamma_E \e}\over (4 \pi)^{2-\e}}\right)^2 M_{2-{\rm loop}}+\cdots\Bigg)
\label{MHVsupL}
\ee
as well, although the determination of $M_{L-{\rm loop}}$ may be quite non-trivial\footnote{It is important to point out here that there is a natural seperation of the $M_{L-{\rm loop}}$ functions into even and odd components.}. In the above we still suppress the tree-level gauge coupling and color structure, worrying only about relative factors between different loop orders.

Before moving on to more non-trivial examples, we point out an important subtlety related to the special case of $n = 3$ in eq. (\ref{MHVsup}). Our experience with scattering amplitudes suggests that we should also be able to define the MHV and anti-MHV three-point superamplitudes, $\a_{3;2}$ and $\bar{\a}_{3;2}$, even though, na\"{i}vely, it would appear that any superamplitude with four Grassmann variables in each term must vanish due to supercharge conservation (Appendix \ref{SWI}). After all, the Grassmann polynomial that expresses supercharge conservation already has eight Grassmann variables in each term. There is a way out, however, if one allows for degenerate kinematics. For three-point kinematics we have $p_1 = - p_2 - p_3$ which implies that $0 = p_1^2 = \spa2.3 \spb3.2$. Making the choice $\spb2.3 = 0$ and $\spa3.2 \neq 0$ leads to the three-point MHV superamplitude (setting $n = 3$ in eq. (\ref{MHVsup})) and making the choice $\spa2.3 = 0$ and $\spb3.2 \neq 0$ will lead us to the anti-MHV three-point superamplitude. Setting all of the holomorphic spinor products to zero in the three-point amplitude forces all of the holomorphic spinors to be proportional to one another. Quantitatively, this means that $\lambda_{\alpha\,\ell} = c_{\ell} \chi_\alpha$ for some spinor $\chi_\alpha$ and coefficients $c_{\ell}$. Consequently, the total supercharge can be written as $Q^a_{~\,\alpha} = \chi_\alpha \sum_{\ell = 1}^3 c_\ell \eta^a_\ell$. The point is that now the $\alpha$ dependence just sits in the spinor $\chi_\alpha$ which pulls out of the sum over $\ell$. Overall factors inside delta functions can't lead to non-trivial constraints and it follows that the overall supercharge conserving delta function is only four-fold in this special case. It is very instructive to realize this discussion explicitly and determine the superspace structure of $\bar{\a}_{3;2}$. 

We start with momentum conservation 
\be
\lambda_{\alpha\,1}\, \tilde{\lambda}_{\dot{\alpha}\,1}+\lambda_{\alpha\,2}\, \tilde{\lambda}_{\dot{\alpha}\,2}+\lambda_{\alpha\,3}\, \tilde{\lambda}_{\dot{\alpha}\,3} = 0
\ee
and project by taking the spinor product of this equation with, say, $\tilde{\lambda}_{\dot{\alpha}\,1}$:
\be
\lambda_{\alpha\,2}\spb2.1 + \lambda_{\alpha\,3}\spb3.1 = 0 \,{\rm .}
\ee
Permuting labels, we can also write
\be
\lambda_{\alpha\,1}\spb1.3 + \lambda_{\alpha\,2}\spb2.3 = 0 \,{\rm .}
\ee
Solving for $\lambda_{\alpha\,1}$ and $\lambda_{\alpha\,3}$ in terms of $\lambda_{\alpha\,2}$, we find
\be
Q^a_{~\,\alpha} = \sum_{\ell = 1}^3 \lambda_{\alpha\,\ell}\,\eta^a_\ell = -\lambda_{\alpha\,2}{\spb2.3 \eta_1^a \over \spb1.3} + \lambda_{\alpha\,2} \eta_2^a -\lambda_{\alpha\,2}{\spb2.1 \eta_3^a \over \spb3.1} = \lambda_{\alpha\,2}\left({\spb2.3 \eta^a_1 + \spb3.1\, \eta^a_2 + \spb1.2 \eta^a_3 \over \spb3.1}\right)
\ee
and the arguments of the last paragraph imply that the superspace structure of $\bar{\a}_{3;2}$ is simply
\bea
\D^{(4)}\left(\spb2.3 \eta^a_1 + \spb3.1\, \eta^a_2 + \spb1.2\, \eta^a_3\right) &=& \left(\spb2.3 \eta^1_1 + \spb3.1\, \eta^1_2 + \spb1.2\, \eta^1_3\right)\left(\spb2.3 \eta^2_1 + \spb3.1\, \eta^2_2 + \spb1.2\, \eta^2_3\right) \times 
\el \times \left(\spb2.3 \eta^3_1 + \spb3.1\, \eta^3_2 + \spb1.2 \,\eta^3_3\right)\left(\spb2.3 \eta^4_1 + \spb3.1\, \eta^4_2 + \spb1.2 \,\eta^4_3\right) 
\elale \prod_{a = 1}^4 \left(\spb2.3 \,\eta^a_1 + \spb3.1\, \eta^a_2 + \spb1.2\, \eta^a_3\right)\,{\rm .}\nn
\eea

Now that the superspace structure of $\bar{\a}_{3;2}$ is fixed, it is a simple matter to match onto, say, the anti-MHV three-gluon amplitude and determine the entire superamplitude. As it stands, we have 
\be
\bar{\a}_{3;2} = \bar{\a}_{3;2}^\prime \prod_{a = 1}^4 \left(\spb2.3 \,\eta^a_1 + \spb3.1\, \eta^a_2 + \spb1.2\, \eta^a_3\right)
\label{partanti3}
\ee
Expanding out both sides of eq. (\ref{partanti3}) for, say, $A^{tree}\left(k_1^{1234}, k_2, k_3\right)$ and extracting the coefficient of $\eta_1^{1234}$, we find
\bea
A^{tree}\left(k_1^{1234}, k_2, k_3\right) &=& \bar{\a}_{3;2}^\prime \prod_{a = 1}^4 \left(\spb2.3\, \eta^a_1 + \spb3.1\, \eta^a_2 + \spb1.2\, \eta^a_3\right)\nonumber \\
{i \spb2.3^4 \over \spb2.3 \spb3.1 \spb1.2} &=& \bar{\a}_{3;2}^\prime \spb2.3^4 \nonumber \\
{i \over \spb2.3 \spb3.1 \spb1.2} &=& \bar{\a}_{3;2}^\prime \,{\rm .}
\eea
Thus, we finally have
\be
\bar{\a}_{3;2} = i {\prod_{a = 1}^4 \left(\spb2.3 \,\eta^a_1 + \spb3.1\, \eta^a_2 + \spb1.2\, \eta^a_3\right) \over \spb1.2 \spb2.3 \spb3.1}\,{\rm .}
\ee

As we will see, the superspace structure of the six-point NMHV superamplitude is in some sense built out of pieces of $\bar{\a}_{3;2}$. The superamplitude $\a_{6;3}$ is of particular interest for us because it represents the desired supersymmetrization of the results derived in Section \ref{gluoncomp}. To proceed, we need to know the superspace structure of $\a_{6;3}$ and which component amplitudes are required to nail it down. Happily, this difficult problem was solved in a recent paper by Elvang, Freedman, and Kiermaier~\cite{EFK}. Here we simply present their formula for $\a_{6;3}$ and refer the reader interested in the derivation to \ref{superNMHV}. The result is
\bea
&&\a_{6;\,3}^{1-{\rm loop}} = {\D^{(8)}(Q^{a\,\A}) \over \spb3.4^4 \spa5.6^4}\Big(A_1^{1-{\rm loop}}\left(p_1^{1234}, p_2, p_3, p_4, p_5^{1234}, p_6^{1234}\right) \prod_{a=1}^4 \left(\spb3.4~\eta_1^a+\spb4.1 ~\eta_3^a +\spb{1}.3~ \eta_4^a\right)
\el+ A_1^{1-{\rm loop}}\left(p_1^{123},p_2^4,p_3,p_4,p_5^{1234},p_6^{1234}\right) \prod_{a=1}^3 \left(\spb3.4~\eta_1^a+\spb4.1 ~\eta_3^a +\spb{1}.3~ \eta_4^a\right)\left(\spb3.4~\eta_2^4+\spb4.2 ~\eta_3^4 +\spb{2}.3~ \eta_4^4\right)
\el+ A_1^{1-{\rm loop}}\left(p_1^{12},p_2^{34},p_3,p_4,p_5^{1234},p_6^{1234}\right)\prod_{a=1}^2 \left(\spb3.4~\eta_1^a+\spb4.1 ~\eta_3^a +\spb{1}.3~ \eta_4^a\right)\prod_{a=3}^4\left(\spb3.4~\eta_2^a+\spb4.2 ~\eta_3^a +\spb{2}.3~ \eta_4^a\right)
\el + A_1^{1-{\rm loop}}\left(p_1^1,p_2^{234},p_3,p_4,p_5^{1234},p_6^{1234}\right) \left(\spb3.4~\eta_1^1+\spb4.1 ~\eta_3^1 +\spb{1}.3~ \eta_4^1\right)\prod_{a=2}^4\left(\spb3.4~\eta_2^a+\spb4.2 ~\eta_3^a +\spb{2}.3~ \eta_4^a\right)
\el + A_1^{1-{\rm loop}}\left(p_1, p_2^{1234}, p_3, p_4, p_5^{1234}, p_6^{1234}\right) \prod_{a=1}^4\left(\spb3.4~\eta_2^a+\spb4.2 ~\eta_3^a +\spb{2}.3~ \eta_4^a\right)\Big)\,{\rm .}\nn
\label{A63}
\eea
It is also well-known that, when used in conjunction with the four dimensional helicity scheme~\cite{4dimHS}, results derived from supersymmetry such as the above hold order-by-order in the dimensional regularization parameter. If we want to use the above formula to give a supersymmetrized version of the all-orders-in-$\e$ six-point NMHV superamplitude, we have to somehow determine the pentagon coefficients of  $A_1^{1-{\rm loop}}\left(p_1^{123},p_2^4,p_3,p_4,p_5^{1234},p_6^{1234}\right)$, $A_1^{1-{\rm loop}}\left(p_1^{12},p_2^{34},p_3,p_4,p_5^{1234},p_6^{1234}\right)$, and  $A_1^{1-{\rm loop}}\left(p_1^1,p_2^{234},p_3,p_4,p_5^{1234},p_6^{1234}\right)$ (the box coefficients for NMHV amplitudes with pairs of scalars or fermions were already computed in~\cite{Risager} or can be deduced from~\cite{Huang}). In this paper we are focused on computing the unknown pentagon coefficients and we will therefore suppress the box contributions to $\Nsym$ amplitudes throughout the rest of this work except when it is desirable to include them for pedagogical purposes.

Now, there are two ways that we could try and go forward. One would be to try and match the superamplitude onto three different pure-gluon amplitudes that we have already computed and solve a linear system to determine the eighteen unknown pentagon coefficients (one system of three equations in three unknowns suffices to determine all eighteen unknowns by virtue of the leading singularity method). Alternatively, we could try and determine the coefficients directly by combining the leading singularity method with $D$ dimensional unitarity, as was done in Section \ref{gluoncomp} for six-gluon amplitudes. Either way we will have to simplify the rather complicated results obtained using BCFW shifts. In this section we try both approaches although it would probably be easier to use the first to determine everything. In Subsection  \ref{ffbarcalc}, we compute amplitudes with fermion/anti-fermion pairs directly because, if we'd like to claim that our approach to $D$ dimensional integrand reconstruction is applicable in principle to one-loop QCD calculations, it is important to see that the calculational method discussed in Section \ref{gluoncomp} can handle amplitudes where some of the external gluons have been replaced by fermions. For the final amplitude with a scalar/anti-scalar pair we obtain the result in Subsection \ref{ssbarcalc} by matching onto the result we derived in Section \ref{gluoncomp} for $A_1^{1-{\rm loop}}\left(p_1^{1234},p_2,p_3^{1234},p_4,p_5^{1234},p_6\right)$.
\subsection{NMHV Amplitudes With a Fermion/Anti-Fermion Pair}
\label{ffbarcalc}
The approach that we use in this subsection to calculate the higher-order in $\e$ pentagon coefficients for $A_1^{1-{\rm loop}}\left(k_1^{1234},k_2^{1234},k_3^{123},k_4^{4},k_5,k_6\right)$ is completely analogous to the approach outlined in Subsection \ref{effgcomp} for six-gluon amplitudes. There are, of course, a few minor differences. The diagrammatics are a bit less obvious (see Figure \ref{ffbardiags}) and some of the three-point tree amplitudes used to reconstruct the one-loop integrands are different. We will need the three-point vertex for a fermion, anti-fermion, and a complex scalar in $D$ dimensions. This is just a Yukawa coupling:
\be
A^{tree}_{\mu^2}(-p_{\bar{f}},k_1^{12},(p-k_1)_f) = i \bar{u}(p)u(p-k_1) \,{\rm .}
\ee
As the attentive reader may have suspected, we are actually going to compute \\$A_1^{1-{\rm loop}}\left(k_1^{1234},k_2^{1234},k_3^{1},k_4^{234},k_5,k_6\right)$ instead of $A_1^{1-{\rm loop}}\left(k_1^{1},k_2^{234},k_3,k_4,k_5^{1234},k_6^{1234}\right)$ to make life as easy as possible. If we compute $A_1^{1-{\rm loop}}\left(k_1^{1234},k_2^{1234},k_3^{1},k_4^{234},k_5,k_6\right)$, it will turn out that we can recycle the four-vertices (eqs. (\ref{mu4tree1}) - (\ref{mu4tree3})) that we used in our six-gluon calculations. All we have to do is rewrite eq. (\ref{A63}) with shifted component amplitudes. Going through the proof in \ref{superNMHV}, we see that this is an entirely straightforward exercise and we arrive at
\FIGURE{
\resizebox{.95\textwidth}{!}{\includegraphics{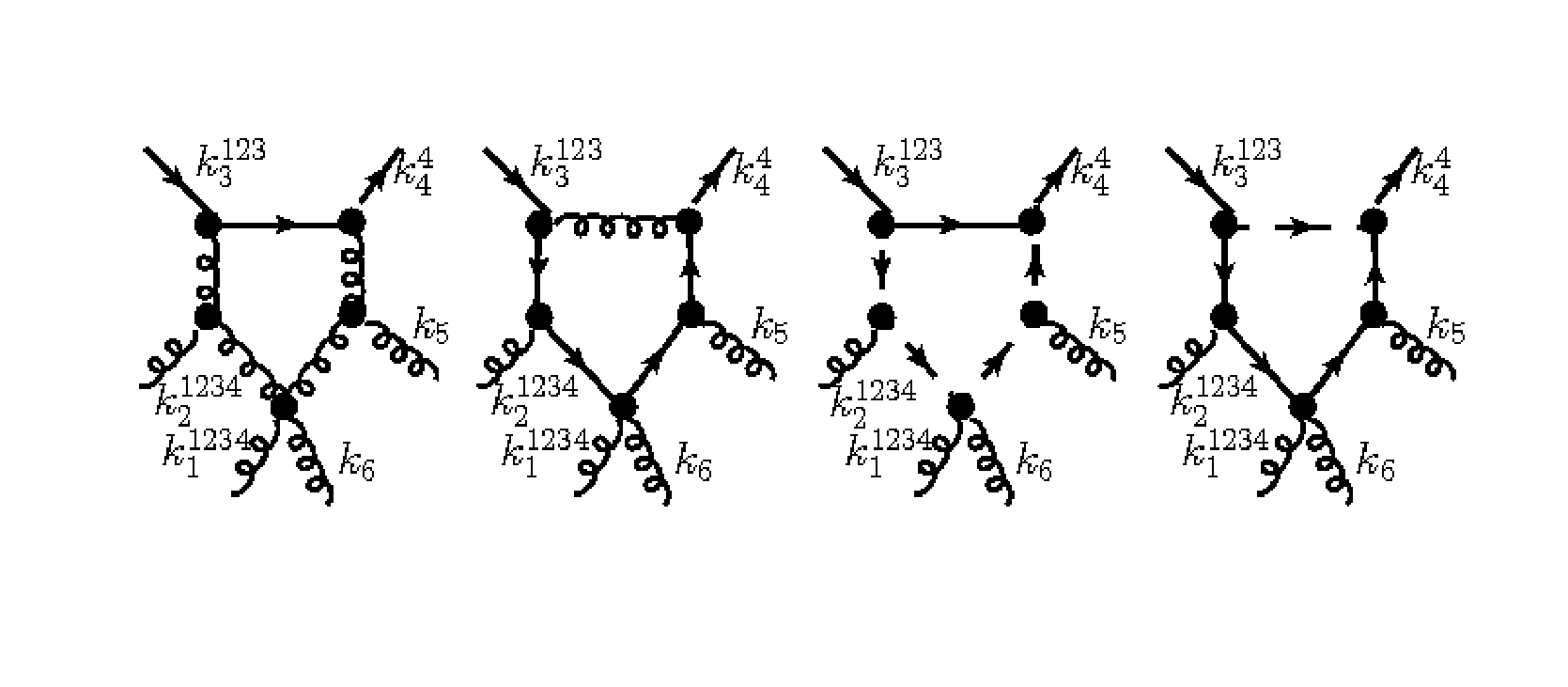}}
\caption{The generalized unitarity diagrams contributing to the $I_5^{(1)}$ pentagon coefficient of $A_1^{1-{\rm loop}}\left(k_1^{1234},k_2^{1234},k_3^{123},k_4^{4},k_5,k_6\right)$.}
\label{ffbardiags}}
\bea
&&\a_{6;\,3}^{1-{\rm loop}} = {\D^{(8)}(Q^{a\,\A}) \over \spb5.6^4 \spa1.2^4}\Big(A_1^{1-{\rm loop}}\left(k_1^{1234}, k_2^{1234}, k_3^{1234}, k_4, k_5, k_6\right) \prod_{a=1}^4 \left(\spb5.6~\eta_3^a+\spb6.3 ~\eta_5^a +\spb{3}.5~ \eta_6^a\right)
\el+ A_1^{1-{\rm loop}}\left(k_1^{1234},k_2^{1234},k_3^{123},k_4^4,k_5,k_6\right) \prod_{a=1}^3 \left(\spb5.6~\eta_3^a+\spb6.3 ~\eta_5^a +\spb{3}.5~ \eta_6^a\right)\left(\spb5.6~\eta_4^4+\spb6.4 ~\eta_5^4 +\spb{4}.5~ \eta_6^4\right)
\el+ A_1^{1-{\rm loop}}\left(k_1^{1234},k_2^{1234},k_3^{12},k_4^{34},k_5,k_6\right)\prod_{a=1}^2 \left(\spb5.6~\eta_3^a+\spb6.3 ~\eta_5^a +\spb{3}.5~ \eta_6^a\right)\prod_{a=3}^4\left(\spb5.6~\eta_4^a+\spb6.4 ~\eta_5^a +\spb{4}.5~ \eta_6^a\right)
\el + A_1^{1-{\rm loop}}\left(k_1^{1234},k_2^{1234},k_3^{1},k_4^{234},k_5,k_6\right) \left(\spb5.6~\eta_3^1+\spb6.3 ~\eta_5^1 +\spb{3}.5~ \eta_6^1\right)\prod_{a=2}^4\left(\spb5.6~\eta_4^a+\spb6.4 ~\eta_5^a +\spb{4}.5~ \eta_6^a\right)
\el + A_1^{1-{\rm loop}}\left(k_1^{1234},k_2^{1234},k_3,k_4^{1234},k_5,k_6\right) \prod_{a=1}^4\left(\spb5.6~\eta_4^a+\spb6.4 ~\eta_5^a +\spb{4}.5~ \eta_6^a\right)\Big)
\label{A63s}
\eea
with very little effort.

Also, when we say compute the amplitude, what we really mean is compute the pentagon coefficient of $I_5^{(1);~D = 6 - 2\e}$. Once this function is determined, the leading singularity equations give the rest of the pentagon coefficients for free. We denote the first diagram in Figure \ref{ffbardiags} $\mathcal{M}_I$, the second $\mathcal{M}_{II}$ and so forth. We present each of these contributions in turn:
\begin{changemargin}{-.6 in}{0 in}
\bea
&&A_{1;\,\mathcal{M}_I}^{1-{\rm loop}}\left(k_1^{1234},k_2^{1234},k_3^{123},k_4^4,k_5,k_6\right)\Big|_{I_5^{(1)}} = -\left({-i \over \sqrt{2}}\right)^2 
\el\pol^\alpha(p_*+k_5+k_6)[4|\gamma_\alpha u(p_*-k_1-k_2-k_3) \bar{u}(p_*-k_1-k_2-k_3)\gamma_\beta |3\rangle\pol^{*\,\beta}(p_*-k_1-k_2)
\el i \sqrt{2}\pol^{*\,\rho_5}(p_*+k_5+k_6) \left(\pol^+(k_5)\cdot (p_*+k_5+k_6)~ g_{\rho_5 \sigma_5}+k_{5\,\rho_5}~\pol^+_{\sigma_5}(k_5)-k_{5\,\sigma_5}~ \pol^+_{\rho_5}(k_5)\right)\pol^{\sigma_5}(p_*+k_6)
\el \pol^{*\,\sigma}(p_*+k_6) {-2 i \over \spab1.{p_*}.1}\Big(\pol^-(k_1)\cdot p_* ~\pol^+(k_6)\cdot p_* ~g_{\rho \sigma}+ \pol^+(k_6)\cdot p_* ~k_{1\,\sigma} \pol^-_\rho(k_1) 
\nonumber\\&&-\pol^+(k_6)\cdot p_* ~k_{1\,\rho}~ \pol^-_\sigma(k_1) + \pol^-(k_1)\cdot p_* ~k_{6\,\sigma}~ \pol^+_{\rho}(k_6) 
\nonumber\\&&- \pol^-(k_1)\cdot p_* ~k_{6\,\rho}~ \pol^+_{\sigma}(k_6) - k_1\cdot k_6 \pol^-_\rho(k_1)\pol^+_\sigma(k_6)\Big)\pol^\rho(p_*-k_1)
\el i \sqrt{2}\pol^{*\,\rho_2}(p_*-k_1) \left(\pol^-(k_2)\cdot (p_*-k_1)~ g_{\rho_2 \sigma_2}+k_{2\,\rho_2}~\pol^-_{\sigma_2}(k_2)-k_{2\,\sigma_2}~ \pol^-_{\rho_2}(k_2)\right)\pol^{\sigma_2}(p_*-k_1-k_2)
\elale {2 i \over \spab1.{p_*}.1} g^{\rho_5 \alpha}[4|\gamma_\alpha (\slashed{p}_*-\slashed{k}_1-\slashed{k}_2-\slashed{k}_3+\mu)\gamma_\beta |3\rangle g^{\sigma_2 \beta}
\el\left(\pol^+(k_5)\cdot (p_*+k_5+k_6)~ g_{\rho_5 \sigma_5}+k_{5\,\rho_5}~\pol^+_{\sigma_5}(k_5)-k_{5\,\sigma_5}~ \pol^+_{\rho_5}(k_5)\right)g^{\sigma_5 \sigma}
\el\Big(\pol^-(k_1)\cdot p_* ~\pol^+(k_6)\cdot p_* ~g_{\rho \sigma}+ \pol^+(k_6)\cdot p_* ~k_{1\,\sigma} \pol^-_\rho(k_1) 
\nonumber\\&&-\pol^+(k_6)\cdot p_* ~k_{1\,\rho}~ \pol^-_\sigma(k_1) + \pol^-(k_1)\cdot p_* ~k_{6\,\sigma}~ \pol^+_{\rho}(k_6) 
\nonumber\\&&- \pol^-(k_1)\cdot p_* ~k_{6\,\rho}~ \pol^+_{\sigma}(k_6) - k_1\cdot k_6 \pol^-_\rho(k_1)\pol^+_\sigma(k_6)\Big)
\el g^{\rho \rho_2} \left(\pol^-(k_2)\cdot (p_*-k_1)~ g_{\rho_2 \sigma_2}+k_{2\,\rho_2}~\pol^-_{\sigma_2}(k_2)-k_{2\,\sigma_2}~ \pol^-_{\rho_2}(k_2)\right)\,{\rm ,}\nn
\label{MI}
\eea
\end{changemargin}
\begin{changemargin}{-.6 in}{0 in}
\bea
&&A_{1;\,\mathcal{M}_{II}}^{1-{\rm loop}}\left(k_1^{1234},k_2^{1234},k_3^{123},k_4^4,k_5,k_6\right)\Big|_{I_5^{(1)}} = \left({-i \over \sqrt{2}}\right)^4 [4|\gamma_\alpha \pol^{*\,\alpha}(p_*-k_1-k_2-k_3)u(p_*+k_5+k_6)\times
\el \times  \bar{u}(p_*+k_5+k_6)\slashed{\pol}^+(k_5)u(p_*+k_6)\left({i (p_*+k_6)\cdot \pol^+(k_6) \over \spab1.{p_*}.1} \bar{u}(p_*+k_6) \slashed{\pol}^-(k_1)u(p_*-k_1)\right)\times
\el \times\bar{u}(p_*-k_1)\slashed{\pol}^-(k_2)u(p_*-k_1-k_2)\bar{u}(p_*-k_1-k_2)\gamma_\beta \pol^{\beta}(p_*-k_1-k_2-k_3) |3\rangle
\elale -{i (p_*+k_6)\cdot \pol^+(k_6) \over 4 \spab1.{p_*}.1} [4|\gamma^\beta(\slashed{p}_*+\slashed{k}_5+\slashed{k}_6+\mu)\slashed{\pol}^+(k_5)(\slashed{p}_*+\slashed{k}_6+\mu)\slashed{\pol}^-(k_1)(\slashed{p}_*-\slashed{k}_1+\mu)\times
\el\times \slashed{\pol}^-(k_2)(\slashed{p}_*-\slashed{k}_1-\slashed{k}_2+\mu)\gamma_\beta |3\rangle \,{\rm ,}\nn
\label{MII}
\eea
\end{changemargin}
\begin{changemargin}{-.6 in}{0 in}
\bea
&&A_{1;\,\mathcal{M}_{III}}^{1-{\rm loop}}\left(k_1^{1234},k_2^{1234},k_3^{123},k_4^4,k_5,k_6\right)\Big|_{I_5^{(1)}} = -(i)^2 [4|u(p_*-k_1-k_2-k_3)\bar{u}(p_*-k_1-k_2-k_3)|3\rangle \times
\el
\left(-\sqrt{2} i (p_*+k_5+k_6)\cdot \pol^+(k_5)\right) {-i \spab1.{p_*}.6^2 \over s_6 \spab1.{p_*}.1} \left(-\sqrt{2} i (p_*-k_1)\cdot \pol^-(k_2)\right)
\elale 2 i [4|\slashed{p}_*-\slashed{k}_1-\slashed{k}_2-\slashed{k}_3+\mu|3\rangle
(p_*+k_5+k_6)\cdot \pol^+(k_5){\spab1.{p_*}.6^2 \over s_6 \spab1.{p_*}.1}(p_*-k_1)\cdot \pol^-(k_2)\,{\rm ,}
\label{MIII}
\eea
\end{changemargin}
\begin{changemargin}{-.6 in}{0 in}
\bea
&&A_{1;\,\mathcal{M}_{IV}}^{1-{\rm loop}}\left(k_1^{1234},k_2^{1234},k_3^{123},k_4^4,k_5,k_6\right)\Big|_{I_5^{(1)}} = (i)^2 [4| u(p_*+k_5+k_6) \left( -{i\over \sqrt{2}}\bar{u}(p_*+k_5+k_6) \slashed{\pol}^+(k_5) u(p_*+k_6)\right) \times
\el {i (p_*+k_6)\cdot \pol^+(k_6)\over \spab1.{p_*}.1}\bar{u}(p_*+k_6)\slashed{\pol}^-(k_1) u(p_*-k_1)\left(-{i\over \sqrt{2}}\bar{u}(p_*-k_1)\slashed{\pol}^-(k_2)u(p_* - k_1 - k_2)\right) \bar{u}(p_*-k_1-k_2)|3\rangle
\el = {i (p_*+k_6)\cdot \pol^+(k_6)\over 2 \spab1.{p_*}.1}[4| (\slashed{p}_*+\slashed{k}_5+\slashed{k}_6)\slashed{\pol}^+(k_5) (\slashed{p}_*+\slashed{k}_6)\slashed{\pol}^-(k_1) (\slashed{p}_*-\slashed{k}_1)\slashed{\pol}^-(k_2)\slashed{p}_*-\slashed{k}_1-\slashed{k}_2)|3\rangle\,{\rm .}
\label{MIV}
\eea
\end{changemargin}
Combining eqs. (\ref{MI}) - (\ref{MIV}) with the appropriate multiplicities, we finally find
\cmb{-.8 in}{0 in}
\bea
&&A_{1}^{1-{\rm loop}}\left(k_1^{1234},k_2^{1234},k_3^{123},k_4^4,k_5,k_6\right)\Big|_{I_5^{(1)}} = A_{1;\,\mathcal{M}_{I}}^{1-{\rm loop}}\left(k_1^{1234},k_2^{1234},k_3^{123},k_4^4,k_5,k_6\right)\Big|_{I_5^{(1)}} 
\el+  A_{1;\,\mathcal{M}_{II}}^{1-{\rm loop}}\left(k_1^{1234},k_2^{1234},k_3^{123},k_4^4,k_5,k_6\right)\Big|_{I_5^{(1)}} 
 + 3 A_{1;\,\mathcal{M}_{III}}^{1-{\rm loop}}\left(k_1^{1234},k_2^{1234},k_3^{123},k_4^4,k_5,k_6\right)\Big|_{I_5^{(1)}} 
 \el+ 3 A_{1;\,\mathcal{M}_{IV}}^{1-{\rm loop}}\left(k_1^{1234},k_2^{1234},k_3^{123},k_4^4,k_5,k_6\right)\Big|_{I_5^{(1)}} \,{\rm .}
 \label{unsimpfermans}
\eea
\cme
As before, we should project the pentagon integrals onto the dual conformal basis using (\ref{PtoDCB}) before attempting to simplify the result. After trying all BCFW shifts, we were able to find a simple formula for the above pentagon coefficient. Suppressing the overall factor of $\e$ from (\ref{PtoDCB}), we find:
\cmb{-.8 in}{0 in}
\bea
&&K_1^{f \bar{f}} = {i \over 2}C_1 {\Big(\spab2.{1+6}.5\spab1.{2+3}.4\spab4.{1+2}.6+\spab5.{1+6}.2\spa1.2\spa2.4\spb4.5\spb5.6\Big)\over \spab2.{1+6}.5 \spab5.{1+6}.2 s_3 s_6}\times 
\el \times \Big(\spab2.{1+6}.5\spab1.{2+3}.4\spab3.{1+2}.6+\spab5.{1+6}.2\spa1.2\spa2.3\spb4.5\spb5.6 \Big)\,{\rm ,}
\label{fermcoef(1)}
\eea
\cme
where the $C_i$ appear in the reduction of a massless scalar hexagon integral to a sum of one mass pentagons and they entered into our six gluon results in Subsection \ref{gresults}.

The $I_5^{(1);~D = 6 - 2\e}$ pentagon coefficient of $A_{1}^{1-{\rm loop}}\left(k_1^{1234},k_2^{1234},k_3^{1},k_4^{234},k_5,k_6\right)\Big|_{I_5^{(1)}}$ can be derived in a completely analogous fashion. We will not describe the calculation in detail because it is extremely similar to that above but the final result is, of course, important and we present it using the similar notation:
\cmb{-.8 in}{0 in}
\bea
&&H_1^{\bar{f} f} = -{i \over 2}C_1 {\Big(\spab2.{1+6}.5\spab1.{2+3}.4\spab4.{1+2}.6+\spab5.{1+6}.2\spa1.2\spa2.4\spb4.5\spb5.6\Big)\over \spab2.{1+6}.5 \spab5.{1+6}.2 s_3 s_6}\times 
\el \times \Big(\spab2.{1+6}.5\spab1.{2+3}.4\spab4.{1+2}.6+\spab5.{1+6}.2\spa1.2\spa2.4\spb3.5\spb5.6 \Big)\,{\rm .}
\label{antifermcoef(1)}
\eea
\cme
In $H_1^{\bar{f} f}$ above, the origin of the overall minus sign is a reflection of the fact that our $D$ dimensional generalized unitarity technique naturally computes $A_{1}^{1-{\rm loop}}\left(k_1^{1234},k_2^{1234},k_3^{4},k_4^{123},k_5,k_6\right)\Big|_{I_5^{(1)}}$, which is off by a minus sign from $A_{1}^{1-{\rm loop}}\left(k_1^{1234},k_2^{1234},k_3^{1},k_4^{234},k_5,k_6\right)\Big|_{I_5^{(1)}} \equiv H_1^{\bar{f} f}$.

Before leaving this subsection, let us say a few more words about how we derived  eqs. (\ref{fermcoef(1)}) and (\ref{antifermcoef(1)}). We treat (\ref{fermcoef(1)}) but (\ref{antifermcoef(1)}) is no more difficult. In fact, (\ref{unsimpfermans}) is particularly easy to simplify down to (\ref{fermcoef(1)}) due to its similarity to $A_{1}^{1-{\rm loop}}\left(k_1^{1234},k_2^{1234},k_3^{1234},k_4,k_5,k_6\right)\Big|_{I_5^{(1)}}$. Comparing eq. (\ref{fermcoef(1)}) above to
\be
K_1 = {i \over 2} C_1 {\Big(\spab2.{1+6}.5 \spab1.{2+3}.4 \spab3.{1+2}.6 +\spab5.{1+6}.2 \spa1.2 \spa2.3 \spb4.5 \spb5.6 \Big)^2\over s_6 s_3 \spab2.{6+1}.5 \spab5.{6+1}.2}
\ee
from Subsection \ref{gresults}, we see that one BCFW shift is particularly helpful in determining the analytic structure of $K_1^{f \bar{f}}$. Suppose we make the shift
\bea
\lambda_3 &&\rightarrow \lambda_3(z) = \lambda_3 + z \lambda_4\nn
\tilde{\lambda}_4 &&\rightarrow \tilde{\lambda}_4(z) = \tilde{\lambda}_4 - z \tilde{\lambda}_3 \,{\rm ,}
\eea
on the unsimplified formula for $K_1^{f\bar{f}}$ given by eq. (\ref{unsimpfermans}). What we will find is that this shift evaluated at a random phase-space point looks like
$$K_1^{f\bar{f}} + K_1 z$$
evaluated numerically at the random point. This immediately tells us to just take one of the factors of 
$$\Big(\spab2.{(1+6)}.5 \spab1.{(2+3)}.4 \spab3.{(1+2)}.6 +\spab5.{(1+6)}.2 \spa1.2 \spa2.3 \spb4.5 \spb5.6 \Big)$$
in $K_1$ and replace $\lambda_3$ with $\lambda_4$ in that factor to get $K_1^{f \bar{f}}$. Of course, there is no guarantee that something like this will work in general, but we will see that we are also able to guess a simple result for the first pentagon coefficient of the scalar/anti-scalar amplitude considered in the next subsection.
\subsection{NMHV Amplitudes With a Scalar/Anti-Scalar Pair}
\label{ssbarcalc}
In this subsection, we use the results that we have so far for $A_1^{1-{\rm loop}}\left(k_1^{1234},k_2^{1234},k_3^{1234},k_4,k_5,k_6\right)$, $A_1^{1-{\rm loop}}\left(k_1^{1234},k_2^{1234},k_3,k_4^{1234},k_5,k_6\right)$, $A_1^{1-{\rm loop}}\left(k_1^{1234},k_2,k_3^{1234},k_4,k_5^{1234},k_6\right)$,\\ $A_1^{1-{\rm loop}}\left(k_1^{1234},k_2^{1234},k_3^{123},k_4^4,k_5,k_6\right)$, and  $A_1^{1-{\rm loop}}\left(k_1^{1234},k_2^{1234},k_3^{1},k_4^{234},k_5,k_6\right)$ and eq. (\ref{A63s}) for $\a_{6;3}$ to deduce an expression for the $I_5^{(1);~D = 6 - 2\e}$ pentagon coefficient of\\ $A_1^{1-{\rm loop}}\left(k_1^{1234},k_2^{1234},k_3^{12},k_4^{34},k_5,k_6\right)$. Clearly, the first step is to expand eq. (\ref{A63s}) and extract the coefficient of $\eta_1^{1234}\eta_3^{1234}\eta_5^{1234}$. Doing this results in the relation 
\bea
&&A_1^{1-{\rm loop}}\left(k_1^{1234}, k_2, k_3^{1234}, k_4, k_5^{1234}, k_6\right) = {\spab1.{3+5}.6^4 \over \spa1.2^4 \spb5.6^4} A_1^{1-{\rm loop}}\left(k_1^{1234}, k_2^{1234}, k_3^{1234}, k_4, k_5, k_6\right)
\el+4{\spab1.{3+5}.6^3 \spa3.1 \spb6.4\over \spa1.2^4 \spb5.6^4} A_1^{1-{\rm loop}}\left(k_1^{1234},k_2^{1234},k_3^{123},k_4^4,k_5,k_6\right) 
\el+ 6{\spab1.{3+5}.6^2 \spa3.1^2 \spb6.4^2\over \spa1.2^4 \spb5.6^4} A_1^{1-{\rm loop}}\left(k_1^{1234},k_2^{1234},k_3^{12},k_4^{34},k_5,k_6\right)
\el + 4{\spab1.{3+5}.6 \spa3.1^3 \spb6.4^3 \over \spa1.2^4 \spb5.6^4} A_1^{1-{\rm loop}}\left(k_1^{1234},k_2^{1234},k_3^{1},k_4^{234},k_5,k_6\right) 
\el +{\spa3.1^4 \spb6.4^4\over \spa1.2^4 \spb5.6^4}A_1^{1-{\rm loop}}\left(k_1^{1234},k_2^{1234},k_3,k_4^{1234},k_5,k_6\right)
\label{scaleqn}
\eea
which can be used to trivially solve for the $I_5^{(1);~D = 6 - 2\e}$ pentagon coefficient of \\$A_1^{1-{\rm loop}}\left(k_1^{1234},k_2^{1234},k_3^{12},k_4^{34},k_5,k_6\right)$.

We can simplify the complicated looking expression that results by using BCFW shifts. By examining the results that we have so far
\cmb{-1.2 in}{0 in}
\bea
K_1 &=& {i \over 2} C_1 {\Big(\spab2.{1+6}.5 \spab1.{2+3}.4 \spab3.{1+2}.6 +\spab5.{1+6}.2 \spa1.2 \spa2.3 \spb4.5 \spb5.6 \Big)^2\over \spab2.{1+6}.5 \spab5.{1+6}.2 s_3 s_6} \\
K_1^{f \bar{f}} &=& {i \over 2}C_1 {\Big(\spab2.{1+6}.5\spab1.{2+3}.4\spab4.{1+2}.6+\spab5.{1+6}.2\spa1.2\spa2.4\spb4.5\spb5.6\Big)\over \spab2.{1+6}.5 \spab5.{1+6}.2 s_3 s_6}\times  \nonumber \\
&& \times \Big(\spab2.{1+6}.5\spab1.{2+3}.4\spab3.{1+2}.6+\spab5.{1+6}.2\spa1.2\spa2.3\spb4.5\spb5.6 \Big)\\
H_1^{\bar{f} f} &=& -{i \over 2}C_1 {\Big(\spab2.{1+6}.5\spab1.{2+3}.4\spab4.{1+2}.6+\spab5.{1+6}.2\spa1.2\spa2.4\spb4.5\spb5.6\Big)\over \spab2.{1+6}.5 \spab5.{1+6}.2 s_3 s_6}\times \nonumber \\
&& \times \Big(\spab2.{1+6}.5\spab1.{2+3}.4\spab4.{1+2}.6+\spab5.{1+6}.2\spa1.2\spa2.4\spb3.5\spb5.6 \Big)\\
H_1 &=& {i \over 2} C_1 {\Big(\spab2.{6+1}.5 \spab1.{2+4}.3 \spab4.{1+2}.6 +\spab5.{6+1}.2 \spa1.2 \spa2.4 \spb3.5 \spb5.6 \Big)^2\over \spab2.{1+6}.5 \spab5.{1+6}.2 s_3 s_6} 
\eea
\cme
we see that a natural guess for the $I_5^{(1);~D = 6 - 2\e}$ pentagon coefficient of \\$A_1^{1-{\rm loop}}\left(k_1^{1234},k_2^{1234},k_3^{12},k_4^{34},k_5,k_6\right)$ is
\be
M_1^{s s^*} = {i \over 2}C_1 {\Big(\spab2.{1+6}.5\spab1.{2+3}.4\spab4.{1+2}.6+\spab5.{1+6}.2\spa1.2\spa2.4\spb4.5\spb5.6\Big)^2\over \spab2.{1+6}.5 \spab5.{1+6}.2 s_3 s_6}
\label{scalguess}
\ee
because it has a factor in common with both $K_1^{f \bar{f}}$ and $H_1^{\bar{f} f}$. In fact, this is {\it almost} right. If we try subtracting eq. (\ref{scalguess}) from the expression obtained by solving eq. (\ref{scaleqn}), we find that what's left over is easily determined using BCFW analysis to be
$$ {i\over 6}C_1 \spa1.2^2 \spb5.6^2$$
which implies that the entire pentagon coefficient is given by
\be
M_1^{s s^*} = {i \over 2}C_1 \left({\Big(\spab2.{1+6}.5\spab1.{2+3}.4\spab4.{1+2}.6+\spab5.{1+6}.2\spa1.2\spa2.4\spb4.5\spb5.6\Big)^2\over \spab2.{1+6}.5 \spab5.{1+6}.2 s_3 s_6} + {1\over 3}\spa1.2^2 \spb5.6^2\right)
\ee
All of the necessary pieces are now in place and we can use them to determine all of the higher-order terms in the $\Nsym$ superamplitude $\a_{6;3}$. We do this in the next subsection.
\subsection{The $\Nsym$ Supersymmetrization of the Six-Point NMHV Amplitudes}
\label{supersymresults}
In this subsection we use the results derived in the last two subsections and some of those derived in Section \ref{gluoncomp} to write down the full form of the higher-order in $\e$ contributions (in the dual conformal basis) to the one-loop planar six-point NMHV superamplitude. We present all of the higher-order pieces of the component scattering amplitudes that appear in eq. (\ref{A63s}) for $\a_{6;3}$. To determine the other pentagon coefficients we first solved the requisite leading singularity equations numerically using our $I_5^{(1);~D = 6 - 2\e}$ pentagon coefficients as inputs. Then, based on the analytical formulas obtained in this section and the last, we found it straightforward to guess appropriate compact formulas for the remaining undetermined coefficients, checking everything numerically against our numbers from the leading singularity equations. If we first define
\bea
K_1 &=& {i \over 2} C_1 {\Big(\spab2.{1+6}.5 \spab1.{2+3}.4 \spab3.{1+2}.6 +\spab5.{1+6}.2 \spa1.2 \spa2.3 \spb4.5 \spb5.6 \Big)^2\over s_6 s_3 \spab2.{6+1}.5 \spab5.{6+1}.2} \nonumber \\ \\
K_2 &=& {i \over 2} C_2 {\spab3.{1+2}.6^2 \spa1.2^2 \spb4.5^2 t_1^2 \over s_1 s_4 \spab3.{1+2}.6 \spab6.{1+2}.3} \\
K_3 &=& {i \over 2} C_3 {\spab1.{2+3}.4^2 \spa2.3^2 \spb5.6^2 t_1^2 \over s_2 s_5 \spab1.{2+3}.4 \spab4.{2+3}.1} \\
K_4 &=& {i \over 2} C_4 {\Big(\spab2.{1+6}.5 \spab1.{2+3}.4 \spab3.{1+2}.6 +\spab5.{1+6}.2 \spa1.2 \spa2.3 \spb4.5 \spb5.6 \Big)^2\over s_6 s_3 \spab2.{6+1}.5 \spab5.{6+1}.2} \nonumber \\ \\
K_5 &=& {i \over 2} C_5 {\spab3.{1+2}.6^2 \spa1.2^2 \spb4.5^2 t_1^2 \over s_1 s_4 \spab3.{1+2}.6 \spab6.{1+2}.3} \\
K_6 &=& {i \over 2} C_6 {\spab1.{2+3}.4^2 \spa2.3^2 \spb5.6^2 t_1^2 \over s_2 s_5 \spab1.{2+3}.4 \spab4.{2+3}.1} \, {\rm ,}
\eea
\bea
K_1^{f \bar{f}} &=& {i \over 2}C_1 {\Big(\spab2.{1+6}.5\spab1.{2+3}.4\spab3.{1+2}.6+\spab5.{1+6}.2\spa1.2\spa2.3\spb4.5\spb5.6 \Big)\over s_6 s_3 \spab2.{6+1}.5 \spab5.{6+1}.2}\times  \nonumber \\
&& \times \Big(\spab2.{1+6}.5\spab1.{2+3}.4\spab4.{1+2}.6+\spab5.{1+6}.2\spa1.2\spa2.4\spb4.5\spb5.6\Big) \\ 
K_2^{f \bar{f}} &=& {i \over 2} C_2 {\Big(\spab3.{1+2}.6 \spa1.2 \spb4.5 t_1\Big)\Big(\spb4.5\spa1.2\spab4.{5+6}.3\spab3.{4+5}.6\Big)\over s_1 s_4 \spab3.{1+2}.6 \spab6.{1+2}.3} \\
K_3^{f \bar{f}} &=& {i \over 2} C_3 {\Big(\spab1.{2+3}.4 \spa2.3 \spb5.6 t_1\Big)\Big(\spb5.6\spa2.3\spab1.{2+3}.4\spab4.{1+2}.3\Big)\over s_2 s_5 \spab1.{2+3}.4 \spab4.{2+3}.1} \\
K_4^{f \bar{f}} &=& {i \over 2}C_4 {\Big(\spab2.{1+6}.5\spab1.{2+3}.4\spab3.{1+2}.6+\spab5.{1+6}.2\spa1.2\spa2.3\spb4.5\spb5.6 \Big)\over s_6 s_3 \spab2.{6+1}.5 \spab5.{6+1}.2}\times  \nonumber \\
&& \times \Big(\spab2.{1+6}.5\spab1.{2+3}.4\spab4.{1+2}.6+\spab5.{1+6}.2\spa1.2\spa2.4\spb4.5\spb5.6\Big) \\ 
K_5^{f \bar{f}} &=& {i \over 2} C_5 {\Big(\spab3.{1+2}.6 \spa1.2 \spb4.5 t_1\Big)\Big(\spb4.5\spa1.2\spab4.{5+6}.3\spab3.{4+5}.6\Big)\over s_1 s_4 \spab3.{1+2}.6 \spab6.{1+2}.3} \\
K_6^{f \bar{f}} &=& {i \over 2} C_6 {\Big(\spab1.{2+3}.4 \spa2.3 \spb5.6 t_1\Big)\Big(\spb5.6\spa2.3\spab1.{2+3}.4\spab4.{1+2}.3\Big)\over s_2 s_5 \spab1.{2+3}.4 \spab4.{2+3}.1} \, {\rm ,}
\eea
\bea
M_1^{s s^*} &=& {i \over 2}C_1 \left({\Big(\spab2.{1+6}.5\spab1.{2+3}.4\spab4.{1+2}.6+\spab5.{1+6}.2\spa1.2\spa2.4\spb4.5\spb5.6\Big)^2\over s_6 s_3 \spab2.{6+1}.5 \spab5.{6+1}.2} + {1\over 3}\spa1.2^2 \spb5.6^2\right) \nn\\ 
M_2^{s s^*} &=& {i \over 2} C_2 \left({\spb4.5^2 \spa1.2^2\spab4.{5+6}.3^2\spab3.{4+5}.6^2 \over s_1 s_4 \spab3.{1+2}.6 \spab6.{1+2}.3}+ {1\over 3}\spa1.2^2 \spb5.6^2\right) \\
M_3^{s s^*} &=& {i \over 2} C_3 \left({\spb5.6^2 \spa2.3^2\spab1.{2+3}.4^2\spab4.{1+2}.3^2 \over  s_2 s_5 \spab1.{2+3}.4 \spab4.{2+3}.1}+ {1\over 3}\spa1.2^2 \spb5.6^2\right) \\
M_4^{s s^*} &=& {i \over 2}C_4 \left({\Big(\spab2.{1+6}.5\spab1.{2+3}.4\spab4.{1+2}.6+\spab5.{1+6}.2\spa1.2\spa2.4\spb4.5\spb5.6\Big)^2\over s_6 s_3 \spab2.{6+1}.5 \spab5.{6+1}.2} + {1\over 3}\spa1.2^2 \spb5.6^2\right) \nn\\ 
M_5^{s s^*} &=& {i \over 2} C_5 \left({\spb4.5^2 \spa1.2^2\spab4.{5+6}.3^2\spab3.{4+5}.6^2 \over s_1 s_4 \spab3.{1+2}.6 \spab6.{1+2}.3}+ {1\over 3}\spa1.2^2 \spb5.6^2\right) \\
M_6^{s s^*} &=& {i \over 2} C_6 \left({\spb5.6^2 \spa2.3^2\spab1.{2+3}.4^2\spab4.{1+2}.3^2 \over  s_2 s_5 \spab1.{2+3}.4 \spab4.{2+3}.1}+ {1\over 3}\spa1.2^2 \spb5.6^2\right) \, {\rm ,}
\eea
\bea
H_1^{\bar{f} f} &=& -{i \over 2}C_1 {\Big(\spab2.{6+1}.5 \spab1.{2+4}.3 \spab4.{1+2}.6 +\spab5.{6+1}.2 \spa1.2 \spa2.4 \spb3.5 \spb5.6 \Big)\over s_6 s_3 \spab2.{6+1}.5 \spab5.{6+1}.2}\times \nonumber \\
&& \times \Big(\spab2.{1+6}.5\spab1.{2+3}.4\spab4.{1+2}.6+\spab5.{1+6}.2\spa1.2\spa2.4\spb4.5\spb5.6\Big)\\
H_2^{\bar{f} f} &=& {i \over 2}C_2 {\spa1.2 \Big(\spab3.{1+2}.6 \spab4.{1+2}.3 \spb5.3 +\spab6.{1+2}.3 \spab4.{1+2}.6 \spb5.6\Big)\over s_1 s_4 \spab3.{1+2}.6 \spab6.{1+2}.3}\times \nonumber \\
&& \times \Big(\spb4.5 \spa1.2\spab4.{5+6}.3\spab3.{4+5}.6\Big) \\
H_3^{\bar{f} f} &=& {i \over 2}C_3 {\spb5.6 \Big(\spab1.{2+3}.4 \spab4.{1+2}.3 \spa2.4 +\spab4.{2+3}.1 \spab1.{2+4}.3 \spa2.1\Big)\over s_2 s_5 \spab4.{2+3}.1 \spab1.{2+3}.4}\times \nonumber \\
&& \times \Big(\spb5.6 \spa2.3\spab1.{2+3}.4\spab4.{1+2}.3\Big) \\
H_4^{\bar{f} f} &=& -{i \over 2}C_4 {\Big(\spab2.{6+1}.5 \spab1.{2+4}.3 \spab4.{1+2}.6 +\spab5.{6+1}.2 \spa1.2 \spa2.4 \spb3.5 \spb5.6 \Big)\over s_6 s_3 \spab2.{6+1}.5 \spab5.{6+1}.2}\times \nonumber \\
&& \times \Big(\spab2.{1+6}.5\spab1.{2+3}.4\spab4.{1+2}.6+\spab5.{1+6}.2\spa1.2\spa2.4\spb4.5\spb5.6\Big)\\
H_5^{\bar{f} f} &=& {i \over 2}C_5 {\spa1.2 \Big(\spab3.{1+2}.6 \spab4.{1+2}.3 \spb5.3 +\spab6.{1+2}.3 \spab4.{1+2}.6 \spb5.6\Big)\over s_1 s_4 \spab3.{1+2}.6 \spab6.{1+2}.3}\times \nonumber \\
&& \times \Big(\spb4.5 \spa1.2\spab4.{5+6}.3\spab3.{4+5}.6\Big) \\
H_6^{\bar{f} f} &=&  {i \over 2}C_6 {\spb5.6 \Big(\spab1.{2+3}.4 \spab4.{1+2}.3 \spa2.4 +\spab4.{2+3}.1 \spab1.{2+4}.3 \spa2.1\Big)\over s_2 s_5 \spab4.{2+3}.1 \spab1.{2+3}.4}\times \nonumber \\
&& \times \Big(\spb5.6 \spa2.3\spab1.{2+3}.4\spab4.{1+2}.3\Big) \,{\rm ,}
\eea
and
\bea
H_1 &=& {i \over 2} C_1 {\Big(\spab2.{6+1}.5 \spab1.{2+4}.3 \spab4.{1+2}.6 +\spab5.{6+1}.2 \spa1.2 \spa2.4 \spb3.5 \spb5.6 \Big)^2\over s_6 s_3 \spab2.{6+1}.5 \spab5.{6+1}.2} \nonumber \\ \\
H_2 &=& {i \over 2} C_2 {\spa1.2^2 \Big(\spab3.{1+2}.6 \spab4.{1+2}.3 \spb5.3 +\spab6.{1+2}.3 \spab4.{1+2}.6 \spb5.6\Big)^2\over s_1 s_4 \spab3.{1+2}.6 \spab6.{1+2}.3} \nonumber \\ \\
H_3 &=& {i \over 2} C_3 {\spb5.6^2 \Big(\spab1.{2+3}.4 \spab4.{1+2}.3 \spa2.4 +\spab4.{2+3}.1 \spab1.{2+4}.3 \spa2.1\Big)^2\over s_2 s_5 \spab4.{2+3}.1 \spab1.{2+3}.4} \nonumber \\ \\
H_4 &=& {i \over 2} C_4 {\Big(\spab2.{6+1}.5 \spab1.{2+4}.3 \spab4.{1+2}.6 +\spab5.{6+1}.2 \spa1.2 \spa2.4 \spb3.5 \spb5.6 \Big)^2\over s_6 s_3 \spab2.{6+1}.5 \spab5.{6+1}.2} \nonumber \\ \\
H_5 &=& {i \over 2} C_5 {\spa1.2^2 \Big(\spab3.{1+2}.6 \spab4.{1+2}.3 \spb5.3 +\spab6.{1+2}.3 \spab4.{1+2}.6 \spb5.6\Big)^2\over s_1 s_4 \spab3.{1+2}.6 \spab6.{1+2}.3} \nonumber \\ \\
H_6 &=& {i \over 2} C_6 {\spb5.6^2 \Big(\spab1.{2+3}.4 \spab4.{1+2}.3 \spa2.4 +\spab4.{2+3}.1 \spab1.{2+4}.3 \spa2.1\Big)^2\over s_2 s_5 \spab4.{2+3}.1 \spab1.{2+3}.4} \nonumber \\
\eea
in what will hopefully be obvious notation given what has been discussed so far, then the final form of our answer reads
\bea
\a_{6;3}^{1-{\rm loop}} =&& \cdots + \e \,{\D^{(8)}(Q^{a\,\A}) \over \spb5.6^4 \spa1.2^4}\sum_{\ell = 1}^6 \Big(K_\ell \prod_{a=1}^4 \left(\spb5.6~\eta_3^a+\spb6.3 ~\eta_5^a +\spb{3}.5~ \eta_6^a\right)
\el+ K_\ell^{f \bar{f}} \prod_{a=1}^3 \left(\spb5.6~\eta_3^a+\spb6.3 ~\eta_5^a +\spb{3}.5~ \eta_6^a\right)\left(\spb5.6~\eta_4^4+\spb6.4 ~\eta_5^4 +\spb{4}.5~ \eta_6^4\right)
\el+ M_\ell^{s s^*}\prod_{a=1}^2 \left(\spb5.6~\eta_3^a+\spb6.3 ~\eta_5^a +\spb{3}.5~ \eta_6^a\right)\prod_{a=3}^4\left(\spb5.6~\eta_4^a+\spb6.4 ~\eta_5^a +\spb{4}.5~ \eta_6^a\right)
\el + H_\ell^{\bar{f} f} \left(\spb5.6~\eta_3^1+\spb6.3 ~\eta_5^1 +\spb{3}.5~ \eta_6^1\right)\prod_{a=2}^4\left(\spb5.6~\eta_4^a+\spb6.4 ~\eta_5^a +\spb{4}.5~ \eta_6^a\right)
\el + H_\ell \prod_{a=1}^4\left(\spb5.6~\eta_4^a+\spb6.4 ~\eta_5^a +\spb{4}.5~ \eta_6^a\right)\Big)\,I_5^{(\ell);~D = 6 - 2 \e}
\label{A63fin}
\eea
where, as usual, we have suppressed the well-known box integral contributions. In Section \ref{WL/MHV} we will see that there is another form for the higher-order in $\e$ contributions to $\a^{1-{\rm loop}}_{6;3}$ that is significantly simpler than eq. (\ref{A63fin}).
\subsection{The Structure of $\a_{n;2}$ At One, Two, and Higher Loops}
\label{BDSsect}
In this subsection we review the ongoing program of research dedicated to understanding the multi-loop structure of the planar MHV superamplitude in $\Nsym$. This program was begun with the seminal paper of Bern, Dixon, Dunbar, and Kosower (BDDK)~\cite{BDDKMHV} which computed all one-loop MHV superamplitudes in $\Nsym$ (as usual, we will only be interested in the planar contributions). Recall the notation used in eq. (\ref{MHVsupL}):
\be
\a_{n;2} = i{{1\over 16} \prod_{a = 1}^4 \sum_{i,j = 1}^n \spa{i}.j \eta^a_i \eta^a_j \over \spa1.2 \spa2.3 \cdots \spa{n}.1}\Bigg(1+\left({g^2 \Nc \mu^{2\e} e^{-\gamma_E \e} \over (4 \pi)^{2-\e}}\right) M_{1-{\rm loop}}+ \left({g^2 \Nc \mu^{2\e} e^{-\gamma_E \e}\over (4 \pi)^{2-\e}}\right)^2 M_{2-{\rm loop}}+\cdots\Bigg) \,{\rm .}
\ee
After $\a_{n;2}^{tree}$ is factored out, the analytic structure at $L$ loops, $M_{L-{\rm loop}}$, can be determined by comparing to, say, $A^{L-{\rm loop}}_1(p_1^{1234},p_2^{1234},p_3,\cdots,p_n)$ modulo the Parke-Taylor amplitude. Although most of the multi-loop $\Nsym$ literature prior to the development of $\Nsym$ on-shell superspace focused on purely gluonic amplitudes, the discussion of \ref{gendisonshell} makes it clear that, in the MHV sector, one can make this choice and still determine the full superamplitude (Of course it is probably more natural to perform all calculations in a way that preserves as many of the supersymmetries as possible~\cite{DHKSgenunit}). In all of the applications that follow, it will be useful to redefine the contribution from the $L$-th loop as follows:
\be
\left({g^2 \Nc \mu^{2\e} e^{-\gamma_E \e} \over (4 \pi)^{2-\e}}\right)^L M_{L-{\rm loop}} = \left({2 g^2 \Nc e^{-\gamma_E \e} \over (4 \pi)^{2-\e}}\right)^L\M^{(L)}(n,t_i^{[r]},\e) = a^L\M^{(L)}(n,t_i^{[r]},\e) \, {\rm ,}
\ee
where we have made the useful definitions
\be
a \equiv {\lambda e^{-\gamma_E \e} \over (4 \pi)^{2-\e}}~~~~{\rm and}~~~~t_i^{[r]} \equiv (p_i + \cdots + p_{i+r-1})^2 \,{\rm .}
\ee
Using this notation, eq. (\ref{MHVsupL}) reads 
\be
\a_{n;2} = i{{1\over 16} \prod_{a = 1}^4 \sum_{i,j = 1}^n \spa{i}.j \eta^a_i \eta^a_j \over \spa1.2 \spa2.3 \cdots \spa{n}.1}\Bigg(1+\sum_{L = 1}^\infty a^L\M^{(L)}(n,t_i^{[r]},\e)\Bigg) \,{\rm .}
\ee

Let us now describe the results of BDDK in~\cite{BDDKMHV} where the structure of $\M^{(1)}(n,t_i^{[r]},\e)$ for all $n$ was determined through $\Ord(\e^0)$. It was found that:
\cmb{0. in}{0 in}
\bea
&&\M^{(1)}(n,t_i^{[r]},\e) =
\el C_\Gamma \sum_{i=1}^{n} \left( -{ 1 \over \e^2 } \bigg(
{ \mu^2  \over -t_i^{[2]} } \bigg)^{\e}
-\sum_{r=2}^{\left[{n\over 2}\right] -1}
\sum_{i=1}^n
  \ln \bigg({ -t_i^{[r]}\over -t_i^{[r+1]} }\bigg)
  \ln \bigg({ -t_{i+1}^{[r]}\over -t_i^{[r+1]} }\bigg) +
D_n\left(t_i^{[r]}\right) + L_n\left(t_i^{[r]}\right) +{ n \pi^2 \over 6 }\right) 
\el+ \Ord(\e)\, {\rm ,}
\label{UnivFunc}
\eea
\cme
where $C_{\G}$ is given by
$$C_\G = {\G(1+\e)\G(1-\e)^2 \over 2 \G(1-2\e)}\,{\rm .}$$
The form of $D_n\left(t_i^{[r]}\right)$ and $L_n\left(t_i^{[r]}\right)$ depends upon whether $n$ is odd or even.
For $n=2m+1$,
$$
D_{2m+1}= -\sum_{r=2}^{m-1} \Bigg( \sum_{i=1}^{n}
\li2 \bigg[ 1- { t_{i}^{[r]} t_{i-1}^{[r+2]}
\over t_{i}^{[r+1]} t_{i-1}^{[r+1]} } \biggr]  \Bigg)\, {\rm ,}
$$
$$
L_{2m+1}= -{ 1\over 2} \sum_{i=1}^n
  \ln \bigg({ -t_{i}^{[m]}\over -t_{i+m+1}^{[m]}  } \bigg)
  \ln \bigg({ -t_{i+1}^{[m]}\over -t_{i+m}^{[m]} } \bigg)\, {\rm ,}
$$
whereas for $n=2m$,
$$
D_{2m}= -\sum_{r=2}^{m-2} \Bigg( \sum_{i=1}^{n}
\li2 \bigg[ 1- { t_{i}^{[r]} t_{i-1}^{[r+2]}
\over t_{i}^{[r+1]} t_{i-1}^{[r+1]} }  \bigg]  \Bigg)
-\sum_{i=1}^{n/2} \li2 \bigg[ 1- { t_{i}^{[m-1]}t_{i-1}^{[m+1]}
\over t_{i}^{[m]}t_{i-1}^{[m]}} \bigg]\, {\rm ,}
$$
$$
L_{2m}=-{1\over 4} \sum_{i=1}^n
  \ln \bigg({ -t_{i}^{[m]}\over -t_{i+m+1}^{[m]}  } \bigg)
  \ln \bigg({ -t_{i+1}^{[m]}\over -t_{i+m}^{[m]} } \bigg)\, {\rm .}
$$
The above only holds for $n \geq 5$. For $n = 4$ we have
\be
\M^{(1)}(4,t_i^{[r]},\e) = C_\G \bigg\{
-{2 \over \e^2} \Big[ \left( {-s\over\mu^2}\right)^{-\e}+ \left({-t\over\mu^2}\right)^{-\e} \Big]
+ \ln^2\left( {-s \over - t} \right) + \pi^2 \bigg\} \,{\rm .}
\label{4ptanal}
\ee

Subsequently, the functions $\M^{(2)}(4,t_i^{[r]},\e)$ and $\M^{(2)}(5,t_i^{[r]},\e)$ were determined through terms of $\Ord(\e^0)$ in~\cite{ABDK} and~\cite{TwoLoopFive} respectively. Remarkably, the following relationships were found:
\bea
\M^{(2)}(4,t_i^{[r]},\e)\Big|_{\Ord(\e^0)} - {1\over 2} \M^{(1)}(4,t_i^{[r]},\e)^2\Big|_{\Ord(\e^0)} &=& \alpha~ \M^{(1)}(4,t_i^{[r]},2\e)\Big|_{\Ord(\e^0)} + \beta \label{4pt2LBDS} \\
\M^{(2)}(5,t_i^{[r]},\e)\Big|_{\Ord(\e^0)} - {1\over 2} \M^{(1)}(5,t_i^{[r]},\e)^2\Big|_{\Ord(\e^0)} &=& \alpha~ \M^{(1)}(5,t_i^{[r]},2\e)\Big|_{\Ord(\e^0)} + \beta  \,{\rm ,}
\label{5pt2LBDS}
\eea
where both sides of the above are only considered through $\Ord(\e^0)$. $\alpha$ and $\beta$ are transcendentality two and four numbers respectively.\footnote{For example, $\zeta(2)$ is transcendentality two and $\zeta(4)$ is transcendentality four. One also speaks of functions carrying transcendentality ({\it e.g.} $\li2[1-{-s \over -t}]$ has transcendentality two).} Generically, $L$-loop planar amplitudes in $\Nsym$ are built out of transcendentality $2 L$ numbers and functions~\cite{KotikovLipatov}. In the above, $\alpha$ and $\beta$ have the transcendentality that they do because both sides of eqs. (\ref{4pt2LBDS}) and (\ref{5pt2LBDS}) are expected to have uniform transcendentality four. 

Given these striking results, Bern, Dixon, and Smirnov proposed~\cite{BDS} the following ansatz for the analytical structure of all planar MHV superamplitudes,
\be
\ln\left(1+\sum_{L = 1}^\infty a^L \M^{(L)}(n,t_i^{[r]},\e)\right) = \sum_{L = 1}^\infty a^L\left(f^{(L)} \M^{(1)}(n,t_i^{[r]},L \e) + C^{(L)}+E^{(L)}(n,\e)\right) \,{\rm ,}
\label{BDSansatz}
\ee
which they checked for $n = 4$ through three loops. In eq. (\ref{BDSansatz}) above, $f^{(L)}$ and $C^{(L)}$ are numbers of the appropriate transcendentality ($2(L-1)$ and $2 L$ respectively) and $E^{(L)}(n,\e)$ contains higher-order in $\e$ contributions that are unimportant because, usually, both sides of (\ref{BDSansatz}) are expanded to some order in $a$ and then higher-order in $\e$ terms are dropped to put all of the $n$ dependence on the right-hand side into the function  $\M^{(1)}(n,t_i^{[r]},L \e)$. Actually, a structure like this is expected in gauge theory on general grounds for the $\e$ pole terms; the IR divergences of planar non-Abelian gauge theory amplitudes are well-understood and known to exponentiate~\cite{Catani,StermanYeomans}. What is really novel about eq. (\ref{BDSansatz}) is that it holds also for the finite terms.

In fact, the so-called BDS ansatz (eq. (\ref{BDSansatz})) is known to be valid to all loop orders if $n = 4 ~{\rm or}~5$~\cite{DHKSward}. We will explain this in Section \ref{WL/MHV} after introducing dual superconformal symmetry. For higher multiplicity, however, life is not so simple. It was proven in~\cite{AMlargen,BLSV1,BDKRSVV} that the BDS ansatz is incomplete at two loops and six points. For this case, which will be the one of primary interest to us, eq. (\ref{BDSansatz}) must be modified:
\be
\M^{(2)}(6,t_i^{[r]},\e)\Big|_{\Ord(\e^0)} - {1\over 2}\M^{(1)}(6,t_i^{[r]},\e)^2\Big|_{\Ord(\e^0)}= \alpha ~\M^{(1)}(6,t_i^{[r]},2\e)\Big|_{\Ord(\e^0)} + \beta + R_{6}^{(2)}\left(t_i^{[r]}\right)\,{\rm ,}
\label{6ptstruct}
\ee
The new term on the right-hand side is called the two-loop, six-point remainder function. It is IR finite and highly constrained. For example, in order to be consistent with the known results for four and five point scattering at two loops, $R_6^{(2)}$ must have vanishing soft and collinear limits in all channels. Also, we know from the discussion above that $R_6^{(2)}$ should be a function of uniform transcendentality four. Furthermore, as we will see in the next section, the remainder function is not an arbitrary function of the $t_i^{[r]}$. In fact, for generic kinematics it is a function of only three independent variables.
\section{Dual Superconformal Symmetry and the Ratio of the Six-Point NMHV and MHV Superamplitudes at Two-Loops}
\label{WL/MHV}
In this section we review developments related to a recently discovered hidden symmetry of the planar $\Nsym$ S-matrix, {\it dual superconformal symmetry}, and we present alternative formulae for the higher-order contributions to $\a_{6;3}^{1-{\rm loop}}$ that manifest the new symmetry as much as possible. The final formula obtained is very simple and is the form of our results used in a recent study of the $\Nsym$ planar NMHV superamplitude at two loops~\cite{KRV}. As will be explained more below, one of the ideas tested in~\cite{KRV} is whether the NMHV superamplitude divided by the MHV superamplitude is dual superconformally invariant, as was proposed earlier in~\cite{DHKSdualconf} by Drummond, Henn, Korchemsky, and Sokatchev. The new results described in this section for $\a_{6;3}^{1-{\rm loop}}$ were obtained in collaboration with one of the authors of~\cite{KRV}, Cristian Vergu. This work has been primarily been about perturbation theory at weak coupling and, therefore, we will describe all developments in perturbation theory even though most of them were first seen non-perturbatively in the strong coupling regime of $\Nsym$ (via the AdS/CFT correspondence). Of course, it would be a shame to completely ignore the strong coupling regime, so we offer a brief account in Appendix \ref{ADS/CFT} for the reader interested in a historical introduction to the ideas discussed in this section.
\subsection{Light-Like Wilson-Loop/MHV Amplitude Correspondence}
\label{WL/MHV2}
Inspired by the AdS/CFT correspondence, a very surprising connection was suggested~\cite{origAldayMald} between two {\it a priori} completely unrelated observables. One of the observables, 
\be
{\a_{n;2} \over \a^{tree}_{n;2}} = 1 + \sum_{L = 1}^\infty a^L \M^{(L)}(n,t_i^{[r]},\e)
\label{analstruc}
\ee
was discussed at length in Subsections \ref{gendisonshell} and \ref{BDSsect}. The other, the expectation value of an $n$-gon (denoted $C_n$) light-like Wilson loop
\be
W[C_n] \equiv {1\over \Nc} \langle0|{\rm Tr}\bigg[P\bigg\{ {\rm exp}\left(i g \oint_{C_n} dx^\nu A^a_\nu(x) t^a\right)\bigg\}\bigg]|0\rangle
\label{WLdef}
\ee
has not been introduced so far, so we will analyze its definition in some detail. Of course, we will also have to understand, at least in principle, how to calculate the set of objects introduced above perturbatively if our goal is to establish a connection between eqs. (\ref{analstruc}) and (\ref{WLdef}). 

Since it may well be the case that the reader is less familiar with Wilson loop expectation values than with scattering amplitudes, we first take a step back and discuss Wilson loops in general. Wilson loops were introduced by Wilson in~\cite{origWilson} in an attempt to better understand the phenomenon of quark confinement in non-Abelian Yang-Mills theory (he had the gauge group $SU(3)_{\rm color}$ in mind). The asymptotic behavior of Wilson loop expectation values as the circumference of the loop goes to infinity tells you whether your gauge theory (in Euclidean space) is confining in the infrared. If we define $A(C)$ to be the area of the surface of minimal area bounded by $C$ and $L(C)$ to be the circumference of $C$, it can be shown~\cite{Polyakovtext} that if we have
\be
W[C] ~~~~\stackrel{L(C)\rightarrow \infty}{\longrightarrow}~~~~ W_0 e^{-k A(C)} \,{\rm ,}
\ee
then the gauge theory is in the confining phase, provided that the contour $C$ is smooth (without cusps) and is space-like back in Minkowski space (see Figure \ref{splike}).
\FIGURE{
\resizebox{.75\textwidth}{!}{\includegraphics{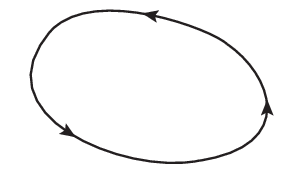}}
\caption{A smooth space-like Wilson loop. The arrows indicate the direction of traversal; if the loop was unoriented it wouldn't be clear how to make sense of the path-ordering in (\ref{WLdef}).}
\label{splike}}

Although the above application is probably what would come to the minds of most researchers if asked about Wilson loops, we have something completely different in mind. The Wilson loop appearing in eq. (\ref{WLdef}) is of a special type: the contour $C_n$ defining it has cusps connected by light-like segments. The expectation value of unions of light-like Wilson lines enter into the calculation of certain universal soft functions in QCD. These soft functions are important because they control the resummation of large logarithms that often appear at the edges of phase-space when one tries to na\"{i}vely compute next-to-leading (or higher) corrections to cross-sections for processes in QCD. As we shall see, $n$-cusp light-like Wilson loop expectation values also play an important in $\Nsym$, but in a rather different way.

It is now time to return to eq. (\ref{WLdef}),
\be
W[C_n] = {1\over \Nc} \langle0|{\rm Tr}\bigg[P\bigg\{ {\rm exp}\left(i g \oint_{C_n} dx^\nu A^a_\nu(x) t^a\right)\bigg\}\bigg]|0\rangle \,{\rm ,}
\ee
and scrutinize everything that enters into the expression on the right-hand side. In the above, the gauge connection, $A_\nu^a$ is contracted with the $SU(\Nc)$ fundamental representation gauge group generators, $t^a$. The quantity $A_\nu^a t^a$ is integrated around the closed contour $C_n$ depicted in Figure \ref{hexWL} (for $n = 6$). Each cusp of $C_n$ is labeled $x_i^\nu$ and the lines between adjacent cusps have lengths $(x_i - x_{i+1})^2 = 0$. The distances between non-adjacent cusps are in general non-zero. If we introduce the notation $x_{i j}^2 = (x_i - x_j)^2$, we have nine distinct non-zero distances for $n = 6$: $\{x_{1 3}^2,\,x_{2 4}^2,\,x_{35}^2,\,x_{46}^2,\,x_{51}^2,\,x_{62}^2,\,x_{14}^2,\,x_{25}^2,\,x_{36}^2\}$. In field theory, when one is faced with evaluating the expectation value of an exponential of field operators, one simply expands the exponential and applies Wick's theorem in the standard way, usually using Feynman diagrams as a book-keeping device. This case is no different, but the appearance of the path-ordering operator, $P$, tells us to order the integrals that we get out of the exponential's Taylor expansion according to how we are traversing the Wilson loop. In fact the only non-commutative structure in the problem are the $t^a$ generator matrices, so the path-ordering in this case is just an ordering on the $SU(\Nc)$ generators that appear in the argument of the exponential. Finally, we have to trace over gauge theory indices to obtain a gauge invariant functional of $C_n$. Suppose we tried to make sense of $W[C_n]$ without the trace:
\be
W^\prime[C_n] = {1\over \Nc} \langle0|P\bigg\{ {\rm exp}\left(i g \oint_{C_n} dx^\nu A^a_\nu(x) t^a\right)\bigg\}|0\rangle \,{\rm .}
\ee
Under a gauge transformation $\Omega$, $A_\nu^a t^a$ becomes $\Omega^{-1} A_\nu^a t^a \Omega+ {i\over g} \Omega^{-1} \partial_\nu \Omega$. This induces a change in $W^\prime[C_n]$,
\be
\Omega^{-1} {1\over \Nc} \langle0|P\bigg\{ {\rm exp}\left(i g \oint_{C_n} dx^\nu A^a_\nu(x) t^a\right)\bigg\}|0\rangle \Omega \,{\rm ,}
\ee
and we see that $W^\prime[C_n]$ is not gauge invariant. This problem is easily fixed by taking the trace over generator matrices and this brings us back to $W[C_n]$.
\FIGURE{
\resizebox{.75\textwidth}{!}{\includegraphics{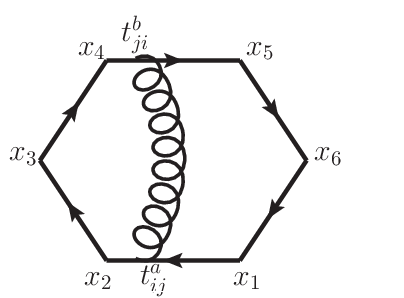}}
\caption{The Feynman diagram for one contribution to $W[C_6]$.}
\label{hexWL}}

Now that we understand how to interpret $W[C_n]$, following~\cite{DKS4pt}, we calculate it to order $\Ord(g^2)$ (lowest non-trivial order). Expanding the path-ordered exponential
gives 
\be
P\bigg\{ {\rm exp}\left(i g \oint_{C_n} dx^\nu A^a_\nu(x) t^a\right)\bigg\} = 1 + i g \oint_{C_n} dx^\nu A^a_\nu(x) t^a +{1\over 2!}(i g)^2 \oint_{C_n}\oint_{C_n}  dx^\rho dy^\sigma A^a_\rho(x) A^b_\sigma(y) t^a_{i j}t^b_{j k} + \cdots ~{\rm .}
\ee 
Truncating the above at $\Ord(g^2)$ and taking its vacuum expectation value gives
\be
1 + {1\over 2!}(i g)^2 \oint_{C_n}\oint_{C_n}  dx^\rho dy^\sigma \langle0|A^a_\rho(x) A^b_\sigma(y)|0\rangle t^a_{i j}t^b_{j k}
\ee
since $\langle0|A^a_\nu(x) t^a|0\rangle = 0$ by virtue of Lorentz invariance. Finally, we take the trace over generator matrices, tack on the overall factor of $1/\Nc$, and obtain $W[C_n]$ through $\Ord(g^2)$:
\be
W[C_n]\Big|_{\Ord(g^2)} = 1 - {g^2\over 2! \Nc} \oint_{C_n}\oint_{C_n}  dx^\rho dy^\sigma \langle0|A^a_\rho(x) A^b_\sigma(y)|0\rangle t^a_{i j}t^b_{j i} \,{\rm .}
\label{WLsetup}
\ee
Since $\langle0|A^a_\rho(x) A^b_\sigma(y)|0\rangle$ is just the well known two-point correlation function for the Yang-Mills field in position space,
\be
\langle0|A^a_\rho(x) A^b_\sigma(y)|0\rangle = {-g_{\rho \sigma} \D^{a b} \mu^{2\e} \pi^\e e^{\gamma_E \e}\over 4 \pi^2(-(x-y)^2)^{1-\e}}
\ee
it is clear that Wilson loop expectation values are conveniently described by Feynman diagrams. For example, if we parametrize our $n$-gon loop,  for $1 \leq i \leq n$, as
\be
\{x^\nu(\tau_i) = x_i^\nu - \tau_i x_{i\,i+1}^\nu,\, y^\nu(\tau_i) = x_i^\nu - \tau_i x_{i\,i+1}^\nu | 0\leq \tau \leq 1\} \,{\rm ,}
\ee
the $\Ord(g^2)$ contribution to $W[C_6]$ shown in Figure \ref{hexWL} can be calculated by integrating over the positions on lines $x_1 - x_2$ and $x_4 - x_5$ where the gluon stretched between them can be absorbed/emitted\footnote{Due to the fact that we have two integrals over the entire closed contour, we pick up a factor of $2!$ (from interchanging the roles of $x^\rho$ and $y^\sigma$) that cancels against the factor of $2!$ in the denominator of eq. (\ref{WLsetup}).}:
\bea
&&-{g^2 \over \Nc} \int_{x_1^\rho}^{x_2^\rho} dx^\rho \int_{x_4^\sigma}^{x_5^\sigma} dy^\sigma {-g_{\rho \sigma} \D^{a b} \mu^{2\e} \pi^\e e^{\gamma_E \e}\over 4 \pi^2(-(x-y)^2)^{1-\e}} t^a_{i j}t^b_{j i}
\elale -{g^2 \over \Nc} \int_{0}^{1} (-d\tau_1 x_{12}^\rho) \int_{0}^{1} (-d\tau_4 x_{45}^\sigma) {-g_{\rho \sigma} \mu^{2\e} \pi^\e e^{\gamma_E \e} \over 4 \pi^2(-(x_1-x_4-\tau_1 x_{12} + \tau_4 x_{45})^2)^{1-\e}} t^a_{i j}t^a_{j i}
\elale {\mu^{2\e}g^2 \pi^\e e^{\gamma_E \e}\over 4 \pi^2 \Nc}\int_{0}^{1} d\tau_1 \int_{0}^{1} d\tau_4 {x_{12}\cdot x_{45} \over (-(x_1-x_4-\tau_1 x_{12} + \tau_4 x_{45})^2)^{1-\e}} C_F \Nc
\elale {\mu^{2\e}g^2 \pi^\e e^{\gamma_E \e} C_F \over 4 \pi^2}\int_{0}^{1} d\tau_1 \int_{0}^{1} d\tau_4 {x_{12}\cdot x_{45} \over (-(x_1-x_4-\tau_1 x_{12} + \tau_4 x_{45})^2)^{1-\e}}
\eea
On general grounds, we expect such a contribution to be a real number for $(x_1-x_4-\tau_1 x_{12} + \tau_4 x_{45})^2 < 0$ and $\e$ sufficiently small, real, and positive. In this paper we will never have to leave the region where these conditions are satisfied.
\FIGURE{
\resizebox{.9\textwidth}{!}{\includegraphics{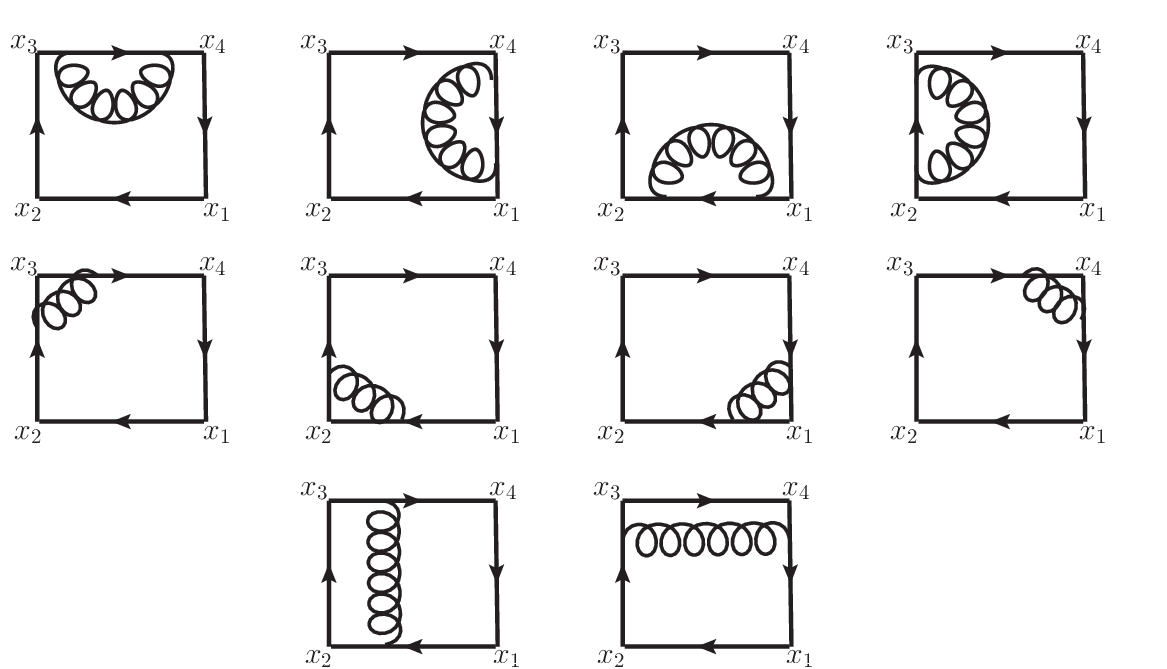}}
\caption{The complete set of Feynman diagrams required to calculate $W[C_4]$ to $\Ord(g^2)$.}
\label{WL4f}}
Of course, for $n = 6$, we will have to add a very large number of topologically distinct contributions in order to obtain a gauge invariant result. It will be simpler and get the point across just as effectively if we follow~\cite{DKS4pt} and calculate $W[C_4]$ in detail to $\Ord{(g^2)}$. The complete set of diagrams for the $\Ord{(g^2)}$ correction to $W[C_4]$ are shown in Figure \ref{WL4f}. In this case the only non-zero invariants are $x_{13}^2$ and $x_{24}^2$. Clearly, the first line of diagrams in Figure \ref{WL4f} vanish once the light-like character of the Wilson loop is taken into account. The second line of diagrams are divergent due to presence of the cusps. These divergences come from the regions of parameter-space where positions of absorption/emission approach a cusp. Such divergences are short distance and therefore ultraviolet in nature. Finally, we will see that the last line of diagrams are finite in four dimensions. If we denote the diagram in class $(\ell)$ that has a gluon stretched between lines $x_i - x_{i+1}$ and $x_j - x_{j+1}$ as $\mathcal{W}_{ij}^{(\ell)}$, 
we have 
\bea
\mathcal{W}_{i i}^{(1)} &=& 0 ~{\rm for~all}~1 \leq i \leq 4 \\
\mathcal{W}_{1 2}^{(2)} &=& \mathcal{W}_{3 4}^{(2)} = -{g^2 C_F e^{\gamma_E \e}(-x_{13}^2 \pi \mu^2)^\e \over 8 \pi^2 \e^2}\\
\mathcal{W}_{2 3}^{(2)} &=& \mathcal{W}_{1 4}^{(2)} = -{g^2 C_F e^{\gamma_E \e}(-x_{24}^2 \pi \mu^2)^\e \over 8 \pi^2 \e^2}\\
\mathcal{W}_{1 3}^{(3)} &=& \mathcal{W}_{2 4}^{(3)} = {g^2 C_F e^{\gamma_E \e}\left(\ln^2\left({x_{13}^2 \over x_{24}^2}\right)+\zeta(2)\right) \over 16 \pi^2} \,{\rm .}
\eea
Taking into account the fact that, in the large $\Nc$ limit,
$$C_F = {\Nc^2-1\over 2\Nc} \longrightarrow {\Nc \over 2} \, {\rm ,}$$
we make the replacement
\be
{g^2 C_F e^{\gamma_E \e} \pi^\e \over  8 \pi^2} \longrightarrow a
\ee
and find that the $\Ord(a)$ analytic structure of $W[C_4]$ is given by~\cite{DKS4pt}
\be
W[C_4]\Big|_{\Ord(a)} = -{1 \over \e^2}\bigg(\left(-x_{13}^2 \mu^2\right)^\e+\left(-x_{24}^2 \mu^2\right)^\e\bigg)+{1\over 2}\left(\ln^2\left({x_{13}^2 \over x_{24}^2}\right)+\pi^2\right) + \Ord(\e)\,{\rm .}
\label{4ptWLfin}
\ee
This is a remarkable result. Recall eq. (\ref{4ptanal}), where we wrote down the one-loop analytic structure of the four-point MHV superamplitude:
\be
\M^{(1)}(4,t_i^{[r]},\e') = C_\G \bigg\{
-{2 \over \e'^2} \Big[ \left( {-s\over\mu'^2}\right)^{-\e'}+ \left({-t\over\mu'^2}\right)^{-\e'} \Big]
+ \ln^2\left( {-s \over - t} \right) + \pi^2 \bigg\} \,{\rm ,}
\label{4ptanal2}
\ee
where $C_\G$ is given by
$$C_\G = {\G(1+\e')\G(1-\e')^2 \over 2 \G(1-2\e')}\,{\rm .}$$
Up to some redefinition of $a$, $\e$, and $\mu$, the above expression for $W[C_4]$ at lowest non-trivial order matches the above formula for $\M^{(1)}(4,t_i^{[r]},\e')$ exactly if we make the identifications
\be
s \leftrightarrow x_{13}^2 ~~~~{\rm and}~~~~ t \leftrightarrow x_{24}^2\,{\rm .} 
\ee
This surprising connection captures the essence of the light-like Wilson loop/MHV amplitude correspondence in planar $\Nsym$. Even more remarkably, the work of~\cite{DKS4pt} generalizes. There is now a large body of evidence for the following relation
\be
\ln \Bigg({\a_{n;2}\over \a_{n;2}^{tree}}\Bigg)\Bigg|_{{\rm finite};\,\Ord(a^L)}  = \ln \bigg(W[C_n]\bigg)\Bigg|_{{\rm finite};\,\Ord(a^L)} + D_n^{(L)}\, {\rm ,}
\label{equiv}
\ee
valid for all multiplicity and for all-loop orders (see Appendix \ref{ADS/CFT} for a bit more discussion). In the above, $D_n^{(L)}$ is transcendentality $2 L$ number. As one might guess from eqs. (\ref{4ptWLfin}) and (\ref{4ptanal2}) there is also a relation between the IR poles on the amplitude side and UV poles on the Wilson loop side. As hinted at above, one must make some non-trivial redefinitions of parameters in order to make this precise. See~\cite{DixonStermanMagnea} for a discussion of the IR poles. 

The key observation is that there is a superconformal symmetry (see Appendix \ref{sconf} if unfamiliar with superconformal symmetry) acting on the Wilson loop in a natural way because it is defined in a configuration space (where the Lagrangian density that possesses this symmetry is constructed). Ultraviolet divergences in the Wilson loop due to the presence of cusps breaks the subgroup of conformal transformations in a controlled fashion. The action of the conformal symmetry is anomalous and one can derive non-perturbatively valid anomalous conformal Ward identities that fix the finite part of $W[C_n]$ that comes from the breaking of the conformal symmetry up to an additive constant at all loop orders~\cite{DHKSward}. What remains must be a function of the conformal cross-ratios. For example, at the six-point level, there are three such cross-ratios
\be
u_1 = {x_{1 3}^2 x_{4 6}^2\over x_{1 4}^2 x_{3 6}^2} \qquad u_2 = {x_{2 4}^2 x_{5 1}^2\over x_{2 5}^2 x_{1 4}^2} \qquad u_3 = {x_{3 5}^2 x_{6 2}^2\over x_{3 6}^2 x_{2 5}^2}\,{\rm ,}
\label{crsrts}
\ee
each of which is invariant under conformal transformations. Now recall that the special conformal transformations can be obtained by conjugating the spatial translations by the conformal inversion operator, $I$ (Appendix \ref{sconf}). Furthermore, it is straightforward to see that $(x_{i j})^{\alpha \dot{\alpha}} = (x_i^\mu - x_j^\mu)(\sigma_\mu)^{\alpha \dot{\alpha}}$ transforms under inversion as:
\be
I[x_{i j}] = x_i^{-1} - x_j^{-1} = -x_j^{-1} (x_i - x_j)x_i^{-1} = -x_j^{-1}x_{i j}x_i^{-1}\,{\rm .}
\ee
Due to the fact that $u_2$ and $u_3$ are obtained by cyclicly permuting $u_1$, we can rest assured that they are conformally invariant if $u_1$ is. We see that
\be
I[u_1] = {I[x_{13}^2] I[x_{46}^2] \over I[x_{1 4}^2] I[x_{3 6}^2]} = {{x_{13}^2 \over x_1^2 x_3^2}{x_{46}^2\over x_4^2 x_6^2} \over {x_{1 4}^2 \over x_1^2 x_4^2} {x_{3 6}^2 \over x_3^2 x_6^2}} = {x_{1 3}^2 x_{4 6}^2\over x_{1 4}^2 x_{3 6}^2}
\ee
and $u_1$ is indeed invariant under inversion. This actually implies the invariance of $u_1$ under the full conformal group, since it is obviously invariant under Poincar\'{e} transformations and dilatations.

We are now in a position to make some remarks about the analytic structure of the $n$-point MHV superamplitudes in $\Nsym$. As we will discuss more in the next subsection, the fact the Wilson loop/MHV amplitude correspondence of eq. (\ref{equiv}) exists implies the existence of a novel hidden symmetry of the planar $\Nsym$ MHV amplitudes through the identification $x_{i}^\mu - x_{i+1}^\mu = p_i^\mu$. This symmetry is ``hidden'' because it cannot have its origin in the Lagrangian (it acts naturally in momentum space). This hidden symmetry is called dual superconformal invariance for reasons that should now be clear. In fact, the dual conformal subgroup already tells us quite a bit of useful information about the analytic structure of the MHV superamplitude. For instance, the reason that the BDS ansatz gives the exact finite part for $n = 4~{\rm or} ~5$ external states is obvious once one understands that the ansatz is just the contribution of the conformal anomaly and that the conformal anomaly is exact for four or five points; due to the light-like nature of the Wilson loops under consideration, no conformally invariant cross-ratios can even be written down for four or five particles in $\Nsym$. In fact, if dual conformal symmetry was not broken by IR divergences, we would expect the full non-perturbative answer to be just a constant times the appropriate tree amplitude.

We can also make precise the arguments of the two-loop six-point remainder function mentioned in Subsection \ref{BDSsect}. Recall eq. (\ref{6ptstruct}) for the analytic structure of $\ln(1+\sum_{L = 1} a^L \M^{(L)}(6,t_i^{[r]},\e))$ expanded up to second order in perturbation theory:
\be
\M^{(2)}(6,t_i^{[r]},\e)\Big|_{\rm finite} - {1\over 2}\M^{(1)}(6,t_i^{[r]},\e)^2\Big|_{\rm finite}= \alpha ~\M^{(1)}(6,t_i^{[r]},2\e)\Big|_{\rm finite} + \beta + R_{6}^{(2)}\left(t_i^{[r]}\right)\,{\rm .}
\ee
Given everything that we have discussed so far, it is clear that the six-point two-loop remainder function
$R_{6}^{(2)}\left(t_i^{[r]}\right)$ must actually be a function of three dual conformally invariant cross-ratios. If we use the dictionary
\cmb{-.2 in}{0 in}
\bea
&&x_{13}^2 \leftrightarrow s_{1} \qquad x_{24}^2 \leftrightarrow s_{2} \qquad x_{35}^2 \leftrightarrow s_{3} \qquad x_{46}^2 \leftrightarrow s_{4}\nonumber \\
x_{15}^2 & \leftrightarrow & s_{5} \qquad x_{26}^2 \leftrightarrow s_{6} \qquad x_{14}^2 \leftrightarrow t_{1}\qquad
x_{25}^2 \leftrightarrow t_{2} \qquad
x_{36}^2 \leftrightarrow t_{3}
\eea
\cme
we see that, from the point of view of dual conformal symmetry, eq. (\ref{crsrts}) becomes
\be
u_1 = {s_1 s_4\over t_1 t_3} \qquad u_2 = {s_2 s_5\over t_2 t_1} \qquad u_3 = {s_3 s_6\over t_3 t_2}
\ee
and we have
\be
R_{6}^{(2)}\left(t_i^{[r]}\right) = R_{6}^{(2)}(u_1,u_2,u_3) \, {\rm .}
\ee
So far, we have really only used the dual conformal subgroup of the dual superconformal symmetry. In the next section we will describe the full dual symmetry group~\cite{DHKSdualconf}.
\subsection{Dual Superconformal Invariance and the Pentagon Coefficients of the Planar $\Nsym$ One-Loop Six-Point NMHV Superamplitude}
\label{DSI}
To realize the dual superconformal generators on their dual superspace~\cite{DHKSdualconf} we introduce variables $\theta_{i\,\alpha}^{a}$ to solve the $\delta^{(8)}$($Q^a_{~\alpha}$) supercharge conservation constraint in much the same way that the $x_{i\,\alpha\dot{\alpha}}$ of the last subsection solve the $\delta^{(4)}$($P_{\alpha \dot{\alpha}}$) momentum conservation constraint. In other words,
\be
\theta_{i\,\alpha}^{a}-\theta_{i+1\,\alpha}^{a} = \lambda_{i\,\alpha}\eta_i^a
\ee
is the supersymmetric complement of the relation
\be
x_{i\,\alpha\dot{\alpha}} - x_{i+1\,\alpha\dot{\alpha}} = \lambda_{i\,\alpha}\tilde{\lambda}_{i\,\dot{\alpha}} \,{\rm .}
\ee
Intuitively, since (dual) superconformal symmetry naturally acts in (momentum) position space and position and momentum are not mutually compatible observables, we expect the algebra of the ordinary superconformal group (see Appendix \ref{sconf})  and the dual superconformal group to be somehow entangled. This intuition is correct; the sketch below shows that there is indeed some overlap between the non-trivial generators of the superconformal (left-hand side) and the dual superconformal (right-hand side) groups:
\bea
~~~~~P_{\alpha \dot{\alpha}}~~~~~&&~~~~~\mc{K}_{\alpha \dot{\alpha}} \nonumber \\
Q^a_{~\alpha}~~~~~~~~~~\bar{Q}_{b\,\dot{\alpha}} &=& \bar{\mc{S}}_{b\,\dot{\alpha}}~~~~~~~~~~\mc{S}^a_{~\alpha} \nonumber \\
S_{a\,\alpha}~~~~~~~~~~\bar{S}^b_{~\dot{\alpha}} &=& \bar{\mc{Q}}^b_{~\dot{\alpha}}~~~~~~~~~~\mc{Q}_{a\,\alpha} \nonumber \\
~~~~~K_{\alpha \dot{\alpha}}~~~~~&&~~~~~\mathcal{P}_{\alpha \dot{\alpha}}
\label{dsalgstruc}
\eea
In the above, the generators $Q^a_{~\alpha}$ and $P_{\alpha \dot{\alpha}}$ on the superconformal side and $\mc{Q}_{a\,\alpha}$ and $\mathcal{P}_{\alpha \dot{\alpha}}$ on the dual superconformal side are actually realized in a pretty trivial fashion and were just included to make the table look more symmetrical:
\bea
Q^a_{~\alpha} &=& \sum_{i = 1}^n \lambda_{i\,\alpha}\eta^a_i\qquad {\rm and} \qquad P_{\alpha \dot{\alpha}} = \sum_{i = 1}^n \lambda_{i\,\alpha}\tilde{\lambda}_{i\,\dot{\alpha}}\nonumber \\
\mc{Q}_{a\,\alpha} &=& \sum_{i = 1}^n {\partial\over \partial \T_{i}^{a\,\alpha}}\qquad {\rm and} \qquad \mathcal{P}_{\alpha \dot{\alpha}} = \sum_{i = 1}^n {\partial\over \partial x_{i}^{~\,\alpha \dot{\alpha}}}\,{\rm .}
\eea
The generators $S_{a\,\alpha}$ and $K_{\alpha \dot{\alpha}}$ on the superconformal side and $\mc{S}^a_{~\alpha}$ and $\mc{K}_{\alpha \dot{\alpha}}$ on the dual superconformal side are a lot more complicated:
\bea
S_{a\,\alpha} &=& \sum_{i = 1}^n {\partial\over \partial \lambda_{i}^{\,~\alpha}}{\partial\over \partial \eta^a_i}\qquad {\rm and} \qquad K_{\alpha \dot{\alpha}} = \sum_{i = 1}^n {\partial\over \partial \lambda_{i}^{\,~\alpha}}{\partial\over \partial \tilde{\lambda}_{i}^{~\,\dot{\alpha}}}\nonumber \\
\mc{S}^a_{~\alpha} &=& \sum_{i = 1}^n \left(-\theta^b_{i\,\alpha}\theta^{a\,\beta}_{i}{\partial\over \partial \T_{i}^{b\,\B}}+x_{i\,\alpha}^{~~\,\dot{\beta}}\theta^{a\,\beta}_{i}{\partial\over \partial x_{i}^{~\,\B \dot{\B}}}+\lambda_{i\,\alpha}\theta^{a\,\gamma}_{i}{\partial\over \partial \lambda_{i}^{\,~\g}}+x_{i+1\,\alpha}^{\,~~~~~\dot{\beta}}\eta^a_i{\partial\over \partial \tilde{\lambda}_{i}^{~\,\dot{\B}}}-\theta^b_{i+1\,\alpha}\eta^a_i{\partial\over \partial \eta^b_i}\right) \qquad {\rm and} \nonumber \\
\mc{K}_{\alpha \dot{\alpha}} &=& \sum_{i = 1}^n \left(x_{i\,\alpha}^{~~\,\dot{\beta}}x_{i\,\dot{\alpha}}^{~~\,\beta}{\partial\over \partial x_{i}^{~\,\B \dot{\B}}}+x_{i\,\dot{\alpha}}^{~~\,\beta}\theta_{i\,\alpha}^b{\partial\over \partial \T_{i}^{b\,\B}}+x_{i\,\dot{\alpha}}^{\,~~\beta}\lambda_{i\,\alpha}{\partial\over \partial \lambda_{i}^{\,~\B}}+x_{i+1\,\alpha}^{~~~~~\,\dot{\beta}}\tilde{\lambda}_{i\,\dot{\alpha}}{\partial\over \partial \tilde{\lambda}_{i}^{~\,\dot{\B}}}+\tilde{\lambda}_{i\,\dot{\alpha}}\theta^b_{i+1\,\alpha}{\partial\over \partial \eta_{i}^b}\right)\,{\rm .}\nn
\eea
Finally, the generators $\bar{Q}_{b\,\dot{\alpha}}$ and $\bar{S}^b_{~\dot{\alpha}}$ on the superconformal side and $\bar{\mc{S}}_{b\,\dot{\alpha}}$ and $\bar{\mc{Q}}^b_{~\dot{\alpha}}$ on the dual superconformal side:
\bea
\bar{Q}_{b\,\dot{\alpha}} &=& \sum_{i = 1}^n \tilde{\lambda}_{i\,\dot{\alpha}}{\partial\over \partial \eta^b_{i}}\qquad {\rm and} \qquad \bar{S}^b_{~\dot{\alpha}} = \sum_{i = 1}^n \eta_{i}^b{\partial\over \partial \tilde{\lambda}_{i}^{~\,\dot{\alpha}}} \label{QbarSbar} \\
\bar{\mc{S}}_{b\,\dot{\alpha}} &=& \sum_{i = 1}^n \left(x_{i\,\dot{\alpha}}^{\,~~\beta}{\partial\over \partial \T_{i}^{b\,\B}}+\tilde{\lambda}_{i\,\dot{\alpha}}{\partial\over \partial \eta^b_{i}}\right)\qquad {\rm and} \qquad \bar{\mc{Q}}^b_{~\dot{\alpha}} = \sum_{i = 1}^n \left(\theta^{b\,\alpha}_{i}{\partial\over \partial x_{i}^{~\,\alpha \dot{\alpha}}}+\eta_{i}^{b}{\partial\over \partial \tilde{\lambda}_{i}^{~\,\dot{\alpha}}}\right)\,{\rm .}\nonumber
\eea
actually match up if we restrict to the on-shell superspace introduced in Section \ref{supercomp} (by ignoring all $\theta_{i\,\alpha}^{a}$ terms).

Of course, if one wants to check explicitly that all the (anti)commutation relations (see Appendix \ref{sconf}) are satisfied, one needs the rest of the representation. The rest of the ordinary and dual superconformal generators are given in Appendix \ref{sconf}. The above discussion was just intended to give the reader some sense of how the ordinary and dual superconformal algebras fit together. Although we will not use it here, it is worth emphasizing that the superconformal and dual superconformal algebras do not commute (this is clear from the form of eq. (\ref{dsalgstruc})). It turns out that their closure is a Yangian~\cite{DHPYangian1,BHMPYangian2}.  It is also worth pointing out that, since the dual superconformal generators are first order differential operators, one may expect them to be better behaved at the quantum level than the usual superconformal generators. It is therefore reasonable to suppose that formulae for one-loop superamplitudes which make the dual superconformal symmetry as manifest as possible will be simpler than those of Section \ref{supercomp} (there the dual superconformal symmetry was hidden).

In~\cite{DHKSdualconf}, Drummond, Henn, Korchemsky, and Sokatchev constructed a set of six dual superconformally invariant functions,

\bea
R_{1 4 6} &=& {\D^{(4)}\left(\spb4.5 \eta_6^a+\spb5.6\eta_4^a+
\spb6.4\eta_5^a\right) \prod_{i = 1}^6 \spa{i}.{i+1}\over t_1 \spa1.2 \spa2.3 \spab1.{5+6}.4 \spab3.{4+5}.6\spb4.5\spb5.6}~~~R_{2 5 1} = {\D^{(4)}\left(\spb5.6 \eta_1^a+\spb6.1\eta_5^a+
\spb1.5\eta_6^a\right) \prod_{i = 1}^6 \spa{i}.{i+1}\over t_2 \spa2.3 \spa3.4 \spab2.{6+1}.5 \spab4.{5+6}.1\spb5.6\spb6.1} \nonumber \\
R_{3 6 2} &=& {\D^{(4)}\left(\spb6.1 \eta_2^a+\spb1.2\eta_6^a+
\spb2.6\eta_1^a\right) \prod_{i = 1}^6 \spa{i}.{i+1}\over t_3 \spa3.4 \spa4.5 \spab3.{1+2}.6 \spab5.{6+1}.2\spb6.1\spb1.2}~~~R_{4 1 3} = {\D^{(4)}\left(\spb1.2 \eta_3^a+\spb2.3\eta_1^a+
\spb3.1\eta_2^a\right) \prod_{i = 1}^6 \spa{i}.{i+1} \over t_1 \spa4.5 \spa5.6 \spab4.{2+3}.1 \spab6.{1+2}.3\spb1.2\spb2.3} \nonumber \\
R_{5 2 4} &=& {\D^{(4)}\left(\spb2.3 \eta_4^a+\spb3.4\eta_2^a+
\spb4.2\eta_3^a\right)\prod_{i = 1}^6 \spa{i}.{i+1} \over t_2 \spa5.6 \spa6.1 \spab5.{3+4}.2 \spab1.{2+3}.4\spb2.3\spb3.4}~~~R_{6 3 5} = {\D^{(4)}\left(\spb3.4 \eta_5^a+\spb4.5\eta_3^a+
\spb5.3\eta_4^a\right) \prod_{i = 1}^6 \spa{i}.{i+1}\over t_3 \spa6.1 \spa1.2 \spab6.{4+5}.3 \spab2.{3+4}.5\spb3.4\spb4.5} \nn
\eea
which they then used to write all of the one-loop box coefficients of $\a^{1-{\rm loop}}_{6;3}$ in a way that meshes well with dual superconformal symmetry. More precisely, they found that they could express all the leading singularities in the computation of the NMHV superamplitude in a manifestly dual superconformally invariant way using $R_{1 4 6}$ and its cyclic permutations. In $\Nsym$ there is a choice of basis (the dual conformal basis introduced in Subsection \ref{GUD}) where the dual superconformal properties of the theory at loop level are as manifest as possible. At the one-loop level, this basis consists of $D = 4 - 2\e$ box integrals and $D = 6 - 2\e$ pentagon integrals. 

The simplicity of the results for boxes suggests that we should try to play the same game for the (now known) pentagon coefficients of the NMHV superamplitude. This is actually not as straightforward as it sounds, due to the fact that the $R_{pqr}$ above do not form a linearly independent set~\cite{DHKSdualconf,DHsuperBCFW}. In fact, for each pentagon topology, it is possible to fit an ansatz of the form
\bea
&&C_i {i \D^{(8)}(Q^{a\,\alpha})\over \spa1.2\spa2.3\spa3.4\spa4.5\spa5.6\spa6.1} \left(z_1^{(i)} R_{413} + z_2^{(i)} R_{524} + z_3^{(i)} R_{635}+ z_4^{(i)} R_{146}+z_5^{(i)} R_{251}+z_6^{(i)} R_{362}\right) \nonumber \\
&=& C_i \a^{tree}_{6;2} \left(z_1^{(i)} R_{413} + z_2^{(i)} R_{524} + z_3^{(i)} R_{635}+ z_4^{(i)} R_{146}+z_5^{(i)} R_{251}+z_6^{(i)} R_{362}\right)
\eea
using just five component amplitudes (for example those used to fix the form of $\a^{1-{\rm loop}}_{6;3}$ in Section \ref{supercomp}). Fortunately, there is an obvious, preferred, maximally symmetric solution: $z_i^{(i)} = z_{i+3}^{(i)}$. For example, for the pentagon coefficient of $I_5^{(5)}$, we set $z_5^{(5)} = z_2^{(5)}$. This choice then forces 
\be
\left(z_1^{(5)}\right)^{\langle~\rangle \leftrightarrow [~]} = z_4^{(5)} \qquad \left(z_3^{(5)}\right)^{\langle~\rangle \leftrightarrow [~]} = z_6^{(5)}
\ee
as well. The other topologies behave in a completely analogous fashion. To simplify the result, it is convenient to work numerically with complex spinors. It is then possible to recognize the origin of the imaginary parts of the $z_i$ as coming from the natural odd six-point invariant
$$\spb1.2\spa2.3\spb3.4\spa4.5\spb5.6\spa6.1-\spa1.2\spb2.3\spa3.4\spb4.5\spa5.6\spb6.1\,{\rm .}$$
We can now determine the rest of the structure by experimenting with real-valued candidate expressions that respect all the constraints of the problem and have the right BCFW shifts in all channels. In the end, we find
\cmb{0 in}{0 in}
\bea
 &&\a^{1-{\rm loop}}_{6;3} =
 \el \cdots + \frac{i}{6} \e ~\a^{tree}_{6;2}\sum_{i = 1}^6 C_i\bigg(
   {1\over 2} \left(2 s_{i+1} s_{i-2} - t_{i} t_{i+1}\right) t_{i-1} \left(R_{i+2\,i-1\,i+1} + R_{i-1\,i+2\,i-2}\right) \nonumber \\
  && - \Big([i \,i+1] \langle i+1\, i+2\rangle [i+2 \,i+3] \langle i+3\, i+4\rangle [i+4 \,i+5] \langle i+5\,i\rangle \nonumber \\
  &&- \langle i\, i+1\rangle [i+1 \,i+2] \langle i+2\, i+3\rangle [i+3\, i+4] \langle i+4\, i+5\rangle [i+5\, i]\Big) \left(R_{i+2 \, i-1\,i+1} - R_{i-1\,i+2\,i-2}\right) \nonumber \\
  && + {1\over 2} \left(2 s_{i-1} s_{i+2} - t_{i-1} t_{i+1}\right) t_{i} \left(R_{i+3\,i\,i+2} + R_{i\,i+3\,i-1}\right) + {1\over 2} \left(2 s_{i+3} s_{i} - t_{i} t_{i-1}\right) t_{i+1} \left(R_{i+1\,i-2\,i} + R_{i-2\,i+1\,i-3}\right) \nonumber \\
  && -\Big([i \,i+1] \langle i+1\, i+2\rangle [i+2 \,i+3] \langle i+3\, i+4\rangle [i+4 \,i+5] \langle i+5\,i\rangle \nonumber \\
  &&- \langle i\, i+1\rangle [i+1 \,i+2] \langle i+2\, i+3\rangle [i+3\, i+4] \langle i+4\, i+5\rangle [i+5\, i]\Big) \left(R_{i+1\,i-2\,i} - R_{i-2\,i+1\,i-3}\right) \bigg) I^{(i),\,D = 6 - 2\e}_5 \,{\rm .} \nn
\label{sd}
\eea
\cme
Remarkably, when written in this form, the pentagon contributions to the one-loop six-point NMHV superamplitude are related by cyclic symmetry. We can use this fact to explain relation (\ref{mystrel}), reproduced below for convenience:
\be
{K_1 \over C_1} = {K_4 \over C_4} \qquad
{K_2 \over C_2} = {K_5 \over C_5} \qquad
{K_3 \over C_3} = {K_6 \over C_6} \, {\rm .}
\ee
Examining eq. (\ref{sd}), it is trivial to see that the only piece of a given pentagon that does not return to itself under $i \rightarrow i+3$, is the $C_i$ coefficient out front. Evidently, relation (\ref{mystrel}) is a property of the full superamplitude because the symmetric choice
$$z_2^{(i)} = z_5^{(i)}$$
in our ansatz was necessary to manifest the cyclic symmetry of the superamplitude; writing the superamplitude in the form given by eq. (\ref{sd}) shows that there are not enough independent R-invariant structures, to support a full $i \rightarrow i+6$ symmetry for the coefficients divided by their $C_i$. That there are only three independent R-invariant structures can be understood as a consequence of parity invariance in the superamplitude; parity acts on R-invariants by shifting their indices from $i$ to $i+3$. 

Now that we have in hand a pretty formula for the pentagon coefficients of $\a^{1-{\rm loop}}_{6;3}$ built out of dual superconformal invariants, it would be nice if there was some application of our result. It is to this that we turn in the next subsection.
\subsection{Ratio of the Six-Point NMHV and MHV Superamplitudes at Two-Loops}
\label{rationfunc}
In~\cite{DHKSdualconf}, Drummond, Henn, Korchemsky, and Sokatchev made an interesting all-loop prediction based on a remarkable one-loop calculation in their paper. They calculated the parity even part of the NMHV ratio function, $R_{{\rm NMHV}} \equiv \a_{6;3}/\a_{6;2}\,$, to $\Ord(a)$ and found a dual superconformally invariant function. This is a non-trivial result because both $\a_{6;3}$ and $\a_{6;2}$ have IR divergences. The universal, helicity-blind structure of the IR divergences guarantees that the NMHV ratio function is finite to all loop orders. However, at loop level the dual superconformal symmetry is anomalous. One way to circumvent this problem might be to write
\be
\a_{6;3} = \a_{6;2} \Big(R_{\rm NMHV} + \Ord(\e)\Big)
\label{ratiofunc}
\ee
to all loop orders and hope that all of the messiness associated with dual superconformal anomalies resides in the $\a_{6;2}$ prefactor.\footnote{Recently, Beisert, Henn, McLoughlin, and Plefka developed a technique to address these anomalies directly by deforming the dual superconformal generators~\cite{BHMPYangian2}.} It is not {\it a priori} clear that the ratio function, $R_{\rm NMHV}$, should have any special properties.. For example, as discussed in~\cite{DHKSdualconf}, it is not obvious that the dual superconformal generator $\bar{\mc{Q}}^a_{~\dot{\alpha}}$ annihilates the ratio function, because this generator (eq. (\ref{QbarSbar})) is  sensitive to the dependence of $R_{\rm NMHV}$ on the dual variables, $x_{i\,\alpha \dot{\alpha}}$, and the dependence of the finite parts of $\a_{6;3}$ and $\a_{6;2}$ on the dual variables is fairly complicated (even at $\Ord(a)$). Therefore it is interesting to check by explicit calculation that $R_{\rm NMHV}$ is given by a dual superconformally invariant function. We have already seen that pulling a factor of $\a_{6;2}^{tree}$ out of $\a_{6;3}^{1-{\rm loop}}$ is a natural operation and simplifies the formula for the one-loop NMHV superamplitude. The question is whether $\a_{6;3}^{1-{\rm loop}}$ simplifies when one factors out the entire one-loop MHV superamplitude.

DHKS carried out this analysis and they found that $R_{\rm NMHV}^{1-{\rm loop}}$ could be expressed in terms of R-invariants and linear combinations of two mass hard, two mass easy, and one mass boxes (see eqs. (\ref{6ptgMHV}) and (\ref{6ptgNMHV})). When evaluated through $\Ord(\e^0)$  (see eqs. (\ref{box1})-(\ref{box2h})), these box integrals give rise to logarithms and dilogarithms. After simplifying all logarithms and dilogarithms, non-trivial cancellations occur and DHKS found the simple dual superconformally invariant result:\footnote{In eq. (\ref{1Lratfunc}) the $\pi^2/3$ factors are inessential and depend on precisely how one defines the analytic structure of the MHV amplitude. We follow the conventions of DHKS in~\cite{DHKSgenunit}.}
\bea
R_{\rm NMHV}^{1-{\rm loop}} &=& {1 \over 4} \sum_{i=1}^6 R_{i\,i+3\,i+5} \bigg(-\ln\left(u_i\right)\ln\left(u_{i+1}\right)+\ln\left(u_{i+1}\right)\ln\left(u_{i+2}\right)+\ln\left(u_{i+2}\right)\ln\left(u_i\right) \nonumber \\
&+&\li2\left(1-u_i\right)+\li2\left(1-u_{i+1}\right)+\li2\left(1-u_{i+2}\right)-{\pi^2\over 3}\bigg) \,{\rm .}
\label{1Lratfunc}
\eea
It is important to note that, in eq. (\ref{1Lratfunc}) above, the index $i$ is understood to be mod 3 for the $u_i$ and mod 6 for the $R_{i\,i+3\,i+5}$. Given the validity of eq. (\ref{ratiofunc}) at one loop and six points, it is reasonable to suspect that something similar will happen at higher loops as well. However, a na\"{i}ve extrapolation from one to higher loops is often dangerous. For example, the BDS ansatz is exact at the one-loop $n$-point level, but is incomplete at two loops and six points, as discussed in \ref{BDSsect}. NMHV configurations first appear at the six-point level and, consequently, the first really non-trivial check of (\ref{ratiofunc}) is at two loops and six points. To this end, Kosower, Roiban, and Vergu recently computed the two-loop six-point NMHV superamplitude and they verified (\ref{ratiofunc}) for the parity even part of the ratio function~\cite{KRV}. Before they could check (\ref{ratiofunc}) at $\Ord(a^2)$, they had to resolve a technical problem related to $\e$ poles induced by $\mu$-integrals at the two-loop level. 

In order to understand the problem we need to recall the discussion of Subsection \ref{gresults} where we introduced $\mu$-integral hexabox integrals. We did not properly define this integral in \ref{gresults} because it was not necessary at the time. The $\mu$-integral hexabox integral depicted in Figure \ref{hexabox} is given by
\cmb{-.8 in}{0 in}
\bea
&&I_{(4,6)}^{(2);\,D = 4 - 2\e}[\mu^2] = \int {dp^{4-2\e}\over (2\pi)^{4-2\e}}  
{1\over p^2 (p-k_2)^2(p-k_1-k_2)^2}\int {dq^4\over (2 \pi)^4}\int {d^{-2\e} \mu \over (2\pi)^{-2\e} }\times
\el
\times{\mu^2 \over ((q+p)^2+2 \vec{\mu}\cdot p-\mu^2)(q^2-\mu^2)((q-k_3)^2-\mu^2)((q-k_3-k_4)^2-\mu^2)((q-k_3-k_4-k_5)^2-\mu^2)((q+k_1+k_2)^2-\mu^2)}\,{\rm .} \nn
\eea
\cme
In this case it turns out that, to leading order, the above integral factorizes~\cite{BDKRSVV} and we can write 
\cmb{0 in}{0 in}
\bea
&I_{(4,6)}^{(2);\,D = 4 - 2\e}[\mu^2]& = I_3^{(2);\,D = 4 - 2\e} I_{6}^{D = 4 - 2\e}[\mu^2] = \left(-{1 \over \e^2}(-s_{1})^{-1-\e}\right)\left(-\e I_6^{D = 6 - 2\e}\right) 
\elale \left(-{1 \over \e^2}(-s_{1})^{-1-\e}\right)\left(-{\e \over 2} \sum_{i = 1}^6 C_i I_5^{(i);\,D = 6 - 2\e}\right) \,{\rm ,}
\label{hexaboxrel}
\eea
\cme
where the last equality follows from eq. (\ref{hexred}). One can check numerically that (\ref{hexaboxrel}) is valid through $\Ord(\e^0)$; the hexabox $\mu$-integral can be evaluated through $\Ord(\e^0)$, apart from trivial factors, is a $1/\e$ pole times a certain linear combination of the finite one-loop functions $I_5^{(i);\,D = 6}$. In our discussion of planar gluon NMHV amplitudes in Section \ref{gluoncomp}, we noted a close connection between the one-loop pentagon coefficients we calculated and appropriate $\mu$-integral hexabox coefficients. For the sake of concreteness, we go back to the particular example discussed in \ref{gresults}, where we wrote down the relationship between the coefficients of $\e \,I_5^{(2);\,D = 6 - 2\e}$ and $I_{(4,6)}^{(2);\,D = 4 - 2\e}[\mu^2]$:
\be
K_2 = {C_2 \over 2 s_1} \mathcal{K}_2 \, {\rm .}
\ee
If we use the above relation to express the $\mathcal{K}_2$ in terms of $K_2$, we find that the contribution from this NMHV $\mu$-integral hexabox to the ratio function at $\Ord(a^2)$ looks like 
\be
\mc{K}_2 I_{(4,6)}^{(2);\,D = 4 - 2\e}[\mu^2] = {(-s_1)^{-\e} K_2 \over \e \,C_2} \sum_{i=1}^6 C_i I_5^{(i);\,D = 6} + \Ord(\e^0)
\label{hexaboxepole}
\ee
To see how this is all related to our one-loop NMHV pentagon coefficients, let us take a step back and remember what we're trying to calculate. Since we want $R_{\rm NMHV}$ to two loops\footnote{In eq. (\ref{ratexp}), $\hat{\a}_{6;3}^{L-{\rm loop}}$ denotes the superamplitude with a factor of $\a_{6;2}^{tree}$ stripped off.} 
\bea
R_{\rm NMHV} &=& {\hat{\a}_{6;3}^{tree} + a \,\hat{\a}_{6;3}^{1-{\rm loop}} + a^2 \hat{\a}_{6;3}^{2-{\rm loop}} +\cdots\over 1 + a\, \M^{(1)}(n,t_i^{[r]},\e)+a^2 \M^{(2)}(n,t_i^{[r]},\e)+\cdots} 
\elale \hat{\a}_{6;3}^{tree} + a \left(\hat{\a}_{6;3}^{1-{\rm loop}}-\hat{\a}_{6;3}^{tree} \M^{(1)}(n,t_i^{[r]},\e)\right) 
\el+ a^2 \left(\hat{\a}_{6;3}^{2-{\rm loop}} - \hat{\a}_{6;3}^{1-{\rm loop}} \M^{(1)}(n,t_i^{[r]},\e) + \hat{\a}_{6;3}^{tree} \M^{(1)}(n,t_i^{[r]},\e)^2 - \hat{\a}_{6;3}^{tree} \M^{(2)}(n,t_i^{[r]},\e)\right) 
\el + \Ord(a^3)\,{\rm ,}\nn
\label{ratexp}
\eea
we see that there are other places for us to look for $1/\e$ poles at $\Ord(a^2)$ besides the actual two loop contributions. It is possible for one-loop contributions of $\Ord(\e)$ to hit the universal soft singular terms (see eq. (\ref{1LIR})) in another one-loop contribution and interfere to produce $1/\e$ singularities. For instance, there will be a contribution of the form
\be
-\left(\e\, K_2 I_5^{(2);\,D = 6 - 2 \e}\right)\left(-{1 \over \e^2}\sum^6_{i=1}\left(-s_{i\,i+1}\right)^{-\e}\right) = {K_2 \sum^6_{i=1}\left(-s_{i\,i+1}\right)^{-\e} \over \e} I_5^{(2);\,D = 6 - 2 \e} + \Ord(\e^0)\nn
\label{pentagonepole}
\ee
coming from the cross-term $-\hat{\a}_{6;3}^{1-{\rm loop}} \M^{(1)}(n,t_i^{[r]},\e)$. This shows that, to obtain all IR divergent contributions to the parity even part of $R_{\rm NMHV}$, the even terms in the one-loop NMHV pentagon coefficients of eq. (\ref{sd}) must be included. Indeed, the authors of~\cite{KRV} have checked at the level of superamplitudes using our results that the even part of $R_{\rm NMHV}$ is dual superconformally invariant. It is now possible to explain why the hexabox coefficients derived by Kosower, Roiban, and Vergu in~\cite{KRV} are so similar to our one-loop pentagon coefficients. A close connection between them is necessary for all of the exotic IR divergent contributions (those that have their origin in $\mu$-integrals) to cancel out in the calculation of the ratio function.

Actually, with a modest amount of additional effort, we can simplify our NMHV pentagon coefficients further and explicitly make contact with the form used by Kosower, Roiban, and Vergu in carrying out their analysis of the two-loop NMHV ratio function. Kosower, Roiban, and Vergu make use of a particular rearrangement of eq. (\ref{sd}). This rearrangement is very nice because, with it, the usual MHV level notions\footnote{At the MHV level, the ``even components'' are simply those terms in the amplitude with no explicit factors of $\pol(i,j,k,\ell)$ and the ``odd components'' are those terms with such factors. We remind the reader that $\pol(i,j,k,\ell)$ was defined in eq. (\ref{epstensor}).} of ``even components'' and ``odd components'' actually make sense in the context of the one-loop NMHV amplitude as well. 

Recall the form of (\ref{sd}) and collect all terms in the above proportional to each R-invariant structure\footnote{There are six such structures: $R_{3 6 2} + R_{6 3 5}$, $R_{4 1 3} + R_{1 4 6}$, $R_{5 2 4} + R_{2 5 1}$, $R_{3 6 2} - R_{6 3 5}$, $R_{4 1 3} - R_{1 4 6}$, and $R_{5 2 4} - R_{2 5 1}$.}:
\bea
 &&\a^{1-{\rm loop}}_{6;3} = \cdots + \frac{i}{6} \e ~\a^{tree}_{6;2}\left\{{\frac{1}{2}}\sum_{i = 1}^6 C_i I^{(i),\,D = 6 - 2\e}_5\right\}\sum_{i = 1}^3 \left(2 s_{i+1} s_{i-2} - t_{i} t_{i+1}\right) t_{i-1} \left(R_{i+2\,i-1\,i+1} + R_{i-1\,i+2\,i-2}\right) \nonumber \\
  && + \frac{i}{6} \e ~\a^{tree}_{6;2}\sum_{i = 1}^3 (-1)^{i} \left( C_i I^{(i),\,D = 6 - 2\e}_5 - C_{i+1} I^{(i+1),\,D = 6 - 2\e}_5 + C_{i-3} I^{(i-3),\,D = 6 - 2\e}_5 - C_{i-2} I^{(i-2),\,D = 6 - 2\e}_5\right)\times
  \el\times \Big(\spb1.2 \spa2.3 \spb3.4 \spa4.5 \spb5.6 \spa6.1 - \spa1.2 \spb2.3 \spa3.4 \spb4.5 \spa5.6 \spb6.1\Big) \left(R_{i+2 \, i-1\,i+1} - R_{i-1\,i+2\,i-2}\right) \,{\rm .}
\label{sd2}
\eea
Using eq. (\ref{hexred}), reproduced below for the convenience of reader,
\be
I^{D = 6 - 2\e}_6 = {1 \over 2}\sum_{i = 1}^6 C_i I^{(i),\,D = 6 - 2\e}_5
\ee
the first line of eq. (\ref{sd2}) can be put into a form that bears a close resemblance to the even components of the higher order pieces of the one-loop MHV superamplitude; it is proportional to the one-loop scalar hexagon integral (see eq. (\ref{6ptMHV})), $I^{D = 6 - 2\e}_6$. 

In fact, a similar simplification is possible for the terms proportional to $R_{i+2 \, i-1\,i+1} - R_{i-1\,i+2\,i-2}$ as well, although it is not at all obvious. We have numerically checked that
\be
C_i = {2(-1)^i \pol(i+1,i+2,i+3,i+4)\over \spb1.2 \spa2.3 \spb3.4 \spa4.5 \spb5.6 \spa6.1 - \spa1.2 \spb2.3 \spa3.4 \spb4.5 \spa5.6 \spb6.1}\,.
\ee
Using this relation we see that the terms in eq. (\ref{sd2}) not directly proportional to $I^{D = 6 - 2\e}_6$ bear a striking resemblance to the odd components of the higher order pieces of the one-loop MHV superamplitude (again, see eq. (\ref{6ptMHV})). Putting everything together, we find
\bea
&&\a^{1-{\rm loop}}_{6;3} = \cdots + \frac{i}{6} \e ~\a^{tree}_{6;2} I^{D = 6 - 2\e}_6 \sum_{i = 1}^3 \left(2 s_{i+1} s_{i-2} - t_{i} t_{i+1}\right) t_{i-1} \left(R_{i+2\,i-1\,i+1} + R_{i-1\,i+2\,i-2}\right) \nonumber \\
  && + \frac{i}{3} \e ~\a^{tree}_{6;2}\sum_{i = 1}^3 \bigg( \pol(i+1,i+2,i+3,i+4) I^{(i),\,D = 6 - 2\e}_5 + \pol(i+2,i+3,i+4,i+5) I^{(i+1),\,D = 6 - 2\e}_5 \nonumber \\
  &&- \pol(i-2,i-1,i,i+1) I^{(i-3),\,D = 6 - 2\e}_5 - \pol(i-1,i,i+1,i+2) I^{(i-2),\,D = 6 - 2\e}_5\bigg)\left(R_{i+2 \, i-1\,i+1} - R_{i-1\,i+2\,i-2}\right)\nn
\label{sdfinal}
\eea
for the higher-order components of the planar one-loop NMHV superamplitude. Eq. (\ref{sdfinal}) is particularly important because it was the form utilized by Kosower, Roiban, and Vergu for their analysis in~\cite{KRV}. It is now clear that, indeed, the notions of even and odd that were used in the context of the planar one-loop MHV superamplitude make sense at the NMHV level as well.
\section{Summary}
\label{sum}
In this work we have discussed several recent developments in the theory of the $\Nsym$ S-matrix. After reviewing some of the most important computational techniques in \ref{revcomp}, we discussed a simple refinement of the $D$ dimensional unitarity technique of Bern and Morgan in \ref{effgcomp}. One notable feature of our approach is that all integrands are reconstructed in $D$ dimensions directly from tree amplitudes without any need for supersymmetric decompositions. While our approach to $D$ dimensional unitarity is probably already familiar to experts in the field, to the best of our knowledge no detailed exposition of the ideas have appeared in print so far.\footnote{A particularly interesting recent study~\cite{BCDHI} further develops $D$ dimensional generalized unitarity along different lines. In future work it would be very nice to make contact with the formalism developed in~\cite{BCDHI}.} We also discuss how our approach to $D$ dimensional unitarity meshes well with the leading singularity method in the context of all-orders-in-$\e$ one-loop $\Nsym$ calculations. In \ref{gresults} we presented simple formulae for the higher-order in $\e$ pentagon coefficients of the planar one-loop six-gluon NMHV amplitudes $A^{1-{\rm loop}}_{1}(k_1^{1234},k_2^{1234},k_3^{1234},k_4,k_5,k_6)$, $A^{1-{\rm loop}}_{1}(k_1^{1234},k_2^{1234},k_3,k_4^{1234},k_5,k_6)$, and $A^{1-{\rm loop}}_{1}(k_1^{1234},k_2,k_3^{1234},k_4,k_5^{1234},k_6)$. Na\"{i}vely, these results may seem rather useless because, if one only cares about the massless $\Nsym$ S-matrix, one never needs the pentagon coefficients.

However, we argue in \ref{gsrel} that, actually, the higher-order in $\e$ pentagon coefficients are useful because they contain non-trivial information about tree-level scattering of massless modes in open superstring theory. After reviewing the non-Abelian Born-Infeld action in \ref{BornInfeld}, we argued in \ref{results} that matrix elements of the non-Abelian Born-Infeld action at $\Ord(\alpha'^2)$ and $\Ord(\alpha'^3)$ can be predicted from all-orders-in-$\e$ $\Nsym$ amplitudes dimensionally shifted to either $D = 8-2\e$ or $D = 10-2\e$. As an amusing by-product of our analysis, we were able to use another close connection between the one-loop all-plus amplitudes in pure Yang-Mills and our stringy corrections at $\Ord(\alpha'^2)$ to understand the vanishing of the all-plus amplitudes when three or more gluons are replaced by photons for $n > 4$.  

At this point, in Section \ref{supercomp}, we explained how to supersymmetrize the results of \ref{gresults}. To this end, we introduced the $\Nsym$ on-shell superspace in \ref{gendisonshell} and discussed some important examples of $\Nsym$ superamplitudes. In \ref{ffbarcalc}, we first explain that, following Elvang, Freedman, and Kiermaier, one can choose the five component amplitudes $A_1^{1-{\rm loop}}\left(k_1^{1234},k_2^{1234},k_3^{1234},k_4,k_5,k_6\right)$, $A_1^{1-{\rm loop}}\left(k_1^{1234},k_2^{1234},k_3^{123},k_4^{4},k_5,k_6\right)$,\\ $A_1^{1-{\rm loop}}\left(k_1^{1234},k_2^{1234},k_3^{12},k_4^{34},k_5,k_6\right)$,$A_1^{1-{\rm loop}}\left(k_1^{1234},k_2^{1234},k_3^{1},k_4^{234},k_5,k_6\right)$, \\and $A_1^{1-{\rm loop}}\left(k_1^{1234},k_2^{1234},k_3,k_4^{1234},k_5,k_6\right)$ and determine the full $\Nsym$ superamplitude in terms of them. We then showed how to extend the methods of \ref{effgcomp} to deal with\\ $A_1^{1-{\rm loop}}\left(k_1^{1234},k_2^{1234},k_3^{123},k_4^{4},k_5,k_6\right)$ and  $A_1^{1-{\rm loop}}\left(k_1^{1234},k_2^{1234},k_3^{1},k_4^{234},k_5,k_6\right)$. It is crucial that our techniques be applicable to amplitudes with external fermions if we want them to be useful for theories with less supersymmetry such as QCD. Finally, in \ref{ssbarcalc} we determine $A_1^{1-{\rm loop}}\left(k_1^{1234},k_2^{1234},k_3^{12},k_4^{34},k_5,k_6\right)$ indirectly and in \ref{supersymresults} we complete the process by collecting our results. We write down for the first time the higher-order pentagon contributions to the six-point NMHV $\Nsym$ superamplitude.

We then show in \ref{DSI} (after reviewing some of the developments that led to the discovery of dual superconformal invariance in \ref{BDSsect} and \ref{WL/MHV2}) that the superamplitude takes on a significantly simpler form if expressed in terms of the R-invariants of Drummond, Henn, Korchemsky, and Sokatchev. Remarkably, in this form, the pentagon coefficients are related by cyclic symmetry. We can understand the greater simplicity of this formula by comparing the explicit operator realization of the ordinary and dual superconformal symmetries. Some of the ordinary superconformal generators are expressed in terms of 2nd-order partial differential operators, whereas {\it all} of the dual superconformal generators are expressed in terms of 1st-order differential operators. This is to be expected since differences of dual variables are just momenta; the dual superconformal symmetry acts naturally in momentum space. As a result, it is not too surprising that, when expressed in terms of R-invariants, the pentagon coefficients look even simpler than those presented in \ref{supersymresults}, where dual superconformal symmetry was obscured. 

Finally, in \ref{rationfunc} we explain the relevance of our results to the study of the dual superconformal properties of (the parity even part of) the two-loop NMHV ratio function in dimensional regularization.  Our higher-order-in-$\e$ pentagon coefficients can interfere with $1/\e^2$ poles in the one-loop MHV superamplitude to produce contributions of order $1/\e$. Thus, the results written down in \ref{DSI} in terms of dual superconformal R-invariants are necessary to produce a finite result for the two-loop NMHV ratio function in on-shell superspace if one is working in dimensional regularization. To this end, we further improve the results presented in \ref{DSI} by seperating them into even and odd components. This decomposition is clearly natural because the results of \ref{DSI} simplify still more. In particular, the even components of the higher order pieces of the one-loop NMHV superamplitude can be rearranged and put into a form where they are actually proportional to the one-loop hexagon integral in $D = 6 - 2 \e$. This feature of the even components was exploited recently in a study of the two-loop NMHV ratio function by Kosower, Roiban, and Vergu~\cite{KRV}.

There has been a tremendous amount of recent progress on the planar $\Nsym$ S-matrix\footnote{Most, if not all, of the works cited here were significantly influenced by the seminal work of Witten~\cite{witten}.}~\cite{newpap1,newpap2,newpap3,newpap4,newpap5,newpap6,newpap7,newpap8,newpap9,newpap10,newpap11,newpap12,newpap13,newpap14,newpap15,newpap16,newpap17,newpap18,newpap19,newpap20,newpap21,newpap22,newpap23,newpap24,newpap25,newpap26,newpap27,newpap28,newpap29,DelDuca1,DelDuca2,DelDuca3,newpap30,newpap32,newpap33,newpap34,newpap35,newpap36}, which, unfortunately, we don't have time to say much about. A recurring theme in recent papers on the subject is the idea that one should be able to learn everything there is to know about the planar $\Nsym$ S-matrix using only four dimensional information. In Subsection \ref{effgcomp} we showed that, generically, four dimensional generalized unitarity cuts are not sufficient to determine one-loop planar $\Nsym$ scattering amplitudes to all orders in the dimensional regularization parameter. Clearly, the analysis of Subsection \ref{results} suggests that the one-loop pentagon coefficents missed by the leading singularity method are important and should be determined independently ({\it e.g.} by using maximal generalized unitarity in $D$ dimensions). However, in the spirit of the recent developments, we should first check whether, perhaps, our predictions for the $\Ord(\alpha'^2)$ and $\Ord(\alpha'^3)$ stringy corrections to $\Nsym$ amplitudes don't really rely on {\it all} pentagon coefficients but only some linear combination thereof.

Recall from the discussion of Subsection \ref{effgcomp} that, at the one-loop $n$-point level, amplitudes computed via the leading singularity method are not uniquely determined to all orders in $\e$ but have
\be
{(n-5)(n-4)(n-3)(n-2)(n-1)\over 120}
\ee
pentagon coefficients that must be determined by some other method. It is conceivable that, after performing the dimension shift operation and summing over all contributions, all the undetermined coefficients actually drop out. In fact, there is evidence that this happens for $\Nsym$ amplitudes dimensionally shifted to $D = 8 - 2\e$; we checked that we could derive the appropriate tree-level stringy corrections at $\Ord(\alpha'^2)$ for both MHV and NMHV $n = 6$  amplitudes, $n = 7$ MHV amplitudes, and $n = 8$ MHV amplitudes using the leading singularity method (without $D$ dimensional unitarity). However, for the $\Ord(\alpha'^3)$ stringy corrections, this no longer works. For example, one can check that the one-loop $\Nsym$ six-point MHV amplitude cannot be used to compute $A^{tree}_{str}\left(k_1^{1234},k_2^{1234},k_3,k_4,k_5,k_6\right)$ at $\Ord(\alpha'^3)$ unless {\it all} pentagon coefficients in the amplitude are determined. Our conclusion is that there are still some questions that can be answered by calculating $\Nsym$ amplitudes to all orders in $\e$ that cannot (at least not obviously) be answered by calculating amplitudes in a framework that requires only four dimensional inputs. 

In this review we have seen that our approach to all-orders one-loop $\Nsym$ scattering amplitudes opens up several interesting avenues of exploration. Besides the connection that we found between stringy corrections and dimensionally shifted one-loop amplitudes, it seems plausible, for example, that a variant of our approach will be the right way to think about computing scattering amplitudes at a generic point in the $\Nsym$ moduli space. It will also be quite interesting to see whether our approach to $D$ dimensional integrand construction is useful for theories with less or no supersymmetry. Our expectation is that the approach advocated here will continue to be relevant and useful for future higher-loop studies of $\Nsym$ and more general amplitudes in dimensional regularization.
\section*{Acknowledgments}
I am grateful to Matt Strassler and Lance Dixon for their guidance, support, and collaboration over the last few years. I would like to thank the theory group at SLAC for its hospitality the last few summers and Carola Berger, Zvi Bern, Darren Forde, Tanju Gleisberg, and Daniel Ma\^{i}tre for enlightening conversations and correspondence. I am also very grateful to Cristian Vergu for enlightening correspondence and collaboration. I am especially grateful to Fernando Alday, Nima Arkani-Hamed, Jacob Bourjaily, Freddy Cachazo, James Drummond, Henriette Elvang, Johannes Henn, Juan Maldacena, David McGady, and Jaroslav Trnka for helping me to better understand exciting recent developments in planar $\Nsym$ gauge theory and related topics. A special acknowledgment goes to Johannes Henn, Karl Landsteiner, Andy O'Bannon, Radu Roiban, Matt Strassler, and Cristian Vergu for providing me with some very useful comments and advice on earlier drafts of this work. In this work, figures were drawn primarily using the Jaxodraw~\cite{JaxD} package. This research was supported in part by the US Department of Energy under contract DE-FG02-96ER40949.
\appendix 
\section{Dimensional Regularization}
\label{dimregs}
In this Appendix, we begin in \ref{IRintro} by giving a cursory review of dimensional regularization, focusing on the regularization of IR divergences, which are the only divergences that appear in the $\Nsym$ theory. In \ref{4DHS} we remind the reader that simply declaring that dimensional regularization will be used to regulate divergences is not meaningful because there are several different variants of dimensional regularization. We describe the salient features of one scheme, called the four dimensional helicity scheme, which is particularly useful for regulating the divergences in maximally supersymmetric gauge theories. In Subsection \ref{muint} we give an explicit derivation of eq. (\ref{PtoDCB}), which played an important role in the body of this work. Finally, in Subsection \ref{IR1Ldiv}, we briefly talk about the general structure of IR divergences in planar one-loop $\Nsym$ scattering amplitudes.
\subsection{Definitions and Principal Applications}
\label{IRintro}
Dimensional regularization was first shown to be a well-defined regulator by 't Hooft and Veltman in~\cite{origtHooftVelt} and in this section we describe the method used by them in their seminal paper\footnote{The reader should be aware that many modern calculations do not use the exact scheme advocated by 't Hooft and Veltman, but variations on the same theme (see {\it e.g.} \ref{4DHS}).}. It is actually quite straightforward to describe the method. The only subtlety is related to defining $\e_{\mu \nu \rho \sigma}$, a point that we will return to later. When one is faced with the evaluation of a divergent scattering amplitude the prescription is to make the replacement $\int {d^4 q\over (2\pi)^4}\rightarrow \mintd{q}{4-2\e}$ in each Feynman integral in the expression for the amplitude derived from Feynman diagrams and simultaneously multiply the answer by a factor $\mu^{2 \e}$. This factor of $\mu^{2 \e}$ is called the unit of mass and its function is to prevent the dimensionality of the scattering amplitude from changing under the above replacement\footnote{The physical meaning of $\mu$ varies depending on the physical meaning of the linear combination of Feynman integrals under consideration.}. The regularization parameter $\e$ is usually understood to be small (but non-zero) and less than one in absolute value. We explain the main features of the method by considering a few basic examples and illustrating how it is used in practice.

Let us first think about regulating an uninteresting toy model like massless $\phi^4$ theory, were there are no complications introduced by external wavefunctions or tensor structures in the numerators of the Feynman integrals. It turns out that everything that we want to have happen happens if we make a couple of well-motivated assumptions about the behavior of the regulated Feynman integrals:
\begin{enumerate}[i.]
\item All of the usual properties that integrals enjoy, such as linearity, still hold for the regulated integrals.
\item The integrals are analytic in the complex $\e$ plane except at isolated non-essential singular points.
\end{enumerate}
Although the first property seems completely trivial, this is actually not at all the case because we're integrating over a space of {\it non-integer} dimension. In fact, attempting to carry out the renormalization procedure for massless $\phi^4$ theory is the perfect way to appreciate this subtlety because one of the integrals that one encounters is the massless one-point function
\be
I_1(0,\e) = i (4 \pi)^{2-\e} \mintd{q}{4-2\e} {1\over q^2}\,{\rm ,}
\label{mlesstad}
\ee
which looks truly pathological. We would be happy if we could simply ignore this integral and, happily, its value is indeed zero in dimensional regularization. It turns out that demanding linearity and uniqueness of the results obtained from a given regulated Feynman integral together with analyticity in $\e$ force (\ref{mlesstad}) to zero~\cite{Collins}. The same conclusion holds for any Feynman integral with a scaleless integrand.

An obvious question is what happens to the one-loop tadpole if we consider {\it massive} $\phi^4$ theory? In this case the answer is not zero and, in fact, the evaluation of this integral will make it clear why analyticity is so crucial for the whole regularization program to work. If we modify \ref{mlesstad} by $p^2 \rightarrow p^2 - m^2$, we get the integral that we want to evaluate. The calculation proceeds in the standard way~\cite{PeskinSchroeder}, first Wick rotating to Euclidean space and then introducing spherical coordinates in $4 - 2 \e$ dimensions, we find
\be
I_1(m^2,\e) = -{\G(1+\e)m^{2-2\e}\over \e (1-\e)}
\label{mtad}
\ee
for $1 < \e < 2$. At first sight, it looks like we've just hit an insurmountable obstacle; the integral $I_1(m^2,\e)$ only converges for $\e$ well away from where we want it (near $\e = 0$). However, this is no problem at all because the function is analytic everywhere except at the isolated points $\e = 0$ and $\e = 1$ and we can easily analytically continue $I_1(m^2,\e)$ to the rest of the complex $\e$-plane. Furthermore, the assumptions of dimensional regularization guarantee that $I_1(m^2,\e)$ has a well-behaved Laurent expansion everywhere in the complex plane. In particular, in a small neighborhood of $\e = 0$ we find
\be
I_1(m^2,\e) = m^2\left(-{1\over \e}+(\gamma_E-1+\ln(m^2)+\Ord(\e)\right)
\ee
The $\ln(m^2)$ factor looks quite peculiar because there is a dimensionful quantity inside a logarithm. In fact, if we kept track of the $\e$ expansion of the unit of mass, we would find a factor $-m^2 \ln(\mu^2)$ which combines together with the $m^2 \ln(m^2)$ in the above to yield a proper, dimensionless logarithm. In this context, the unit of mass, $\mu$, plays the role of the renormalization scale.

In order to discuss dimensional reg in a more non-trivial example involving vector and fermion fields, we have to understand how to modify the Feynman rules for these species of fields in a way that is consistent with the prescription given above for scalar integrals (we must verify that our modification doesn't spoil gauge invariance). The shift to $4-2\e$ dimensions is applied to everything in the problem;  the Dirac matrices, the metric, the external momenta, the external spinor and vector wavefunctions (when applicable), and the epsilon tensor $\e_{\mu \nu \rho \sigma}$ should all make sense in $4-2\e$ if the program is to work\footnote{Actually, as we will see in \ref{4DHS}, this is an unnecessarily stringent requirement; in fact, the loop momenta are the only vectors that have to be treated in $4-2\e$ dimensions.}. Everything generalizes to $4 - 2 \e$ dimensions straightforwardly except $\e_{\mu \nu \rho \sigma}$ and the external wavefunctions. 't Hooft and Veltman reasoned that it had to be sensible to allow the wavefunctions of the external particles to live in exactly four dimensions if helicity amplitudes are to make sense but they had to think harder about the Levi-Civita pseudotensor. Making sense of  $\e_{\mu \nu \rho \sigma}$ in $4 - 2 \e$ dimensions was the main technical problem 't Hooft and Veltman had to solve in order to show that dimensional regularization is well-defined.

The solution they came up with was to leave the definition of $\e_{\mu \nu \rho \sigma}$ alone. This is okay so long as you treat objects built out of $\e_{\mu \nu \rho \sigma}$, like $\gamma^5$, with special care. A consistent treatment of $\gamma^5$ is particularly important because it enters into the calculation of the axial anomaly. Suppose we have some $4-2\e$ dimensional momentum, $\ell_\nu$, contracted into $\gamma^\mu$ and that we want to commute $\slashed{\ell}$ past $\gamma^5$. To do this in a consistent fashion you must first divide $\ell$ up into a four dimensional component and a $-2 \e$ dimensional component: 
\be
\ell = \ell_\parallel +\mu \,{\rm .}
\ee
Then, taking into account the fact that $\gamma^5$ is defined to be an intrinsically four dimensional object, we see that
\be
\{\ell_\parallel, \gamma^5\} = 0\,\,\,\,\,\,\,\,\,{\rm as ~usual}
\ee
\be
{\rm but}\,\,\,\,\,\,\,\,\,\,\,\,\,\,\,\,\,\,[\mu, \gamma^5] = 0 ~{\rm .}
\ee
This prescription for $\gamma^5$ allowed 't Hooft and Veltman to rederive the results of ABJ for the axial anomaly~\cite{Adler, BellJackiw} in a different way.

Now that we have addressed all of the subtleties associated with the construction of a consistent regularization scheme for non-SUSY models\footnote{The scheme described above will not work for models with unbroken supersymmetry. See \ref{4DHS} for a description of the four dimensional helicity scheme, a more sophisticated variant of dimensional regularization, which, in particular, works in the context of $\Nsym$ super Yang-Mills.}, it is very natural to wonder whether IR divergences are also regulated by $\e$ in massless theories or if the method is only suitable for carrying out the renomalization procedure. In fact, dimensional regularization can be used to regulate all physical singularities that crop up in scattering amplitude calculations. We have developed the method to the point where we can consider a realistic example. The one-loop vertex diagram in QED furnishes a nice one because it is very simple and yet has both UV and IR divergences which manifest themselves as poles in $\e$. We begin by evaluating the graph of Figure \ref{selfint} using the appropriate Feynman rules:

\FIGURE{
\resizebox{.95\textwidth}{!}{\includegraphics{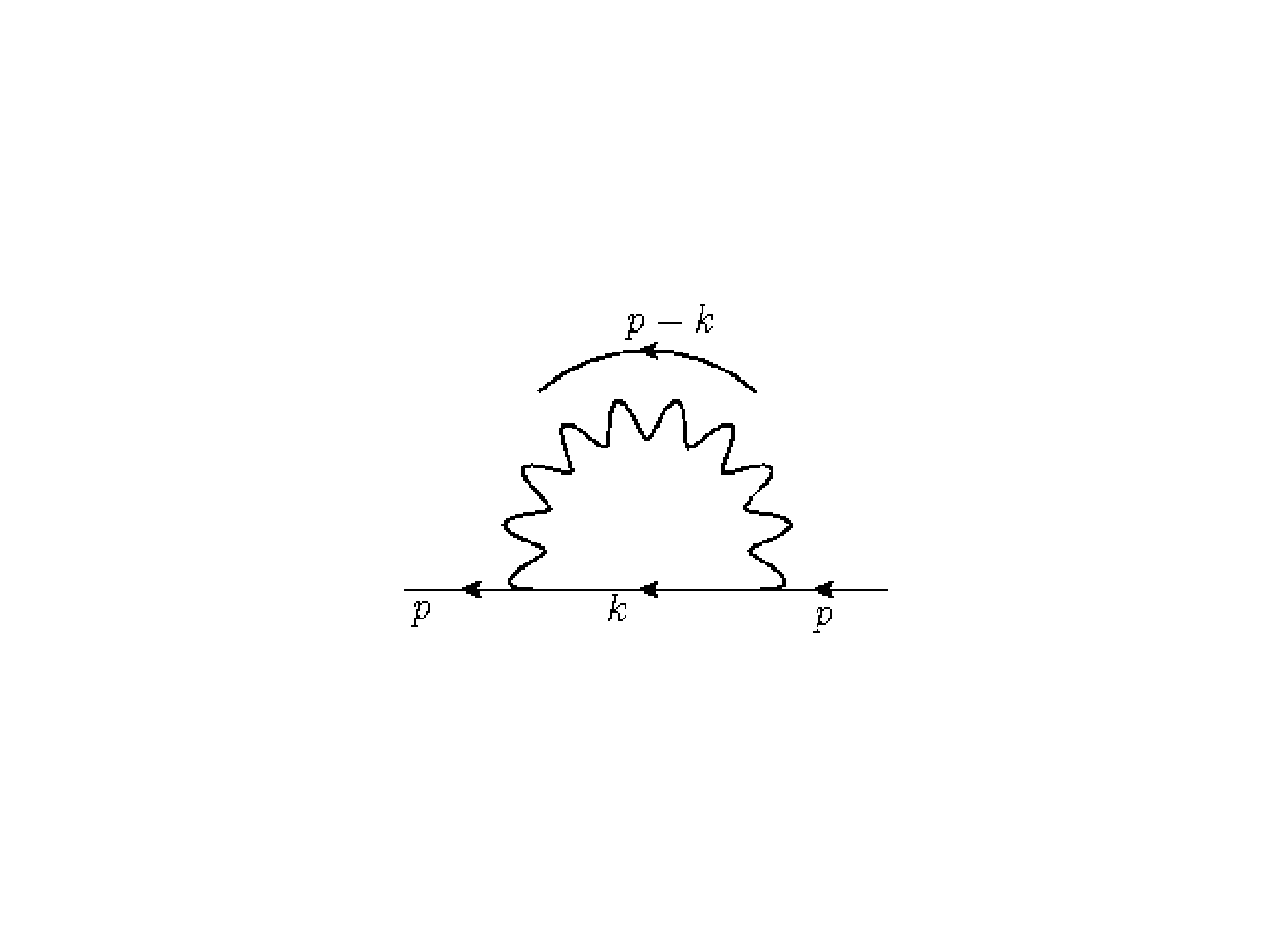}}
\caption{The one-loop fermion self-energy graph in QED.}
\label{selfint}}
\be
\Sigma_1 = - i e^2 \mu^{2\e}\int {d^{4-2\e}k \over (2\pi)^{4-2\e}}\gamma^\mu {\slashed{k}+m \over k^2-m^2} \gamma_\mu {1 \over (p-k)^2}\,{\rm,}
\ee
where we have suppressed the external spinor states because they are irrelevant for us. At the outset, it may be hard to see why this example requires more care than the massive tadpole integral treated above. Recall, however, that $\gamma^\mu \gamma_\mu = 4$ must be replaced with $\gamma^\mu \gamma_\mu = 4-2\e$ and this also modifies the identity $\gamma^\mu \slashed{k} \gamma_\mu = -2 \slashed{k}$ to $\gamma^\mu \slashed{k} \gamma_\mu = -(2-2\e) \slashed{k}$. After applying this identity, Feynman parametrizing, and shifting the integration variable to $q = k - (1-x)p$ one finds
\bea
&& \Sigma_1 = i e^2 \mu^{2\e}\int_0^1 dx \int {d^{4-2\e}q \over (2\pi)^{4-2\e}}{(2-2\e)(\slashed{q}+(1-x) \slashed{p})-m (4-2\e) \over (x (q+(1-x) p)^2 + (1-x)(q+x p)^2-m^2)^2} 
\elale
i e^2 \mu^{2\e}\int_0^1 dx \int {d^{4-2\e}q \over (2\pi)^{4-2\e}}{(2-2\e)(1-x)\slashed{p} -m (4-2\e)\over(q^2+x (1-x) p^2-m^2 x)^2}\,{\rm.}
\label{rawselfen}
\eea
If we now consider a Taylor series expansion of the above about $\slashed{p} = m$. It is very well-known~\cite{PeskinSchroeder} that the first non-zero term\footnote{The integral at $\slashed{p} = m$ is trivially zero, since it is just the integral of a dimensionless function.}, ${d\Sigma_1 \over d\slashed{p}}\Big|_{\slashed{p} = m}$, is related to the mass renormalization of the electron via 
\be
Z_2 = \left(1-{d\Sigma_1 \over d\slashed{p}}\Big|_{\slashed{p} = m}\right)^{-1} \,{\rm .}
\ee
Following~\cite{SirlinIR}, we can separate ${d\Sigma_1 \over d\slashed{p}}\Big|_{\slashed{p} = m}$ into two pieces, one of which converges in the IR but diverges in the UV and one of which converges in the UV but diverges in the IR:
\bea
{d\Sigma_1 \over d\slashed{p}}\Big|_{\slashed{p} = m} &=& 2 i e^2 \mu^{2\e}(1-\e)\int_0^1 dx \int {d^{4-2\e}q \over (2\pi)^{4-2\e}}{(1-x)\over(q^2-m^2 x^2)^2} 
\el
+ 8 i e^2 m^2 \mu^{2\e}\int_0^1 dx \int {d^{4-2\e}q \over (2\pi)^{4-2\e}}{x(1-x)(1+x(1-\e))\over(q^2-m^2 x^2)^3} \, {\rm .}
\eea
The integrals over $q$ and $x$ can be performed without difficulty and we get
\bea
{d\Sigma_1 \over d\slashed{p}}\Big|_{\slashed{p} = m} &=& -\G(\e){2 e^2 (1-\e)\over (4\pi)^{2-\e}}\left({\mu \over m}\right)^{2\e} \int_0^1 x^{-2\e} (1-x) dx
\el
+{4 e^2 \G(1+\e)\over (4\pi)^{2-\e}}\left({\mu \over m}\right)^{2\e} \int_0^1 x^{-1-2\e} (1-x)(1+x(1-\e)) dx
\elale
-\G(\e){e^2 \over (4\pi)^{2-\e} (1-2\e)}\left({\mu \over m}\right)^{2\e}-{2 e^2 \G(1+\e)(1-\e)\over (4\pi)^{2-\e}\e(1-2\e)}\left({\mu \over m}\right)^{2\e}\, {\rm .}
\label{selffinal}
\eea
The first line of (\ref{selffinal}) has a Feynman parameter integral that makes sense at $\e = 0$ but, nevertheless, this contribution has a UV divergence coming from the factor $\G(\e)$ out front of the integral. This is a general feature of UV divergences; they manifest themselves as special (but analytic) functions singular at $\e = 0$ that appear after performing loop integrations that are divergent by power-counting. Once again, the unit of mass is going to play the role of the renormalization scale. Note that the first Feynman parameter integral in (\ref{selffinal}) converges for ${\rm Re}(\e) < 1/2$ and the second converges for ${\rm Re}(\e) < 0$.

From this rich example we draw the conclusion that, in fact, dimensional regularization regularizes IR divergences as well as the more familiar UV ones. Even in complicated cases one can separate out the UV divergences first and use them to renormalize the amplitude because the UV divergences should be manifest after integrating out the loop momenta. The IR divergences typically only appear after integrating out some Feynman parameters as well and, therefore, should be dealt with after the UV divergences have been removed via the renormalization procedure. 
\subsection{The Four Dimensional Helicity Scheme}
\label{4DHS}
The four dimensional helicity scheme is a variant of dimensional regularization introduced to simplify the renormalization of supersymmetric gauge theories. In this Subsection, we discuss its salient features and contrast it to the so-called 't Hooft-Veltman scheme discussed in Subsection \ref{IRintro}. The four dimensional helicity scheme is the variant of dimensional regularization that we implicitly work in throughout the main text. The criterion that one uses to decide if a regulator is appropriate for a given quantum field theory is whether the proposed regulator preserves all the symmetries of the model. Despite the many successes of the 't Hooft-Veltman scheme in the Standard Model, it is not an appropriate regulator for supersymmetric models because examples exist (see {\it e.g.} \cite{DRTJones}) where its use explicitly violates certain supersymmetric Ward identities.

In 2002 the four dimensional helicity scheme~\cite{4dimHS} was proposed\footnote{Very recently, calculations were presented in~\cite{Kilgore} which imply that the four dimensional helicity scheme is not a generally applicable renormalization scheme in the way that, say, dimensional reduction is. We are not in a position to evaluate the claims made by the author of~\cite{Kilgore} at the present time. Regardless, nothing in the present work is affected by the discussion in~\cite{Kilgore} because UV divergences are absent in maximally supersymmetric gauge theory.} as a variant of dimensional regularization fully consistent with supersymmetry to all orders in perturbation theory. As the name suggests, all external momenta and wavefunctions are kept in four dimensions; only the loop momenta are continued to $D$ dimensions. The rules for objects built out of $\e_{\mu \nu \rho \sigma}$ are the same in the 't Hooft-Veltman scheme. The main insight of~\cite{4dimHS} was that one must introduce an additional scale, called the spin dimension, which is taken to be the dimension in which the wavefunctions of all virtual particles circulating in loops live. If supersymmetry is to be preserved, the spin dimension, $D_s$, must be treated as follows.
\begin{enumerate}[i.]
\item{Perform all index contractions as if $D_s > D > 4$.}
\item{After the amplitude is a function only of the loop momenta, external momenta, external wavefunctions, $D$, and $D_s$, set $D_s = 4$.}
\end{enumerate}
It is useful to note that if $D_s$ is set to $D$ we recover the 't Hooft-Veltman scheme.
\subsection{A Useful Integral Reduction Identity Involving Dimensionally-Shifted Integrals at the One-Loop Level}
\label{muint}
In this Subsection we derive eq. (\ref{PtoDCB}) explicitly to supplement the streamlined discussion of Subsection \ref{GUD}. This exercise should also help the reader understand why the coefficients of the pentagon integrals in the dimensionally shifted basis defined in \ref{GUD} contain an explicit factor of $\e$, whereas the box integrals in it do not.
We begin with the dimensionally regulated one-loop integrals of eq. (\ref{IntDef}): 
\be
I_n^{D=4-2 \e}
= i (-1)^{n+1} (4\pi)^{2-\e} \int 
    {d^{4-2\e} \ell \over (2\pi)^{4-2\e} }
  { 1 \over \ell^2 \ldots
    (\ell-\sum_{i=1}^{n-1} K_i )^2 } \,{\rm .}
\label{IntegralDef}
\ee
It has been known for a very long time how to write down an expression for (\ref{IntegralDef}) as a Feynman parameter integral~\cite{PeskinSchroeder}:
\be
I_n^{D=4-2 \e}
= \G(n-2+\e) \int_0^1 d^n x_i {\D\left(1-\sum_{i=1}^n x_i\right) \over \left(\left(\sum_{i=1}^n x_i p_{i-1}\right)^2-\sum_{i=1}^n x_i p_{i-1}^2\right)^{n-2+\e}} \, {\rm ,}
\ee
where $p_i = \sum_{j=1}^{i} K_j$. There is, however, a particularly nice, symmetric way of rewriting this expression~\cite{oneloopdimreg}. The above formula collapses to\footnote{One can verify this relation directly after eliminating one of the variables through the relation $\sum_{i=1}^n x_i = 1$ on both sides of the equation.}
\be
I_n^{D=4-2 \e}
= \G(n-2+\e) \int_0^1 d^n x_i {\D\left(1-\sum_{i=1}^n x_i\right) \over \left(\sum_{i,j=1}^n p_{i-1}\cdot p_{j-1}x_i x_j\right)^{n-2+\e}} \, {\rm .}
\ee 
This representation of $I_n^{D=4-2 \e}$ will enter into our derivation of eq. (\ref{PtoDCB}). The idea is to evaluate the same integral in two different ways. 

Following~\cite{oneloopdimreg}, we define
\be
I_n^{D=4-2 \e}[\ell^2] \equiv i (-1)^{n+1} (4\pi)^{2-\e} \int 
    {d^{4-2\e} \ell \over (2\pi)^{4-2\e} }
  { \ell^2 \over \ell^2 \ldots
    (\ell-\sum_{i=1}^{n-1} K_i )^2 } \, {\rm .}
\ee
Of course, this integral can be trivially reduced by canceling the numerator against the first propagator denominator. 
\be
I_n^{D=4-2 \e}[\ell^2] = -I_{n-1}^{D=4-2 \e} \, {\rm .}
\label{easyway}
\ee
However, we are also free to evaluate it as a Feynman parameter integral. Going through the usual Feynman parametrization procedure we find something of the form
\bea
&&I_n^{D=4-2 \e}[\ell^2] = i (-1)^{n+1} (4\pi)^{2-\e} \G(n) \int_0^1 d^n x_i \D\left(1-\sum_{i=1}^n x_i\right) \times
\el
\times \int 
    {d^{4-2\e} q \over (2\pi)^{4-2\e} }
  { q^2 + \sum_{i,j=1}^n p_{i-1}\cdot p_{j-1}x_i x_j +\sum_{i=1}^n x_i p_{i-1}^2\over \left(q^2 - \sum_{i,j=1}^n p_{i-1}\cdot p_{j-1}x_i x_j\right)^{n-2+\e}} \, {\rm .}
\eea
These integrals are easily carried out by using the standard formulae~\cite{PeskinSchroeder}
\bea
 &&i (-1)^{n+1} (4\pi)^{2-\e} \int  {d^{4-2\e} q \over (2\pi)^{4-2\e} } {1\over (q^2-\Delta)^n} = {\G(n-2+\e)\over \G(n)\Delta^{n-2+\e}}
 \el
{\rm and}~ i (-1)^{n+1} (4\pi)^{2-\e} \int  {d^{4-2\e} q \over (2\pi)^{4-2\e} } {q^2\over (q^2-\Delta)^n} = -{(2-\e)\G(n-3+\e)\over \G(n)\Delta^{n-3+\e}} \, {\rm .}
\eea
$I_n^{D=4-2 \e}[\ell^2]$ becomes 
\begin{changemargin}{-.6 in}{0 in}
\bea
&&I_n^{D=4-2 \e}[\ell^2] = \int_0^1 d^n x_i \D\left(1-\sum_{i=1}^n x_i\right) \Bigg(-{(2-\e)\G(n-3+\e)\over \left(\sum_{i,j=1}^n p_{i-1}\cdot p_{j-1}x_i x_j\right)^{n-3+\e}} 
\el
+ {\G(n-2+\e) \sum_{i,j=1}^n p_{i-1}\cdot p_{j-1}x_i x_j \over \left(\sum_{i,j=1}^n p_{i-1}\cdot p_{j-1}x_i x_j\right)^{n-2+\e}} + {\G(n-2+\e) \sum_{i=1}^n x_i p_{i-1}^2 \over \left(\sum_{i,j=1}^n p_{i-1}\cdot p_{j-1}x_i x_j\right)^{n-2+\e}}\Bigg)\,{\rm .}
\eea
\end{changemargin}
This simplifies nicely:
\begin{changemargin}{-.6 in}{0 in}
\bea
&&I_n^{D=4-2 \e}[\ell^2] = \int_0^1 d^n x_i \D\left(1-\sum_{i=1}^n x_i\right)\Bigg(-{(2-\e)\G(n-2+(\e-1))\over \left(\sum_{i,j=1}^n p_{i-1}\cdot p_{j-1}x_i x_j\right)^{n-2+(\e-1)}}
\el + {(n-3+\e)\G(n-2+(\e-1)) \over \left(\sum_{i,j=1}^n p_{i-1}\cdot p_{j-1}x_i x_j\right)^{n-2+(\e-1)}} + {\G(n-2+\e) \sum_{i=1}^n x_i p_{i-1}^2 \over \left(\sum_{i,j=1}^n p_{i-1}\cdot p_{j-1}x_i x_j\right)^{n-2+\e}}\Bigg)
\elale
\int_0^1 d^n x_i \D\left(1-\sum_{i=1}^n x_i\right) \left({(n-5+2\e)\G(n-2+(\e-1)) \over \left(\sum_{i,j=1}^n p_{i-1}\cdot p_{j-1}x_i x_j\right)^{n-2+(\e-1)}} + {\G(n-2+\e) \sum_{i=1}^n x_i p_{i-1}^2 \over \left(\sum_{i,j=1}^n p_{i-1}\cdot p_{j-1}x_i x_j\right)^{n-2+\e}}\right)
\elale
(n-5+2\e)I_n^{D=6-2 \e} + \sum_{i=1}^n p_{i-1}^2 I_n^{D=4-2 \e}[x_i]\,{\rm .}
\label{hardway}
\eea
\end{changemargin}
Equating the last line of (\ref{hardway}) with the right-hand side of(\ref{easyway}),
\be
-I_{n-1}^{D=4-2 \e} = (n-5+2\e)I_n^{D=6-2 \e} + \sum_{i=1}^n p_{i-1}^2 I_n^{D=4-2 \e}[x_i]
\ee
we finally obtain a non-trivial relation between scalar integrals. 

In fact, all of the above analysis goes through unchanged if $I_{n}^{D=4-2 \e}[\ell^2]$ is replaced by $I_{n}^{D=4-2 \e}[(\ell-p_{i-1})^2]$, allowing us to derive a total of $n$ relations that can be written in a unified way as
\be
-I_{n-1}^{(i);\,D=4-2 \e} = (n-5+2\e)I_n^{D=6-2 \e} + 2 \sum_{j=1}^n S_{i j} I_n^{D=4-2 \e}[x_i]\,{\rm ,}
\ee
where we have introduced the daughter-integral notation (which first appeared in Subsection \ref{GU4}) and the matrix $S_{i j}$ defined as
\bea
&& S_{i j} = -{1 \over 2}~ (p_i + ... + p_{j-1})^2,~~~~~ i \neq j
\el 
S_{i j} = 0,~~~~~~~~~~~~~~~~~~~~~~~~~~~~~~~i = j \, ,
\eea
where both $i$ and $j$ are to be taken mod $n$. Solving for $I_n^{D=4-2\e}[x_i]$, we obtain
\be
I_n^{D=4-2\e}[x_i] = {1\over 2}\bigg[\sum_{j=1}^n S_{i j}^{-1} I_{n-1}^{(j),~D=4-2\e} + (n-5+2\e)C_i I_n^{D=6-2 \e} \bigg]\,{\rm ,}
\label{interrel}
\ee
where $C_i = \sum_{j=1}^n S_{i j}^{-1}$. Finally, we can exploit the identity $\sum_{i=1}^n x_i = 1$ and sum over the index $i$ in the above. This yields
\be
I_n^{D=4-2\e} = {1\over 2}\bigg[\sum_{j=1}^n C_j I_{n-1}^{(j),~D=4-2\e} + (n-5+2\e)C_0 I_n^{D=6-2 \e} \bigg]\,{\rm ,}
\label{finalrel}
\ee
where $C_0 = \sum_{i=1}^n C_i$. This is the final form of our desired relation.

This formula for $n = 5$, 
\be
I_5^{D=4-2\e} = {1\over 2}\bigg[\sum_{j=1}^5 C_j I_{4}^{(j),~D=4-2\e} + 2\e C_0 I_5^{D=6-2 \e} \bigg]\,{\rm ,}
\label{5ptrel}
\ee
turns out to be very useful in the analysis of one-loop $\Nsym$ amplitudes in dimensional regularization because the five-point scalar integral is related to a linear combination of four-point scalar integrals plus a five-point integral in $D=6-2\e$ dimensions that has an explicit factor of $\e$ out front. Furthermore, it turns out that the $D=6-2\e$ scalar integral has no poles in $\e$. This then implies that eq. (\ref{5ptrel}) corresponds to a special case that relates the five-point integral to four-point integrals, up to $\Ord(\e)$ contributions that can be neglected if one is only interested in computing one-loop amplitudes to $\Ord(\e^0)$. For us, this relation provides a convenient way to separate higher order in $\e$ contributions from those that contribute only through $\Ord(\e^0)$.

Eq. (\ref{finalrel}) for $n = 6$ also appears throughout the main text (recall eq. (\ref{hexred})). To see the utility of (\ref{finalrel}) for this value of $n$, one needs to know something more about the matrix $S_{i j}$ and its rank. In deriving eq. (\ref{interrel}), we implicitly assume that $S_{i j}$ is invertible. Actually this is only a valid assumption for $n \leq 6$ and $n = 6$ is the borderline case. It is well-known that, beginning at the six-point level, additional non-linear constraints on scattering processes exist coming from the fact that it is no longer possible to find an $n-1$ dimensional linearly independent subset of the $n$ external momenta~\cite{oneloopdimreg}. 

To be more concrete, let us specialize to $n = 6$ and count the degrees of freedom for external momenta in $D = 4$. The sum of all momenta is zero by construction, so clearly at most five of the external momenta are linearly independent. However, it must be the case that any one of these five momenta can be expressed as a linear combination of the other four, simply because the vector space that we're working in is four dimensional. More precisely, we have the six relations
\be
{\rm Det}(k_i \cdot k_j)_{r} = 0 {\rm ,}
\label{Gram}
\ee
where the $r$ subscript is to be interpreted as an instruction to delete the $r$-th column of the matrix $(k_i \cdot k_j)$. It turns out that the changing the value of $r$ doesn't change the left-hand side of (\ref{Gram}) and, therefore, all six  equations give the same constraint\footnote{{\it A priori} (\ref{Gram}) could have been identically satisfied. It turns out that this is not the case and, as a result, there is a non-trivial constraint on the kinematics.} on the kinematics. The object $C_0$ is proportional to ${\rm Det}(k_i \cdot k_j)_{r}$ and therefore can be set equal to zero. This results in
\be
I_6^{D=4-2\e} = {1\over 2}\sum_{i=1}^6 C_i I_5^{(i),~D=4-2\e}
\ee
a special case of (\ref{finalrel}) for $n = 6$.
\subsection{IR Structure of One-Loop Planar Amplitudes in $\Nsym$}
\label{IR1Ldiv}
In this subsection, we review the results of references~\cite{Kunszt} and~\cite{GieleGlover} where all possible IR divergences at one loop in massless gauge theories were classified. Actually, the one-loop IR divergences in $\Nsym$ are a little bit simpler than in the general case. In general, one expects two distinct epsilon pole structures at one loop: poles that have their origin in purely soft or soft-collinear virtual particles and poles that have their origin in purely collinear virtual particles. The purely collinear singularities are governed by terms that have the schematic form~\cite{GieleGlover}
\be
{1 \over \e} \left({\mu^2 \over -\tilde{s}}\right)^\e A^{tree}\left(k_1^{h_1},~\cdots,~k_{n}^{h_n}\right)\, ,
\ee
where $\tilde{s}$ is some kinematic scale. However, there are clearly no divergences of this form in the integral basis of eqs.~\ref{box0}-\ref{box3} (valid for planar $\Nsym$ through $\Ord(\e^0)$). Rather, one sees divergences of the schematic form
\be
{1 \over \e^2} \left({\mu^2 \over -\tilde{s}}\right)^\e A^{tree}\left(k_1^{h_1},~\cdots,~k_{n}^{h_n}\right)
\ee
in those integral functions, which correspond to soft-collinear and soft singularities. We conclude that the virtual IR divergences in planar $\Nsym$ one-loop scattering processes have their origin in soft or soft-collinear virtual particles connecting pairs of adjacent external lines\footnote{If we did not restrict ourselves to planar contributions, then the virtual particles could connect non-adjacent external lines as well.}. Quantitatively, if we take the index $i$ in what follows to be mod $n$, we have
\cmb{-.6 in}{0 in}
\bea
&&A^{1-{\rm loop}}_1 \left(k_1^{h_1},~\cdots,~k_{n}^{h_n}\right)\Bigg|_{\rm singular} = 
\el -{1 \over \e^2}{g^2 \Nc \mu^{2\e} e^{-\gamma_E \e}\over (4\pi)^{2-\e}} {\G(1+\e)\G^2(1-\e) \over \G(1-2\e)}\sum^n_{i=1}\left({\mu^2 \over -s_{i\,i+1}}\right)^\e A^{tree}\left(k_1^{h_1},~\cdots,~k_{n}^{h_n}\right)
\label{1LIR}
\eea
\cme
for color-ordered partial amplitudes in the Euclidean kinematical region (defined in Subsection \ref{GU4}).
\section{$\Nsym$ Superconformal Symmetry}
\label{sconf}
In this work we study the Yang-Mills theory based on the four-dimensional $\Nsym$ supersymmetric extension of the Poincar\'{e} group. This extension, called the $\Nsym$ superconformal group, is an example of a {\it Lie supergroup}, a generalization of a Lie group that possesses a $\mathbb{Z}_2$ graded Lie algebra. The $\Nsym$ superconformal group is normally discussed in the context of the Lagrange density of the $\Nsym$ gauge theory. For our purposes, we are more interested in fleshing out the discussion of Subsection \ref{DSI}. In \ref{DSI} particular realizations of both the ordinary and the closely related dual $\Nsym$ superconformal symmetries were discussed in the context of an on-shell chiral superspace construction. It turns out that, in classifying the symmetries of superamplitudes on this chiral on-shell superspace, one actually needs to consider the action of the generators of the {\it centrally extended} $\Nsym$ superconformal group. We begin by briefly describing each (ordinary) symmetry operation. 

Of course, it is not so easy to give the reader a feeling for the fermionic symmetries. Consequently, we argue by analogy to the appropriate even (under the $\mathbb{Z}_2$ grading) cases when discussing the symmetries associated with odd generators. In the end, we are mostly interested in representations of the appropriate Lie superalgebras on the on-shell superspace. Therefore, in the second part of this appendix, we write down the $\Nsym$ superconformal and dual superconformal algebras and give explicit representations thereof.

First, we remind the reader that the Poincar\'{e} group by itself is nothing but the isometry group of Minkowski space. As such it contains
\bea
{\rm spacetime ~translations:}&&~~~~~~~~~~x'^\mu = x^\mu + r^\mu ~~~~~{\rm and}\\
{\rm spacetime ~rotations:}&&~~~~~~~~~~x'^\mu = M^\mu_\nu x^\nu \,{\rm .}
\eea
Since there are four coordinates to translate in, three pairing of coordinate axes ($\{x,y\}$, $\{x,z\}$, and $\{y,z\}$) to define spatial rotations in, and three spatial directions to boost in, the dimension of the Poincar\'{e} group is ten. In this appendix, we will follow the conventions used in the main text and label generators using spinor notation. Spatial translations are generated by the momentum operator, $\mathbb{P}_{\alpha \dot{\alpha}}$, and spacetime rotations are generated by $\mathbb{M}_{\alpha \beta}$ and $\bar{\mathbb{M}}_{\dot{\alpha} \dot{\beta}}$. 

Now, suppose that one adds four fermionic directions to $\mathbb{R}^{1,3}$ labeled by $a$ for $a \in \{1,2,3,4\}$. One certainly expects any well-behaved theory to be invariant under the full isometry group of the space on which the theory sits. We clearly have to allow for translations along the new fermionic directions
\be
{\rm spacespace ~translations:}~~~~~~~~~~\theta'^a_\mu = \theta^a_\mu + \eta^a_\mu
\ee
in addition to the spacetime translations discussed above. Superspace translations are generated by the so-called supercharges, $\mathbb{Q}^a_{~\,\alpha}$ and $\bar{\mathbb{Q}}_{a\,\dot{\alpha}}$. There are sixteen of these fermionic generators in all because there are four fermionic coordinate axes and the supercharges carry spacetime indices as well. 

The $SU(4)_R$ symmetry discussed extensively in the main text fits neatly into this picture: the R-symmetry acts by rotating the supercharges into one other. As is well-known, $SU(4)_R$ has fifteen generators, $\mathbb{R}^a_{~\,b}$. On on-shell superspace, $\mathbb{R}^a_{~\,b}$ is realized as
\be
R^a_{~\,b} = \sum_{i = 1}^n \left(\eta_i^a{\partial \over \partial \eta_{i}^b}-{1 \over 4}\D_b^{\,~a}\eta^c_i{\partial \over \partial \eta_{i}^c}\right)\, ;
\ee
the trace part by itself is not a symmetry of the theory. However, in attempting to implement $\Nsym$ supersymmetry on on-shell scattering amplitudes, one discovers that it {\it is} possible to build an additional symmetry generator by appropriately modifying the generator of the trace part that we were initially tempted to discard. This new generator, $\mathbb{Z}$, is called the central charge of the Lie superalgebra due to the fact that it commutes with all of the other generators and is related to the helicity quantum number of on-shell superamplitudes. On on-shell superspace, the generator of the central charge,
\be
Z = \sum_{i = 1}^n \left(1 + {1\over 2}\lambda_i^{~\,\A}{\partial \over \partial \lambda_i^{~\,\A}}-{1\over 2}\tilde{\lambda}_i^{~\,\dot{\alpha}}{\partial \over \partial \tilde{\lambda}_i^{~\,\dot{\alpha}}}-{1\over 2}\eta_i^a{\partial \over \partial \eta_i^a}\right)\,,
\ee
is identified with one minus the helicity operator summed over states~\cite{DHPYangian1}:
\be
Z = \sum_{i = 1}^n \left(1-h_i\right)\,{\rm .}
\ee
By construction, each superfield $\Phi(p,\eta)$ has helicity $+ 1$ (see eq. (\ref{supwvfun})). Therefore, $Z$ annihilates all on-shell superamplitudes and it follows that there is indeed an additional bosonic symmetry as claimed. Clearly, the construction of $Z$ is tied up with the chirality of the on-shell superspace since we arbitrarily chose to work with superfields of helicity $+ 1$, rather than $-1$.

Now, there is a natural extension of the Poincar\'{e} group that provides five additional bosonic generators. What we are alluding to is the well-known conformal group which, in addition to the ten dimensional Poincar\'{e} group, consists of 
\bea
{\rm dilatations:}&&~~~~~~~x'^\mu = \alpha\, x^\mu~~~~~{\rm and}\\
{\rm special~conformal~transformations:}&&~~~~~~~x'^\mu = {x^\mu - r^\mu x^2 \over 1- 2 r \cdot x + r^2 x^2} \,{\rm .}
\eea
The dilatation operation, generated by $\mathbb{D}$, is just a rescaling of the coordinates and, at the level of operators, it measures the classical scaling dimension. The special conformal transformations, generated by $\mathbb{K}_{\alpha \dot{\alpha}}$, are a bit more difficult to understand, as their action on Minkowski space looks rather complicated. A nice way to proceed is as follows. If we introduce the discrete operation of conformal inversion
\be
{\rm inversion:}~~~~~~~x'^\mu = {x^\mu \over x^2} \equiv I[x^\mu] \,{\rm ,}
\ee
it turns out~\cite{DHKSdualconf} that one can think of the special conformal symmetries as being generated by an inversion, a translation, and another inversion applied in succession:
\be
\mathbb{K}_{\alpha \dot{\alpha}} = I\, \mathbb{P}_{\alpha \dot{\alpha}} \,I \,{\rm .}
\label{invK}
\ee  

Na\"{i}vely one might think that we have now successfully identified all generators. However, it turns out that we are still missing the analogs of special conformal transformations along the fermionic directions~\cite{Minwalla}. Indeed, we can identify sixteen new fermionic generators, the generators of the special supersymmetry transformations, along the lines of eq. (\ref{invK}):
\be
\mathbb{S}_a^{\alpha} = I\, \bar{\mathbb{Q}}_{a\,\dot{\alpha}}\,I~~~~{\rm and}~~~~\bar{\mathbb{S}}^{a\,\dot{\alpha}} = I\, \mathbb{Q}^a_{~\,\alpha}\,I \,{\rm .}
\ee
Now that we have actually succeeded in identifying all symmetry generators, we can give explicit forms for them and write down the (anti)commutation relations that they ought to satisfy.
\subsection{The Ordinary and Dual $\Nsym$ Superconformal Algebras and Differential Operator Representations Thereof}
\label{scalg}
We first present, in spinor notation, the non-trivial (anti)commutation relations of the ordinary $\Nsym$ superconformal algebra:
\bea
&&\left[ \mathbb{D}, \mathbb{P}_{\alpha \dot{\alpha}} \right] = \mathbb{P}_{\alpha \dot{\alpha}} \qquad ~~~~~\left[ \mathbb{D}, \mathbb{K}_{\alpha \dot{\alpha}} \right] = -\mathbb{K}_{\alpha \dot{\alpha}} 
\el
\left[ \mathbb{D}, \mathbb{Q}^a_{~\,\alpha} \right] = {1\over 2} \mathbb{Q}^a_{~\,\alpha} \qquad ~~\left[ \mathbb{D}, \bar{\mathbb{Q}}_{a\,\dot{\alpha}} \right] = {1\over 2}\bar{\mathbb{Q}}_{a\,\dot{\alpha}}
\el
\left[ \mathbb{D}, \mathbb{S}_{a\,\alpha} \right] = -{1\over 2} \mathbb{S}_{a\,\alpha} \qquad ~\left[ \mathbb{D}, \bar{\mathbb{S}}^a_{~\,\dot{\alpha}} \right] = -{1\over 2}\bar{\mathbb{S}}^a_{~\,\dot{\alpha}}
\el
\left[ \mathbb{K}_{\alpha \dot{\alpha}}, \mathbb{Q}^a_{~\,\beta} \right] = \e_{\A \B}\bar{\mathbb{S}}^a_{~\,\dot{\alpha}} \qquad ~\left[ \mathbb{K}_{\alpha \dot{\alpha}}, \bar{\mathbb{Q}}_{a\,\dot{\beta}} \right] = \e_{\dot{\A} \dot{\B}} \mathbb{S}_{a\,\alpha}
\el
\left[ \mathbb{K}_{\alpha \dot{\alpha}}, \mathbb{P}_{\beta \dot{\beta}} \right] = \e_{\A \B}\e_{\dot{\A}\dot{\B}} \mb{D}+{1
\over 2}\e_{\dot{\A}\dot{\B}}\mb{M}_{\A \B}+{1
\over 2}\e_{\A \B}\bar{\mb{M}}_{\dot{\A}\dot{\B}}
\el
\left[ \mathbb{P}_{\A \dot{\A}}, \mathbb{S}_{a\,\B} \right] = \e_{\A \B} \bar{\mathbb{Q}}_{a\,\dot{\alpha}} \qquad ~~\left[ \mathbb{P}_{\A \dot{\A}}, \bar{\mathbb{S}}^a_{~\,\dot{\B}} \right] = \e_{\dot{\A} \dot{\B}}\mathbb{Q}^a_{~\,\alpha}
\el
\left[ \mathbb{M}_{\A \B}, \mathbb{M}_{\g \D} \right] = \e_{\A \g} \mathbb{M}_{\B \D} + \e_{\B \g} \mathbb{M}_{\A \D}+\e_{\A \D} \mathbb{M}_{\B \g}+\e_{\B \D} \mathbb{M}_{\A \g}
\el
\left[ \bar{\mathbb{M}}_{\dot{\A} \dot{\B}}, \bar{\mathbb{M}}_{\dot{\g} \dot{\D}} \right] = \e_{\dot{\A} \dot{\g}} \bar{\mathbb{M}}_{\dot{\B} \dot{\D}} + \e_{\dot{\B} \dot{\g}} \bar{\mathbb{M}}_{\dot{\A} \dot{\D}}+\e_{\dot{\A} \dot{\D}} \bar{\mathbb{M}}_{\dot{\B} \dot{\g}}+\e_{\dot{\B} \dot{\D}} \bar{\mathbb{M}}_{\dot{\A} \dot{\g}}
\el
\left[ \mathbb{M}_{\A \B}, \mathbb{S}_{a\,\g} \right] = \e_{\B \g} \mathbb{S}_{a\,\A}+\e_{\A \g} \mathbb{S}_{a\,\B} \qquad ~~\left[ \bar{\mathbb{M}}_{\dot{\A} \dot{\B}}, \bar{\mathbb{S}}^a_{~\,\dot{\g}} \right] = \e_{\dot{\A} \dot{\g}}\bar{\mathbb{S}}^a_{~\,\dot{\B}}+\e_{\dot{\B} \dot{\g}}\bar{\mathbb{S}}^a_{~\,\dot{\A}}
\el
\left[ \mathbb{M}_{\A \B}, \mathbb{Q}^a_{~\,\g} \right] = \e_{\B \g} \mathbb{Q}^a_{~\,\A}+\e_{\A \g} \mathbb{Q}^a_{~\,\B} \qquad \left[ \bar{\mathbb{M}}_{\dot{\A} \dot{\B}}, \bar{\mathbb{Q}}_{a\,\dot{\g}} \right] = \e_{\dot{\A} \dot{\g}}\bar{\mathbb{Q}}_{a\,\dot{\B}}+\e_{\dot{\B} \dot{\g}}\bar{\mathbb{Q}}_{a\,\dot{\A}}
\el
\left[ \mathbb{M}_{\A \B},\mathbb{K}_{\g \dot{\g}}\right] = \e_{\B \g} \mathbb{K}_{\A \dot{\g}}+\e_{\A \g} \mathbb{K}_{\B \dot{\g}} \qquad ~\left[ \bar{\mathbb{M}}_{\dot{\A} \dot{\B}}, \mathbb{K}_{\g \dot{\g}} \right] = \e_{\dot{\A} \dot{\g}}\mathbb{K}_{\g \dot{\B}}+\e_{\dot{\B} \dot{\g}}\mathbb{K}_{\g \dot{\A}}
\el
\left[ \mathbb{M}_{\A \B},\mathbb{P}_{\g \dot{\g}}\right] = \e_{\B \g} \mathbb{P}_{\A \dot{\g}}+\e_{\A \g} \mathbb{P}_{\B \dot{\g}} \qquad ~~~\left[ \bar{\mathbb{M}}_{\dot{\A} \dot{\B}}, \mathbb{P}_{\g \dot{\g}} \right] = \e_{\dot{\A} \dot{\g}}\mathbb{P}_{\g \dot{\B}}+\e_{\dot{\B} \dot{\g}}\mathbb{P}_{\g \dot{\A}}
\el
\left[ \mb{R}^a_{~\,b},\mb{R}^c_{~\,d} \right] = \D_b^{~\,c}\mb{R}^a_{~\,d}-\D_d^{~\,a}\mb{R}^c_{~\,b}
\el
\left[ \mb{R}^a_{~\,b},\mathbb{Q}^c_{~\,\A} \right] = \D_b^{~\,c}\mathbb{Q}^a_{~\,\A} - {1\over 4}\D_b^{~\,a}\mathbb{Q}^c_{~\,\A} \qquad \left[ \mb{R}^a_{~\,b},\bar{\mathbb{Q}}_{c\,\dot{\A}} \right] = -\left(\D_c^{~\,a}\bar{\mathbb{Q}}_{b\,\dot{\A}} - {1\over 4}\D_b^{~\,a}\bar{\mathbb{Q}}_{c\,\dot{\A}}\right)
\el
\left[ \mb{R}^a_{~\,b},\mathbb{S}_{c\,\A} \right] = -\left(\D_c^{~\,a}\mathbb{S}_{b\,\A} - {1\over 4}\D_b^{~\,a}\mathbb{S}_{c\,\A}\right) \qquad \left[ \mb{R}^a_{~\,b},\bar{\mathbb{S}}^c_{~\,\dot{\A}} \right] = \D_b^{~\,c}\bar{\mathbb{S}}^a_{~\,\dot{\A}} - {1\over 4}\D_b^{~\,a}\bar{\mathbb{S}}^c_{~\,\dot{\A}}
\el
\left\{\mathbb{Q}^a_{~\,\A},\bar{\mathbb{Q}}_{b\,\dot{\A}}\right\} = \D_b^{~\,a} \mb{P}_{\A \dot{\A}} \qquad \left\{\mathbb{S}_{a\,\A},\bar{\mathbb{S}}^b_{~\,\dot{\A}}\right\} = \D_a^{~\,b} \mb{K}_{\A \dot{\A}}
\el
\left\{\mathbb{S}_{a\,\A},\mathbb{Q}^b_{~\,\B}\right\} = {1
\over 2}\D_a^{~\,b} \mb{M}_{\A \B} -\e_{\A \B} \mb{R}^b_{~\,a}+{1 \over 2}\e_{\A \B}\D_a^{~\,b}\left(\mb{D}+\mb{Z}\right)
\el
\left\{\bar{\mathbb{S}}^a_{~\,\dot{\A}},\bar{\mathbb{Q}}_{b\,\dot{\B}}\right\} = {1
\over 2}\D_b^{~\,a} \bar{\mb{M}}_{\dot{\A} \dot{\B}} +\e_{\dot{\A} \dot{\B}} \mb{R}^a_{~\,b}+{1 \over 2}\e_{\dot{\A} \dot{\B}}\D_b^{~\,a}\left(\mb{D}-\mb{Z}\right)
\label{commalg}
\eea
Our focus in this work is on the differential operator representation of the above superalgebra on on-shell superspace (discussed in \ref{DSI}). For a supermatrix representation of the $\Nsym$ superconformal algebra we refer the interested reader to~\cite{DHPYangian1}.  The representation that we present acts on the on-shell superspace of \ref{gendisonshell}:
\bea
P_{\A \dot{\A}} &=& \sum_{i = 1}^n \lambda_{i\,\A} \tilde{\lambda}_{i \,\dot{\A}} \qquad \qquad \qquad K_{\A \dot{\A}} = \sum_{i = 1}^n {\partial \over \partial \lambda_i^{~\,\A}}{\partial \over \partial \tilde{\lambda}_i^{~\,\dot{\alpha}}} 
\nonumber \\
M_{\A \B} &=&  \sum_{i = 1}^n \left(\lambda_{i\,\alpha}{\partial \over \partial \lambda_i^{~\,\B}}+\lambda_{i\,\B}{\partial \over \partial \lambda_i^{~\,\A}}\right) \qquad \bar{M}_{\dot{\A} \dot{\B}} =  \sum_{i = 1}^n \left(\tilde{\lambda}_{i\,\dot{\A}}{\partial \over \partial \tilde{\lambda}_i^{~\,\dot{\B}}}+\tilde{\lambda}_{i\,\dot{\B}}{\partial \over \partial \tilde{\lambda}_i^{~\,\dot{\alpha}}}\right)
\nonumber \\
D &=&  \sum_{i = 1}^n \left({1 \over 2}\lambda_i^{~\,\alpha}{\partial \over \partial \lambda_i^{~\,\A}}+{1 \over 2}\tilde{\lambda}_i^{~\,\dot{\A}}{\partial \over \partial \tilde{\lambda}_i^{~\,\dot{\alpha}}}+1\right)\qquad R^a_{~\,b} = \sum_{i = 1}^n \left(\eta_i^a{\partial \over \partial \eta_{i}^b}-{1 \over 4}\D_b^{\,~a}\eta^c_i{\partial \over \partial \eta_{i}^c}\right)
\nonumber \\
Q^a_{~\,\alpha}&=&  \sum_{i = 1}^n \lambda_{i\,\alpha}\eta_i^a \qquad \bar{Q}_{a\,\dot{\A}} = \sum_{i = 1}^n \tilde{\lambda}_{i\,\dot{\alpha}} {\partial \over \partial \eta_{i}^a}\qquad S_{a\,\A} = \sum_{i = 1}^n {\partial \over \partial \lambda_i^{~\,\A}} {\partial \over \partial \eta_{i}^a} \qquad \bar{S}^a_{~\,\dot{\alpha}} =  \sum_{i = 1}^n \eta^a_i {\partial \over \partial \tilde{\lambda}_i^{~\,\dot{\alpha}}}
\nonumber \\
Z &=& \sum_{i = 1}^n \left(1 + {1\over 2}\lambda_i^{~\,\A}{\partial \over \partial \lambda_i^{~\,\A}}-{1\over 2}\tilde{\lambda}_i^{~\,\dot{\alpha}}{\partial \over \partial \tilde{\lambda}_i^{~\,\dot{\alpha}}}-{1\over 2}\eta_i^a{\partial \over \partial \eta_i^a} \right)
\eea
We have painstakingly checked (using the conventions of Penrose and Rindler~\cite{PenroseRindler} for raising and lowering spinor indices) that this representation satisfies the above (anti)commutation relations. 

It's important to note that the above superalgebra is not appropriate for the dual superconformal symmetry discussed in \ref{DSI}; to write down the dual superconformal algebra one should take (\ref{commalg}), swap the $SU(4)_R$ chiralities of all operators ({\it e.g.} $\mathbb{Q}^a_{~\,\A}$ becomes $\mathbb{Q}_{a\,\A}$), and then appropriately adjust the (anti)commutation relations involving $\mb{R}^a_{~\,b}$. Using the explicit expressions given in \ref{DSI}, we find:
\bea
&&\left[ \mathbb{D}, \mathbb{P}_{\alpha \dot{\alpha}} \right] = \mathbb{P}_{\alpha \dot{\alpha}} \qquad ~~~~~\left[ \mathbb{D}, \mathbb{K}_{\alpha \dot{\alpha}} \right] = -\mathbb{K}_{\alpha \dot{\alpha}} 
\el
\left[ \mathbb{D}, \mathbb{Q}_{a\,\alpha} \right] = {1\over 2} \mathbb{Q}_{a\,\alpha} \qquad ~~\left[ \mathbb{D}, \bar{\mathbb{Q}}^a_{~\,\dot{\alpha}} \right] = {1\over 2}\bar{\mathbb{Q}}^a_{~\,\dot{\alpha}}
\el
\left[ \mathbb{D}, \mathbb{S}^a_{~\,\alpha} \right] = -{1\over 2} \mathbb{S}^a_{~\,\alpha} \qquad ~\left[ \mathbb{D}, \bar{\mathbb{S}}_{a\,\dot{\alpha}} \right] = -{1\over 2}\bar{\mathbb{S}}_{a\,\dot{\alpha}}
\el
\left[ \mathbb{K}_{\alpha \dot{\alpha}}, \mathbb{Q}_{a\,\beta} \right] = \e_{\A \B}\bar{\mathbb{S}}_{a\,\dot{\alpha}} \qquad ~\left[ \mathbb{K}_{\alpha \dot{\alpha}}, \bar{\mathbb{Q}}^a_{~\,\dot{\beta}} \right] = \e_{\dot{\A} \dot{\B}} \mathbb{S}^a_{~\,\alpha}
\el
\left[ \mathbb{K}_{\alpha \dot{\alpha}}, \mathbb{P}_{\beta \dot{\beta}} \right] = \e_{\A \B}\e_{\dot{\A}\dot{\B}} \mb{D}+{1
\over 2}\e_{\dot{\A}\dot{\B}}\mb{M}_{\A \B}+{1
\over 2}\e_{\A \B}\bar{\mb{M}}_{\dot{\A}\dot{\B}}
\el
\left[ \mathbb{P}_{\A \dot{\A}}, \mathbb{S}^a_{~\,\B} \right] = \e_{\A \B} \bar{\mathbb{Q}}^a_{~\,\dot{\alpha}} \qquad ~~\left[ \mathbb{P}_{\A \dot{\A}}, \bar{\mathbb{S}}_{a\,\dot{\B}} \right] = \e_{\dot{\A} \dot{\B}}\mathbb{Q}_{a\,\alpha}
\el
\left[ \mathbb{M}_{\A \B}, \mathbb{M}_{\g \D} \right] = \e_{\A \g} \mathbb{M}_{\B \D} + \e_{\B \g} \mathbb{M}_{\A \D}+\e_{\A \D} \mathbb{M}_{\B \g}+\e_{\B \D} \mathbb{M}_{\A \g}
\el
\left[ \bar{\mathbb{M}}_{\dot{\A} \dot{\B}}, \bar{\mathbb{M}}_{\dot{\g} \dot{\D}} \right] = \e_{\dot{\A} \dot{\g}} \bar{\mathbb{M}}_{\dot{\B} \dot{\D}} + \e_{\dot{\B} \dot{\g}} \bar{\mathbb{M}}_{\dot{\A} \dot{\D}}+\e_{\dot{\A} \dot{\D}} \bar{\mathbb{M}}_{\dot{\B} \dot{\g}}+\e_{\dot{\B} \dot{\D}} \bar{\mathbb{M}}_{\dot{\A} \dot{\g}}
\el
\left[ \mathbb{M}_{\A \B}, \mathbb{S}^a_{~\,\g} \right] = \e_{\B \g} \mathbb{S}^a_{~\,\A}+\e_{\A \g} \mathbb{S}^a_{~\,\B} \qquad ~~\left[ \bar{\mathbb{M}}_{\dot{\A} \dot{\B}}, \bar{\mathbb{S}}_{a\,\dot{\g}} \right] = \e_{\dot{\A} \dot{\g}}\bar{\mathbb{S}}_{a\,\dot{\B}}+\e_{\dot{\B} \dot{\g}}\bar{\mathbb{S}}_{a\,\dot{\A}}
\el
\left[ \mathbb{M}_{\A \B}, \mathbb{Q}_{a\,\g} \right] = \e_{\B \g} \mathbb{Q}_{a\,\A}+\e_{\A \g} \mathbb{Q}_{a\,\B} \qquad \left[ \bar{\mathbb{M}}_{\dot{\A} \dot{\B}}, \bar{\mathbb{Q}}^a_{~\,\dot{\g}} \right] = \e_{\dot{\A} \dot{\g}}\bar{\mathbb{Q}}^a_{~\,\dot{\B}}+\e_{\dot{\B} \dot{\g}}\bar{\mathbb{Q}}^a_{~\,\dot{\A}}
\el
\left[ \mathbb{M}_{\A \B},\mathbb{K}_{\g \dot{\g}}\right] = \e_{\B \g} \mathbb{K}_{\A \dot{\g}}+\e_{\A \g} \mathbb{K}_{\B \dot{\g}} \qquad ~\left[ \bar{\mathbb{M}}_{\dot{\A} \dot{\B}}, \mathbb{K}_{\g \dot{\g}} \right] = \e_{\dot{\A} \dot{\g}}\mathbb{K}_{\g \dot{\B}}+\e_{\dot{\B} \dot{\g}}\mathbb{K}_{\g \dot{\A}}
\el
\left[ \mathbb{M}_{\A \B},\mathbb{P}_{\g \dot{\g}}\right] = \e_{\B \g} \mathbb{P}_{\A \dot{\g}}+\e_{\A \g} \mathbb{P}_{\B \dot{\g}} \qquad ~~~\left[ \bar{\mathbb{M}}_{\dot{\A} \dot{\B}}, \mathbb{P}_{\g \dot{\g}} \right] = \e_{\dot{\A} \dot{\g}}\mathbb{P}_{\g \dot{\B}}+\e_{\dot{\B} \dot{\g}}\mathbb{P}_{\g \dot{\A}}
\el
\left[ \mb{R}^a_{~\,b},\mb{R}^c_{~\,d} \right] = \D_b^{~\,c}\mb{R}^a_{~\,d}-\D_d^{~\,a}\mb{R}^c_{~\,b}
\el
\left[ \mb{R}^a_{~\,b},\mathbb{Q}_{c\,\A} \right] = -\left(\D_c^{~\,a}\mathbb{Q}_{b\,\A} - {1\over 4}\D_b^{~\,a}\mathbb{Q}_{c\,\A}\right) \qquad \left[ \mb{R}^a_{~\,b},\bar{\mathbb{Q}}^c_{~\,\dot{\A}} \right] = \D_b^{~\,c}\bar{\mathbb{Q}}^a_{~\,\dot{\A}} - {1\over 4}\D_b^{~\,a}\bar{\mathbb{Q}}^c_{~\,\dot{\A}}
\el
\left[ \mb{R}^a_{~\,b},\mathbb{S}^c_{~\,\A} \right] = \D_b^{~\,c}\mathbb{S}^a_{~\,\A} - {1\over 4}\D_b^{~\,a}\mathbb{S}^c_{~\,\A} \qquad \left[ \mb{R}^a_{~\,b},\bar{\mathbb{S}}_{c\,\dot{\A}} \right] = -\left(\D_c^{~\,a}\bar{\mathbb{S}}_{b\,\dot{\A}} - {1\over 4}\D_b^{~\,a}\bar{\mathbb{S}}_{c\,\dot{\A}}\right)
\el
\left\{\mathbb{Q}_{a\,\A},\bar{\mathbb{Q}}^b_{~\,\dot{\A}}\right\} = \D_a^{~\,b} \mb{P}_{\A \dot{\A}} \qquad \left\{\mathbb{S}^a_{~\,\A},\bar{\mathbb{S}}_{b\,\dot{\A}}\right\} = \D_b^{~\,a} \mb{K}_{\A \dot{\A}}
\el
\left\{\mathbb{S}^a_{~\,\A},\mathbb{Q}_{b\,\B}\right\} = {1
\over 2}\D_b^{~\,a} \mb{M}_{\A \B} +\e_{\A \B} \mb{R}^a_{~\,b}+{1 \over 2}\e_{\A \B}\D_b^{~\,a}\left(\mb{D}+\mb{Z}\right)
\el
\left\{\bar{\mathbb{S}}_{a\,\dot{\A}},\bar{\mathbb{Q}}^b_{~\,\dot{\B}}\right\} = {1
\over 2}\D_a^{~\,b} \bar{\mb{M}}_{\dot{\A} \dot{\B}} - \e_{\dot{\A} \dot{\B}} \mb{R}^b_{~\,a}+{1 \over 2}\e_{\dot{\A} \dot{\B}}\D_a^{~\,b}\left(\mb{D}-\mb{Z}\right) \,.
\eea
For the sake of completeness, we collect the dual superconformal generators here as well:
\cmb{-.4 in}{0 in}
\bea
\mc{Q}_{a\,\alpha} &=& \sum_{i = 1}^n {\partial\over \partial \T_{i}^{a\,\A}} ~~~ \mathcal{P}_{\alpha \dot{\alpha}} = \sum_{i = 1}^n {\partial\over \partial x_{i}^{~\,\A\dot{\A}}} ~~~
\bar{\mc{S}}_{b\,\dot{\alpha}} = \sum_{i = 1}^n \left(x_{i\,\dot{\alpha}}^{\,~~\beta}{\partial\over \partial \T_{i}^{b\,\B}}+\tilde{\lambda}_{i\,\dot{\alpha}}{\partial\over \partial \eta^b_{i}}\right) ~~~ \bar{\mc{Q}}^b_{~\dot{\alpha}} = \sum_{i = 1}^n \left(\theta^{b\,\alpha}_{i}{\partial\over \partial x_{i}^{~\,\A\dot{\A}}}+\eta_{i}^{b}{\partial\over \partial \tilde{\lambda}_{i}^{~\,\dot{\A}}}\right)\nn
\mc{S}^a_{~\alpha} &=& \sum_{i = 1}^n \left(-\theta^b_{i\,\alpha}\theta^{a\,\beta}_{i}{\partial\over \partial \T_{i}^{b\,\B}}+x_{i\,\alpha}^{~~\,\dot{\beta}}\theta^{a\,\beta}_{i}{\partial\over \partial x_{i}^{~\,\B \dot{\B}}}+\lambda_{i\,\alpha}\theta^{a\,\gamma}_{i}{\partial\over \partial \lambda_{i}^{\,~\g}}+x_{i+1\,\alpha}^{\,~~~~~\dot{\beta}}\eta^a_i{\partial\over \partial \tilde{\lambda}_{i}^{~\,\dot{\B}}}-\theta^b_{i+1\,\alpha}\eta^a_i{\partial\over \partial \eta^b_i}\right)  \nonumber \\
\mc{K}_{\alpha \dot{\alpha}} &=& \sum_{i = 1}^n \left(x_{i\,\alpha}^{~~\,\dot{\beta}}x_{i\,\dot{\alpha}}^{~~\,\beta}{\partial\over \partial x_{i}^{~\,\B \dot{\B}}}+x_{i\,\dot{\alpha}}^{~~\,\beta}\theta_{i\,\alpha}^b{\partial\over \partial \T_{i}^{b\,\B}}+x_{i\,\dot{\alpha}}^{\,~~\beta}\lambda_{i\,\alpha}{\partial\over \partial \lambda_{i}^{\,~\B}}+x_{i+1\,\alpha}^{~~~~~\,\dot{\beta}}\tilde{\lambda}_{i\,\dot{\alpha}}{\partial\over \partial \tilde{\lambda}_{i}^{~\,\dot{\B}}}+\tilde{\lambda}_{i\,\dot{\alpha}}\theta^b_{i+1\,\alpha}{\partial\over \partial \eta_{i}^b}\right)\nn
\mc{M}_{\A \B} &=& \sum_{i = 1}^n \left(x_{i\,\alpha}^{~~\,\dot{\alpha}}{\partial\over \partial x_{i}^{~\,\B \dot{\A}}}+x_{i\,\beta}^{~~\,\dot{\alpha}}{\partial\over \partial x_{i}^{~\,\alpha \dot{\A}}}+\theta_{i\,\alpha}^{a}{\partial\over \partial \T_{i}^{a\,\B}}+\theta_{i\,\beta}^{a}{\partial\over \partial \T_{i}^{a\,\alpha}}+\lambda_{i\,\alpha}{\partial\over \partial \lambda_{i}^{\,~\B}}+\lambda_{i\,\beta}{\partial\over \partial \lambda_{i}^{\,~\alpha}}\right)\nonumber \\
\bar{\mc{M}}_{\dot{\A}\dot{\B}} &=& \sum_{i = 1}^n \left(x_{i\,~~\dot{\alpha}}^{~\,\alpha}{\partial\over \partial x_{i}^{~\,\alpha \dot{\B}}}+x_{i\,~~\dot{\beta}}^{~\,\alpha}{\partial\over \partial x_{i}^{~\,\alpha \dot{\alpha}}}+\tilde{\lambda}_{i\,\dot{\alpha}}{\partial\over \partial \tilde{\lambda}_{i}^{~\,\dot{\B}}}+\tilde{\lambda}_{i\,\dot{\beta}}{\partial\over \partial \tilde{\lambda}_{i}^{~\,\dot{\alpha}}}\right)\nonumber \\
\mc{R}^a_{~b} &=& \sum_{i = 1}^n \left(\T_i^{a\,\alpha}{\partial\over \partial \T_{i}^{a\,\A}}+\eta_i^a {\partial\over \partial \eta^b_{i}}-{1\over 4}\D_{b}^{~\,a}\T_{i}^{c\,\alpha}{\partial\over \partial \T_{i}^{c\,\A}}-{1\over 4}\D_{b}^{~\,a}\eta_i^c{\partial\over \partial \eta^c_i}\right)\nonumber \\
\mc{D} &=& -\sum_{i = 1}^n \left(x_{i}^{~\,\alpha \dot{\alpha}}{\partial\over \partial x_{i}^{~\,\alpha \dot{\alpha}}}+{1\over 2}\T_{i}^{a\,\alpha}{\partial\over \partial \T_{i}^{a\,\alpha}}+{1\over 2}\lambda_{i}^{\,~\alpha}{\partial\over \partial \lambda_{i}^{\,~\alpha}}+{1\over 2}\tilde{\lambda}_{i}^{~\,\dot{\alpha}}{\partial\over \partial \tilde{\lambda}_{i}^{~\,\dot{\alpha}}}\right)\nonumber \\
\mc{Z} &=& -{1\over 2}\sum_{i = 1}^n \left(\lambda_{i}^{\,~\alpha}{\partial\over \partial \lambda_{i}^{\,~\alpha}}-\tilde{\lambda}_{i}^{\,~\dot{\alpha}}{\partial\over \partial \tilde{\lambda}_{i}^{\,~\dot{\alpha}}}-\eta^a_i{\partial\over \partial \eta_{i}^a}\right)\,.
\eea
\cme
\section{Supersymmetric Ward Identities and Their Solutions}
\label{SWI}
In this Appendix we derive $\Nsym$ {\it supersymmetric Ward identities} ($\Nsym$ SWI), linear relations between on-shell scattering amplitudes that are consequences of the action of $\Nsym$ supersymmetry on the space of states. After warming up with the very simple (but very important) example of the relations $\mathcal{A}\left(k_1,k_2,\cdots, k_n\right) = 0$ and $\mathcal{A}\left(k_1^{1234},k_2,\cdots, k_n\right) = 0$, which are true even non-perturbatively in any supersymmetric gauge theory, we study the much more non-trivial case of the six-point NMHV superamplitude. Recently, it has been shown that, for general superamplitudes, a solution to the complete set of $\Nsym$ supersymmetric Ward identities can be found using ideas from representation theory. We illustrate this for the six-point NMHV superamplitude in \ref{superNMHV}.
\subsection{$\mathcal{A}\left(k_1,k_2,\cdots, k_n\right) = 0$ and $\mathcal{A}\left(k_1^{1234},k_2,\cdots, k_n\right) = 0$}
\label{all+0}
Before we can begin, we have to say a few more words about notation. Throughout most of this paper it has mattered relatively little what conventions are used for the evaluation of spinor products. There is a good numerical program for the evaluation of spinor products, S@M~\cite{S@M}, and, therefore, all the results presented in the main text can be checked without worrying too much about the underlying conventions. However, for the sake of completeness, we now present one convention that meshes well with the $\Nsym$ on-shell superspace of Subsection \ref{gendisonshell}. We define the holomorphic spinor product
of $\lambda_\alpha = (\lambda_1,\lambda_2)$ and $\chi_\alpha = (\chi_1,\chi_2)$ as
\be
\spa{\lambda}.{\chi} = \lambda_\B \chi^\B = \e_{\A \B} \lambda^\A \chi^\B \equiv {\rm Det}\left(\begin{array}{cc} \lambda^1 & \chi^1 \\ \lambda^2 & \chi^2 \end{array}\right)
\ee

This definition will be helpful in analyzing the consequences of supercharge conservation, from which $\mathcal{A}\left(k_1,k_2,\cdots, k_n\right) = 0$ and $\mathcal{A}\left(k_1^{1234},k_2,\cdots, k_n\right) = 0$ follow trivially. First recall that, implicitly, there is always an overall four-momentum conserving delta function out front of any scattering amplitude. This factor is so trivial that one normally doesn't even write it. It is, however, an extremely important part of the scattering amplitude. Because of it the total momentum operator $P_{\A \dot{\A}}$, which acts multiplicatively, annihilates the amplitude. In $\Nsym$ supersymmetric gauge theory, half of the supercharge operators are realized in a completely analogous fashion on the superamplitudes (that is to say, multiplicatively). In this paper, we implement $Q^{a\,\alpha}$ multiplicatively:
\be
Q^{a\,\alpha}~\a(\Phi_1,\cdots,\Phi_n) = \left( \sum_{\ell=1}^n \lambda^\A_{\ell}~\eta^a_\ell \right) ~\a(\Phi_1,\cdots,\Phi_n) = 0\,.
\ee
For this to make sense, it must be possible to pull an overall factor of 
\be
\prod_{\alpha = 1}^2 \prod_{a = 1}^4 \delta\left(\sum_{\ell=1}^n \lambda^\alpha_\ell~\eta^a_\ell\right) = \delta^{(8)} \left(\sum_{\ell=1}^n \lambda^\alpha_\ell~\eta^a_\ell\right)
\ee
out of each superamplitude; eq. (\ref{kexpans}) becomes
\begin{equation}
    \a(\Phi_1,\cdots, \Phi_n) = \delta^{(8)} \left(\sum_{\ell=1}^n \lambda^\alpha_\ell~\eta^a_\ell\right)\left(\hat{\a}_{n;2} + \hat{\a}_{n;3} + \cdots + \hat{\a}_{n;\, n-2}\right) \,{\rm .}
    \label{kexpansD}
\end{equation}

Now, a useful simplification of the supercharge conserving delta function follows from our definition of the holomorphic spinor product. Suppose we fix the $SU(4)_R$ index and consider the product
\be
\delta\left(\sum_{\ell=1}^n \lambda^1_\ell~\eta^a_\ell\right)\delta\left(\sum_{\ell=1}^n \lambda^2_\ell~\eta^a_\ell\right) = \left(\sum_{\ell=1}^n \lambda^1_\ell~\eta^a_\ell\right)\left(\sum_{\ell=1}^n \lambda^2_\ell~\eta^a_\ell\right)\,{\rm .}
\ee
Let us rearrange the product in the following way. First, suppose that we select two of the external momenta indexed by $i$ and $j$. For now we assume that $i < j$. If we expand the above product and collect all terms involving both $i$ and $j$, we get
\be
\lambda^1_i \lambda^2_j \eta^a_i \eta^a_j + \lambda^1_j \lambda^2_i \eta^a_j \eta^a_i = (\lambda^1_i \lambda^2_j - \lambda^1_j \lambda^2_i) \eta^a_i \eta^a_j = \spa{i}.j \eta^a_i \eta^a_j \,{\rm .}
\ee
We can easily drop the assumption that $i > j$ if we write
\be
\spa{i}.j \eta^a_i \eta^a_j = {1\over 2}\left(\spa{i}.j \eta^a_i \eta^a_j + \spa{j}.i \eta^a_j \eta^a_i\right) \,{\rm .}
\ee
Finally, we have 
\be
\left(\sum_{\ell=1}^n \lambda^1_\ell~\eta^a_\ell\right)\left(\sum_{\ell=1}^n \lambda^2_\ell~\eta^a_\ell\right) = {1\over 2}\sum_{i,j = 1}^n \spa{i}.j \eta^a_i \eta^a_j
\ee
and we have a more useful expression for our eight-fold Grassmann delta function:
\be
\delta^{(8)} \left(\sum_{\ell=1}^n \lambda^\alpha_\ell~\eta^a_\ell\right) = {1\over 16}\prod_{a = 1}^4 \sum_{i,j = 1}^n \spa{i}.j \eta^a_i \eta^a_j
\ee

From the above expression for $\delta^{(8)}\left(Q^{a\,\A}\right)$, we see that, for non-degenerate kinematics, a superamplitude must have a $k$-charge of at least two to be consistent with supercharge conservation. Since $\mathcal{A}\left(k_1,k_2,\cdots, k_n\right)$ belongs to the $k$-charge zero sector and $\mathcal{A}\left(k_1^{1234},k_2,\cdots, k_n\right)$ to the $k$-charge one sector, these amplitudes and all other amplitudes related to them by $\Nsym$ supersymmetry vanish for generic kinematical configurations. However, we remind the reader that one can have non-vanishing $k = 1$ amplitudes if one allows all spinors to be proportional to one another. As was demonstrated in \ref{gendisonshell}, this construction is important in its own right because it allows one to define the three-point MHV and anti-MHV superamplitudes. 

The study of supersymmetric Ward identities has a long history (see {\it e.g.}~\cite{GP, GPvN, DSWI, BDDPR}) and culminated recently in~\cite{EFK} where Elvang, Freedman, and Kiermaier presented a complete solution to the $\Nsym$ SWI. In other words, as discussed in Subsection \ref{gendisonshell}, it is now possible to express any $\Nsym$ superamplitude as a linear combination of a minimal number of linearly independent component amplitudes. To illustrate the power of the new techniques, we derive eq. (\ref{A63}) for the six-point NMHV superamplitude in the next subsection.
\subsection{$\mathcal{A}_{6;3}$}
\label{superNMHV}
In~\cite{EFK}, Elvang, Freedman, and Kiermaier (EFK) found a useful general solution to the $\Nsym$ SWI using only the constraints coming from ordinary $\Nsym$ supersymmetry and $SU(4)_R$ invariance. Their main new observation was that one can profit by demanding that on-shell superamplitudes be manifestly $SU(4)_R$ invariant. Then, using representation theory, they described the solution for superamplitudes in the $(k,\,n)$ sector. In this subsection we focus on $(3,\,6)$ sector because the resulting formula is what we use in the main text. 

The basic approach used by EFK was to start with an ansatz and then impose the constraints of $\Nsym$ supersymmetry and $SU(4)_R$ invariance on it. The starting point for their ansatz is Nair's formula~\cite{Nair} for the MHV tree-level superamplitudes, derived in eq. (\ref{MHVsup}) and reproduced here for convenience
\be
\a_{n;\,2} = i {{1\over 16}\prod_{a = 1}^4 \sum_{i,j = 1}^n \spa{i}.j \eta^a_i \eta^a_j \over \spa1.2 \spa2.3 \cdots \spa{n}.1}
\label{MHVsuper2}
\ee
The moral of Subsection \ref{all+0} above is that all $\Nsym$ superamplitudes are naturally written with a factor of $\D^{(8)}\left( Q^{a\,\alpha} \right)$ pulled out front. It therefore follows that the superspace structure of an arbitrary superamplitude will be of the form 
\be
\a_{n;\,k} = \D^{(8)}\left( Q^{a\,\alpha} \right)\Big(~\cdots~\Big)\,{\rm .}
\ee
As noted in Subsection \ref{gendisonshell}, the NMHV superamplitude $\a_{6;3}$ requires a total of twelve Grassmann variables. Since eight variables are already present in the supercharge conserving delta function, we need just four more from some other source. The EFK ansatz for $\a_{6;3}$ is 
\be
\a_{6;\,3}^{\rm EFK} = \D^{(8)}\left( Q^{a\,\alpha} \right)\sum_{i,j,k,l = 1}^6 q_{i j k l}~ \eta^1_i \eta^2_j \eta^3_k \eta^4_l \,{\rm .}
\label{EFK}
\ee

A familiar result from the representation theory of $SU(N)$ is that one can learn all about the algebra of $SU(N)$ simply by studying the $SU(2)$ subalgebras inside of it (see {\it e.g.}~\cite{Georgi}). In the same vein, we can impose $SU(4)_R$ invariance on (\ref{EFK}) by demanding that (\ref{EFK}) be invariant under infinitesimal $SU(2)_R$ rotations acting on any pair of $SU(4)_R$ indices. For example, an infinitesimal $\sigma_1$ rotation of the pair $(1,\,2)$ parametrized by $\theta$ acts as
\be
\left(\begin{array}{cc} 0 & \theta \\ \theta & 0 \end{array}\right) \left(\begin{array}{c} \eta_i^1 \\ \eta_i^2 \end{array}\right) = \theta \left(\begin{array}{c} \eta_i^2 \\ \eta_i^1 \end{array}\right)~{\rm .}
\ee
For short, we can write $\delta_{R_{(1,\,2)}} \eta^1_i = \theta \eta^2_i$ etc. Now suppose we apply the $\delta_{R_{(1,\,2)}}$ operation to (\ref{EFK}). We already know that the $\D^{(8)}$ term is $SU(4)_R$ invariant by itself, so we focus on the other term and find:
\bea
&& \D_{R_{(1,\,2)}}\left(\sum_{i,j,k,l = 1}^6 q_{i j k l} ~\eta^1_i \eta^2_j \eta^3_k \eta^4_l\right) = \theta \sum_{i,j,k,l = 1}^6 q_{i j k l} (\eta^2_i \eta^2_j \eta^3_k \eta^4_l + \eta^1_i \eta^1_j \eta^3_k \eta^4_l) ~~~{\rm and} \\
&& \D_{R_{(1,\,2)}}\left(\sum_{i,j,k,l = 1}^6 q_{j i k l} ~\eta^1_j \eta^2_i \eta^3_k \eta^4_l\right) = \theta \sum_{i,j,k,l = 1}^6 q_{j i k l} (\eta^2_j \eta^2_i \eta^3_k \eta^4_l + \eta^1_j \eta^1_i \eta^3_k \eta^4_l) \, {\rm .}
\eea
If we insist that the right-hand sides of the above two equations be $SU(2)_R$ invariant, then they must vanish. Comparing, we see that $q_{i j k l} = q_{j i k l}$. The exact same argument can be applied to each pair of $SU(4)_R$ indices and it therefore follows that $q_{i j k l}$ must be symmetric in each pair of indices. In what follows, we will often need to refer to the part of $\a_{6;\,3}^{\rm EFK}$ that was acted on non-trivial by $SU(4)_R$ so we define:
\be
P_{\rm NMHV} \equiv \sum_{i,j,k,l = 1}^6 q_{i j k l}~ \eta^1_i \eta^2_j \eta^3_k \eta^4_l
\label{PNMHVq}
\ee

To impose supersymmetry on their NMHV ansatz, EFK used the eight spinor supercharges contracted with arbitrary spinors $\e_\alpha$ and $\tilde{\e}_{\dot{\alpha}}$
\be
Q^a = \sum_{i = 1}^6 \spa\e.i \eta^a_i~~~~{\rm and}~~~~\bar{Q}_a = \sum_{i = 1}^6 \spb\e.i {\partial \over \partial \eta^a_i}
\ee
and demanded that they all annihilate $P_{\rm NMHV}$ (the $\D^{(8)}$ term is, of course, already annihilated by all supercharges). As we shall see shortly, it pays to start with the $\bar{Q}_a$. Before going further, it is convenient to explicitly eliminate $\eta_5^a$ and $\eta_6^a$ using the $\D^{(8)}$ term in $\a_{6;\,3}^{\rm EFK}$. This is accomplished simply by taking the inner product of the equation enforced by the delta function,
\be
Q^{a\,\alpha} = \sum_{i=1}^6 |i\rangle \eta_i^a = 0 \, {\rm ,}
\ee
with $\langle 5 |$, then again with $\langle 6 |$, and then finally solving a system of two equations in two unknowns. Explicit expressions for $\eta_5^a$ and $\eta_6^a$ in terms of the other four Grassmann variables (with $SU(4)_R$ index $a$) result. We find:
\be
\eta^a_5 = -\sum_{i=1}^4 {\spa{i}.6 \over \spa5.6} \eta_i^a ~~~{\rm and}~~~\eta^a_6 = \sum_{i=1}^4 {\spa{i}.5 \over \spa5.6} \eta^a_i \,{\rm .}
\ee
Plugging these results back into eq. (\ref{PNMHVq}) gives
\be
P_{\rm NMHV} = {1 \over \spa5.6^4} \sum^{4}_{i,j,k,l=1} c_{ijkl} ~\eta_i^1\eta_j^2\eta_k^3\eta_l^4 \,{\rm ,}
\label{PNMHVc}
\ee
where the $c_{i j k l}$ are some linear combination of the $q_{i j k l}$. The precise relationship between the $q_{i j k l}$ and the $c_{i j k l}$ is not important, but note that the total symmetry of the $q_{i j k l}$ implies that the $c_{i j k l}$ must be totally symmetric as well.

Suppose we start by investigating the constraints resulting from the action of $\bar{Q}_1$ on $P_{\rm NMHV}$:
\bea
\bar{Q}_1 P_{\rm NMHV} &=& {1 \over \spa5.6^4} \sum_{r,i,j,k,l=1}^4 c_{i j k l} \spb{\e}.r {\partial \over \partial \eta^1_r} (\eta_i^1 \eta_j^2 \eta_k^3 \eta_l^4)
\elale {1 \over \spa5.6^4} \sum_{i,j,k,l=1}^4 c_{i j k l} \spb{\e}.i \eta_j^2 \eta_k^3 \eta_l^4
\elale {1 \over \spa5.6^4} \sum_{j,k,l=1}^4 \left( \sum_{i=1}^4 c_{i j k l} \spb{\e}.i\right) \eta_j^2 \eta_k^3 \eta_l^4 \, {\rm .}
\label{QP}
\eea
Since we demand that $\bar{Q}_1 P_{\rm NMHV} = 0$, the quantity in parentheses on the last line of (\ref{QP}) must vanish. Using the resulting relation allows us to eliminate $c_{3 j k l}$ and $c_{4 j k l}$ from the sum over $i$ in eq. (\ref{QP}). We find
\begin{changemargin}{-1.0 in}{0 in}
\bea
P_{NMHV} &=& {1\over \spa5.6^4} \left( \sum_{j,k,l=1}^4\sum_{i=1}^2 c_{ijkl}~\eta_i^1 \eta_j^2 \eta_k^3 \eta_l^4 + \sum_{j,k,l=1}^4 (c_{3jkl} ~\eta_3^1+c_{4jkl} ~\eta_4^1)\eta_j^2 \eta_k^3 \eta_l^4 \right)
\elale {1\over \spa5.6^4} \left(\sum_{j,k,l=1}^4\sum_{i=1}^2 c_{ijkl}~\eta_i^1 \eta_j^2 \eta_k^3 \eta_l^4 + \sum_{j,k,l=1}^4 \Bigg(\left(-\sum_{i=1}^2 {\spb{i}.4\over \spb3.4} c_{ijkl}\right) ~\eta_3^1+\left(\sum_{i=1}^2 {\spb{i}.3\over \spb3.4} c_{ijkl}\right) ~\eta_4^1\Bigg)\eta_j^2 \eta_k^3 \eta_l^4 \right)
\elale {1\over \spa5.6^4 \spb3.4} \sum_{j,k,l=1}^4\sum_{i=1}^2 c_{ijkl} (\spb3.4~\eta_i^1+\spb4.i ~\eta_3^1 +\spb{i}.3~ \eta_4^1)\eta_j^2 \eta_k^3 \eta_l^4\, {\rm .}\nn
\eea
\end{changemargin}
Very similar reasoning applies to $j,\,k,\,{\rm and}\,l$. If we define
\begin{changemargin}{-1 in}{0 in}
\be
X_{ijkl} = (\spb3.4~\eta_i^1+\spb4.i ~\eta_3^1 +\spb{i}.3~ \eta_4^1)(\spb3.4~\eta_j^2+\spb4.j ~\eta_3^2 +\spb{j}.3~ \eta_4^2)(\spb3.4~\eta_k^3+\spb4.k ~\eta_3^3 +\spb{k}.3~ \eta_4^3)(\spb3.4~\eta_l^4+\spb4.l~\eta_3^4 +\spb{l}.3~ \eta_4^4)
\label{Xpoly}
\ee
\end{changemargin}
we eventually arrive at
\bea
&& P_{\rm NMHV} = \sum_{i,j,l,k=1}^2 c_{ijkl} {X_{ijkl} \over \spb3.4^4 \spa5.6^4} ~~~{\rm and}
\el \a_{6;\,3}^{\rm EFK} = \D^{(8)}(Q^{a\,\A})\sum_{i,j,l,k=1}^2 c_{ijkl} {X_{ijkl} \over \spb3.4^4 \spa5.6^4}
\eea
$X_{ijkl}$ has a superspace structure that closely resembles that of the three-point anti-MHV superamplitude and is annihilated by the $Q^a$ supercharges. This is why it was smart to start with the $\bar{Q}_a$ supersymmetries; after solving the $\bar{Q}_a$ Ward identities, the $Q^a$ Ward identities are satisfied automatically. 

Our final task is to identify the $c_{ijkl}$ with particular components of $\a_{6;\,3}$. Following EFK, we first exploit the total symmetry of $c_{ijkl}$ to write 
\be
\a_{6;\,3}^{\rm EFK} = {\D^{(8)}(Q^{a\,\alpha}) \over \spb3.4^4 \spa5.6^4}\sum_{1\leq i \leq j \leq k \leq l \leq 2} c_{ijkl} X_{(i j k l)}\,{\rm ,}
\ee
where $X_{(ijkl)}$ represents the sum over all distinct arrangements of the fixed indices $i,\,j,\,k,\,l$. For example, $X_{(1112)} = X_{1112}+X_{1121}+X_{1211}+X_{2111}$. The number of distinct entries in $X_{ijkl}$ is given by the appropriate multinomial coefficient. It is now a straightforward matter to extract a particular component NMHV amplitude from the superamplitude. For example, the partial amplitude $A\left(k_1^{1234}, k_2, k_3, k_4, k_5^{1234}, k_6^{1234}\right)$ is obtained by collecting all terms proportional to $\eta_1^1\eta_1^2\eta_1^3\eta_1^4\eta_5^1\eta_5^2\eta_5^3\eta_5^4\eta_6^1\eta_6^2\eta_6^3\eta_6^4$. The variables $\eta_5^1\eta_5^2\eta_5^3\eta_5^4\eta_6^1\eta_6^2\eta_6^3\eta_6^4$ eat the factor ${\D^{(8)}(Q^{a\,\alpha}) / \spa5.6^4}$. By staring at eq. (\ref{Xpoly}), we see that the last four Grassmann variables must come from the $X_{(1111)}$ term in the sum because $\spb3.4^4 \eta^1_1 \eta^2_1 \eta^3_1 \eta^4_1$ is the only contribution with all four $\eta_1^a$. So, in the end, we conclude that $c_{1111} = A\left(k_1^{1234}, k_2, k_3, k_4, k_5^{1234}, k_6^{1234}\right)$ and this suggests a strategy to determine the remaining four $c_{ijkl}$: $c_{1112},\,c_{1122},\,c_{1222},\,{\rm and}\,c_{2222}$. The idea is to first take the  eight Grassmann variables
$$\eta_5^1\eta_5^2\eta_5^3\eta_5^4\eta_6^1\eta_6^2\eta_6^3\eta_6^4$$
and, as above, use them to eat the factor of ${\D^{(8)}(Q^{a\,\alpha}) / \spa5.6^4}$. Then the trick is to select helicity states for $k_1$ and $k_2$ that smoothly interpolate between $A\left(k_1^{1234}, k_2, k_3, k_4, k_5^{1234}, k_6^{1234}\right)$ (for $c_{1111}$) and $A\left(k_1, k_2^{1234}, k_3, k_4, k_5^{1234}, k_6^{1234}\right)$ (for $c_{2222}$). In other words,
\bea
&&c_{1112} = A\left(k_1^{123},k_2^4,k_3,k_4,k_5^{1234},k_6^{1234}\right)
\el c_{1122} = A\left(k_1^{12},k_2^{34},k_3,k_4,k_5^{1234},k_6^{1234}\right)
\el c_{1222} = A\left(k_1^{1},k_2^{234},k_3,k_4,k_5^{1234},k_6^{1234}\right) \,{\rm .}
\eea

At long last, we arrive at a new form for the superamplitude $\a_{6;\,3}$, expressed as a linear combination of five component amplitudes:
\begin{changemargin}{-.4 in}{0 in}
\bea
&&\a_{6;\,3} = {\D^{(8)}(Q^{a\,\alpha}) \over \spb3.4^4 \spa5.6^4}\Big(A\left(k_1^{1234}, k_2, k_3, k_4, k_5^{1234}, k_6^{1234}\right) X_{(1111)}+ A\left(k_1^{123},k_2^4,k_3,k_4,k_5^{1234},k_6^{1234}\right) X_{(1112)} 
\el+ A\left(k_1^{12},k_2^{34},k_3,k_4,k_5^{1234},k_6^{1234}\right) X_{(1122)} + A\left(k_1^{1},k_2^{234},k_3,k_4,k_5^{1234},k_6^{1234}\right) X_{(1222)} 
\el+ A\left(k_1, k_2^{1234}, k_3, k_4, k_5^{1234}, k_6^{1234}\right) X_{(2222)}\Big)\,{\rm ,}\nn
\eea
\end{changemargin}
where the $X_{(ijkl)}$ are sums over distinguishable permutations of the $X_{ijkl}$ (for fixed indices $i,\,j,\,k,\,l$) defined by eq. (\ref{Xpoly}). If we expand the above we arrive at eq. (\ref{A63}):
\bea
&&\a_{6;\,3} = {\D^{(8)}(Q^{a\,\alpha}) \over \spb3.4^4 \spa5.6^4}\Big(A\left(k_1^{1234}, k_2, k_3, k_4, k_5^{1234}, k_6^{1234}\right) \prod_{a=1}^4 \left(\spb3.4~\eta_1^a+\spb4.1 ~\eta_3^a +\spb{1}.3~ \eta_4^a\right)
\el+ A\left(k_1^{123},k_2^4,k_3,k_4,k_5^{1234},k_6^{1234}\right) \prod_{a=1}^3 \left(\spb3.4~\eta_1^a+\spb4.1 ~\eta_3^a +\spb{1}.3~ \eta_4^a\right)\left(\spb3.4~\eta_2^4+\spb4.2 ~\eta_3^4 +\spb{2}.3~ \eta_4^4\right)
\el+ A\left(k_1^{12},k_2^{34},k_3,k_4,k_5^{1234},k_6^{1234}\right)\prod_{a=1}^2 \left(\spb3.4~\eta_1^a+\spb4.1 ~\eta_3^a +\spb{1}.3~ \eta_4^a\right)\prod_{a=3}^4\left(\spb3.4~\eta_2^a+\spb4.2 ~\eta_3^a +\spb{2}.3~ \eta_4^a\right)
\el + A\left(k_1^{1},k_2^{234},k_3,k_4,k_5^{1234},k_6^{1234}\right) \left(\spb3.4~\eta_1^1+\spb4.1 ~\eta_3^1 +\spb{1}.3~ \eta_4^1\right)\prod_{a=2}^4\left(\spb3.4~\eta_2^a+\spb4.2 ~\eta_3^a +\spb{2}.3~ \eta_4^a\right)
\el + A\left(k_1, k_2^{1234}, k_3, k_4, k_5^{1234}, k_6^{1234}\right) \prod_{a=1}^4\left(\spb3.4~\eta_2^a+\spb4.2 ~\eta_3^a +\spb{2}.3~ \eta_4^a\right)\Big)\,{\rm .}\nn
\eea
\section{The AdS/CFT Correspondence, the Strong-Coupling Form of the Four-Gluon Amplitude, and Hidden Symmetries of Planar $\Nsym$}
\label{ADS/CFT}
In this appendix, we first describe in \ref{ADS} the general idea of the Anti-De Sitter/Conformal Field Theory (AdS/CFT) correspondence (in its original incarnation) between $\Nsym$ SYM and type IIB superstring theory on $AdS_5 \times S^5$. Then, in \ref{4gluestrong}, we turn to a specific application, namely the computation of the planar $\Nsym$ four-gluon scattering amplitudes at strong coupling. The derivation of even this simplest of the simple MHV amplitudes becomes quite technical at strong coupling, so we present a sketch of the calculation and proceed rather quickly to the final results. We feel that this is justified, so long as we take care to omit only technical details that play no direct role in this paper. The intent of this particular appendix is not to provide rigorous justifications but, rather, to help the less expert reader better appreciate the history and background behind the relevant developments discussed at greater length in the main text. Of particular interest is the way in which the strong coupling calculation described in this appendix led researchers to two previously unknown hidden symmetries of the large $\Nc$ S-matrix, dual superconformal invariance and fermionic T-duality.
\subsection{The AdS/CFT Correspondence for $\Nsym$ Super Yang-Mills}
\label{ADS}
One can most easily understand the AdS/CFT correspondence by thinking about the physics of $N_c$ coincident $D_3$-branes in type IIB superstring theory~\cite{MAGOO}. $D_3$-branes were introduced in Subsection \ref{BornInfeld} in the context of the non-Abelian Born-Infeld action. Recall that, in the low energy limit, the non-Abelian Born-Infeld action describes the interactions of the massless modes of open superstrings\footnote{As before, the non-Abelian Born-Infeld action is not sensitive to the details of the string construction.} terminating on a stack of $N_c$ coincident $D_3$-branes. The end result is a $U(\Nc)$ $\Nsym$ super Yang-Mills theory living on the (3+1 dimensional) world-volume of the $D_3$-branes plus stringy corrections of $\Ord(\alpha'^2)$ and higher. Quite remarkably, it was proposed in~\cite{origMald} that, if desired, one can describe the low energy physics of open superstrings ending on these $N_c$ coincident $D_3$-branes by instead considering the massless modes of closed type IIB superstrings in the geometry very close to the $D_3$-brane stack (the $D_3$-branes back-react on $\mb{R}^{1,9}$ to produce a non-trivial geometry). A great deal of evidence for this proposal has accumulated over the years but, for now, we will simply try to give the reader an intuitive feel for how the AdS/CFT correspondence (in its weakest form) works. In this subsection we closely follow the discussion, notation, and conventions of~\cite{Andysthesis}.

One massless mode associated with the closed superstring is, of course, the graviton. If we declare that we are only interested in the dynamics of the graviton near the stack of $D_3$-branes, we can replace the closed string sector of the full type IIB superstring theory with the field equations for the graviton derived from the classical Lagrangian of $\mathcal{N}=2$ supergravity~\cite{DHFrev}. At the level of the supergravity action, the dynamical field associated with the graviton is the spacetime metric. In flat space superstring theory one has the Lorentz group in ten dimensional spacetime, $SO(1,9)$. Inserting $\Nc$ $D_3$-branes breaks this group down to an $SO(1,3)$ along the world-volume directions of the $D$-branes and an $SO(6)$ rotating the other six directions into each other. It turns out that there is a unique metric\footnote{In this appendix we adopt the conventions used throughout most of the string theory literature. In particular, we use diag$\{-1,1,1,1\}$ for the Minkowski metric in Appendix \ref{ADS/CFT} whereas, throughout the rest of this paper, diag$\{1,-1,-1,-1\}$ is employed.} that satisfies the supergravity equations of motion and is consistent with this symmetry breaking pattern:
\be
{1\over \sqrt{1+{4 \pi g_s \Nc \alpha'^2 \over r^4}}}\Big(-dx_0^2 + dx_1^2 + dx_2^2 + dx_3^2\Big)+\sqrt{1+{4 \pi g_s \Nc \alpha'^2 \over r^4}}\Big(dr^2 + r^2 d\Omega^2_5\Big)\,{\rm ,}
\ee
where $r$ is the distance from the stack of $D_3$-branes. Clearly, $r$ can be thought of as the radius of an $S_5$ for which $d\Omega^2_5$ is the measure of solid angle. This metric has a nice small $r$ limit\footnote{Although it is doesn't fit neatly into the main line of our discussion, we briefly comment on what happens to the Ramond-Ramond five-form field strength, $F^{\rm R R}_5$, in the small $r$ limit. It turns out~\cite{MAGOO} that the supergravity solution has a Ramond-Ramond five-form flux $\Nc$ through the $S_5$ part of the $AdS_5 \times S_5$ that emerges in this limit. In other words, $F^{\rm R R}_5$ is such that $\int_{S_5} F^{\rm R R}_5 = \Nc$.}. That is to say, in the neighborhood of the $\Nc$ $D_3$-branes, the metric reduces to
\bea
&&{r^2 \over \sqrt{4 \pi g_s \Nc \alpha'^2}}\Big(-dx_0^2 + dx_1^2 + dx_2^2 + dx_3^2\Big)+{\sqrt{4 \pi g_s \Nc \alpha'^2}\over r^2}\Big(dr^2 + r^2 d\Omega^2_5\Big)
\elale {\sqrt{4 \pi g_s \Nc \alpha'^2}\over r^2} dr^2+{r^2 \over \sqrt{4 \pi g_s \Nc \alpha'^2}}\Big(-dx_0^2 + dx_1^2 + dx_2^2 + dx_3^2\Big)+\sqrt{4 \pi g_s \Nc \alpha'^2} d\Omega^2_5 \,{\rm ,}\nn
\label{ADS5}
\eea
which is nothing but the metric of $AdS_5 \times S_5$. To summarize, Maldacena's proposal is that there exist two very different descriptions of the same physics. In its weakest form, we have $U(\Nc)$ $\Nsym$ super Yang-Mills theory in $3 + 1$ dimensional Minkowski space (in some limit to be determined below) on one side and on the other we have type IIB supergravity in an $AdS_5 \times S_5$ background.  

Unless we understand when the curvature corrections on the supergravity side start to matter, the proposed duality won't be of much use to us, since we don't know much about quantum gravity in strongly curved backgrounds. Another important technical point we must address is what regime of the $U(\Nc)$ $\Nsym$ field theory we are probing if we ignore the fluctuations in the  Ramond-Ramond fields on the supergravity side. Basically, what we want is for the radius of $AdS_5$ to be sufficiently large relative to the string scale that the spacetime we are working in is weakly curved. In symbols, we have
\be
\sqrt{4 \pi g_s \Nc} >> 1\,{\rm ,}
\ee
where we have cancelled a factor of $\alpha'$ on both sides of the above. Now, suppose that we insist on $g_s << 1$ so that $\mathcal{N}=2$ supergravity is solidly within the perturbative regime (the only regime we understand fairly well). At first sight it would appear that the above criterion cannot be satisfied if the supergravity theory is in the perturbative regime. However, $\Nc$ need not be a small parameter the way it is in QCD. Suppose we define a new coupling constant, $\lambda \equiv 4 \pi g_s \Nc$. We can first take $\Nc \rightarrow \infty$ and $g_s \rightarrow 0$ in such a way that the product $g_s N_c$ is constant. After rewriting everything in terms of $\lambda$, we {\it then} take $\lambda$ large. This procedure lets us keep the supergravity theory perturbative while at the same time allowing us to satisfy the above inequality. Now, from the low energy open string description (the non-Abelian Born-Infeld action), we can make the identification
\be
g^2 = 2 \pi g_s \,{\rm .}
\ee
The regime that we know how to work with on the $AdS$ side corresponds to taking the large $\Nc$ limit in the $U(\Nc)$ $\Nsym$ theory and then going to strong 't Hooft coupling. Thus, we are led to the remarkable conclusion that weakly coupled type IIB supergravity in an $AdS_5 \times S_5$ background can give us analytical control over planar $\Nsym$ with gauge group  $U(\Nc)$ in the strong coupling regime. It is important to note that, at least for the application discussed below, the gauge group is effectively $SU(\Nc)$ due to the fact that an overall $U(1)$ decouples from the $D_3$-brane dynamics (see {\it e.g.}~\cite{Dieter}).

One obvious check of the proposed duality is that the symmetries of the supergravity theory on $AdS_5 \times S_5$ match the well-known symmetries of $\Nsym$. $\Nsym$ has a $SO(4,2)$ conformal symmetry and an internal $SO(6)$ R-symmetry\footnote{In principle we should be more careful. For example, strictly speaking, we should write $SU(4)_R$ instead of $SO(6)_R$ because $SU(4)_R$ is the spin cover of $SO(6)_R$. However, for the purposes of this appendix, it suffices to identify Lie groups that have isomorphic Lie algebras.}. It is easy to see that $SO(6)$ R-symmetry matches the $SO(6)$ isometry group of $S_5$. What about the $SO(4,2)$ conformal symmetry? If we try to embed $AdS_5$ in $\mathbb{R}^6$ we arrive at
\be
-X_0^2-X_1^2+X_2^2+X_3^2+X_4^2+X_5^2 = -\sqrt{\lambda}\alpha'
\ee
for the induced metric. In these global coordinates it is clear that the isometry group of $AdS_5$ is $SO(4,2)$ and everything works out nicely. It is worth pointing out that, in most AdS/CFT analyses, the $S_5$ factor is more or less ignored. This is because it corresponds to the internal $SO(6)_R$ symmetry of $\Nsym$ which plays only a peripheral role in many problems.

Though we have made it clear what the AdS/CFT correspondence is supposed to do for us, we do not yet know precisely how to go from one picture to another. In fact, there is no systematic procedure; each class of observables must be treated on a case-by-case basis. To illustrate how this might work we consider the particular example of Wilson loops. These observables are fairly well understood objects in $\Nsym$ field theory but we would like to give a dual gravitational description of them. With this description in hand, we will be able to use AdS/CFT duality to analytically calculate Wilson loops at strong coupling in the $\Nsym$ theory (at least in principle). First let us look at a cartoon of $AdS_5$:
\FIGURE{
\resizebox{0.75\textwidth}{!}{\includegraphics{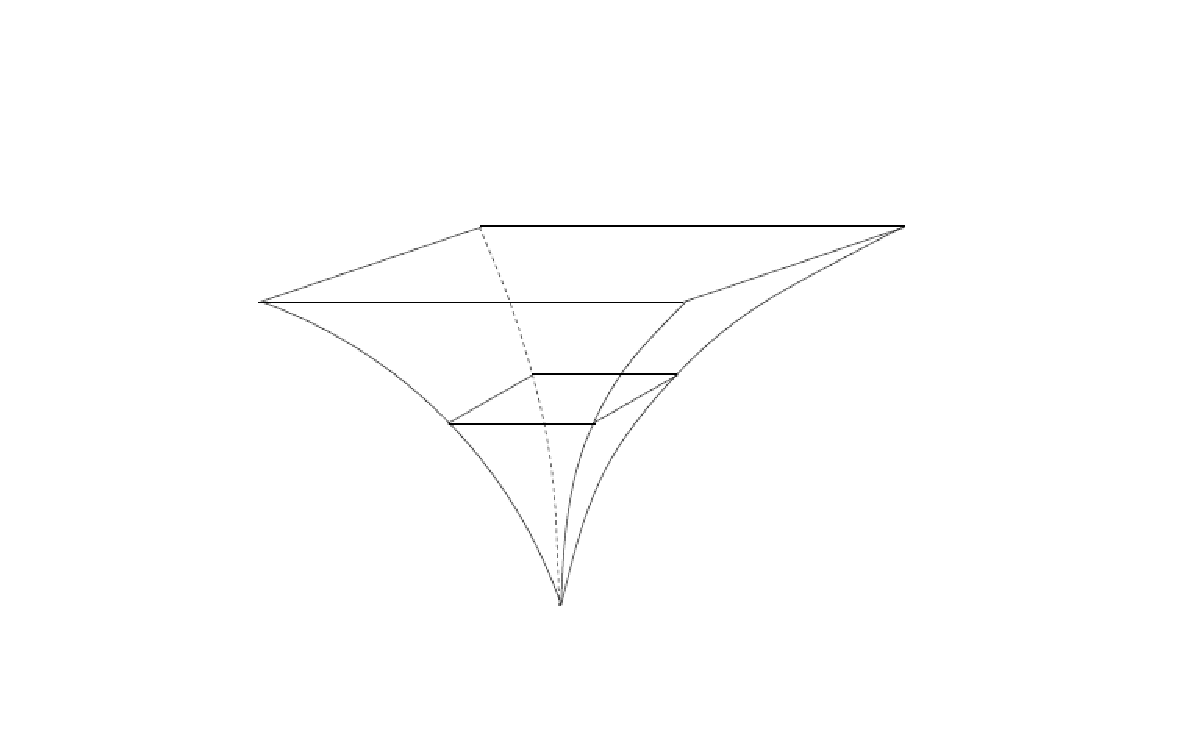}}
\caption{A cartoon of $AdS$ space.}
\label{ADScart}}
Intuitively, the parameter $r$ in eq. (\ref{ADS5}) measures ``depth'' in $AdS_5$ (vertical direction in Figure \ref{ADScart}). Each fixed-$r$ cross-section in Figure \ref{ADScart} is flat $3 + 1$ dimensional Minkowski space times a conformal factor.  As $r \rightarrow \infty$, the metric becomes conformally equivalent to ordinary Minkowski spacetime and this is known as the boundary of $AdS_5$. In Figure \ref{ADScart} the top is the boundary and the bottom ($r = 0$) corresponds to being deep inside $AdS_5$. At $r = 0$ there is a coordinate singularity in the $AdS_5$ metric (called the Poincar\'{e} horizon). This is an artifact of the metric we used, eq. (\ref{ADS5}), which is usually called the Poincar\'{e} patch.\footnote{In global coordinates this singularity disappears.}

Finally, we remark that UV physics in large $\Nc$ $\Nsym$ field theory maps to physics near the boundary in $AdS_5$ and, conversely, IR physics in the field theory maps to physics deep inside $AdS_5$ near $r = 0$. In other words, the $r$ variable in the $AdS_5$ metric can be thought of as an energy scale in the dual $\Nsym$ field theory~\cite{Bianchi}. Actually, one needs to be a bit careful because there is another commonly used form of the Poincar\'{e} metric
\be
{R^2 \over z^2} (dz^2 + dx_\mu dx^\mu) \,{\rm ,}
\label{nuADSmet}
\ee
where the boundary is at $z = 0$ and the Poincar\'{e} horizon (IR in the dual field theory) is at $z = \infty$. In fact, this is the form that we will use in the next subsection.

Unfortunately, a detailed discussion of~\cite{MaldacenaWL}, the original work on Wilson loops in $AdS_5$, would take us too far afield. Nevertheless, it is instructive to outline the $AdS_5$ description for $\Nsym$ Wilson loops. In what follows we will primarily be interested in light-like Wilson loop expectation values but the basic principle is essentially the same for any Wilson loop expectation value. Recall from Subsection \ref{WL/MHV2} that a Wilson loop expectation value is defined as
\be
W[C] = {1\over \Nc} \langle0|{\rm Tr}\bigg[P\bigg\{ {\rm exp}\left(i g \oint_{C} dx^\nu A^a_\nu(x) t^a\right)\bigg\}\bigg]|0\rangle
\ee
for some closed contour, $\mathcal{C}$. The spirit of the AdS/CFT correspondence can be summarized by the statement that it is holographic. The use of the word ``holography'' comes from the similarity of the AdS/CFT correspondence to the way in which holograms are generated\footnote{In holography three dimensional images are produced by light reflecting off of the boundary of the object being imaged. Through AdS/CFT, one has control over a theory of quantum gravity living on the whole of $AdS_5$ if one understands a conformal quantum field theory on the boundary of $AdS_5$.}. In this particular example, if one puts a contour, $\mathcal{C}$, on the boundary of $AdS_5$ and wants to compute the corresponding Wilson loop expectation value, it is very natural to consider a string world-sheet in $AdS_5$ that has the desired contour as its {\it boundary} (see Figure \ref{WLcart}). A trivial but important observation is that there is gravity in the bulk and objects in $AdS_5$ tend to fall towards the horizon. This implies that the  world-sheet bounding the Wilson loop will inevitably sag into the bulk until it reaches equilibrium. Then, the Wilson loop expectation value is simply computed by making a saddle-point approximation to the partition function of the string world-sheet bounding $\mathcal{C}$ (it's not clear that this makes sense in anything other than Euclidean signature). In conclusion, up to some undetermined normalization that must be fixed by other considerations, we have the following prescription for the evaluation of planar $\Nsym$ Wilson loop expectation values via AdS/CFT:
\be
\langle W_{\mathcal{C}}\rangle = A ~e^{-S_{cl}[\bar{\chi}]}\,{\rm ,}
\ee
where $\bar{\chi}$ is shorthand for the solution (or family of solutions) of the equations of motion that minimizes the classical relativistic string action\footnote{This action, the Nambu-Goto action, is very well-known and will be defined in the next subsection.} with $AdS_5$ target space, subject to the above mentioned boundary condition. We will see more explicitly how this works in practice in the next subsection.
\FIGURE{
\resizebox{0.75\textwidth}{!}{\includegraphics{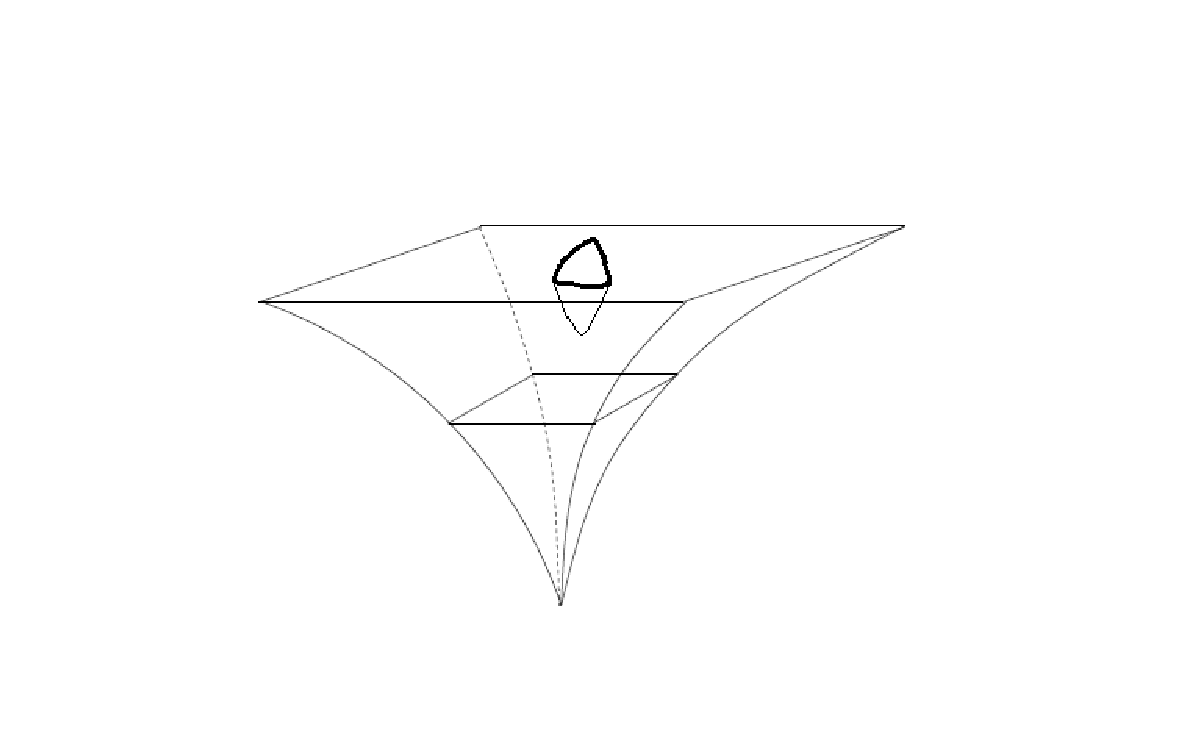}}
\caption{Setup for the computation of a Wilson loop expectation value in $AdS_5$.}
\label{WLcart}}
\newpage
\subsection{Planar Four-Gluon Scattering at Strong Coupling and Large $\Nc$ Hidden Symmetries of the $\Nsym$ S-Matrix}
\label{4gluestrong}
In a remarkable paper~\cite{origAldayMald}, Alday and Maldacena explained how to calculate planar gluon scattering amplitudes at large 't Hooft coupling and explicitly constructed the solution in the four-point case. Strictly speaking, their solution is only valid for scattering in the eikonal limit, but for the simple case of four gluons there is no essential loss of generality. Before attempting to describe the scattering of open superstrings in $AdS_5$ it is very instructive to first analyze something simpler. By carrying out the analysis for flat space bosonic string theory we can gain some intuition about what sort of answer to expect when we redo the calculation in $AdS_5$. We shall consider a scattering amplitude with four tachyons, the lowest-lying states in the bosonic string spectrum to avoid unnecessary complications introduced by gluon polarization vectors. Analyzing the flat-space bosonic string theory result will, in our opinion, make the subsequent discussion of open superstring scattering in $AdS_5$ much easier to follow.

Before we discuss the eikonal scattering of bosonic strings we give a very brief review of the two most widely used action functionals for bosonic strings. These are the Nambu-Goto action and the Polyakov action. The Nambu-Goto action was proposed~\cite{Goto} and has the form:
\be
S_{NG} = {1\over 2 \pi \alpha'}\int_{\mathcal{W}} d\sigma d\tau \sqrt{{\rm det}\Big[ \partial_\alpha \chi^\mu \partial_\beta \chi^\nu g_{\mu \nu}\Big]} \,{\rm .}
\label{NGA}
\ee
Here $g_{\mu \nu}$ is the target space metric and the derivatives are with respect to the coordinates, $\sigma$ and $\tau$, of the two dimensional world-sheet, $\mathcal{W}$. In the context of bosonic string theory, the target space is $\mathbb{R}^{26}$, but the Nambu-Goto action is applicable to classical relativistic strings propagating in curved spacetimes ({\it e.g.} $AdS_5$) as well. 

Some years later, Polyakov~\cite{Polyakov} reformulated the Nambu-Goto action:
\be
S_P = {1 \over 4 \pi \alpha'} \int_{\mathcal{W}}d\sigma d\tau \sqrt{-{\rm det}(h_{a b})} h^{a b} g_{\mu \nu} \partial_a \chi^\mu \partial_b \chi^\nu
\label{Polyact}
\ee
He introduced an auxiliary metric, $h_{a b}$, that one can eliminate through the equations of motion (if desired) to recover the Nambu-Goto action.  The utility of $h_{a b}$ is that it facilitates a path integral formulation of bosonic string theory~\cite{Polyakov} and allows one to write perturbative scattering amplitudes as functional integrals. For example, the tree-level four-tachyon string partial amplitude that will be of interest to us is given by 
\be
A_{tach}^{tree}(k_1, k_2, k_3, k_4) = \int {d[\chi_\mu]d[h^{a b}]\over \mathcal{Z}}~ e^{-S_P} \prod_{i = 1}^4 \mathcal{V}_i(\chi_\mu) \,{\rm ,}
\label{tachtree}
\ee
where the $\mathcal{V}_i(\chi_\mu)$ are the tachyon vertex operators\footnote{In string theory one uses the so-called ``operator-state correspondence'' to compute correlation functions. This just means that the operators used to describe states look like external states themselves.} given by 
\be
\mathcal{V}_i(\chi_\mu) = \int_{\partial \mathcal{W}} dz_j e^{i k_j^\mu \chi_\mu(z_j)} \,{\rm ,}
\label{vertexop}
\ee
where the integral in (\ref{vertexop}) appears because each tachyon can be inserted anywhere on the boundary of the composite world-sheet. Clearly, we are free to choose the order in which we insert the  tachyons on the boundary. Here, we choose the ordering $\{z_1, \,z_2,\,z_3,\,z_4\}$ corresponding to a Chan-Paton factor ${\rm Tr}\{T^{a_1}T^{a_2}T^{a_3}T^{a_4}\}$.
 
The starting pointing for our tree-level string theory partial amplitude is a string diagram, where four string world-sheets merge to form a single, composite world-sheet. Now, the string world-sheet enjoys a full $SL(2,\mathbb{R})$ conformal invariance~\cite{Becker}. This $SL(2,\mathbb{R})$ together with the invariance of the world-sheet under general coordinate transformations puts severe constraints on the string amplitude. For example, it follows from elementary complex analysis that there exists a Schwarz-Christoffel transformation which maps the composite world-sheet to the half plane with a singular point for each external state on the boundary  (illustrated in Figure \ref{worldsheet}). 

\FIGURE{
\resizebox{0.65\textwidth}{!}{\includegraphics{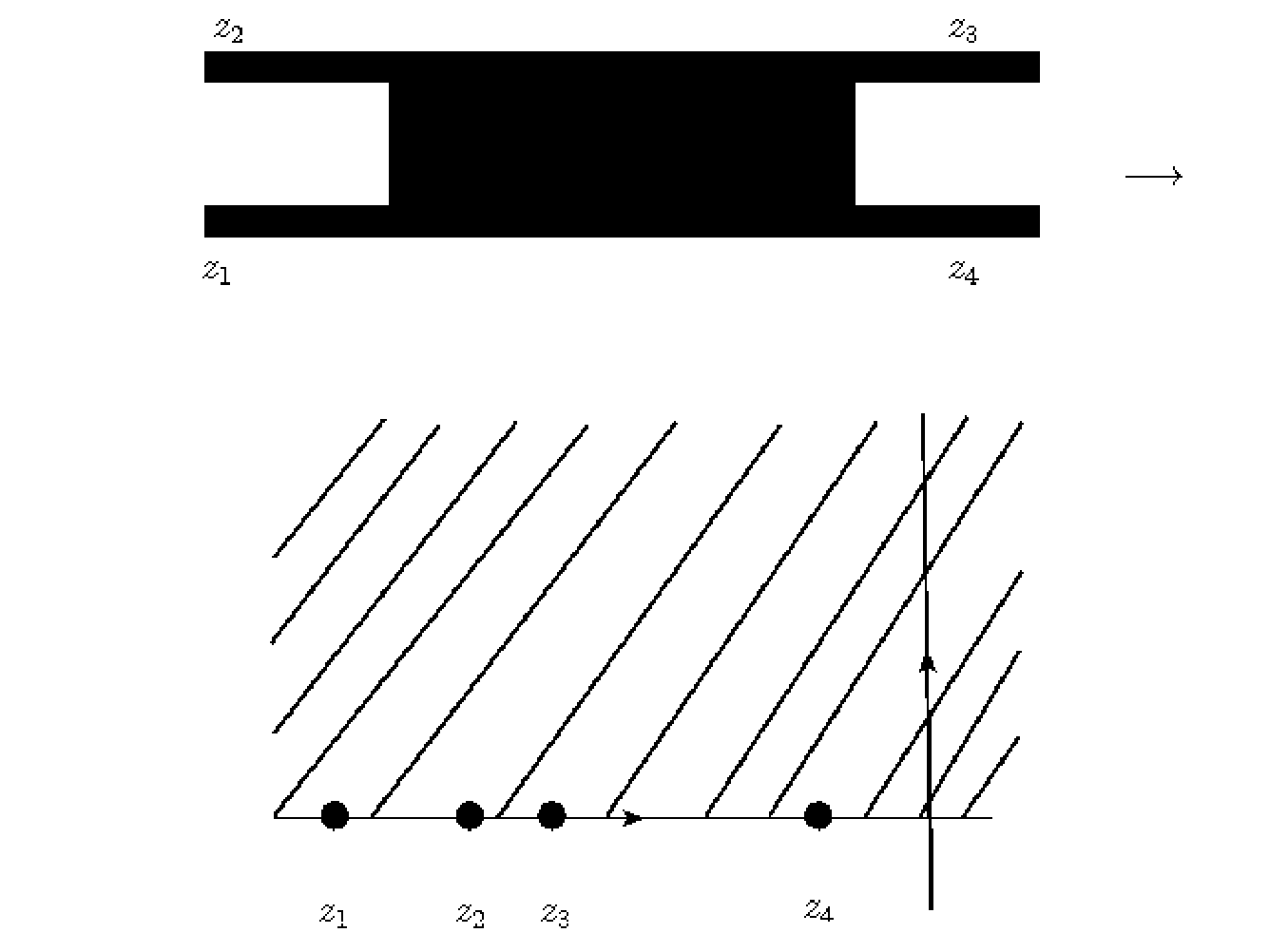}}
\caption{The tree-level open string diagram of interest can be mapped to the upper half plane with four operator insertions by a Schwarz-Christoffel transformation.}
\label{worldsheet}}

The stringy formula of eq. (\ref{tachtree}) for the tree-level tachyon amplitude looks very complicated but taking into account the symmetries of the problem will allow us to bypass most of the necessary integrations. To start, we gauge-fix all available symmetries. The invariance of the action under general coordinate transformations allows us to choose $h^{a b}$ to be the Euclidean metric on $\mathbb{R}^2$. Gauge-fixing $SL(2,\mathbb{R})$ is a bit more involved because of the non-trivial Jacobian that results. To avoid a lengthy digression, suffice it to say that the three generators of $SL(2,\mathbb{R})$ allow us to fix $z_1 = 0,\,z_3 = 1,~{\rm and}\,z_4 = \infty$ and the interested reader can find the resulting Jacobian worked out in~\cite{Green}. Actually, ignoring the Jacobian is fine; if we kept it would just cancel against terms containing $z_4$ that na\"{i}vely look dangerous due to the fact that we gauge-fixed $z_4$ to infinity. Gauge-fixing (\ref{tachtree}) makes it look much more tractable:
\begin{changemargin}{0 in}{0 in}
\bea
&&A_{tach}^{tree}\left(k_1, k_2, k_3, k_4\right)
\el \int_{-\infty}^{\infty} dz_4\int_{-\infty}^{z_4} dz_3\int_{-\infty}^{z_3} dz_2 \int_{-\infty}^{z_2} dz_1 \int{d[\chi_\mu]d[h^{a b}]\over \mathcal{Z}}~ e^{-{1 \over 4 \pi \alpha'} \int_{\mathcal{W}}d\sigma d\tau \sqrt{-{\rm det}(h_{a b})} h^{a b} \partial_a \chi^\mu \partial_b \chi_\mu} ~ e^{\sum_j i k_j^\mu \chi_\mu(z_j)} \nonumber \\
&\approx & \int_{0}^{1} dz_2 \int d[\chi_\mu]~ e^{-{1 \over 4 \pi \alpha'} \int_{\mathcal{W}}d\sigma d\tau ~\chi^\mu \big(-\partial^2\big) \chi_\mu} ~ e^{\sum_j i k_j^\mu \chi_\mu(z_j)} \,{\rm .}
\eea
\end{changemargin}
The functional integral over $\chi_\mu$ in the above is trivially evaluated (it is Gaussian) by changing variables $\chi_\mu(z_i) = \chi'_\mu(z_i) + i \sum_j G(|z_i - z_j|) k^j_\mu$ and using the fact that $G(|z_i - z_j|) = -2 \alpha' \ln(|z_i - z_j|)$ is the two dimensional Green's function for the equation $-\partial^2 \chi = 0$. Carrying out this functional calculus\footnote{The alert reader will notice that there is a divergence in the Green's function at $z_i = z_j$ and there is no obvious reason why $i \neq j$. We gloss over this technical problem because, in the case of superstrings (which is what we will ultimately be interested in), the tachyon mode is lifted to a massless, gluon mode and this problem goes away by virtue of the on-shell condition $k_i^2 = 0$.} and using the gauge conditions for the $z_i$ (and throwing away pieces proportional to logs of $z_4$) results in 
\be
A_{tach}^{tree}\left(k_1, k_2, k_3, k_4\right)\approx \int_{0}^{1} dz_2 ~e^{\alpha' \big(k_1 \cdot k_2 \ln(z_2)+k_2 \cdot k_3 \ln(1-z_2)\big)}\,{\rm .}
\label{fintachint}
\ee
Now if we work in the eikonal limit (high energy, fixed angle scattering), Gross and Mende~\cite{GrossMende} observed that string amplitudes are well-approximated by a saddle-point estimate. In other words, the integral in (\ref{fintachint}) can be approximated by extremizing the argument of the exponential and then evaluating the exponential on this solution. The argument is extremized at 
\be
\bar{z}_2 = {k_1 \cdot k_2 \over k_1 \cdot k_2 + k_2 \cdot k_3}
\ee
and we finally obtain 
\be
A_{tach}^{tree}\left(k_1, k_2, k_3, k_4\right)\approx A_0 e^{\alpha' \Big(k_1 \cdot k_2 \ln\Big({k_1 \cdot k_2 \over k_1 \cdot k_2 + k_2 \cdot k_3}\Big)+k_2 \cdot k_3 \ln\Big({k_2 \cdot k_3 \over k_1 \cdot k_2 + k_2 \cdot k_3}\Big)\Big)}\,{\rm .}
\label{fintach}
\ee
for the partial amplitude. Note that this solution only makes sense in the {\it Euclidean} region where both $k_1 \cdot k_2$ and $k_2 \cdot k_3$ are negative. Although the details will of course be somewhat different, it is reasonable to expect the eikonal scattering of open superstrings in $AdS_5$ to have the same basic exponential structure that we saw with the bosonic string in flat space. 

So far, it is probably not clear why we went to the trouble of deriving eq. (\ref{fintach}). The important point is that, if the basic exponential structure of (\ref{fintach}) carries over to classical strings on $AdS_5$, we can guess using the AdS/CFT correspondence that the partial amplitude $A\left(k_1^{1234},k_2^{1234},k_3,k_4\right)$ in $\Nsym$ gauge theory at strong coupling is of the form $A^{tree}\left(k_1^{1234},k_2^{1234},k_3,k_4\right) e^{-S(s,\,t)}$. Conjecturing this form could probably be dismissed as overly speculative if it was not for the work of Bern, Dixon, and Smirnov in~\cite{BDS}. BDS found that the finite parts of multi-loop contributions to $A\left(k_1^{1234},k_2^{1234},k_3,k_4\right)/A_{4;\,\spa{1}.{2}}^{\textrm{\scriptsize{MHV}}}$ appeared to be exponentiating and, based on an explicit three loop calculation, they formulated an all-orders ansatz for the four-point amplitude. Essentially, they predicted
\be
A^{strong}\left(k_1^{1234},k_2^{1234},k_3,k_4\right) = A_{4;\,\spa{1}.{2}}^{\textrm{\scriptsize{MHV}}} e^{B_{\rm{IR}}(s,t,\lambda,\e) + {\gamma_c(\lambda) \over 4} \big({1\over 2} \ln(t/s)^2 + {2 \pi^2 \over 3} \big) + C(\lambda)}
\label{BDSpredict}
\ee
which is quite remarkable, since the finite part of $A_1^{1-{\rm loop}}\left(k_1^{1234},k_2^{1234},k_3,k_4\right)/A_{4;\,\spa{1}.{2}}^{\textrm{\scriptsize{MHV}}}$ is proportional to $\ln(t/s)^2 + \pi^2$~\cite{BDDKMHV}; roughly speaking BDS conjectured that the complete, non-perturbative dependence of $A\left(k_1^{1234},k_2^{1234},k_3,k_4\right)/A_{4;\,\spa{1}.{2}}^{\textrm{\scriptsize{MHV}}}$ on $s$ and $t$ is fixed by the one-loop quantum corrections. In (\ref{BDSpredict}), $B_{\rm{IR}}(s,t,\lambda,\e)$ refers to the IR divergent terms and $C(\lambda)$ to terms that are not predicted by the ansatz. $B_{\rm{IR}}(s,t,\lambda,\e)$ is actually pretty well understood (see {\it e.g.}~\cite{Dixonstrongrev} for a recent review). $C(\lambda)$ exists because it is generally not possible to predict finite terms that have no dependence on $s$ or $t$. Finally, we comment on the physical interpretation of $\gamma_c(\lambda)$. This constant, called the cusp anomalous dimension plays a central role in the general theory of IR divergences in gauge theories. At each order in perturbation theory, the contribution to the cusp anomalous dimension at that order controls the most $1/\e^2$ singular contributions~\cite{CVthesis}.

Remarkably, Alday and Maldacena~\cite{origAldayMald} were able to determine the analog of eq. (\ref{fintach}) for four-point superstring scattering in $AdS_5$ and found complete agreement with the prediction of BDS. Their paper utilized a novel mapping of $AdS_5$ to $AdS_5$, which, roughly speaking, interchanged the IR and the UV. In some sense, this mapping, was every bit as important as the final formula (\ref{BDSpredict}) because it allowed researchers to uncover previously unknown hidden symmetries of the large $\Nc$ S-matrix. 

To begin, we'd like to visualize the scattering process in $AdS_5$. Unfortunately, the composite world-sheet is much harder to draw in AdS space than it was in flat space (the left-hand side of Figure \ref{worldsheet} was very simple). The picture to have in mind is as follows. Recall the general shape of $AdS_5$ drawn in Figure \ref{ADScart} and the corresponding metric of eq. (\ref{ADS5}). In that picture, the IR is at the bottom and the UV is at the top. With the metric of (\ref{nuADSmet}) Figure \ref{ADScart} is still the right cartoon; the only difference is the now the top corresponds to $z = 0$ and the bottom corresponds to $z = \infty$. Recall also that each fixed-$z$ cross-section of the Figure corresponds to $\mathbb{R}^{1,3}$ times a conformal factor. This means that, if we want to, we can insert a $D_3$-brane at any fixed value of $z$. In fact, we need to insert a $D_3$-brane stack near $z = \infty$ for our open superstrings to end on. This makes sense because, as we know, gluon scattering amplitudes have IR divergences and open superstrings need $D$-branes to terminate on. Thus, one should visualize a stack of $D_3$-branes at the bottom of Figure \ref{ADScart} with four interacting open strings attached to the $D_3$-branes.

In fact, Alday and Maldacena argued that the stringy scattering process are naturally forced into the eikonal regime where the semi-classical approximation becomes applicable. They found the boundary conditions described in the above paragraph a bit hard to work with directly and, consequently, tried changing variables from
\be
{R^2\over z^2} (dz^2 + dx_\mu dx^\mu)
\label{startmet}
\ee
to some other spacetime where, hopefully, the scattering problem would map to something more tractable. They succeeded in a very remarkable way: they discovered a change of variables that maps the scattering problem in $AdS_5$ to the calculation of a light-like Wilson loop {\it again} in $AdS_5$. It turns out that the right variable change to try is defined through the relations
\be
z = R^2/r(\sigma,\tau) ~~~~~~ \partial_\tau y^\mu = i {r(\sigma,\tau)^2 \over R^2} \partial_\sigma x^\mu(\sigma,\tau) ~~~~~~ \partial_\sigma y^\mu = -i{r(\sigma,\tau)^2 \over R^2} \partial_\tau x^\mu(\sigma,\tau)) \,{\rm .}
\label{tdual}
\ee
Though it is not obvious, this non-local coordinate transformation carries $AdS_5$ to $AdS_5$; eq. (\ref{startmet}) maps to
\be
{R^2\over r^2} (dr^2 + dy_\mu dy^\mu) \,{\rm ,}
\label{Tdualmet}
\ee
where now all variables implicitly depend on the world-sheet coordinates $\sigma$ and $\tau$. What this means is that now we must seek a classical world-sheet that sits inside $AdS_5$ and extremizes the Nambu-Goto action.

In fact, the existence of the above mapping is related to the existence of hidden symmetries of $\Nsym$ in the large $\Nc$ limit. After applying (\ref{tdual}), the $D_3$-brane stack that we inserted in the IR near the Poincar\'{e} horizon gets mapped to a stack in the UV near the boundary, $r = 0$. The composite world-sheet now hangs from the boundary of $AdS_5$, not unlike the situation depicted in Figure \ref{WLcart}. The transformation performed on the $x^\mu$ variables can be used to figure out how to interpret the boundary of world-sheet in the new coordinates. Starting with a displacement in the new $\mathbb{R}^{1,3}$, $y_\mu(\tau,\, \sigma_f)-y_\mu(\tau,\,\sigma_i)$, we have
\be
y_\mu(\tau,\,\sigma_f)-y_\mu(\tau,\,\sigma_i) = \int_{\sigma_i}^{\sigma_f} d\sigma \partial_\sigma y_\mu = \int_{\sigma_i}^{\sigma_f} d\sigma  \Big(-i{r^2 \over R^2} \partial_\tau x^\mu\Big) = 2 \pi \alpha' \int_{\sigma_i}^{\sigma_f} d\sigma  \mathcal{P}_\mu^\tau = 2 \pi \alpha' p_\mu \,{\rm ,}
\ee
where we have recognized $-i/(2\pi\alpha') (r/R)^2 \partial_\tau x^\mu$ as the canonical momentum density conjugate to $x^\mu$. In the original $AdS_5$ where four open superstrings scatter, each gluon vertex operator carried one of $p_1^\mu,\,p_2^\mu,\,p_3^\mu,\,{\rm or}\,p_4^\mu$. These momenta satisfy the constraints
\be
p_i^2 = 0 ~~~{\rm and}~~~\sum_{i = 1}^4 p_i^\mu = 0 {\rm .}
\ee
In the new coordinates these conditions evidently imply that each external state now corresponds to a light-like line segment in $\mathbb{R}^{1,3}$ and, furthermore, that the sum of all four light-like line segments form a closed loop. Thus, we see that the apparent similarity between the configuration of the composite world-sheet in the $\{r(\sigma,\tau),\,y^\mu(\sigma,\tau)\}$ variables and the setup for the calculation of a Wilson loop in $AdS_5$ is not accidental. In fact, they are in one-to-one correspondence. Finally, we conclude that the calculation of the four-point MHV amplitude at strong coupling in planar $\Nsym$ boils down to the calculation of a four-sided light-like Wilson loop in a dual $AdS_5$.

This correspondence allowed various authors~\cite{DKS4pt,BM,BRTW} to make several bold conjectures that are now on very solid ground. First,~\cite{DKS4pt} observed that a correspondence between light-like Wilson loops (in position space) and MHV scattering amplitudes (in momentum space) holds at {\it weak} coupling, implying that there is a dual conformal invariance acting on {\it momentum} invariants, since light-like Wilson loops are conformally invariant objects in position space. Later, it was understood in~\cite{DHKSdualconf} that this dual conformal invariance is actually more naturally treated as a dual superconformal invariance, which we discuss in detail in \ref{WL/MHV} and \ref{sconf}. A important point, recently clarified in~\cite{DrummondFerro}, is whether there exists a symmetry (beyond dual superconformal invariance) responsible for the light-like Wilson loop/MHV amplitude duality. Two different groups~\cite{BM,BRTW} established the existence of a new symmetry for strings in $AdS_5 \times S_5$, called fermionic T-duality. Although these papers were major achievements, a fully field theoretic description ({\it i.e.} one that obviously extends to the weak coupling regime of $\Nsym$) of the new symmetry was lacking until the recent work of~\cite{DrummondFerro}. The authors of~\cite{DrummondFerro} proved that fermionic T-duality is a symmetry of the field theory and acts by interchanging ordinary superconformal invariance and dual superconformal invariance, confirming the hypothesis advanced earlier in~\cite{BM}.
\newpage
\bibliographystyle{JHEP}
\bibliography{JHEPexample2}
\end{document}